\documentclass[sigconf]{acmart}
\usepackage{xspace,balance,tabularx,multirow}
\usepackage{flushend}
\usepackage{tikz}
\usepackage{pgfplots}
\usetikzlibrary{patterns}
\pgfplotsset{compat=1.16}
\usepackage{subcaption}
\usepackage[ruled, vlined, linesnumbered]{algorithm2e}
\usepackage{xcolor}
\usepackage{colortbl}
\usepackage{hyperref}
\usepackage{bbold}
\SetKwComment{Comment}{$\triangleright$\ }{}
\usepackage{enumitem}
\usepackage[aboveskip=-4pt]{subcaption}

\AtBeginDocument{%
  \providecommand\BibTeX{{%
    \normalfont B\kern-0.5em{\scshape i\kern-0.25em b}\kern-0.8em\TeX}}}

\acmConference[SIGMOD '23]{Proceedings of the 2023 International Conference on Management of Data}{June 18-23, 2023}{Seattle, USA}
\acmBooktitle{Proceedings of the 2023 International Conference on Management of Data (SIGMOD ’23), June 18-23, 2023}
\acmPrice{15.00}
\acmISBN{978-1-4503-XXXX-X/18/06}

\newcommand{\rc}[1]{{{#1}}}

\makeatletter
\newcommand*\bigcdot{\mathpalette\bigcdot@{.5}}
\newcommand*\bigcdot@[2]{\mathbin{\vcenter{\hbox{\scalebox{#2}{$\m@th#1\bullet$}}}}}
\makeatother

\def\header{\vspace{1mm} \noindent}

\newtheorem{proposition}{Proposition}

\SetKw{Break}{break}

\newcommand{\ie}{{i.e.},\xspace}
\newcommand{\eg}{{e.g.},\xspace}

\newcommand{\stitle}[1]{\vspace*{0.5em}\noindent{\bf #1.\/}}

\usepackage{pifont}
\newcommand{\fbego}{{\em FbEgo}\xspace}
\newcommand{\twego}{{\em TwEgo}\xspace}
\newcommand{\wiki}{{\em Wiki-ii}\xspace}
\newcommand{\physic}{{\em Physician}\xspace}
\newcommand{\filmtrust}{{\em FilmTrust}\xspace}
\newcommand{\scinet}{{\em SciNet}\xspace}
\newcommand{\amazon}{{\em Amazon}\xspace}
\newcommand{\itzerofour}{{\em It-2004}\xspace}
\newcommand{\youtube}{{\em Youtube}\xspace}
\newcommand{\orkut}{{\em Orkut}\xspace}
\newcommand{\dblp}{{\em DBLP}\xspace}
\newcommand{\twitter}{{\em Twitter}\xspace}

\newcommand{\nd}{ND}
\newcommand{\ulcv}{ULCV}

\newcommand{\ar}{AR}

\newcommand{\DM}{\mathbf{D}\xspace}
\newcommand{\XM}{\mathbf{X}\xspace}

\newcommand{\YM}{\mathbf{Y}\xspace}

\newcommand{\LM}{\mathbf{L}\xspace}

\newcommand{\bwdpushsn}{\textsf{GBP}\xspace}
\newcommand{\bwdpushsnb}{\textsf{{GBP}}\xspace}
\newcommand{\fwdpushsn}{\textsf{GFP}\xspace}
\newcommand{\fwdpushsnb}{\textsf{{GFP}}\xspace}
\newcommand{\pprdegname}{$\mathsf{DPPR}$\xspace}
\newcommand{\pprdegnameb}{$\boldsymbol{\mathsf{DPPR}}$\xspace}
\newcommand{\pprdistname}{$\mathsf{PDist}$\xspace}
\newcommand{\pprdistnameb}{$\boldsymbol{\mathsf{PDist}}$\xspace}
\newcommand{\pprviz}{$\mathsf{PPRviz}$\xspace}
\newcommand{\pprvizb}{$\boldsymbol{\mathsf{PPRviz}}$\xspace}
\newcommand{\taupush}{$\mathsf{Tau}$-$\mathsf{Push}$\xspace}
\newcommand{\taupushb}{$\boldsymbol{\mathsf{Tau}}$-$\boldsymbol{\mathsf{Push}}$\xspace}

\newcommand{\louvainplus}{$\mathsf{Louvain+}$\xspace}
\newcommand{\louvainplusb}{$\boldsymbol{\mathsf{Louvain+}}$\xspace}
\newcommand{\dnpr}{$\mathsf{DPR}$\xspace}
\newcommand{\dnprb}{$\boldsymbol{\mathsf{DPR}}$\xspace}

\newcommand{\pprdist}{\boldsymbol{\Delta}\xspace}
\newcommand{\pprdeg}{\pi_d\xspace}
\newcommand{\minleaf}{\gamma}
\newcommand{\leaf}{F}

\newcommand{\forasn}{$\mathsf{GFRA}$\xspace}
\newcommand{\forasnb}{$\boldsymbol{\mathsf{GFRA}}$\xspace}
\newcommand{\fora}{$\mathsf{FORA}$\xspace}
\newcommand{\foraplus}{$\mathsf{FORA+}$\xspace}
\newcommand{\forab}{$\boldsymbol{\mathsf{FORA}}$\xspace}
\newcommand{\fwdpush}{$\mathsf{Forward}$-$\mathsf{Push}$\xspace}
\newcommand{\fwdpushb}{$\boldsymbol{\mathsf{Forward}}$-$\boldsymbol{\mathsf{Push}}$\xspace}
\newcommand{\bwdpush}{$\mathsf{Backward}$-$\mathsf{Push}$\xspace}

\newcommand{\poweriter}{$\mathsf{PI}$\xspace}
\newcommand{\resacc}{$\mathsf{ResAcc}$\xspace}
\newcommand{\louvain}{$\mathsf{Louvain}$\xspace}

\newcommand{\linlog}{$\mathsf{LinLog}$\xspace}
\newcommand{\forceatlas}{$\mathsf{ForceAtlas}$\xspace}
\newcommand{\fr}{$\mathsf{FR}$\xspace}
\newcommand{\pivotmds}{$\mathsf{PMDS}$\xspace}
\newcommand{\mds}{$\mathsf{CMDS}$\xspace}
\newcommand{\kadraw}{$\mathsf{KDraw}$\xspace}
\newcommand{\openord}{$\mathsf{OpenOrd}$\xspace}
\newcommand{\kadrawb}{$\boldsymbol{\mathsf{KDraw}}$\xspace}
\newcommand{\openordb}{$\boldsymbol{\mathsf{OpenOrd}}$\xspace}
\newcommand{\simrank}{$\mathsf{SimRank}$\xspace}
\newcommand{\gf}{$\mathsf{GFactor}$\xspace}

\newcommand{\leemb}{$\mathsf{LapEig}$\xspace}
\newcommand{\lle}{$\mathsf{LLE}$\xspace}
\newcommand{\nodevec}{$\mathsf{Node2vec}$\xspace}
\newcommand{\sdne}{$\mathsf{SDNE}$\xspace}

\newcommand{\frb}{$\boldsymbol{\mathsf{FR}}$\xspace}

\newlength\lengtha 

\definecolor{forestgreen}{RGB}{34, 139, 34}

\definecolor{RYB1}{RGB}{192, 128, 255}
\definecolor{RYB2}{RGB}{255, 192, 32}
\definecolor{RYB3}{RGB}{139, 0, 0}
\definecolor{RYB4}{RGB}{0, 128, 255}

\newenvironment{customlegend}[1][]{%
    \begingroup
    \csname pgfplots@init@cleared@structures\endcsname
    \pgfplotsset{#1}%
}{%
    \csname pgfplots@createlegend\endcsname
    \endgroup
}%

\def\addlegendimage{\csname pgfplots@addlegendimage\endcsname}

\makeatletter
\newcommand\footnoteref[1]{\protected@xdef\@thefnmark{\ref{#1}}\@footnotemark}
\makeatother

\let\oldnl\nl
\newcommand{\nonl}{\renewcommand{\nl}{\let\nl\oldnl}}

\newcommand{\trref}{Appendix~\ref{sec:appendix}}

\begin{document}

\title{Effective and Efficient PageRank-based Positioning for Graph Visualization}
\subtitle{[Technical Report]}


\author{Shiqi Zhang}
\affiliation{%
  \institution{National University of Singapore}
  \country{Singapore}
}
\affiliation{%
  \institution{Southern University of Science and Technology}
  \country{China}
}
\email{s-zhang@comp.nus.edu.sg}

\author{Renchi Yang}
\authornote{This work was done while at National University of Singapore.}
\affiliation{%
  \institution{Hong Kong Baptist University}
  \country{China}
}
\email{renchi@hkbu.edu.hk}

\author{Xiaokui Xiao}
\affiliation{%
  \institution{National University of Singapore}
  \country{Singapore}
}
\email{xkxiao@nus.edu.sg}




\author{Xiao Yan}
\email{yanx@sustech.edu.cn}
\author{Bo Tang}
\email{tangb3@sustech.edu.cn}
\affiliation{%
  \institution{Southern University of Science and Technology}
  \country{China}
}



\renewcommand{\shortauthors}{Shiqi Zhang et al.}


\begin{abstract}
Graph visualization is a vital component in many real-world applications (\eg social network analysis, web mining, and bioinformatics) that enables users to unearth crucial insights from complex data. Lying in the core of graph visualization is the node distance measure, which determines how the nodes are placed on the screen. A favorable node distance measure should be informative in reflecting the full structural information between nodes and effective in optimizing visual aesthetics. However, existing node distance measures yield sub-par visualization quality as they fall short of these requirements. Moreover, most existing measures are computationally inefficient, incurring a long response time when visualizing large graphs. To overcome such deficiencies, we propose a new node distance measure, \pprdistname, geared towards graph visualization by exploiting a well-known node proximity measure, \textit{personalized PageRank}. Moreover, we propose an efficient algorithm \taupush for estimating \pprdistname under both single- and multi-level visualization settings. With several carefully-designed techniques, \taupush offers non-trivial theoretical guarantees for estimation accuracy and computation complexity. Extensive experiments show that our proposal significantly outperforms 13 state-of-the-art graph visualization solutions on 12 real-world graphs in terms of both efficiency and effectiveness (including aesthetic criteria and user feedback). In particular, our proposal can interactively produce satisfactory visualizations within one second for billion-edge graphs.
\end{abstract}

\ccsdesc[500]{Human-centered computing~Graph drawings}
\ccsdesc[500]{Theory of computation~Graph algorithms analysis}
\ccsdesc[500]{Theory of computation~Random walks and Markov chains}

\keywords{Graph Visualization; Personalized PageRank; Approximate Algorithm}


\maketitle

\section{Introduction}\label{sec:intro}
Graph visualization is an effective way to help users comprehend and analyze complex relational data (\eg social and biological networks). It has been identified as one of the most popular and challenging graph processing tasks in graph database systems and graph libraries~\cite{sahu2017ubiquity}. In practice, graph visualization finds extensive applications, such as analyzing protein interactions in the organism~\cite{agapito2013visualization}, studying scholars' co-authoring behaviors~\cite{rodrigues2015gmine}, and capturing the massive hyperlinks among web pages~\cite{bikakis2016graphvizdb}. This motivates a plethora of graph visualization solutions~\cite{eades1984heuristic,fruchterman1991graph,noack2005energy,martin2011openord,jacomy2014forceatlas2,torgerson1952multidimensional,kamada1989algorithm,gansner2004graph,brandes2006eigensolver,gansner2012maxent,meyerhenke2017drawing,rodrigues2015gmine,abello2006ask,archambault2008grouseflocks,hu2005efficient} and softwares~\cite{bastian2009gephi,auber2004tulip,de2018exploratory,shannon2003cytoscape}.

A central problem in graph visualization is calculating an effective layout, \ie the coordinate of each node on the screen, which seeks to place closely-related nodes close and unrelated nodes far apart based on a node distance measure calculated from the graph.
In most existing solutions, classic node distance measures, including the shortest distance \cite{torgerson1952multidimensional,kamada1989algorithm,gansner2004graph,brandes2006eigensolver,gansner2012maxent,meyerhenke2017drawing,shi2009himap} and the direct linkage \cite{eades1984heuristic,fruchterman1991graph,noack2005energy,martin2011openord,jacomy2014forceatlas2,archambault2008grouseflocks,abello2006ask}, are widely used.
However, such distance measures overlook the high-order structure information of the graph and fail to optimize some critical visual features, leading to sub-par visualizations (\eg node overlapping and edge distortion). For example, the classic stress method \cite{gansner2004graph} determines node positions by employing unrecognizable shortest distances, resulting in severe node overlapping.

Another long-standing challenge is visualizing large graphs. When handling large graphs, most existing solutions~\cite{meyerhenke2017drawing,martin2011openord,rodrigues2015gmine,shi2009himap,abello2006ask,archambault2008grouseflocks,hu2005efficient} adopt the \textit{multi-level} scheme to avoid the poor readability and prohibitive overhead caused by visualizing all nodes in a \textit{single-level} fashion. Specifically, the multi-level strategy organizes the nodes of the input graph $G$ into a tree $\mathcal{H}$, such that (i) each leaf in $\mathcal{H}$ is a node in $G$, and (ii) each non-leaf node in $\mathcal{H}$, referred to as a {\it supernode}, has only a small number of children. 
The user can navigate through $\mathcal{H}$ and visualize any set $\mathcal{S}$ of nodes or supernodes that have the same parent. 
Utilizing the multi-level scheme, existing solutions still struggle to cope with large graphs as they entail significant overheads to compute the node distances.
The reason is that these methods (\eg \cite{meyerhenke2017drawing,martin2011openord}) require calculating all pairwise distances (\eg shortest distance) for the leaf nodes in the subtrees of two supernodes $\mathcal{V}_i$ and $\mathcal{V}_j$ before determining the distance between $\mathcal{V}_i$ and $\mathcal{V}_j$; otherwise, the visualization of supernodes will be uninformative in reflecting the underlying graph structure, which results in poor visualization quality.

To address the aforementioned challenges in effectiveness and efficiency, we propose a new graph-theoretic node distance measure dedicated to enhancing visualization quality, referred to as \pprdistname. 
\pprdistname takes inspiration from {\it personalized PageRank (PPR)}~\cite{page1999pagerank}, a proximity measure quantifying the connectivity from a source node to a target node via random walks in a graph. Considering the requirements of graph visualization, \pprdistname incorporates degree and symmetry information into PPR, and ameliorates PPR with transformation and truncation.
Through such optimizations, \pprdistname circumvents the visual issues (\eg node overlapping and edge distortion) that may be caused by PPR, while accurately preserving the graph structure (\eg degree and high-order proximity information). Notably, our analysis shows that \pprdistname offers non-trivial visualization quality guarantees in terms of two widely-used aesthetic criteria.

Unfortunately, computing \pprdistname for supernodes is challenging as it involves a multitude of leaf nodes underlying the supernodes. There are a plethora of approaches for PPR computation in the literature \cite{page1999pagerank,wang2017fora,lin2020index,wang2019efficient}, but they focus on single-source or top-$k$ PPR queries rather than arduous all-pair queries in the case of \pprdistname. As a result, these approaches are inefficient when adopted for \pprdistname computation due to redundant graph traversal operations and random walk simulations.  
To this end, we propose \taupush, an efficient solution for computing approximate \pprdistname. Compared to the PPR computation methods, \taupush achieves superior time complexity and empirical efficiency, while retaining strong accuracy guarantees. Under the hood, \taupush adopts a filter-refinement paradigm accommodating three carefully-designed techniques. First, \taupush computes a rough estimation of each \pprdistname by \textit{grouped forward graph traversal}. Next, \taupush identifies a set of failed target nodes $\mathcal{T}$ by leveraging the global PageRank of the graph. Finally, \taupush refines the \pprdistname estimation of such nodes by performing a handful of graph traversal operations backward from $\mathcal{T}$. Note that our \pprdistname and its computation algorithm \taupush are not limited to single- and multi-level graph visualizations, and could underpin other scenarios, including graph query result visualization~\cite{bhowmick2020aurora} and incremental graph exploration~\cite{herman2000graph}. 

Based on \pprdistname and \taupush, we create our graph visualization solution \pprviz and experimentally evaluate \pprviz against 13 state-of-the-art methods using 12 real-world graphs. 
Our empirical results demonstrate that \pprviz outperforms the competitors in both effectiveness and efficiency. First and foremost, \pprviz obtains considerably better visualization quality in terms of {aesthetic metrics and user satisfaction} than the competitors.
Moreover, \pprviz runs at a fraction of the computational cost of the competing methods for both pre-processing and online visualization.
For instance, \pprviz takes only 1 second to {visualize the (super) node sets} from a graph with 3 billion edges, whereas none of the competitors can produce a visualization within 1000 seconds.

To summarize, this paper makes the following contributions:
\begin{itemize}[topsep=2pt,itemsep=1pt,parsep=0pt,partopsep=0pt,leftmargin=11pt]
    \item We propose \pprdistname, a new node distance measure that not only fully captures the structural information of the input graph but also optimizes the aesthetic metrics.
    \item We devise \taupush, an algorithm for efficient \pprdistname computation with three tailored techniques, which enables responsive visualizations even on billion-edge graphs.
    \item We conduct extensive experiments to demonstrate the superiority of \pprdistname and \taupush over the state-of-the-art graph visualization methods in both effectiveness and efficiency.
\end{itemize}

\section{Preliminaries}\label{sec:gvp}
In this section, we first elaborate on the procedure of graph visualization and then introduce the multi-level mechanism for visualizing large graphs. Finally, we discuss how to assess the visualization quality.
\rc{Table~\ref{tab:notations} lists the frequently used notations in our paper.}

\begin{table}[!t]
\centering
\renewcommand{\arraystretch}{1.1}
\begin{footnotesize}
\caption{Frequently used notations.}
\vspace{-2mm} 
\label{tab:notations}
\rc{
	\begin{tabular}{|p{0.5in}|p{2.5in}|}
		\hline
		{Notation} &  {Description}\\
		\hline
		$G$=$(V,E)$   & A graph $G$ with node set $V$ and edge set $E$.\\
		\hline
		$n, m$   & The numbers of nodes and edges in $G$, respectively.\\
        \hline
        $d(v_i)$   & The out-degree of node $v_i$ in $G$.\\
        \hline
        $\XM$   & The position matrix of nodes in $G$.\\
        \hline
        $\mathcal{S}, k$  & The level-($\ell$+1) supernode, the number of level-$\ell$ children in $\mathcal{S}$.\\
        \hline
        $\mathcal{V}_i, \leaf(\mathcal{V}_i)$  & The level-$\ell$ supernode, the set of leaf nodes in $\mathcal{V}_i$.\\
        \hline
        $\pi, \alpha$  & The PPR value, the restart probability in the random walk.\\
        \hline
        $\pprdist[i,j]$ & The \pprdistname value of node pair $(v_i,v_j)$ defined in Eq. \eqref{eq:pprdist}.\\
        \hline
        $\pprdeg(v_i,v_j)$ & The \pprdegname value of node $v_j$ w.r.t.\ node $v_i$ defined in Eq. \eqref{eq:pprdist}.\\
        \hline
        $\pprdeg(\mathcal{V}_i,\mathcal{V}_j)$  & The \pprdegname value of supernode pair $(\mathcal{V}_i,\mathcal{V}_j)$ defined in Eq. \eqref{eq:spprdeg}.\\
        \hline
        $\hat{\pi}_d(v_i,v_j)$  & The approximate \pprdegname of node $v_j$ w.r.t.\ node $v_i$ (see Eq. \eqref{eq:fwd-invariant}).\\
        \hline
        $r(v_i,v_j)$ & The residue of node $v_j$ w.r.t.\ node $v_i$ (see Eq. \eqref{eq:fwd-invariant}).\\
        \hline
        $r_{max}, r^{b}_{max}$ & The forward and backward residue thresholds, respectively. \\
        \hline
        $\epsilon, \delta$  & The approximation parameters in Definition~\ref{def:appro-pprdeg}.\\
        \hline
        $\tau_i$  & The \dnpr of level-$\ell$ supernode $\mathcal{V}_i$ in Eq.\eqref{eq:dnpr-dnppr}.\\
        \hline
	\end{tabular}
	}
\end{footnotesize}
\vspace{-2mm}
\end{table}

\subsection{Graph Visualization} \label{sec:formulation}
Let $G=(V,E)$ be a graph, where $V$ is a set of $n$ nodes and $E$ is a set of $m$ edges. {We assume that $G$ is a directed and homogeneous graph, where the nodes and edges have no labels/attributes. We consider the task of visualizing the input graph $G$ on the two-dimensional Euclidean space, where the graph $G$ is laid out based on a position matrix $\XM \in \mathbb{R}^{n\times 2}$ and $\XM[i]\in \mathbb{R}^{2}$ (\ie the $i$-{th} row of $\XM$) records the coordinate of node $v_i \in V$ on the screen. 
Following prior visualization methods~\cite{eades1984heuristic,fruchterman1991graph,martin2011openord,jacomy2014forceatlas2,kamada1989algorithm,gansner2004graph,brandes2006eigensolver,gansner2012maxent}, we represent each node with a circle by placing the center of the circle on the coordinate of the node, and draw each directed edge using a straight arrowhead line. For undirected edges, we omit the arrowhead to avoid a cluttered display. Thus, $\|\XM[i]-\XM[j]\|$ is the distance between node $v_i$ and $v_j$ on the screen and $l(v_i,v_j)=\|\XM[i]-\XM[j]\|$ is the length of edge $(v_i,v_j)\in E$.} 

Given an input graph $G$, graph visualization typically consists of two phases: (i) \textit{distance matrix computation} and (ii) \textit{position matrix embedding}.
In the first phase, a specific node distance matrix $\DM \in \mathbb{R}^{n\times n}$ is computed, in which $\DM[i,j]$ reflects the graph-theoretic distance between nodes $v_i$ and $v_j$. For example, \cite{fruchterman1991graph,noack2005energy,jacomy2014forceatlas2} directly employ the adjacent matrix as $\DM$ and \cite{kamada1989algorithm,gansner2004graph,gansner2012maxent} use the all-pair shortest distances as $\DM$.
In the second phase, the distance matrix $\DM$ is converted to a position matrix $\XM$, such that the on-screen distance $\|\XM[i]-\XM[j]\|$ of node pair $(v_i,v_j)$ is close to $\DM[i,j]$ for all node pairs $(v_i,v_j) \in V\times V$. 
In literature, there exist several optimization techniques that transform the distance matrix into the position matrix, \eg gradient descent~\cite{gansner2012maxent}, simulated annealing~\cite{davidson1996drawing}, and stress majorization~\cite{gansner2004graph}.

\subsection{Multi-level Mechanism}\label{sec:prelim-multi}
Directly visualizing all nodes in a large graph usually results in a giant hairball with little discernible structure information, due to the sheer numbers of nodes and edges in the layout~\cite{gibson2013survey}. Thus, most existing methods~\cite{rodrigues2015gmine,shi2009himap,abello2006ask,archambault2008grouseflocks} and softwares~\cite{auber2004tulip,de2018exploratory,shannon2003cytoscape} visualize large graphs in an interactive and \textit{multi-level} manner, so as to cut down the number of nodes in each drawing.
As surveyed in \cite{herman2000graph,von2011visual}, multi-level methods consist of two phases: (i) \textit{preprocessing} and (ii) \textit{interactive visualization}.
In the preprocessing phase, a {\it supergraph} hierarchy is constructed such that nodes of the graph $G$ are organized into a tree $\mathcal{H}$, where (i) each leaf is a node in $G$, and (ii) each non-leaf node, referred to as a {\it supernode}, contains $k$ children. For convenience, we say that each leaf node is at level-0, and that each supernode $\mathcal{V}_i$ is at level-$(\ell+1)$ if its children are at level-$\ell$ ($\ell \ge 0$). In the interactive visualization phase, users can select any supernode $\mathcal{S}$ at level-$(\ell+1)$, and ask for a visualization of the $k$ children of $\mathcal{S}$, where the corresponding position matrix $\XM \in \mathbb{R}^{k\times 2}$ is derived following the visualization procedure described in the preceding section. 
For example, if $\mathcal{S}$ consists of leaf children, then multi-level methods visualize the subgraph of $G$ induced by the nodes in $\mathcal{S}$. On the other hand, if $\mathcal{S}$ consists of supernode children, then they visualize a high-level graph where (i) each node represents a supernode $\mathcal{V}_i$ in $\mathcal{S}$ and (ii) each edge connects from supernode $\mathcal{V}_i$ to supernode $\mathcal{V}_j$ if $G$ contains an edge from a leaf node in the sub-tree of $\mathcal{V}_i$ to another leaf node in the sub-tree of $\mathcal{V}_j$.
Notice that the size constraint $k$ is conducive to curtailing visual clutter~\cite{duncan1998balanced} and can be configured by users according to their needs.
Throughout this paper, we refer to visualizing the entire graph on one single layout (resp.\ multiple levels of layouts) as single-level (resp.\ multi-level) visualization.

Although the supergraph hierarchy can be constructed offline and the time complexity of position matrix embedding is reduced to $O(k^3)$~\cite{gansner2004graph} for the children of $\mathcal{S}$, existing multi-level methods still incur high costs for the distance matrix computation of the children in $\mathcal{S}$, especially on graphs with many nodes and edges.
A keen reader may propose to directly compute the distances of the children in $\mathcal{S}$ by treating $\mathcal{S}$ as a stand-alone graph. However, doing so overlooks the underlying structures of $\mathcal{S}$, and thus impairs visualization quality as pinpointed in \cite{martin2011openord,meyerhenke2017drawing}.
Instead, a canonical approach adopted by existing methods~\cite{meyerhenke2017drawing,sokal1958statistical,martin2011openord} is to measure the distance between two supernodes $\mathcal{V}_i$ and $\mathcal{V}_j$ based on the average distance between every node pair $v_i$ and $v_j$, where $v_i$ (resp.\ $v_j$) is a leaf node in $\mathcal{V}_i$ (resp.\ $\mathcal{V}_j$). In other words, this approach requires computing $k^{\ell}\times k^{\ell}$ pairs of distances for two level-$\ell$ supernodes, which poses a challenge in terms of computation efficiency.

\subsection{Visualization Quality Assessment} 
Intuitively, a high-quality graph visualization should not only reflect the topological information of the input graph but also have good \textit{readability}~\cite{gibson2013survey,battista1998graph}. In other words, the position matrix $\XM$ in the Euclidean space should accurately reflect the structure of $G$, in the sense that well-connected nodes (resp.\ poorly-connected nodes) should be placed close (resp.\ far apart) to each other.
As for good readability, it means that the layout should avoid negative visual artifacts such as nodes overlapping with each other or edges with drastically different lengths. In relation to this, there exist several aesthetic criteria in the literature that quantify the readability of graph layouts. In this paper, we adopt the two most commonly-used aesthetic metrics~\cite{noack2007unified,klammler2018aesthetic,purchase2002metrics,taylor2005applying,purchase2002empirical,bennett2007aesthetics}, \ie \textit{node distribution (ND)} and \textit{uniform length coefficient variance (ULCV)}, defined as follows.

\begin{definition}[\nd]\label{def:node-distri} 
For a position matrix $\XM$, the node distribution is \nd$\textstyle (\XM)=\sum\limits_{i< j}\frac{1}{||\XM[i]-\XM[j]||^2}$.
\end{definition}

\begin{definition}[\ulcv]\label{def:edge-var}
For a position matrix $\XM$, let $l_{\sigma}$ (resp.\ $l_{\mu}$) be the standard deviation (resp.\ mean) of the edge lengths. The uniform length coefficient variance is \ulcv$(\XM)=l_{\sigma}/l_{\mu}$.
\end{definition}

\nd{} is the summation of the reciprocals of the squared distance between node pairs in the graph layout. A large \nd{} score indicates the existence of visual clutter in the layout. In particular, overlapping nodes (occupying the same node positions) lead to an infinite \nd{} score.
\ulcv{} measures the skewness of edge lengths. A large \ulcv{} indicates that some edges are considerably longer or shorter than others, in which case the layout tends to look distorted.

\section{PPR-based Node Distance}\label{sec:pds}
This section presents our node distance measure \pprdistname for graph visualization. We first formally define \pprdistname for the case of single-level visualization and then extend it to multi-level visualization.

\subsection{Single-level \pprdistnameb}\label{sec:pdsdef}
\stitle{Definition}
\pprdistname is formulated based on {\it personalized PageRank (PPR)}, a node proximity measure defined as follows. Given a directed graph $G=(V,E)$, two nodes $v_i,v_j\in V$, and a {\it restart probability} $\alpha$, the PPR $\pi(v_i,v_j)$ from $v_i$ to $v_j$ is defined as the probability that a {\em random walk with restart} (RWR) \cite{tong2006fast} originating from $v_i$ would end at $v_j$. Specifically, an RWR starts from $v_i$, and at each step, it either (i) terminates at the current node with probability $\alpha$, or (ii) with the remaining $1-\alpha$ probability, navigates to a random out-neighbor of the current node. Intuitively, a large PPR $\pi(v_i,v_j)$ indicates that numerous paths exist from $v_i$ to $v_j$; in other words, $v_i$ is well connected to $v_j$. Based on PPR, we define \pprdistname as follows.
\begin{definition}[\pprdistname]\label{def:pprdist}
Let $\pprdist \in \mathbb{R}^{n \times n}$ be the \pprdistname matrix for all node pairs in a graph $G$ and $\pprdist[i,j]$ be the \pprdistname between nodes $v_{i}$ and $v_{j}$. We define $\pprdist[i,j]$ as
\begin{equation}\label{eq:pprdist}
\pprdist[i,j]=\min{\Big(\max\Big(1-\log{\left(\pi_d(v_i,v_j)+\pi_d(v_j,v_i)\right)}, 2\Big),2\log{n}\Big)},
\end{equation}
where $\pi_d(v_i,v_j)=\pi(v_i,v_j)\cdot d(v_i)$ denotes the {\em degree-normalized PPR} (\pprdegname) from $v_{i}$ to $v_{j}$, and $d(v_i)$ is the out-degree of $v_i$.
\end{definition}

The intuition behind \pprdistname is that if node pair $(v_i,v_j)$ has a high PPR value $\pi(v_i,v_j)$, then its \pprdistname $\Delta[i, j]$ should be small. As such, \pprdistname ensures that well-connected nodes are placed closely in a visualization. Moreover, \pprdistname also accounts for the out-degrees of $v_i$ and $v_j$ in its definition to overcome the inherent limitation of PPR for visualization. To explain, we consider the example in Fig.~\ref{fig:graph-example}, which shows a graph with nodes $v_0\textrm{--}v_9$, as well as the PPR and \pprdistname values for node pairs $(v_0,v_8)$, $(v_2,v_0)$, and $(v_6,v_9)$. Observe that $\pi(v_2,v_0)=0.11<\pi(v_6,v_9)=0.44$, even though node pairs $(v_2,v_0)$ and $(v_6,v_9)$ are both directly connected via an edge.  
This indicates that for adjacent node pairs in a graph, their PPR values could vary considerably. 
As a consequence, directly transforming PPR values into node distances would bring about a large variance in edge lengths in the visualization. 
To alleviate this issue, in our formulation of \pprdistname, we scale each $\pi(v_i,v_j)$ by multiplying it with the out-degree of the source node $v_i$. It has been shown in previous work \cite{yang2020homogeneous} that such a scaling contributes to an accurate measure of the strength of connections between nodes. 
In Fig.~\ref{fig:graph-example}, the \pprdistname of $(v_2,v_0)$ is relatively close to that of $(v_6,v_9)$, which is more consistent with the fact that $v_2$ and $v_6$ are direct neighbors of $v_0$ and $v_9$, respectively. Moreover, the \pprdistname of $(v_0, v_8)$ is large, reflecting the fact that $v_0$ is far away from $v_8$ in the input graph.

\begin{table}[t]
    \hspace{2mm}
    \begin{minipage}{0.30\linewidth}
		\centering
		\includegraphics[width=0.8\linewidth]{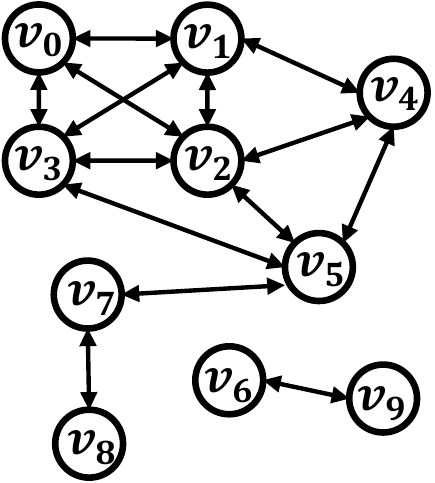}
	\end{minipage}
	\begin{minipage}{0.6\linewidth}
		\centering
    	\renewcommand{\arraystretch}{1.1}
    \begin{scriptsize}
    \resizebox{0.7\columnwidth}{!}{%
    	\begin{tabular}{|c|c|c|}\hline
                     & PPR &  \pprdistname  \\  \hline   
                    $(v_0,v_8)$ & {0.01}  & 3.73\\  \hline
                    $(v_2,v_0)$ & 0.11  & {0.92}\\  \hline
                    $(v_6,v_9)$ & 0.44  & {1.12}\\  \hline
        \end{tabular}
    }
    \end{scriptsize}
	\end{minipage}
	\hspace{4mm}
	\hfill
	\vspace{-2mm}
	\captionof{figure}{PPR and \pprdistnameb in a toy graph.}
	\label{fig:graph-example}
	\vspace{-2mm}
\end{table}

To make the node distance symmetric, we formulate \pprdistname based on  $\pi_d(v_i,v_j)+\pi_d(v_j,v_i)$, since $\pi_d(v_i,v_j) \ne \pi_d(v_j,v_i)$ in general.
Furthermore, \pprdistname takes the inverse of the natural logarithm of \pprdegname $\pi_d(v_i,v_j)+\pi_d(v_j,v_i)$ and imposes an empirical truncation to narrow down the distance variance, thereby ensuring the distances between nodes lie in a reasonable range that can be well fitted into a canvas.
In particular, the lower and upper bounds of \pprdistname are set to 2 and $2\log{n}$ for precluding node overlapping and blank space issues, respectively. The rationale of choosing $2\log{n}$ as the upper bound is that the average PPR value between a node pair in a graph with $n$ nodes is $1/n$ ~\cite{wang2017fora,shi2019realtime} and \pprdistname should focus on reflecting the proximity information of node pairs with above-average PPR values (\ie relatively high relevance).

\stitle{Bounds for aesthetic criteria}
The following two theorems\footnote{We refer interested readers to~\trref{} for all proofs.} establish the worst-case upper bounds in terms of both \nd{} and \ulcv{} (see Definitions \ref{def:node-distri}--\ref{def:edge-var}), when utilizing \pprdistname for visualization. 

\begin{theorem}\label{thm:pprd-nd}
Given the \pprdistname matrix $\pprdist$ of a graph $G$,
suppose that $||\XM[i]-\XM[j]||=\pprdist[i,j]$ $\forall{v_i,v_j}\in V$, then ND$(\XM)\leq 0.215e\cdot m+0.0175n^2$, where $e$ is the Euler constant.
\end{theorem}

\begin{theorem}\label{thm:pprd-ulcv}
Given the \pprdistname matrix $\pprdist$ of a graph $G$,
suppose that $||\XM[i]-\XM[j]||=\pprdist[i,j]$ $\forall{v_i,v_j}\in V$ and the restart probability $\textstyle\alpha \leq \tfrac{1}{2}-\sqrt{\tfrac{1}{4}-\tfrac{1}{2e}}$, then  ULCV$\textstyle(\XM)\leq\tfrac{\left(\log{\tfrac{1}{{2\alpha(1-\alpha)}}}-1\right)}{4}$.
\end{theorem}

Note that the upper bound of the restart probability $\alpha$ in Theorem~\ref{thm:pprd-ulcv} (which is approximately 0.243) is not restrictive, as $\alpha$ is usually set to 0.15 or 0.2~\cite{wang2020personalized,wu2021unifying,wang2016hubppr,lofgren2016personalized}.

\begin{figure}[!t]
\centering
\begin{small}
  \begin{tabular}{ccc}
    \includegraphics[width=0.3\columnwidth]{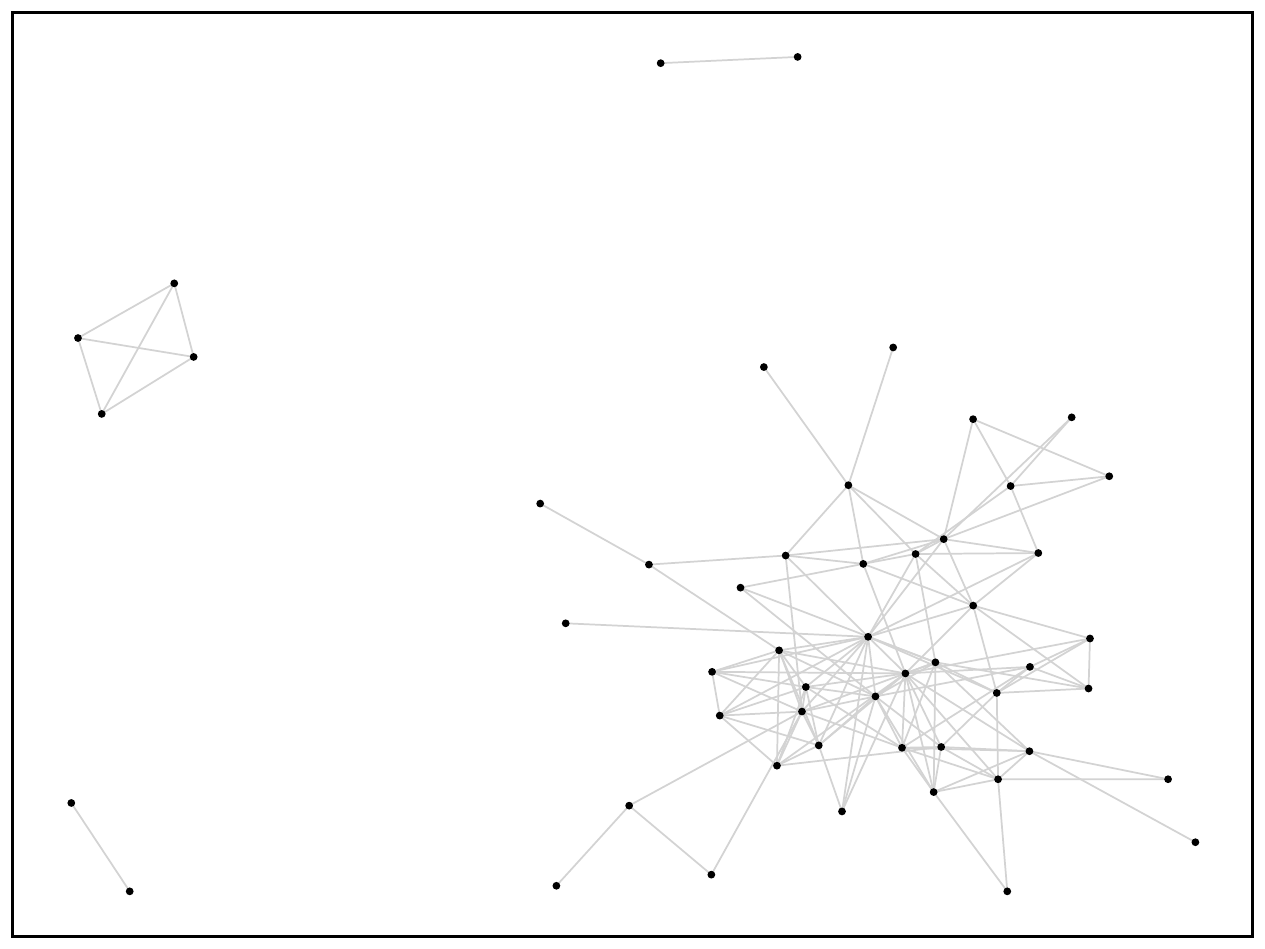}
     &
    \includegraphics[width=0.3\columnwidth]{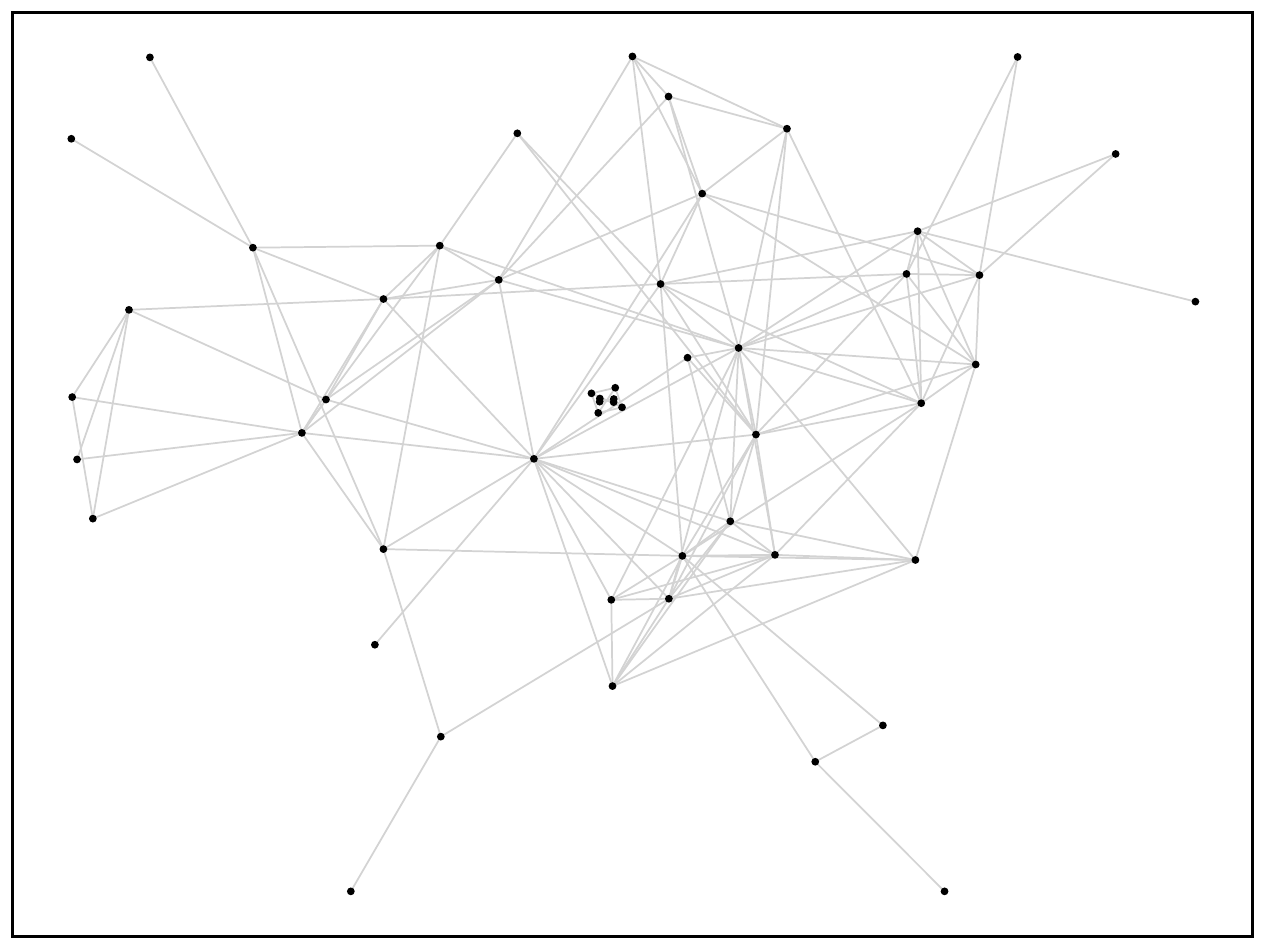}
     &
    \includegraphics[width=0.3\columnwidth]{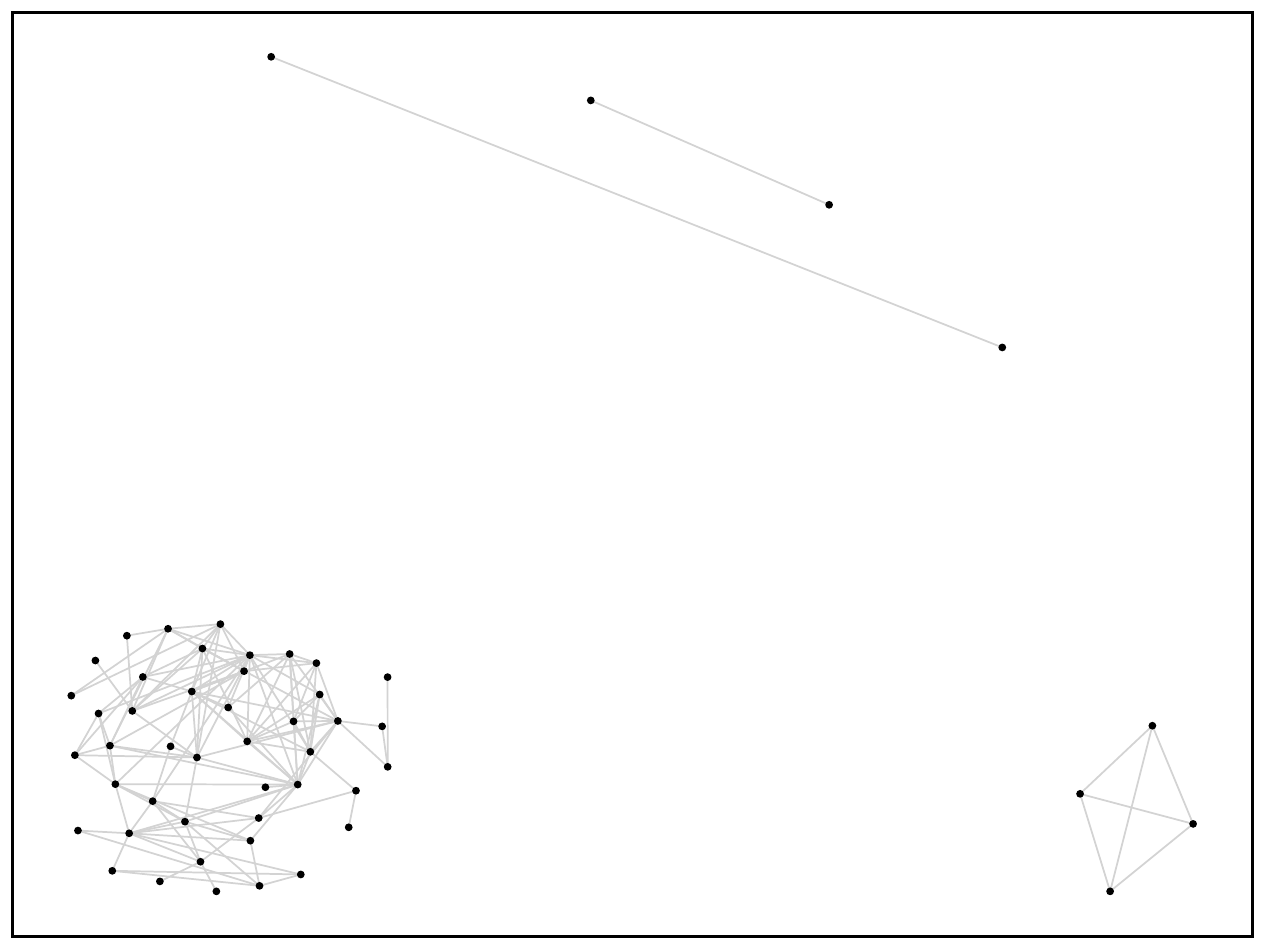}
    \\
    \footnotesize{\nd{}=2.4E+03, \ulcv{}=0.39} & \footnotesize{\nd{}=2.0E+04, \ulcv{}=0.46} & \footnotesize{\nd{}=6.2E+03, \ulcv{}=0.75}
    \\
    (a) \pprdistname  &   (b) Shortest distance~\cite{gansner2004graph}
    & (c) \simrank~\cite{jeh2002simrank}
    \\
  \end{tabular}
\end{small}
\vspace{-2mm}
\caption{Visualization results of \pprdistnameb and two other node distances measures on the \fbego graph. Smaller aesthetic scores indicate better quality.}\label{fig:case-study}
\vspace{-2mm}
\end{figure}

\stitle{A case study} 
To verify the effectiveness of \pprdistname, we use \pprdistname and two classic node distance measures (\ie the shortest distance and \simrank-based distance) to visualize the graph \fbego via the same position matrix embedding algorithm~\cite{gansner2004graph}.
Note that the \simrank-based distance is obtained by plugging \simrank~\cite{jeh2002simrank} into Eq.~\eqref{eq:pprdist}.
Fig.~\ref{fig:case-study} displays the visualization results of the three distance measures.
It can be observed that \pprdistname yields a high-quality visualization result, which organizes the graph into a well-connected cluster and three cliques. In comparison, the widely-used shortest distance measure~\cite{gansner2004graph,gansner2012maxent,brandes2006eigensolver,meyerhenke2017drawing} suffers from severe node overlapping and edge distortion issues. 
Regarding the visualization derived from \simrank-based distances, the edges of the cliques are distorted, as \simrank assigns node pairs within the large connected component high similarity scores but $0$ similarity scores to the 2-cliques.
Besides the qualitative results, we also report \nd{} and \ulcv{} scores in Fig.~\ref{fig:case-study}, and we can find that \pprdistname considerably outperforms the shortest distance and the \simrank-based distance in terms of both metrics, validating the superior aesthetic performance of \pprdistname as proved in Theorem~\ref{thm:pprd-nd} and Theorem~\ref{thm:pprd-ulcv}.

\subsection{Multi-level \pprdistnameb}\label{sec:lpds}

\stitle{Definition}
{Given a user-specified level-$(\ell+1)$ supernode $\mathcal{S}$, multi-level visualization requires calculating the level-$\ell$ \pprdistname between the $k$ level-$\ell$ children supernodes of $\mathcal{S}$.} Recall from Definition~\ref{def:pprdist} that a linchpin functioned in single-level \pprdistname is the \pprdegname between two nodes. Thus, we extend the \pprdegname in Definition~\ref{def:pprdist} to level-$\ell$ \pprdegname (Definition~\ref{def:spprdeg}), and level-$\ell$ \pprdistname can be derived accordingly by plugging Eq.~\eqref{eq:spprdeg} into Eq.~\eqref{eq:pprdist}. 

\begin{definition}[Level-$\ell$ \pprdegname]\label{def:spprdeg} For two level-$\ell$ supernodes $\mathcal{V}_i$ and $\mathcal{V}_j$, denote the set of leaf nodes in $\mathcal{V}_i$ as $\leaf(\mathcal{V}_i)$, the level-$\ell$ \pprdegname $\pprdeg(\mathcal{V}_i, \mathcal{V}_j)$ of $\mathcal{V}_j$ w.r.t.\ $\mathcal{V}_i$ is defined as
\begin{equation}\label{eq:spprdeg}
\pprdeg(\mathcal{V}_i, \mathcal{V}_j) = \frac{\sum\limits_{v_s\in \leaf(\mathcal{V}_i), v_t \in \leaf(\mathcal{V}_j)}\pi_d(v_s,v_t)}{|\leaf(\mathcal{V}_i)|\cdot|\leaf(\mathcal{V}_j)|},
\end{equation}
where $\pi_d(v_s,v_t)$ is \pprdegname of $v_t$ w.r.t.\ $v_s$.
\end{definition}
Intuitively, the level-$\ell$ \pprdegname measures the connectivity from supernode $\mathcal{V}_i$ to $\mathcal{V}_j$ by taking the average of the \pprdegname values from the leaf nodes in $\mathcal{V}_i$ to those in $\mathcal{V}_j$. The idea of measuring the proximity between two supernodes by summarizing the structure of the underlying leaf nodes is also adopted in~\cite{van2008visualizing,martin2011openord,sokal1958statistical}.~{In particular, ~\cite{martin2011openord,sokal1958statistical} also consider the average of all pairwise distances of the leaf nodes.}

Compared with our level-$\ell$ \pprdegname, a simple and straightforward way is to take the level-$\ell$ children of supernode $\mathcal{S}$ as a weighted graph and compute the \pprdegname on it (called W-\pprdegname). More precisely, the graph is constructed by treating each supernode $\mathcal{V}_i\in \mathcal{S}$ as a node and merging the edges between supernodes $\mathcal{V}_i$ and $\mathcal{V}_j$ in $\mathcal{S}$ as a weighted edge. Although W-\pprdegname can be computed very efficiently as it requires only an $O(k^2)$ cost, it ignores the micro-structures of each supernode and the paths containing nodes outside $\mathcal{S}$, resulting in sub-par visualization quality. To exemplify, we consider Fig.~\ref{fig:supergraph-example}, which shows a graph with nodes $v_0\textrm{--}v_5$ and a level-2 supernode $\mathcal{S}$ with level-1 supernodes $\mathcal{V}_0, \mathcal{V}_1, \mathcal{V}_2$, as well as the level-$\ell$ \pprdegname ($\ell$-\pprdegname in short) and W-\pprdegname values for supernode pairs $(\mathcal{V}_1,\mathcal{V}_0)$, $(\mathcal{V}_2,\mathcal{V}_0)$, and $(\mathcal{V}_2,\mathcal{V}_1)$. Intuitively, for the source supernode $\mathcal{V}_2$, $\mathcal{V}_1$ has better connectivity to it than $\mathcal{V}_0$ as the children of $\mathcal{V}_2$ and $\mathcal{V}_0$ share one common neighbor $v_2$, whereas the children of $\mathcal{V}_2$ and $\mathcal{V}_1$ have two common neighbors $v_3$ and $v_4$.
From the table in Fig.~\ref{fig:supergraph-example}, we observe that the W-\pprdegname values of $(\mathcal{V}_2,\mathcal{V}_0)$ and $(\mathcal{V}_2,\mathcal{V}_1)$ are the same, which is counter-intuitive. In contrast, level-$\ell$ \pprdegname addresses this issue and accurately captures the structure of the original graph.                           
    
\begin{table}[!t]
\begin{minipage}{0.3\linewidth}
		\centering
		\includegraphics[width=0.8\columnwidth]{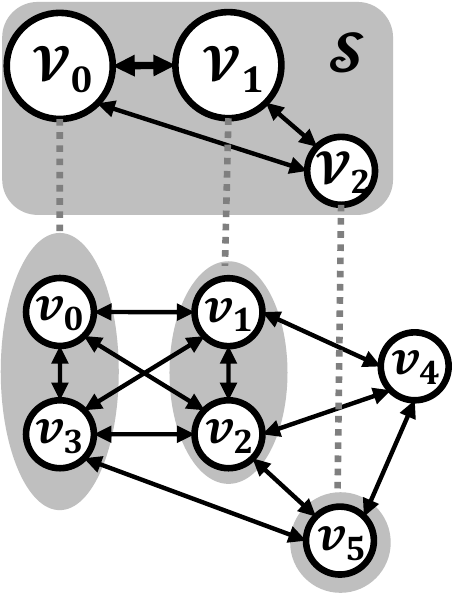}
	\end{minipage}
	\begin{minipage}{0.65\linewidth}
		\centering
		\renewcommand{\arraystretch}{1.1}
\begin{footnotesize}
\begin{tabular}{|c|c|c|}
\hline
     & $\ell$-\pprdegname &  W-\pprdegname  \\ \hline
$(\mathcal{V}_1,\mathcal{V}_0)$ & { 0.56}  & 0.74        \\ \hline
$(\mathcal{V}_2,\mathcal{V}_0)$ & { {0.39}} &   0.69  \\ \hline
$(\mathcal{V}_2,\mathcal{V}_1)$ & { {0.50}} &  0.69   \\ \hline
\end{tabular}
\end{footnotesize}
	\end{minipage}
	\vspace{-2mm}
	\captionof{figure}{A toy example of level-$\boldsymbol{\ell}$ \pprdegnameb.}
	\label{fig:supergraph-example}
\vspace{-2mm}
\end{table}

\subsection{Efficiency Challenges}\label{sec:multilevel-challenge}

As per Eq.\eqref{eq:pprdist} and Eq.\eqref{eq:spprdeg}, the level-$\ell$ \pprdistname between $\mathcal{V}_i$ and $\mathcal{V}_j$ requires computing the exact PPR values for all pairs of leaf nodes in $\leaf(\mathcal{V}_i)\times \leaf(\mathcal{V}_j)$, where $\leaf(\mathcal{V}_i)$ (resp.\ $\leaf(\mathcal{V}_j)$) signifies the set of leaf nodes of $\mathcal{V}_i$ (resp.\ $\mathcal{V}_j$). As pointed out in prior work \cite{yang2020homogeneous}, exact PPR computation is prohibitive as it entails summing up an infinite series. An option is to compute the near-exact result by performing power iterations (\poweriter)~\cite{page1999pagerank} until the absolute error of PPR is less than $10^{-9}$ (the precision of float). That said, \rc{we still need to invoke \poweriter for each of the $O(k^{\ell+1})$ leaf nodes in the specified level-$(\ell+1)$ supernode $\mathcal{S}$, where each invocation has $O(m)$ time complexity~\cite{page1999pagerank}.} As a result, the near-exact computation of the level-$\ell$ \pprdistname matrix incurs a high cost of $O(k^{\ell+1}m)$, which may be $O(mn)$ in the worst case (\ie for the highest level supergraph).

To alleviate the above-said efficiency issue, we resort to computing approximate level-$\ell$ \pprdistname. Towards this end, we define the concept of $(\epsilon,\delta)-$approximate level-$\ell$ \pprdegname.

\begin{definition}[$(\epsilon,\delta)-$approximate level-$\ell$ \pprdegname]\label{def:appro-pprdeg}
Let $\epsilon$ and $\delta$ be two constants, for any two supernodes $\mathcal{V}_i, \mathcal{V}_j\in \mathcal{S}$ and $\mathcal{V}_i \neq \mathcal{V}_j$, $\widehat{\pi}_d(\mathcal{V}_i, \mathcal{V}_j)$ is an $(\epsilon,\delta)-$approximation of level-$\ell$ \pprdegname $\pprdeg(\mathcal{V}_i, \mathcal{V}_j)$ if it satisfies the following conditions.
\begin{itemize}[leftmargin=*]
    \item If $\pi_d(\mathcal{V}_i, \mathcal{V}_j)< \delta$, $|\widehat{\pi}_d(\mathcal{V}_i, \mathcal{V}_j)-\pi_d(\mathcal{V}_i, \mathcal{V}_j)| \leq \epsilon \cdot \delta$.
    \item If $\pi_d(\mathcal{V}_i, \mathcal{V}_j)\ge \delta$, $|\widehat{\pi}_d(\mathcal{V}_i, \mathcal{V}_j)-\pi_d(\mathcal{V}_i, \mathcal{V}_j)| \leq \epsilon \cdot \pi_d(\mathcal{V}_i, \mathcal{V}_j)$.
\end{itemize}
\end{definition}

The following Lemma~\ref{lem:appro-relation} shows that we can convert the $(\epsilon,\delta)$-approximate level-$\ell$ \pprdegname into an approximate level-$\ell$ \pprdistname. Accordingly, we focus on computing $(\epsilon,\delta)-$approximate level-$\ell$ \pprdegname in the rest of the paper.

\begin{lemma}\label{lem:appro-relation}
Given constants $\theta$ and $\sigma$, by setting $\delta = \frac{e^{1-\sigma}}{2}$ and $\epsilon = 1-\left(\frac{1}{e^{2}}\right)^{\theta}$, $(\epsilon,\delta)$-approximate level-$\ell$ \pprdegname results in an approximate level-$\ell$ \pprdistname $\widehat{\pprdist}[i,j]$ satisfying
\begin{itemize}[leftmargin=*]
    \item If $\pprdist[i,j]< \sigma$, $|\pprdist[i,j]-\widehat{\pprdist}[i,j]| \leq \theta\cdot {\pprdist[i,j]}$,
    \item If $\pprdist[i,j]\geq \sigma$, $|\pprdist[i,j]-\widehat{\pprdist}[i,j]| \leq \theta\cdot\sigma$ .
\end{itemize}
\end{lemma}

To compute $(\epsilon,\delta)-$approximate level-$\ell$ \pprdegname values w.r.t.\ a source supernode $\mathcal{V}_i$, a straightforward approach is to utilize existing single source PPR (SSPPR) approximation methods~\cite{fogaras2005towards,andersen2006local,wang2017fora,wu2021unifying,hou2021massively,shi2019realtime}. Specifically, the state-of-the-art methods~\cite{wang2017fora,shi2019realtime,wang2019efficient,lin2020index,luo2019baton,wu2021unifying} are built upon \fwdpush~\cite{andersen2006local}. Given a source leaf node $v_i$, \fwdpush maintains an estimated \pprdegname $\widehat{\pi}_d(v_i,v_j)$ and a residue value $r(v_i,v_j)$ for each node $v_j$, where $r(v_i,v_i)$ is set to $d(v_i)$ and all $\widehat{\pi}_d(v_i,v_j)$ and other residue $r(v_i,v_j)$ are set to 0 initially. 
Then, \fwdpush starts a deterministic graph traversal from source node $v_i$, and at each step converts $\alpha$ portion of the residue of current node $v_j$ (\ie $r(v_i,v_j)$) into its estimated \pprdegname value $\widehat{\pi}_d(v_i,v_j)$ and distributes the remaining $1-\alpha$ part to $v_j$'s {\em out-neighbors} evenly. For convenience, we refer to the operation of distributing a fraction of node $v_j$'s residue to one of its out-neighbors as a {\em push} operation.

\begin{figure}[!t]
\centering
\begin{small}
\includegraphics[width=0.9\columnwidth]{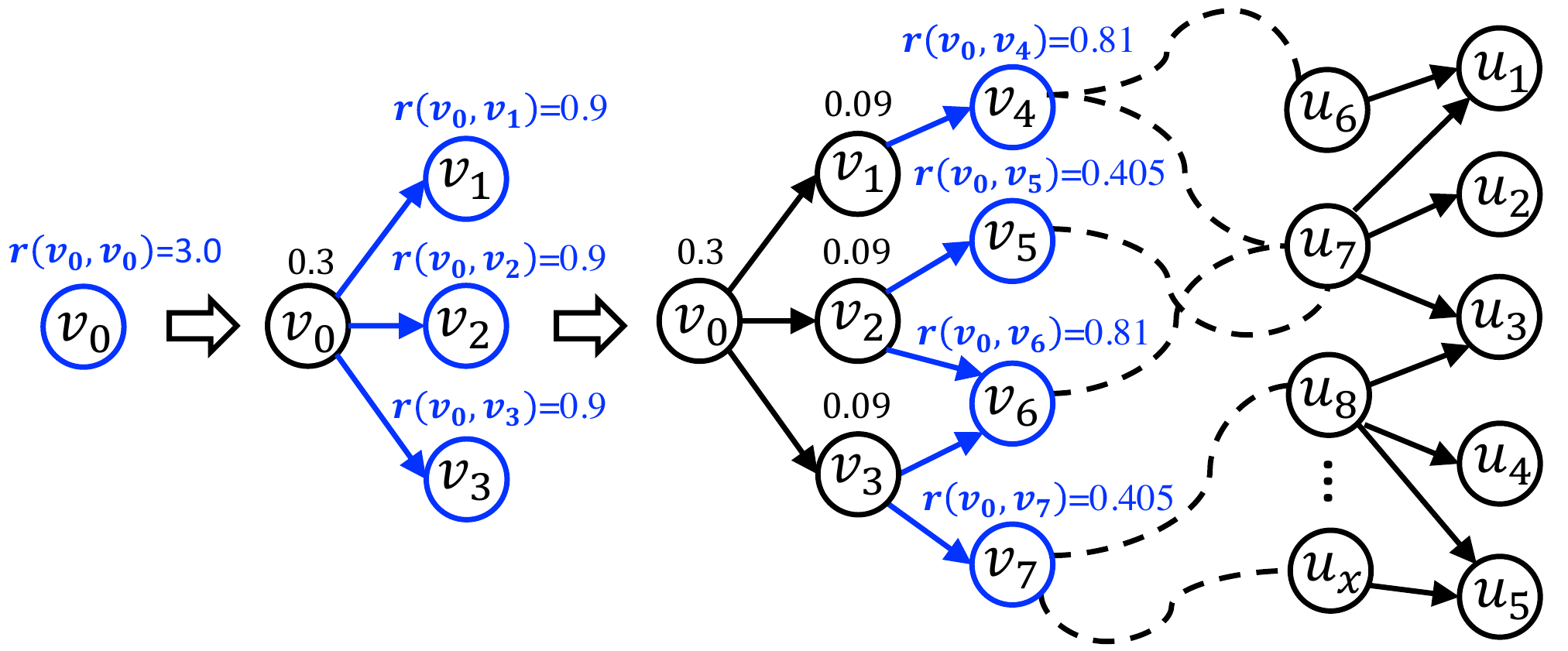}
\end{small}
\vspace{-2mm}
\caption{An example of \fwdpushb.}\label{fig:push-example}
\vspace{-2mm}
\end{figure}

\stitle{A running example}
To exemplify, suppose that the restart probability $\alpha=0.1$ and \fwdpush is performed starting from node $v_0$ in Fig.~\ref{fig:push-example}. Correspondingly, $v_0$ first converts $\alpha=0.1$ fraction of initial residue $r(v_0,v_0)=d(v_0)=3.0$ to $\widehat{\pi}(v_0,v_0)=0.3$, and then evenly pushes the remaining residue (\ie $2.7$) to each of its out-neighbors, resulting in $r(v_0,v_1)=r(v_0,v_2)=r(v_0,v_3)=0.9$. After that, nodes $v_1,v_2$, and $v_3$ continue such push operations and lead to $r(v_0,v_4)=r(v_0,v_6)=0.81$ for nodes $v_4,v_6$ and $r(v_0,v_5)=r(v_0,v_7)=0.405$ for nodes $v_5,v_7$.

In particular, the following invariant~\cite{andersen2006local} holds during the course of push operations:                 
\begin{equation}\label{eq:fwd-invariant}
\pi_d(v_i,v_j)=\widehat{\pi}_d(v_i,v_j)+\sum_{v_k\in V} \frac{1}{d(v_k)}\cdot r(v_i,v_k)\cdot \pi_d(v_k,v_j),
\end{equation}
where $\widehat{\pi}_d(v_i,v_j)$ is an estimation of $\pi_d(v_i,v_j)$ and the other term $\textstyle \sum_{v_k\in V}{\frac{1}{d(v_k)}\cdot r(v_i,v_k)\cdot \pi_d(v_k,v_j)}$ can be regarded as the estimation error. Intuitively, an exact \pprdegname can be obtained when the residue of each node is eventually depleted, \ie $r(v_i,v_k)=0$ for all ${v_k}\in V$. In practice, \fwdpush computes approximate \pprdegname by conducting pushes until every residue value is less than a given threshold $r_{max}$, referred to as the {\em forward residue threshold}.
When $r_{max}$ is small, \fwdpush incurs significant computational overhead, as a large number of push operations are required to deplete residues. To address this issue, most of the state-of-the-art methods~\cite{wang2017fora,shi2019realtime,wang2019efficient,lin2020index,luo2019baton} follow the two-phase paradigm proposed in \fora~\cite{wang2017fora}. In particular, they first invoke \fwdpush~\cite{andersen2006local} with early termination conditions to derive rough approximations of \pprdegname values, and then refine results by exploiting random walk samplings~\cite{fogaras2005towards} to estimate the error term in Eq.~\eqref{eq:fwd-invariant}.
To illustrate, consider the r.h.s.\ graph in Fig.~\ref{fig:push-example}. Instead of always using pushes, \fora-based methods would terminate at nodes $v_4$--$v_7$ and simulate random walks from $v_4$--$v_7$ as per their residues to probe the far-reaching nodes.

\begin{figure}[!t]
\centering
\begin{small}
\includegraphics[width=0.9\columnwidth]{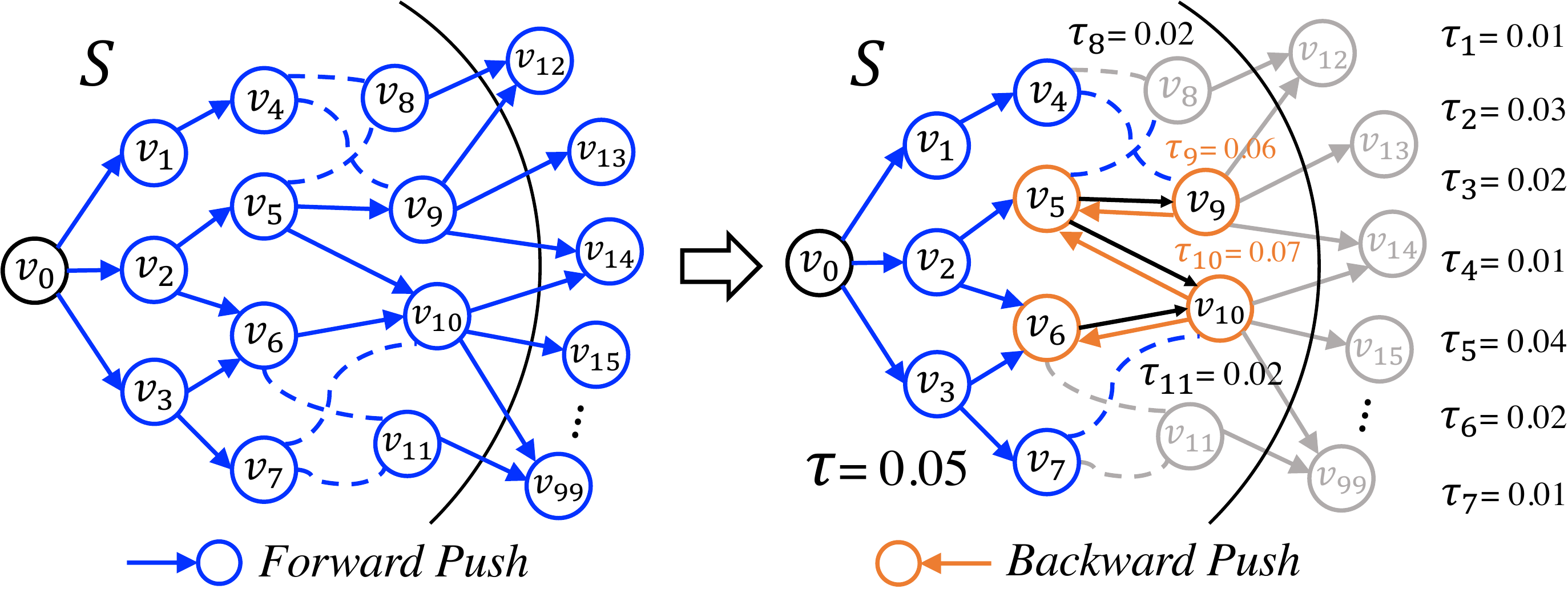}
\end{small}
\vspace{-2mm}
\caption{
\rc{Bidirectional push in \taupushb.}
} \label{fig:taupush-example}
\vspace{-2mm}
\end{figure}

Unfortunately, \fora-based solutions still entail considerable computational overheads if applied directly to \pprdegname approximation for two reasons. First, although SSPPR approximation solutions can improve the efficiency of single-source \pprdegname computation, we need $O(k^{\ell+1})$ (up to $n$ in the worst case) single-source \pprdegname queries for $O(k^{\ell+1})$ leaf nodes in the specified level-$(\ell+1)$ supernode $\mathcal{S}$, which is costly when $\ell$ is large. 
Second, most push operations or random walk samples in such methods are futile in the context of \pprdegname computation.
This is due to the fact that these methods aim to obtain accurate \pprdegname estimates of \textit{all} nodes w.r.t.\ the source. However, the majority of visited nodes by such operations are not in the supernode $\mathcal{S}$ that we are interested in visualizing. Notably, at level-1, roughly $\frac{n-k}{n}$ of nodes located outside $\mathcal{S}$, where $k$ is a small integer (\eg 25) and $n$ can be up to millions.
To illustrate, consider the example in Fig.~\ref{fig:taupush-example}. For simplicity, we assume that supernode $\mathcal{S}$ is at level-1, and it contains 12 leaf nodes $v_0$--$v_{11}$. As shown in the l.h.s.\ graph in Fig.~\ref{fig:taupush-example}, \fwdpush iteratively performs push operations until the estimated \pprdegname values of all nodes, including $88$ nodes $v_{12}$--$v_{99}$ outside $\mathcal{S}$, satisfied the desired approximation.

\section{\taupushb Algorithm}\label{sec:taupush}
In this section, we present \taupush, the efficient algorithm for estimating level-$\ell$ \pprdistname via computing $(\epsilon,\delta)-$approximate level-$\ell$ \pprdegname.
At a high level, \taupush overcomes the limitations of \fora-based methods through (i) a filter-refinement paradigm, which eliminates redundant push operations or random walks without affecting the accuracy guarantee, and (ii) a grouped push strategy, which reduces the number of leaf-level invocations from $O(k^{\ell+1})$ to $O(k)$.
In what follows, we first illustrate the main idea of \taupush in Section~\ref{sec:mainidea}, followed by the algorithmic details of two subroutines of \taupush in Section~\ref{sec:implement}. We further provide theoretical results of \taupush in Section~\ref{sec:analysis}.

\subsection{Main Idea and Algorithm}\label{sec:mainidea}
\taupush estimates level-$\ell$ \pprdegname values by pushing in a bidirectional fashion. In particular, \taupush first performs a small number of forward push operations from the source $\mathcal{V}_i$ to derive an accurate \pprdegname for each target node in the vicinity of $\mathcal{V}_i$. Next, it conducts a handful of pushes reversely along in-neighbors, referred to as \textit{backward push} operations, from the rest of target nodes in $\mathcal{S}$ (\ie the nodes far-reaching from $\mathcal{V}_i$), to obtain their approximate \pprdegname values. To explain the backward push, we consider the r.h.s.\ graph in Fig.~\ref{fig:taupush-example}.
Given $\alpha=0.1$ and node $v_{10}$ with initial residue $r(v_{10},v_{10})=1$, we will convert $0.1$ fraction of $r(v_{10},v_{10})$ to its estimated \pprdegname and propagate the remaining $0.9$ to its in-neighbors $v_5$ and $v_6$, which receive $\frac{0.9}{d(v_5)}=0.45$ and $\frac{0.9}{d(v_6)}=0.9$, respectively. 
By combining forward and backward pushes, we can avoid the issue of ``pushing too deeply'' mentioned in Section~\ref{sec:multilevel-challenge}, and hence, prune excessive pushes/samples for nodes outside $\mathcal{S}$, \eg nodes $v_{12}$--$v_{99}$ in Fig.~\ref{fig:taupush-example}. 
In other words, \taupush filters out the majority of nodes whose approximate \pprdegname values obtained by \fwdpush are sufficiently accurate, and only refines the \pprdegname estimation for each remaining node using backward pushes.

\begin{figure}[!t]
\centering
\begin{small}
\includegraphics[width=0.9\linewidth]{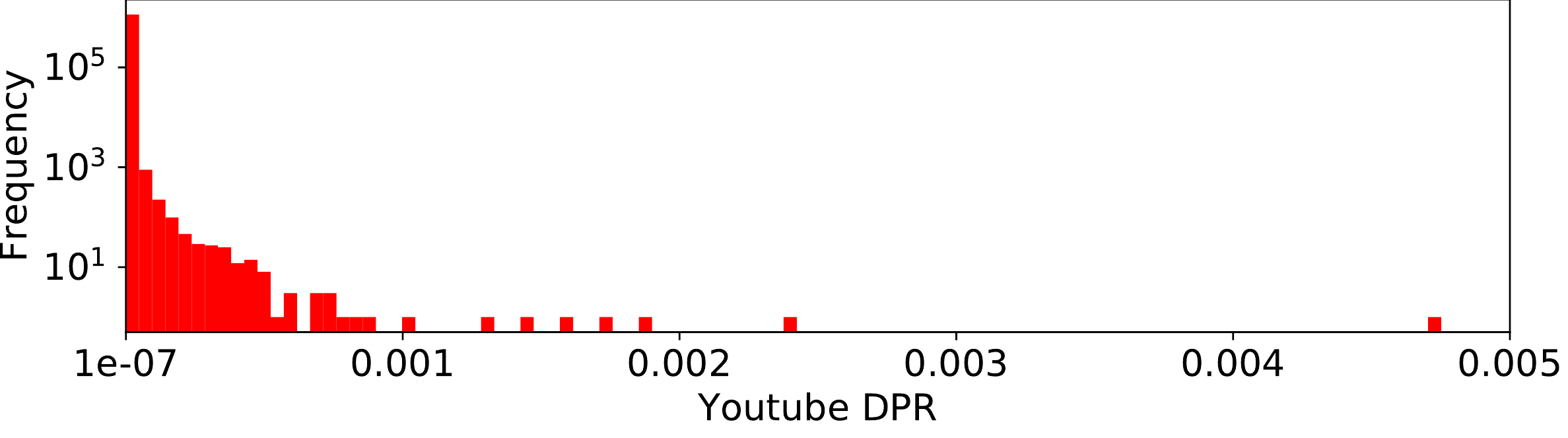}
\end{small}
\vspace{-2mm}
\caption{The distributions of \dnprb on \youtube.} \label{fig:dnpr-youtube}
\vspace{-2mm}
\end{figure}

The linchpin to enabling the aforementioned filter-refinement optimization is the identification of nodes that requires backward pushes. In relation to this, we introduce the notion of {\em degree-normalized PageRank} (\dnpr). For each supernode $\mathcal{V}_j$, its \dnpr is defined as
\begin{equation}\label{eq:dnpr-dnppr}
\tau_j = \frac{1}{m\cdot|\leaf(\mathcal{V}_j)|}{\sum_{v_t\in \leaf(\mathcal{V}_j)}\sum_{v_k\in V}{\pi_d(v_k,v_t)}},
\end{equation}
where $\pi_d(v_k,v_j)$ is the \pprdegname of $v_j$ w.r.t.\ $v_k$ and $\leaf(\mathcal{V}_j)$ signifies the set of leaf nodes in $\mathcal{V}_j$. In particular, when $\mathcal{V}_j$ is a leaf node $v_j$ (\ie $\leaf(\mathcal{V}_j)=v_j$), its \dnpr $\tau_j$ is the summation of \pprdegname values pertinent to $v_j$, which indicates the global importance of $v_j$ in the entire graph.
Building on our theoretical analysis, if the residue threshold $r_{max}$ used in \fwdpush is set properly, the approximate \pprdegname values of nodes $v_j$ with $\tau_j$ less than the given \dnpr threshold $\tau$ are guaranteed to be $(\epsilon,\delta)$-approximate (see Lemma \ref{lem:fpsn-correct}). 
For example, the r.h.s.\ graph in Fig.~\ref{fig:taupush-example} shows that, given $\tau=0.05$, nodes $v_1$--$v_8$ and $v_{11}$ can be filtered out as their \dnpr values are less than $\tau$, and hence, we only need to conduct backward pushes from $v_9$ and $v_{10}$.
We observe that \dnpr values of real-world graphs often follow the power law distribution, where only a few nodes have large values and the remaining nodes have nominal values.
For instance, as shown in Fig.~\ref{fig:dnpr-youtube}, the \dnpr values of over $99.98\%$ of nodes on the well-known \youtube graph are less than $0.001$, while the largest \dnpr is about $0.005$. Hence, by checking \dnpr and choosing a proper $\tau$, we can exclude the great majority of nodes in $\mathcal{S}$ for further refinement, thereby reducing the computational cost. 
It is worth mentioning that the \dnpr of each leaf node can be efficiently calculated in the preprocessing stage and used as the input to \taupush.

Notice that the supernode $\mathcal{S}$ considered above is a level-$1$ supernode with $k$ leaf nodes. As for any level-$(\ell+1)$ supernode $\mathcal{S}$ containing $O(k^{\ell+1})$ leaf nodes, we need to invoke \fwdpush $O(k^{\ell+1})$ times for the computation of all-pair level-$\ell$ \pprdegname values in $\mathcal{S}$, which is rather time-consuming when $\ell$ is large. To alleviate this issue, we design Group Forward-Push (\fwdpushsn), an improved and optimized \fwdpush equipped with a grouped push strategy: it starts push operations from all leaf nodes in the level-$\ell$ supernode simultaneously instead of separately. Analogously, we also propose Group Backward-Push (\bwdpushsn) for accelerating backward push operations in level-$\ell$ supernodes.

\stitle{Algorithm}
Algorithm~\ref{alg:taupush} shows the pseudo-code of \taupush, which consists of two main phases: \fwdpushsn and \bwdpushsn.
Specifically, given a supernode $\mathcal{S}$ and constant $k$, \taupush starts by setting \dnpr threshold $\tau$ as ${1}/{\sqrt{k\cdot n}}$ and forward residue threshold as
\begin{equation}\label{eq:rmax}
r_{max}=\frac{\epsilon\cdot \delta}{m\cdot\tau}
\end{equation}
at Lines 1-2. 
The setting of $\tau$ strikes a good balance between forward and backward phases, leading to an optimized worst-case time complexity (as analyzed later in Section \ref{sec:analysis}).
Next, for each supernode $\mathcal{V}_i\in \mathcal{S}$, \taupush invokes \fwdpushsn with residue threshold $r_{max}$ from $\mathcal{V}_i$ to derive $\widehat{\pi}_d(\mathcal{V}_i,\mathcal{V}_j)$, a rough estimation of \pprdegname for each supernode $\mathcal{V}_j\in\mathcal{S}$ (Lines 3-4).
Subsequently, \taupush calculates the residue threshold $r^b_{max}$ used in \bwdpushsn as per the following equation (Line 5):
\begin{equation}\label{eq:brmax}
r^b_{max}=\frac{\epsilon\cdot\delta}{\max_{\mathcal{V}_i\in \mathcal{S}\backslash\mathcal{V}_j}\frac{\sum_{v_s\in \leaf(\mathcal{V}_i)}{d(v_s)}}{|\leaf(\mathcal{V}_i)|}},
\end{equation}
where the denominator signifies the maximum average degree of leaf nodes in each supernode $\mathcal{V}_i$ of $\mathcal{S}$. \taupush identifies all supernodes $\mathcal{V}_j$ satisfying \dnpr $\tau_j > \tau$ and utilizes \bwdpushsn with residue threshold $r^b_{max}$ to refine the approximate \pprdegname value of every supernode pair $\mathcal{V}_i,\mathcal{V}_j$ in $\mathcal{S}$ (Lines 6-7). Eventually, for each supernode pair $\mathcal{V}_i,\mathcal{V}_j\in \mathcal{S}$, the level-$\ell$ approximate \pprdegname is converted to the desired level-$\ell$ approximate \pprdistname (Lines 8-9).

\begin{algorithm}[!t]
\KwIn{{Graph $G$, supernode $\mathcal{S}$ and constant $k$.}}
\KwOut{{Estimated $\widehat{\pprdist}[i,j],$\ $ \forall{\mathcal{V}_i,\mathcal{V}_j}\in \mathcal{S}$.}}
Set \dnpr threshold $\tau \gets {1}/{\sqrt{k\cdot n}}$\; 
Set forward residue threshold $r_{max}$ by Eq. \eqref{eq:rmax}\; 
\For{each supernode $\mathcal{V}_i \in \mathcal{S}$}{
\ $\forall{\mathcal{V}_j \in \mathcal{S}},\ \widehat{\pi}_d(\mathcal{V}_i,\mathcal{V}_j) \gets$\fwdpushsn$(G,\mathcal{S}, \mathcal{V}_i, r_{max})$\;    
}
Set backward residue threshold $r^b_{max}$ according to Eq. \eqref{eq:brmax}\;
\For{each supernode $\mathcal{V}_j \in \mathcal{S}$\ such that\ $\tau_j > \tau$}{
    \ $\forall{\mathcal{V}_i \in \mathcal{S}},\ \widehat{\pi}_d(\mathcal{V}_i,\mathcal{V}_j) \gets$\bwdpushsn$(G,\mathcal{S}, \mathcal{V}_j,r^b_{max})$\;  
}
\For{each supernode pair $\mathcal{V}_i,\mathcal{V}_j \in \mathcal{S}$}{
{Convert \pprdegname $\widehat{\pi}_d(\mathcal{V}_i,\mathcal{V}_j)$ to \pprdistname $\widehat{\pprdist}[i,j]$ by Eq.\eqref{eq:pprdist}\;}
}
\caption{\taupush}
\label{alg:taupush}
\end{algorithm}

\begin{algorithm}[!t]
\KwIn{Graph $G$, supernode $\mathcal{S}$, source $\mathcal{V}_i$, threshold $r_{max}$.}
\KwOut{Estimated \pprdegname $\widehat{\pi}_d(\mathcal{V}_i,\mathcal{V}_j),$\ $ \forall{\mathcal{V}_j}\in \mathcal{S}$.}
Initialize approximate \pprdegname $\widehat{\pi}_d(\mathcal{V}_i,\mathcal{V}_j)\gets 0$\ $\forall{\mathcal{V}_j}\in \mathcal{S}$\;
Initialize residues $r(\mathcal{V}_i,v_j) \gets \frac{d(v_j)}{|\leaf(\mathcal{V}_i)|}$\ $\forall{v_{j}}\in \leaf(\mathcal{V}_i)$\;
\While{$\exists v_k\in V \ \textrm{such that}\ r(\mathcal{V}_i,v_k)>d(v_k)\cdot r_{max}$}{
    \If{$v_k$ is in any supernode $\mathcal{V}_j\in \mathcal{S}$}{
    $\widehat{\pi}_d(\mathcal{V}_i,\mathcal{V}_j)\gets \widehat{\pi}_d(\mathcal{V}_i,\mathcal{V}_j) + \frac{\alpha\cdot r(\mathcal{V}_i,v_k)}{|\leaf(\mathcal{V}_j)|}$\;
    }
    \For{each out-neighbor ${v_{j}}$ of $v_k$}{
    \ $r(\mathcal{V}_i,v_j)\gets r(\mathcal{V}_i,v_j)+(1-\alpha)\cdot \frac{r(\mathcal{V}_i,v_k)}{d(v_k)}$\;
    }
    \ $r(\mathcal{V}_i,v_k)\gets 0$\;
}
\caption{\fwdpushsn}
\label{alg:fwdpushsn}
\end{algorithm}

\subsection{\fwdpushsn and \bwdpushsn}\label{sec:implement}

In the following, we elaborate on the algorithmic details of \fwdpushsn and \bwdpushsn used in \taupush.

\stitle{\fwdpushsnb}
Algorithm~\ref{alg:fwdpushsn} shows the pseudo-code of \fwdpushsn. 
Akin to \fwdpush, \fwdpushsn maintains two lists of variables in the course of pushes from source supernode $\mathcal{V}_i$: (i) the estimated \pprdegname $\widehat{\pi}_d(\mathcal{V}_i,\mathcal{V}_j)$, $\forall{\mathcal{V}_j}\in \mathcal{S}$, and (ii) the residue $r(\mathcal{V}_i,v_k)$, $\forall{v_k}\in V$. Initially, all variables are set to 0 except that $r(\mathcal{V}_i,v_j)$ is set to $\frac{d(v_j)}{|\leaf(\mathcal{V}_i)|}$ for each node $v_j$ in the leaf node set $\leaf(\mathcal{V}_i)$ of $\mathcal{V}_i$ (Lines 1-2). After that, \fwdpushsn starts an iterative process to traverse $G$ from nodes with non-zero residues, \ie nodes in $\leaf(\mathcal{V}_i)$. In particular, in each iteration, it inspects the residue of each node in $V$ to identify a node $v_k$ whose residue is greater than $d(v_k)\cdot r_{max}$. If such a node $v_k$ exists, \fwdpushsn first adds $\frac{\alpha\cdot r(\mathcal{V}_i,v_k)}{|\leaf(\mathcal{V}_j)|}$ to the approximate \pprdegname $\widehat{\pi}_d(\mathcal{V}_i,\mathcal{V}_j)$, where $\mathcal{V}_j$ is the supernode containing $v_k$ (Lines 4-5). Then, it evenly distributes the remaining $(1-\alpha)$ fraction of residue to the out-neighbors of $v_k$ (Lines 6-7). Afterwards, \fwdpushsn resets the residue $r(\mathcal{V}_i,v_k)$ to $0$ (Line 8) and proceeds to next iteration.
Lemma \ref{lem:fpsn-correct} indicates the correctness of Algorithm~\ref{alg:fwdpushsn}.
\begin{lemma}\label{lem:fpsn-correct} 
Given a source supernode $\mathcal{V}_i\in\mathcal{S}$ and a \dnpr threshold $\tau$,
by setting $\textstyle{r_{max}}$ as in Eq. \eqref{eq:rmax},
\fwdpushsn returns $(\epsilon,\delta)-$approximate level-$\ell$ \pprdegname $\widehat{\pi}_d(\mathcal{V}_i,\mathcal{V}_j)$ for $\mathcal{V}_j\in\mathcal{S}$ with \dnpr $\tau_j\leq \tau$.
\end{lemma}

\stitle{\bwdpushsnb}
\bwdpushsn can be regarded as the backward counterpart of \fwdpushsn. 
As shown in Algorithm \ref{alg:bwdpushsn}, \bwdpushsn initializes residue value $r(v_i,\mathcal{V}_j)$ of each leaf node $v_i$ in target supernode $\mathcal{V}_j$ to ${1}/{|\leaf(\mathcal{V}_j)|}$ (Line 2). Distinct from \fwdpush, \bwdpushsn conducts the graph traversal from the target supernode $\mathcal{V}_j$, following the incoming edges of each node. To be more precise, \bwdpushsn evenly pushes $(1-\alpha)$ fraction of residue $r(v_k,\mathcal{V}_j)$ to the in-neighbors of current node $v_k$ (Lines 6-7), rather than out-neighbors in \fwdpushsn. In addition, another two minor differences are: (i) the residue threshold $r^b_{max}$ is defined as Eq. \eqref{eq:brmax} (Line 3), and (ii) in each iteration, \bwdpushsn increases the estimated \pprdegname $\widehat{\pi}_d(\mathcal{V}_i,\mathcal{V}_j)$ by $\frac{ \alpha\cdot d(v_k)\cdot r(v_k,\mathcal{V}_j)}{|\leaf(\mathcal{V}_i)|}$, where $\mathcal{V}_i$ is the supernode consisting of current node $v_k$ (Lines 4-5). Lemma \ref{lem:bpsn-correct} proves the correctness of \bwdpushsn. 

\begin{lemma}\label{lem:bpsn-correct}
Given a target supernode $\mathcal{V}_j\in\mathcal{S}$ and a threshold $\textstyle{r^b_{max}}$ in Eq. \eqref{eq:brmax}, \bwdpushsn returns the $(\epsilon,\delta)$-approximate level-$\ell$ \pprdegname $\widehat{\pi}_d(\mathcal{V}_i,\mathcal{V}_j)$ for each source supernode $\mathcal{V}_i\in\mathcal{S}$ and $\mathcal{V}_i \neq \mathcal{V}_j$.
\end{lemma}

\subsection{Theoretical Results}\label{sec:analysis}

\begin{algorithm}[!t]
\KwIn{Graph $G$, supernode $\mathcal{S}$, target $\mathcal{V}_j$, threshold $r^b_{max}$.}
\KwOut{Estimated \pprdegname $\widehat{\pi}_d(\mathcal{V}_i,\mathcal{V}_j),$\ $ \forall{\mathcal{V}_i}\in \mathcal{S}$.}
\rc{
Initialize approximate \pprdegname $\widehat{\pi}(\mathcal{V}_i,\mathcal{V}_j) \gets 0 $\ $ \forall{\mathcal{V}_{i}}\in \mathcal{S}$\;
Initialize residues $r(v_i,\mathcal{V}_j) \gets {1}/{|\leaf(\mathcal{V}_j)|},$\ $\forall{v_{i}}\in \leaf(\mathcal{V}_j)$\;
    \While{$\exists v_k\in V \ \textrm{such that}\ r(v_k,\mathcal{V}_j)>r^b_{max}$}{
        \If{$v_k$ is in any supernode $\mathcal{V}_i\in \mathcal{S}$}{
            $\widehat{\pi}_d(\mathcal{V}_i,\mathcal{V}_j)\gets \widehat{\pi}_d(\mathcal{V}_i,\mathcal{V}_j) + \frac{ \alpha\cdot d(v_k)\cdot r(v_k,\mathcal{V}_j)}{|\leaf(\mathcal{V}_i)|}$\;
        }
        \For{each in-neighbor $v_i$ of $v_k$}{
        $r(v_i,\mathcal{V}_j) \gets r(v_i,\mathcal{V}_j) + (1-\alpha)\cdot \frac{r(v_k,\mathcal{V}_j)}{d(v_i)}$\;
        }
        $r(v_k,\mathcal{V}_j)\gets 0$\;
    }
}
\caption{\bwdpushsn}
\label{alg:bwdpushsn}
\end{algorithm}

\stitle{Correctness}
Combining Lemmata~\ref{lem:fpsn-correct} and~\ref{lem:bpsn-correct} leads to Theorem \ref{thm:taupush-correct}, which establishes the correctness of \taupush (Algorithm \ref{alg:taupush}).
\begin{theorem}\label{thm:taupush-correct}
For any user-selected supernode $\mathcal{S}$ and threshold $\tau$, by setting $r_{max}$ and $r^b_{max}$ as in Algorithm \ref{alg:taupush}, respectively, Algorithm~\ref{alg:taupush}
returns $(\epsilon,\delta)-$approximate level-$\ell$ \pprdegname $\widehat{\pi}_d(\mathcal{V}_i,\mathcal{V}_j)$ for $\mathcal{V}_i,\mathcal{V}_j\in\mathcal{S}$ and $\mathcal{V}_i \neq \mathcal{V}_j$.
\end{theorem}

\stitle{Time complexity}
The worst-case time complexity of \taupush is 
$\textstyle{O\left(\frac{k\cdot n\cdot m\cdot \tau}{\epsilon\cdot\delta} + \frac{m}{\tau\cdot\epsilon\cdot\delta}\right)}$.
By setting
$\textstyle\tau={1}/{\sqrt{k\cdot n}}$, the above complexity is minimized to $\textstyle O\left(\frac{k\cdot m}{\epsilon}\cdot \sqrt{k\cdot n}\right)$. By employing \fwdpushsn only, the worst-case complexity is $O\left(\frac{k^2\cdot n\cdot m}{\epsilon}\right)$, which is $\sqrt{k\cdot n}$ times slower than \taupush.
The expected complexity of \taupush for a randomly selected supernode $\mathcal{S}$ is $O\left(\textstyle\sum_{\mathcal{V}_i \in \mathcal{S}}\frac{d(\mathcal{V}_i)}{|\leaf(\mathcal{V}_i)|}\cdot\frac{k\cdot m}{\epsilon\cdot\delta\cdot n}\right)$.

\stitle{Indexing scheme}
We first store $O(n)$ \dnpr values as the index, which can be efficiently pre-computed in a similar way to global PageRank by setting the $k$-th entry in the initial global PageRank as $\frac{d(v_k)}{m}$.
Notice that \bwdpushsn is only conducted from target $\mathcal{V}_j$ with $\tau_j> \tau={1}/{\sqrt{k\cdot n}}$ and is independent of the query supernode $\mathcal{S}$. Hence, the index space is $O\left(k\cdot\sqrt{k\cdot n}\right)$ since \bwdpushsn estimates \pprdegname of $O\left(\sqrt{k\cdot n}\right)$ target nodes w.r.t.\ $O(k)$ source supernodes in $\mathcal{S}$. Overall, the index space of \taupush is $O\left(n+k\cdot\sqrt{k\cdot n}\right)$.

\stitle{{\taupushb vs. \forab}}
We compare \taupush with \fora in terms of time complexity and index space for a randomly selected supernode $\mathcal{S}$. 
As summarized in Table~\ref{tab:complex-compare}, \taupush improves the time complexity of \fora by $\frac{n\cdot\sqrt{k\cdot m}}{k^3\cdot\log{n}}$, where $k$ is the number of supernodes/nodes to be visualized and is usually small as discussed in Section~\ref{sec:prelim-multi}. For example, \taupush is about four orders of magnitude faster than \fora on the \youtube graph with 1 million nodes and 3 million edges.
Besides that, we observe that the indexing space of \fora is usually $3$--$5\times$ larger than that of \taupush in the experiments.

\begin{table}[!t]
\caption{Index space and time complexity of methods for approximate level-$\boldsymbol{\ell}$ \pprdistnameb computation.}
\label{tab:complex-compare}
\vspace{-2mm}
\renewcommand{\arraystretch}{1.1}
\begin{footnotesize}
\begin{tabular}{|c|c|c|}
\hline
                                & \fora & \taupush  \\ \hline
{Indexing space}                                                            & $O\left(\frac{\log{n}\cdot\sqrt{k\cdot m}}{\epsilon}\right)$          &    $O\left(n+k\cdot\sqrt{n\cdot k}\right)$          \\ \hline
{Time complexity}  & $O\left(\frac{m\cdot\sqrt{k\cdot m}}{\epsilon}\right)$   & ${O\left(\frac{k^3\cdot(\log{n})^2}{\epsilon}\right)}$                  \\ \hline
\end{tabular}
\end{footnotesize}
\vspace{-2mm}
\end{table}

\begin{figure*}[!t]
    \centering
    \includegraphics[width=\linewidth]{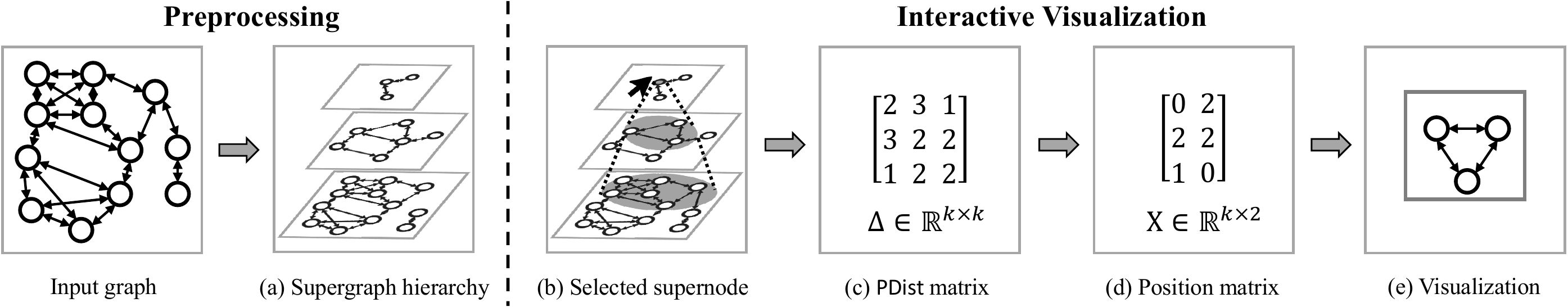}
	\vspace{-6mm}
    \caption{Procedure of \pprvizb.} \label{fig:pprviz}
    \vspace{-2mm}
\end{figure*}

\section{\pprvizb}\label{sec:pprviz}
This section presents \pprviz, our framework for static graph visualization.
Fig.~\ref{fig:pprviz} illustrates the procedure of \pprviz, which consists of two phases, \ie (i) preprocessing and (ii) interactive visualization.
In the preprocessing phase, it first constructs a supergraph hierarchy for the given graph $G$, then builds the index for the proposed \taupush. It is worth noting that the preprocessing phase is conducted only once for any queries on the given $G$.
In the interactive visualization phase, \pprviz contains two stages, namely \pprdistname matrix computation and position matrix embedding. As the index schema of \taupush has been explained in Section~\ref{sec:analysis}, we focus on illustrating the remaining stages as follows.
\begin{itemize} [leftmargin=*]
\item {\bf Supergraph hierarchy construction}: 
\pprviz first constructs a supergraph hierarchy of the input graph in the preprocessing step~\cite{abello2006ask,shi2009himap,archambault2008grouseflocks}.
Take Fig.~\ref{fig:pprviz}(a) as an example. The bottom layer is the input graph, whose non-overlapping node partitions are organized as level-1 supernodes in the middle layer. Similarly, the clustering procedure is repeated to generate the top layer. We employ the \louvain algorithm~\cite{blondel2008fast} to construct the supergraph hierarchy. Specifically, \louvain first treats all level-$\ell$ supernodes and their relationships as a new graph and assigns each supernode to a stand-alone partition. After that, it builds the partitions of level-$\ell$ supernodes (\ie supernodes at level-$(\ell+1)$) by iteratively moving each supernode to the neighbor's partition with maximum modularity improvement, where the moduarlity~\cite{newman2006modularity} measures the density of links inside partitions as compared to links between partitions.
Following the size requirement in Section~\ref{sec:prelim-multi}, we adapt \louvain under the constraint that each supernode $\mathcal{S}$ (resp.\ the coarsest supergraph) should have at most $k$ children (resp.\ supernodes) and call this adaption \louvainplus. Note that $k$ can be configured by users according to their needs.
\item {\bf \pprdistnameb matrix computation}: During interactive visualization, a user can select a level-$(\ell+1)$ supernode $\mathcal{S}$ (marked by an arrow in Fig.~\ref{fig:pprviz}(b)). To visualize the intra-structure of $\mathcal{S}$, \ie the shaded area in the middle layer, \pprviz leverages \taupush in Algorithm~\ref{alg:taupush} to calculate the \pprdistname matrix $\pprdist \in \mathbb{R}^{k \times k}$ for the $k$ children of $S$ (Fig.~\ref{fig:pprviz}(c)), which reflects the theoretical graph distance between the children by utilizing information of the related nodes in the input graph, \ie the shaded area in the bottom layer of Fig.~\ref{fig:pprviz}(b).
\item {\bf Position matrix embedding}: Given the \pprdistname matrix $\pprdist \in \mathbb{R}^{k \times k}$, \pprviz converts it into a position matrix $\XM \in \mathbb{R}^{k \times 2}$ (see Fig.~\ref{fig:pprviz}(d)), where $\XM[i]\in\mathbb{R}^2$ represents the two-dimensional coordinate of the $i$-th child node on the screen. In particular, the position matrix embedding step solves the following optimization problem~\cite{gansner2004graph}:
\begin{equation}\label{eq:loss}
\arg \min\limits_{\XM} \ L(\XM | \pprdist) = \sum\limits_{i<j}\left(1-\frac{||\XM[i]-\XM[j]||}{\pprdist[i,j]}\right)^2.
\end{equation}
Intuitively, it aims to ensure that the Euclidean distance $||\XM[i]-\XM[j]||$ derived from the position matrix $\XM$ is close to the \pprdistname $\pprdist[i,j]$. Towards this end, \pprviz employs the standard method for solving Eq.~\eqref{eq:loss}, namely, the {\it stress majorization} technique~\cite{gansner2004graph}. 
The time complexity of this method is $O(k^3)$~\cite{gansner2004graph}, which is insignificant as $k$ is usually small to avoid visual clutter~\cite{huang2009measuring}.
\end{itemize}

\stitle{Applications and future works}
The proposed framework \pprviz is not limited to multi-level visualization and is applicable for visualizing static homogeneous graphs in other scenarios. For example, \pprviz can visualize the entire graph or the subgraph returned by the graph query in a single-level fashion~\cite{bhowmick2020aurora}, where \pprviz skips the supergraph hierarchy construction stage and sets $k=n$ for the visualization phase.
Furthermore, \pprviz can be employed for incremental exploration~\cite{herman2000graph}, where users can move a focal area over the entire graph to explore the subgraph they are interested in. 
Notice that \pprviz can easily cope with the small dynamic graphs (including subgraphs, motifs, etc.) returned by database queries because our proposed \taupush algorithm is highly efficient and can be real-time responsive when a new visualization request is made.
Regarding the scenarios on large dynamic or attributed graphs, even though existing solutions~\cite{zhuang2019dynamo,wang2017revisiting} are applicable for the supergraph hierarchy construction and position embedding stages of \pprviz, the proposed \pprdistname or corresponding estimation algorithm \taupush fails to extend to these scenarios trivially.
In particular, under the dynamic setting, it is still unclear how to efficiently estimate the proposed \pprdistname with a rigorous theoretical accuracy guarantee. In addition, \pprviz can be extended to visualize attributed or knowledge graphs by treating node/edge attributes as additional nodes~\cite{xu2009semi}. However, such a method requires a new distance measure considering both topologies and attributes. These challenges motivate us to design new approaches in future works.

\section{Related Work}\label{sec:relatedwork}

\stitle{Graph visualization}
Conventional single-level graph visualization methods have been extensively studied~\cite{hu2015visualizing,gibson2013survey,von2011visual,herman2000graph} and can be classified into two main categories: (i) \emph{force-directed methods}, \eg \fr~\cite{fruchterman1991graph}, \linlog~\cite{noack2005energy}, \forceatlas~\cite{jacomy2014forceatlas2} and others~\cite{eades1984heuristic,martin2011openord,jacomy2014forceatlas2,chung2012finding}; and (ii) \emph{stress methods}, \eg \mds~\cite{gansner2004graph}, \pivotmds~\cite{brandes2006eigensolver} and others~\cite{torgerson1952multidimensional,kamada1989algorithm,gansner2012maxent,meyerhenke2017drawing}. In particular, force-directed methods model a graph as a force system, where adjacent nodes attract each other and all nodes repulse each other. The position matrix is derived by minimizing the composite forces in the entire system. Stress methods utilize the shortest distance as the node distance to guide node placement. They embed the shortest distance matrix into a position matrix by optimization techniques, \eg gradient descent~\cite{gansner2012maxent} and stress majorization~\cite{gansner2004graph}. To boost efficiency, many optimizations are incorporated, \eg grid-based partitioning~\cite{fruchterman1991graph,martin2011openord}, quad-tree~\cite{noack2005energy,jacomy2014forceatlas2}, and dimensionality reduction~\cite{brandes2006eigensolver}. Moreover, as surveyed by~\cite{goyal2018graph}, graph embedding methods, \eg \gf~\cite{ahmed2013distributed}, \sdne~\cite{wang2016structural}, \leemb~\cite{belkin2003laplacian}, \lle~\cite{roweis2000nonlinear} and \nodevec~\cite{grover2016node2vec}, can also be applied for visualization by treating the embedding matrix with dimension being 2 as the position matrix~\cite{perozzi2014deepwalk,tang2015line,goyal2018graph}. Multi-level methods~\cite{meyerhenke2017drawing,martin2011openord,rodrigues2015gmine,shi2009himap,abello2006ask,archambault2008grouseflocks,hu2005efficient} mainly focus on designing graph clustering algorithms, and the single-level methods are directly employed to layout each cluster. For example, \openord~\cite{martin2011openord} clusters nodes based on their Euclidean distances in the graph layout and uses \fr for visualization; \kadraw~\cite{meyerhenke2017drawing} applies label propagation for clustering and uses~\cite{gansner2012maxent} for visualization; GrouseFlocks~\cite{archambault2008grouseflocks} groups nodes based on certain graph structures. Compared with our \pprdistname distance matrix, the distance measures employed in existing methods fail to preserve the topological information comprehensively and hence dampen visualization quality. Specifically, force-directed methods only consider the direct links in the graph. Stress methods consider the shortest path from the source node to the target node but ignore other intricate paths. A related work~\cite{chung2012finding} uses the force-directed method for visualization after determining the edge weights by PPR; however, this method is still inherently a force-directed method, hence suffering from the corresponding limitations.

\stitle{PPR computation} 
The efficient computation of PPR has been extensively studied~\cite{fogaras2005towards,andersen2006local,wang2017fora,jung2017bepi,wang2016hubppr,lofgren2016personalized,lofgren2013personalized,shin2015bear,wu2021unifying,shi2019realtime,yang2020homogeneous,wang2020personalized,yoon2018tpa,hou2021massively,lin2020index,wang2019efficient}. Among these works, BEAR~\cite{shin2015bear} and BePI~\cite{jung2017bepi} improve the power iteration method~\cite{page1999pagerank} and achieve high efficiency by indexing several large matrices. However, the index space limits their applicability to large graphs. BiPPR~\cite{lofgren2016personalized} and HubPPR~\cite{wang2016hubppr} combine random walks~\cite{fogaras2005towards} with \bwdpush~\cite{lofgren2013personalized}, and are subsequently improved by \fora~\cite{wang2017fora}. However, \fora and its improved solutions~\cite{hou2021massively,wu2021unifying,wang2019efficient,lin2020index} suffer from numerous push operations and ineffective sample problems as illustrated in Section~\ref{sec:lpds}. This is because our level-$\ell$ \pprdistname computation aims at the aggregated PPRs between two clusters, which is essentially different from PPR computation for a single source node. Another line of work computes PPR using the idea of particle filtering~\cite{gallo2020personalized}, which performs a deterministic graph traversal from a set of source nodes. Nevertheless, it is non-trivial to determine the initial particle distribution for level-$\ell$ \pprdegname, and it does not offer quality guarantees. In contrast, \taupush can significantly outperform existing visualization and PPR solutions in efficiency with quality guarantees.

\section{Experiment Evaluation}\label{sec:exp}
We introduce the experiment settings in Section~\ref{sec:exp-setting}. 
Section~\ref{sec:exp-quality} evaluates the visualization quality of our \pprdistname and the competing methods. Sections~\ref{sec:exp-time} and~\ref{sec:exp-dnppr} compare the efficiency of \taupush against the visualization and PPR-based solutions.
Interested readers are referred to~\trref{} for additional results and analysis on visualization and \taupush's variants.

\subsection{Experiment Settings}\label{sec:exp-setting}

\stitle{Competitors and parameter settings}
We compare \pprviz with 13 representative graph visualization methods from different categories:
(i) 3 \textit{single-level force-directed methods}: \fr~\cite{fruchterman1991graph}, \linlog~\cite{noack2005energy}, \forceatlas~\cite{jacomy2014forceatlas2}; (ii) 2 \textit{single-level stress methods}: \mds~\cite{gansner2004graph}, \pivotmds~\cite{brandes2006eigensolver}; (iii) 5 \textit{single-level graph embedding methods}: \gf~\cite{ahmed2013distributed}, \sdne~\cite{wang2016structural}, \leemb~\cite{belkin2003laplacian}, \lle~\cite{roweis2000nonlinear}, \nodevec~\cite{grover2016node2vec}; (iv) {\textit{an SimRank-based adaptation} mentioned in Section~\ref{sec:pdsdef}}; (v) 2 \textit{multi-level methods}: \openord~\cite{martin2011openord} and \kadraw~\cite{meyerhenke2017drawing}.
We follow the parameter settings of all competitors as recommended in their respective papers. 
For \pprviz, we set the maximum number of nodes in a cluster to 25 (\ie $k = 25$) as suggested in \cite{huang2009measuring}. For a fair comparison, we follow~\cite{archambault2008grouseflocks,rodrigues2015gmine} and modify \openord and \kadraw such that only the partial view of the clusters $\mathcal{S}$ in the zoom-in path is visualized. Since \openord does not allow cluster size constraint and \kadraw employs a complicated method to determine cluster size, we set the maximum number of supernodes in the coarsest supergraph (instead of all levels) to $k$.
We observe that \pprviz usually shows more nodes than \openord and \kadraw in a visualization, and hence the efficiency of \pprviz is not caused by processing fewer nodes. 
Additionally, we compare our proposed \taupush against 4 PPR approximation solutions \poweriter~\cite{page1999pagerank}, \fora~\cite{wang2017fora}, \foraplus~\cite{wang2019efficient} and \resacc~\cite{lin2020index}.
For all methods, {we set $\epsilon=1-1/e$ and $\delta=1/(10k)$ by default.} 
\rc{For \fora-based competitors \fora, \foraplus, and \resacc, we set the initial residue $r(v_i,v_i)=d(v_i)$ for the source each $v_i$ and follow the settings of other parameters as suggested in the original papers, where the correctness of \pprdistname estimation is still satisfied.} The experiment source code and the implementation of \pprviz are available at \url{https://github.com/jeremyzhangsq/PPRviz-reproducibility/}.

\begin{table}[!t]
\centering
\caption{Dataset statistics (${K\!=\!10^3, M\!=\!10^6, B\!=\!10^9}$)}\label{tab:dataset}
\vspace{-2mm}
\renewcommand{\arraystretch}{1.1}
\begin{footnotesize}
\begin{tabular}{|c|c|c|c|}
\hline
 Dataset   & $n$   & $m$    & Description                                       \\ \hline
 \twego      & 23 & 52  & Ego network\cite{snapnets}  \\ \hline
 \fbego     & 52 & 146 & Ego network\cite{snapnets}                \\ \hline
 \wiki     & 186 & 632 & Authorship network\cite{konect}                \\ \hline
 \physic     & 241 & 1.8K & Social network\cite{konect}                \\ \hline
 \filmtrust    & 874  & 2.6K   & User trust network\cite{konect}                      \\ \hline
 \scinet      & 1.5K & 5.4K  & Collaboration network\cite{konect}       \\ \hline
 \amazon     & 334.9K & 1.9M & Product network \cite{snapnets}           \\ \hline
 \youtube      & 1.1M & 6.0M & Social network \cite{snapnets}           \\ \hline
 \orkut      & 3.1M & 234.4M & Social network \cite{snapnets}           \\ \hline
 \dblp        & 5.4M & 17.2M & Collaboration network \cite{konect}           \\ \hline
 \itzerofour     & 41.3M & 2.3B & Crawled network \cite{BoVWFI}           \\ \hline
 \twitter     & 41.7M & 3.0B & Social network \cite{kwak2010twitter}           \\ \hline
\end{tabular}
\end{footnotesize}
\vspace{-2mm}
\end{table}

\stitle{Datasets and performance metrics}
We use the 12 real-world graphs in Table~\ref{tab:dataset} for the experiments.
We generate visualizations in a single-level fashion on 6 smaller graphs, and use \nd{} and \ulcv{} to evaluate the visualization quality of \pprviz and the competitors.
For a fair comparison, we follow NetworkX~\cite{hagberg2008exploring} and normalize each layout to the same scale.
The 6 larger graphs are used to evaluate visualization efficiency, on which we report the \textit{response time} and \textit{total preprocessing time}. For the single-level methods, the response time is the time to visualize the entire graph. For \pprviz and the multi-level methods, the response time is the average visualization time for the children of each supernode over 100 random zoom-in paths. Each path starts with the supergraph on the highest level (corresponds to the entire graph) and randomly selects a supernode in each level until reaching level-$0$ (\ie the original graph) to simulate interactive exploration. The preprocessing time is the time taken before visualization. 
We terminate a method if its response time (resp.\ preprocessing time) exceeds 1000 seconds (resp.\ 12 hours). 
All experiments are conducted on a Linux machine with Intel Xeon(R) Gold 6240@2.60GHz CPU and 80GB RAM in single-thread mode. Note that the memory utilized by \pprviz is comparable to that used by storing the input graph.

\begin{table*}[!t]
\vspace{-2mm}
\centering
\caption{\nd{} of \pprvizb and the baselines, the best in bold and the second best in italic, $\boldsymbol\infty$ indicates infinity.}
\label{tab:metrics-nd}
\vspace{-3mm}
\renewcommand{\arraystretch}{1.1}
\begin{footnotesize}
\resizebox{1\linewidth}{!}{%
\begin{tabular}{|c|c|c|c|c|c|c|c|c|c|c|c|c|}
\hline
           & \pprviz          & \openord/\fr     & \linlog & \forceatlas & \mds     & \pivotmds & \gf     & \sdne    & \leemb   & \lle     & \nodevec & \simrank         \\ \hline
\twego     & \textit{2.1E+02} & \textbf{1.2E+02} & 1.1E+03 & 1.8E+03     & 1.2E+03  & $\infty$  & 3.1E+08 & $\infty$ & $\infty$ & 4.6E+02  & 1.1E+04  & 5.2E+02          \\
\fbego     & \textit{2.4E+03} & \textbf{1.1E+03} & 9.5E+03 & 1.3E+04     & 2.0E+04  & $\infty$  & 3.6E+12 & $\infty$ & $\infty$ & 3.9E+07  & 1.2E+05  & 6.2E+03          \\
\wiki      & \textbf{2.7E+04} & \textit{2.7E+04} & 1.4E+05 & 8.1E+04     & 4.9E+04  & $\infty$  & 9.2E+11 & $\infty$ & $\infty$ & 7.5E+29  & 2.5E+06  & 2.7E+04          \\
\physic    & \textbf{6.7E+04} & \textit{8.7E+04} & 7.6E+05 & 8.2E+05     & 1.5E+05  & $\infty$  & 2.5E+10 & $\infty$ & $\infty$ & 4.0E+09  & 9.4E+07  & 1.1E+05          \\
\filmtrust & \textbf{9.1E+05} & 7.1E+06          & 3.2E+08 & 1.4E+07     & $\infty$ & $\infty$  & 1.2E+17 & $\infty$ & $\infty$ & 1.4E+10  & 9.6E+07  & \textit{2.9E+06} \\
\scinet    & \textbf{2.0E+06} & 6.5E+12          & 2.3E+09 & 1.9E+08     & 9.9E+12  & $\infty$  & 1.1E+17 & $\infty$ & $\infty$ & $\infty$ & 6.6E+07  & \textit{2.2E+06} \\ \hline
\end{tabular}
}
\end{footnotesize}
\vspace{0mm}
\end{table*}

\begin{table*}[!t]
\centering
\caption{\ulcv{} of \pprvizb and the baselines, the best in bold and the second best in italic, ``-'' indicates undefined.}
\label{tab:metrics-ulcv}
\vspace{-3mm}
\renewcommand{\arraystretch}{1.1}
\begin{footnotesize}
\resizebox{1\linewidth}{!}{%
\begin{tabular}{|c|c|c|c|c|c|c|c|c|c|c|c|c|}
\hline
           & \pprviz       & \openord/\fr  & \linlog & \forceatlas & \mds & \pivotmds     & \gf  & \sdne & \leemb & \lle & \nodevec & \simrank \\ \hline
\twego     & \textbf{0.22} & 0.35          & 0.57    & 0.37        & 0.40 & \textit{0.23} & 0.45 & 1.96  & 1.15   & 0.46 & 0.80     & 0.84     \\
\fbego     & \textbf{0.39} & \textit{0.42} & 0.67    & 0.49        & 0.46 & 0.45          & 0.91 & 0.94  & 0.98   & 0.77 & 0.96     & 0.75     \\
\wiki      & \textbf{0.35} & \textit{0.41} & 1.09    & 0.64        & 0.62 & 0.78          & 0.62 & 0.94  & 1.04   & 1.27 & 0.86     & 0.53     \\
\physic    & \textbf{0.45} & 0.53          & 0.90    & 0.55        & 0.80 & \textit{0.47} & 0.95 & 1.67  & 1.02   & 0.77 & 1.41     & 0.53     \\
\filmtrust & \textbf{0.48} & \textit{0.54} & 1.99    & 0.96        & 1.05 & 0.69          & 0.64 & 1.31  & 1.70   & 0.87 & 0.89     & 1.78     \\
\scinet    & \textbf{0.34} & 0.77          & 4.70    & 1.52        & 1.74 & \textit{0.74} & 0.86 & 1.72  & 1.26   & -    & 1.32     & 1.98     \\ \hline
\end{tabular}
}
\end{footnotesize}
\vspace{0mm}
\end{table*}

\subsection{Visualization Quality}\label{sec:exp-quality}
We evaluate the visualization quality of \pprviz and the competitors quantitatively via two popular aesthetic metrics, and qualitatively via a user study and a case study on 6 smaller graphs.

\subsubsection{\bf Aesthetic Metrics}
\label{sec:exp-metrics}
We report the \nd{} and \ulcv{} scores of all approaches in Table~\ref{tab:metrics-nd} and Table~\ref{tab:metrics-ulcv}, respectively.
Since \openord applies \fr to visualize each supergraph, we treat them as one method and report their results in one column. The results of \kadraw are omitted as it fails to process graphs with multiple connected components. 
We use ``-'' and $\infty$ to indicate undefined and infinity scores, respectively. 
Table~\ref{tab:metrics-nd} shows that \pprviz consistently outperforms the competitors in \nd{} on all graphs except \twego and \fbego, where \pprviz has comparable performance to the best method \fr. Note that \fr achieves the smallest \nd{} scores on \twego and \fbego because it places nodes of the largest connected component far apart from each other, which causes the edge distortion issue. In particular, the \nd{} scores of \forceatlas, \fr and \linlog are 5, 2, and 3 orders of magnitude larger than \pprviz on \scinet, respectively. 
Regarding stress methods, we find that \mds and \pivotmds yield infinite \nd{} scores, indicating severe node overlapping issues.
This is because they use the shortest distance between two nodes as the node distance, which only takes a few discrete values and thus fails to distinguish from different node pairs.
Furthermore, \pivotmds computes the position of a non-pivot node as the weighted combination of its connected pivot nodes. Thus, degree-one non-pivot nodes connected to the same pivot will share the same position. The graph embedding methods, especially \sdne and \leemb, have worse \nd{} scores than \pprviz and other competitors, as embedding-based methods are specially designed for machine learning tasks without considering visualization quality.
From Table~\ref{tab:metrics-ulcv}, we can see that \pprviz always performs the best in \ulcv{}. For instance, on \scinet, \pprviz is 14$\times$ better than \linlog. The superior performance of \pprviz is attributed to our carefully designed \pprdistname.

\begin{figure}[!t]
\centering
\begin{small}
\begin{tikzpicture}
    \begin{customlegend}[legend columns=5,
        legend entries={\pprviz,\mds,\fr,\nodevec,\simrank},
        legend columns=-1,
        area legend,
        legend style={at={(0.45,1.15)},anchor=north,draw=none,font=\scriptsize,column sep=0.15cm}]
        \addlegendimage{color=black,fill=red} 
        \addlegendimage{color=black,fill=pink} 
        \addlegendimage{color=black,fill=orange} 
        \addlegendimage{color=black,pattern color=violet,pattern=dots} 
        \addlegendimage{color=black,pattern color=teal,pattern=north west lines} 
    \end{customlegend}
\end{tikzpicture}
\\[-\lineskip]
\vspace{-1mm}
\hspace{-2mm}
\subfloat[T1]{
\begin{tikzpicture}[scale=1]
\begin{axis}[
    height=\columnwidth/2.4,
    width=\columnwidth/1.6,
    ybar=1.0pt,
    bar width=0.08cm,
    enlarge x limits=true,
    ylabel={\em frequency},
    xlabel= {\em rank},
    xmin=0.8, xmax=5.2,
    xtick={1,2,3,4,5},
    xticklabel style = {font=\footnotesize},
    ymin=1,
    ymax=150,
    ytick={1,10,100},
    yticklabels={$10^{0}$,$10^1$,$10^2$},
    ymode=log,
    yticklabel style = {font=\footnotesize},
    log basis y={10},
    every axis y label/.style={at={(current axis.north west)},right=5mm,above=0mm},
    legend style={at={(0.02,0.98)},anchor=north west,cells={anchor=west},font=\tiny}
    ]
\addplot [color=black,fill=red] coordinates {
(1,	87)
(2,	50)
(3,	34)
(4,	3)
(5,	3)
}; 
\addplot [color=black,fill=pink] coordinates {
(1,	26)
(2,	40)
(3,	57)
(4,	36)
(5,	17)
}; 
\addplot [color=black,fill=orange] coordinates {
(1,	44)
(2,	41)
(3,	39)
(4,	42)
(5,	9)
}; 
\addplot [color=black,pattern color=violet,pattern=dots] coordinates {
(1,	5)
(2,	2)
(3,	8)
(4,	26)
(5,	134)
};
\addplot [color=black,pattern color=teal,pattern=north west lines] coordinates {
(1,	18)
(2,	42)
(3,	36)
(4,	67)
(5,	11)
};
\end{axis}
\end{tikzpicture}\vspace{2mm}\hspace{-1mm}%
}
\subfloat[T2]{
\begin{tikzpicture}[scale=1]
\begin{axis}[
    height=\columnwidth/2.4,
    width=\columnwidth/1.6,
    ybar=1.0pt,
    bar width=0.08cm,
    enlarge x limits=true,
    ylabel={\em frequency},
    xlabel= {\em rank},
    xmin=0.8, xmax=5.2,
    xtick={1,2,3,4,5},
    xticklabel style = {font=\footnotesize},
    ymin=1,
    ymax=150,
    ytick={1,10,100},
    yticklabels={$10^{0}$,$10^1$,$10^2$},
    ymode=log,
    yticklabel style = {font=\footnotesize},
    log basis y={10},
    every axis y label/.style={at={(current axis.north west)},right=5mm,above=0mm},
    legend style={at={(0.02,0.98)},anchor=north west,cells={anchor=west},font=\tiny}
    ]
\addplot [color=black,fill=red] coordinates {
(1,	98)
(2,	45)
(3,	24)
(4,	8)
(5,	5) 
};
\addplot [color=black,fill=pink] coordinates {
(1,	12)
(2,	34)
(3,	42)
(4,	48)
(5,	43)
}; 
\addplot [color=black,fill=orange] coordinates {
(1,	28)
(2,	28)
(3,	73)
(4,	33)
(5,	17)
}; 
\addplot [color=black,pattern color=violet,pattern=dots] coordinates {
(1,	14)
(2,	7)
(3,	16)
(4,	42)
(5,	100)
};
\addplot [color=black,pattern color=teal,pattern=north west lines] coordinates {
(1,	28)
(2,	65)
(3,	24)
(4,	48)
(5,	14)
};
\end{axis}
\end{tikzpicture}\vspace{2mm}\hspace{-1mm}%
}
\vspace{-1mm}
\end{small}
\vspace{-3mm}
\caption{\rc{Results of T1 and T2, frequency of selected ranking.}} \label{fig:user-study-t12}
\vspace{-2mm}
\end{figure}
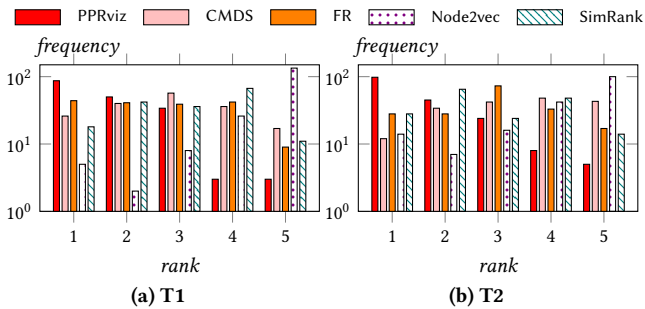

\begin{table}[!t]
\centering
\caption{Results of T3, frequency of being selected.}
\vspace{-2mm}
\label{tab:user-study-t3}
\renewcommand{\arraystretch}{1.1}
\begin{small}
\begin{tabular}{|c|c|c|c|}
\hline
    & {\taupush}        & {\poweriter}   & No difference    \\ \hline
Frequency   & 54             & 43           & \textbf{83}               \\ \hline
\end{tabular}
\end{small}
\vspace{-4mm}
\end{table}


\begin{figure*}[!t]
\vspace{-2mm}
\begin{small}
  \begin{tabular}{cccccc}
    \includegraphics[width=0.14\textwidth]{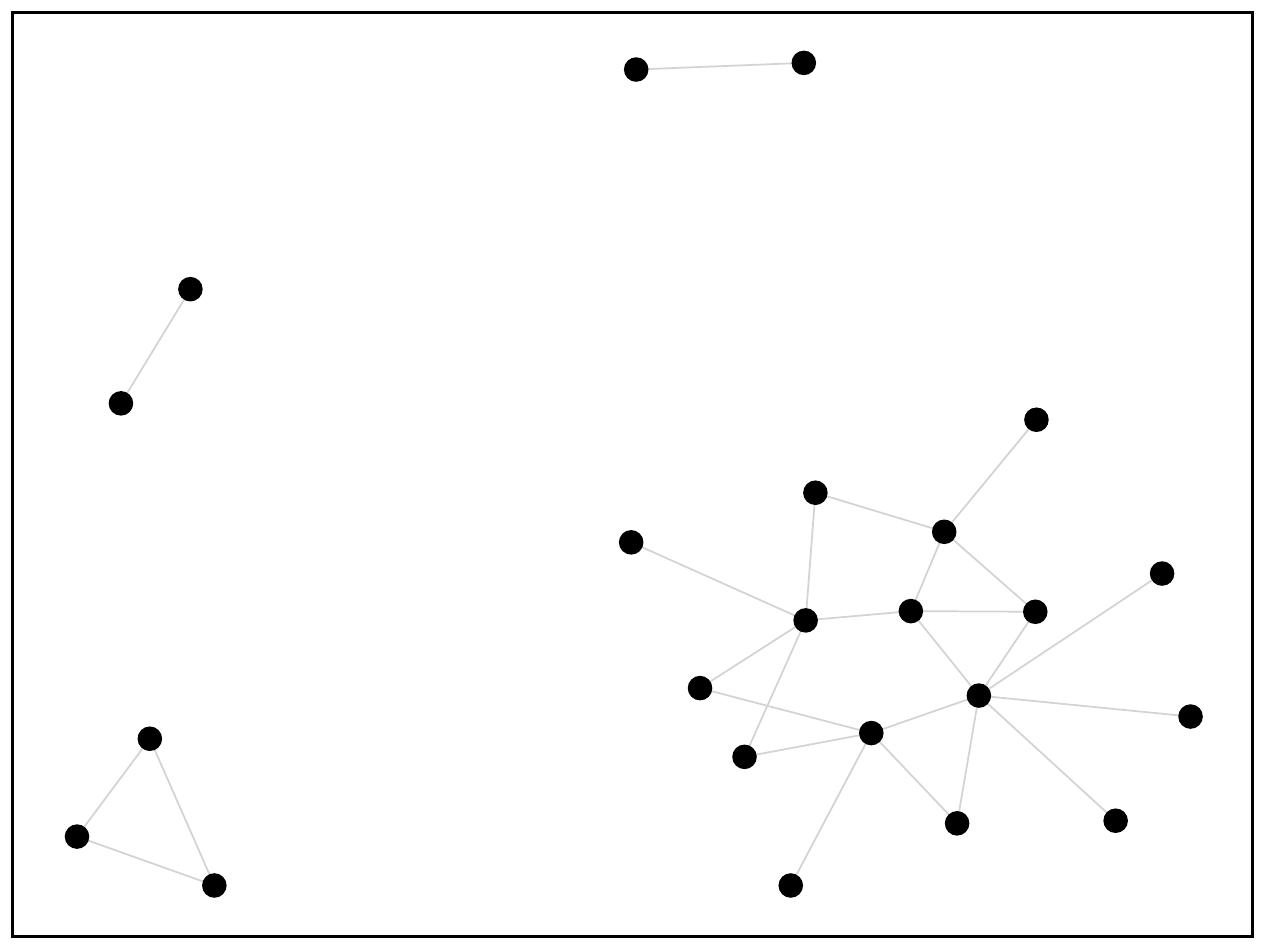}
    &
    \includegraphics[width=0.14\textwidth]{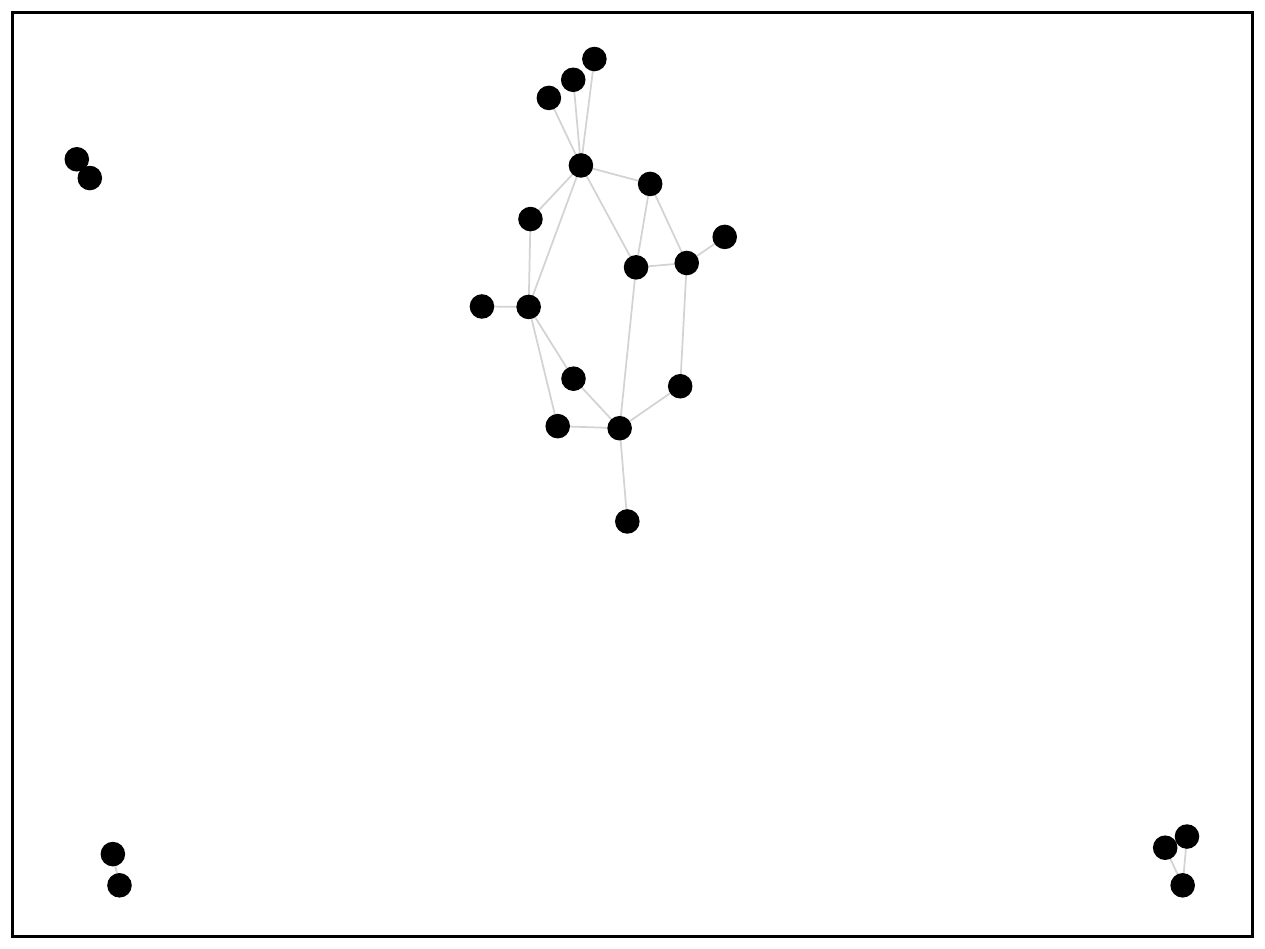}
    &
    \includegraphics[width=0.14\textwidth]{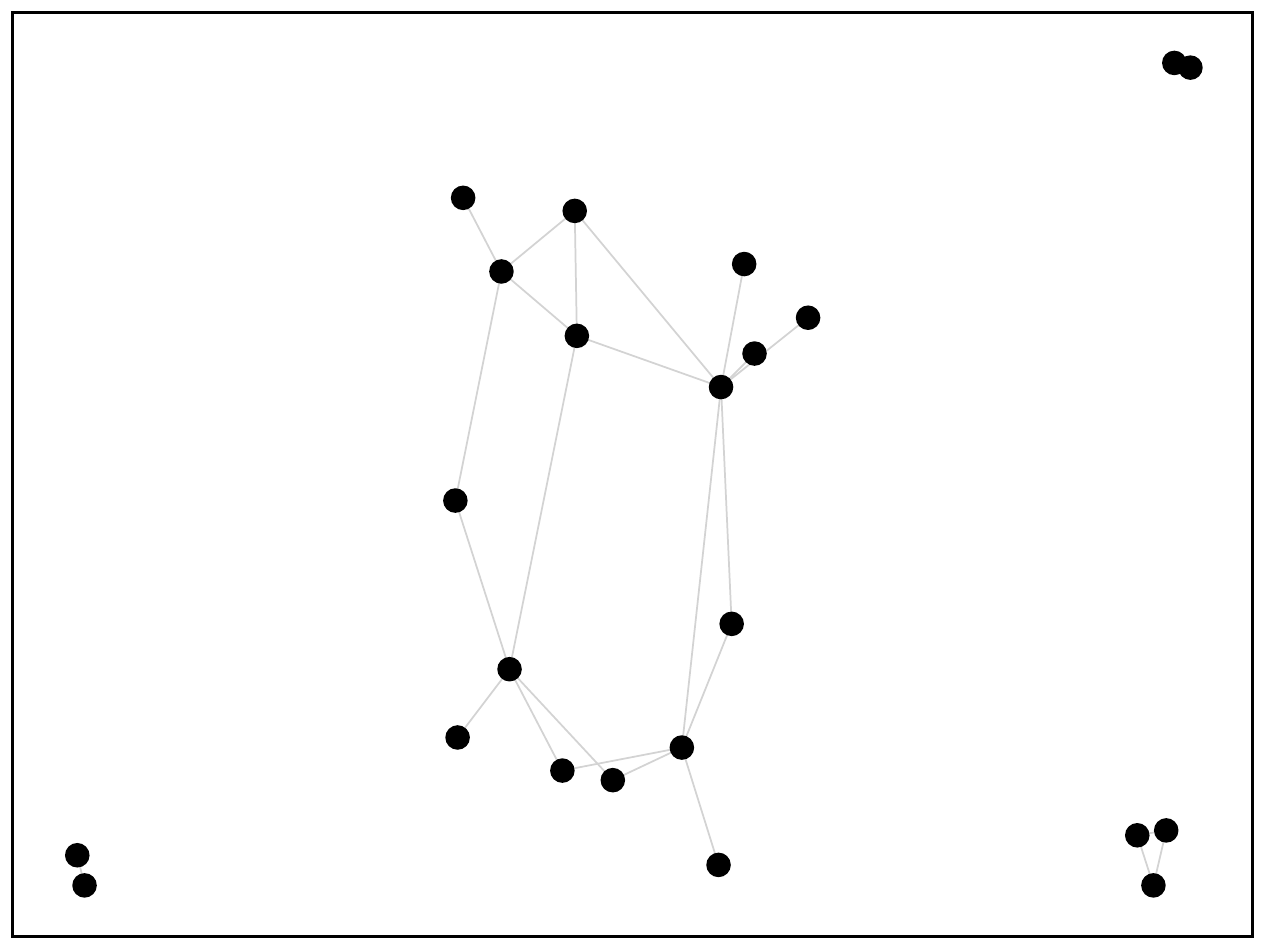}
    &
    \includegraphics[width=0.14\textwidth]{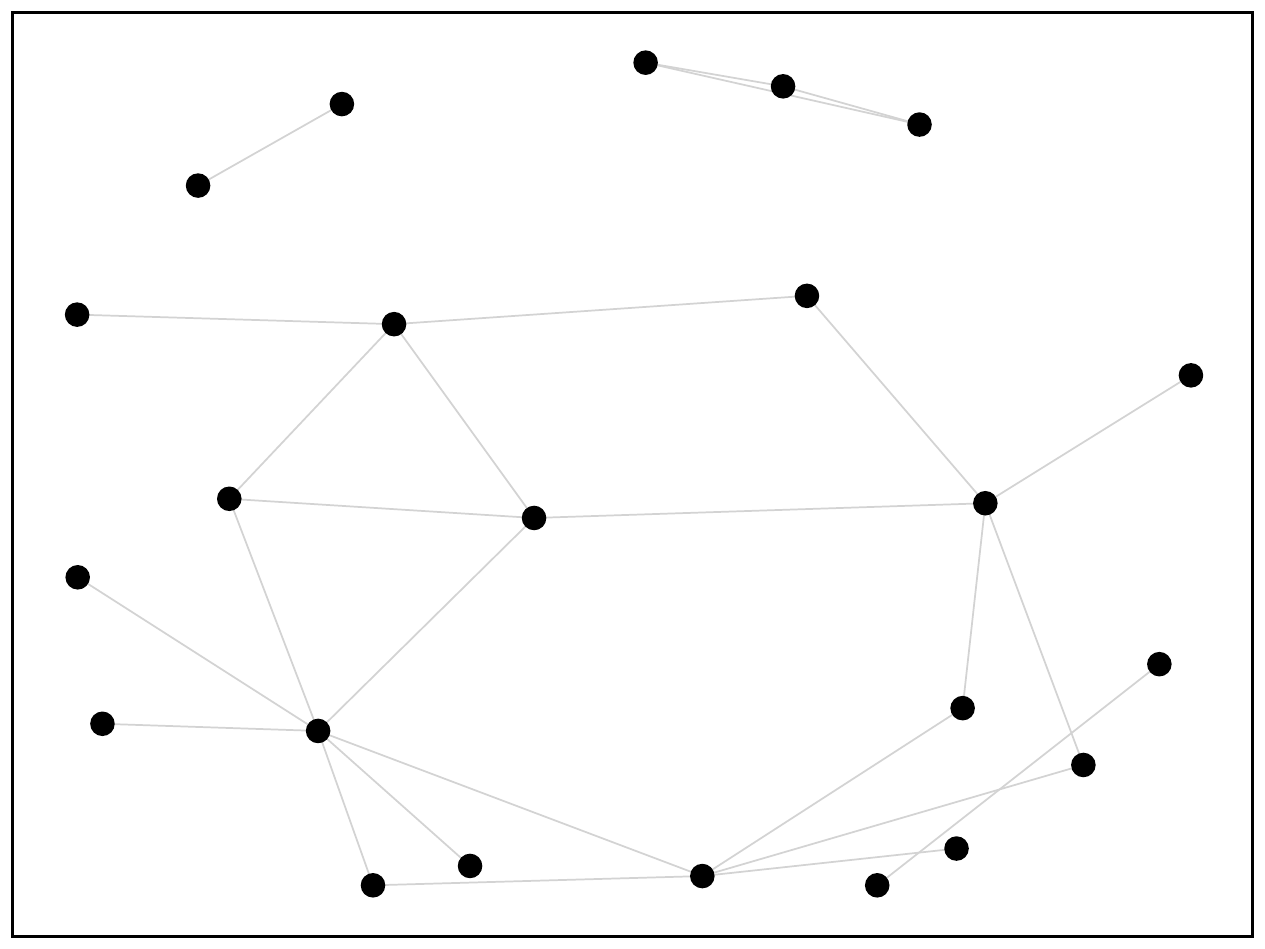}
    &
    \includegraphics[width=0.14\textwidth]{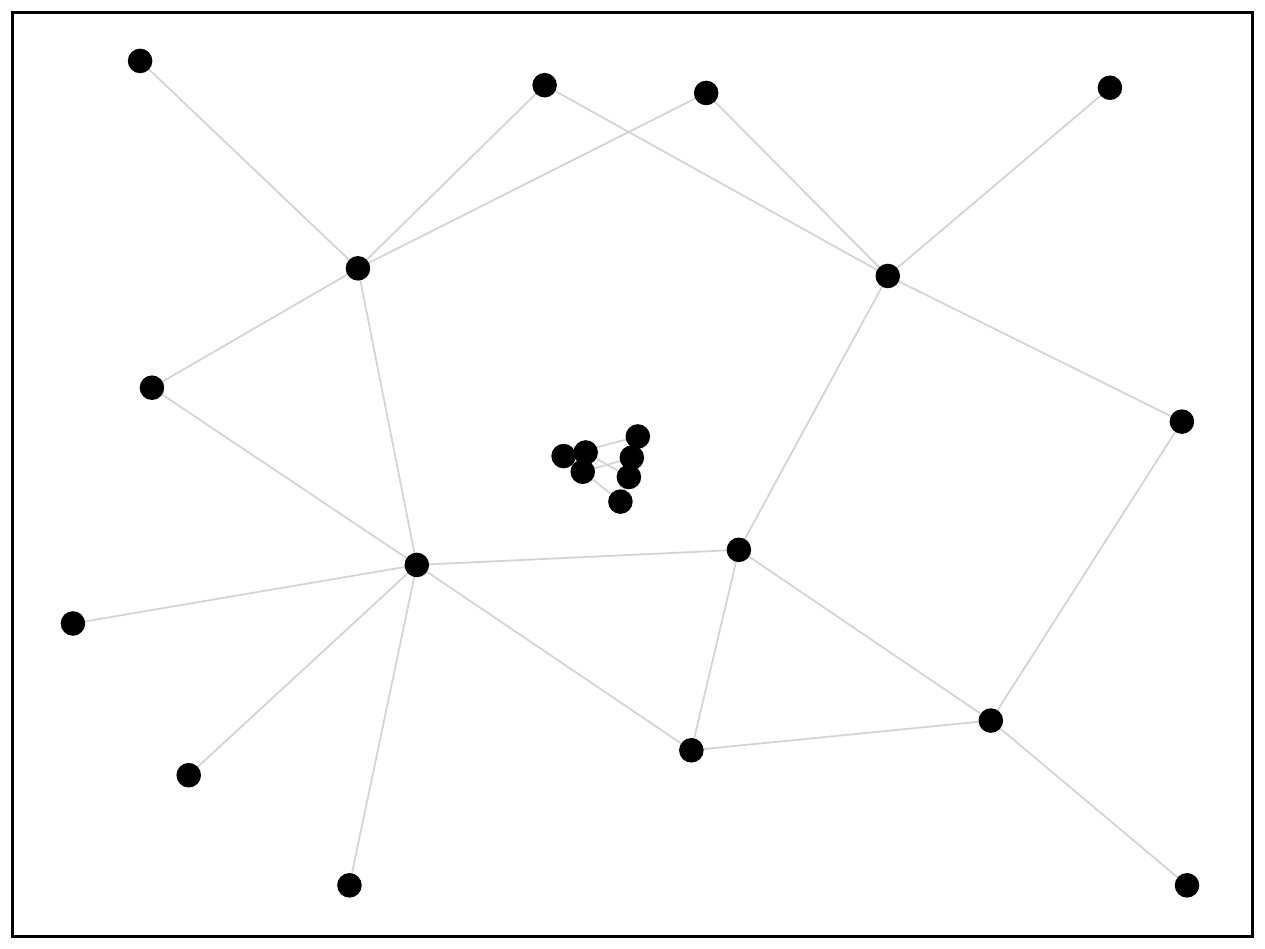}
    &
    \includegraphics[width=0.14\textwidth]{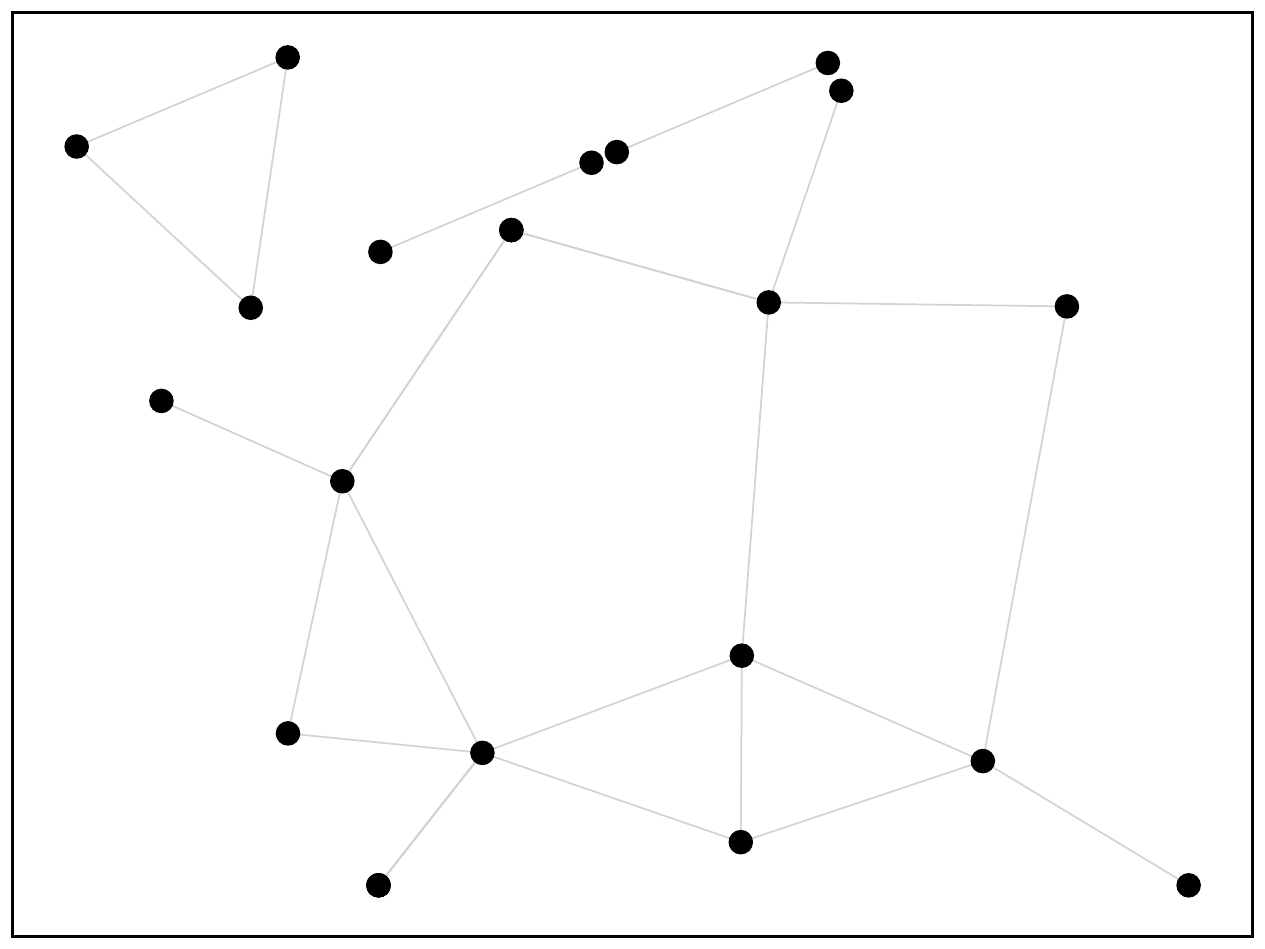}
    \\
    (a) \pprviz (Ours)  & (b) $\blacklozenge$~\forceatlas~\cite{jacomy2014forceatlas2} & (c) $\blacklozenge$~\linlog~\cite{noack2005energy} & (d) $\blacklozenge$~\fr~\cite{fruchterman1991graph} & (e) $\blacktriangle$~ \mds~\cite{gansner2004graph} & (f) $\blacktriangle$~ \pivotmds~\cite{brandes2006eigensolver} \vspace{2mm} \\
    \includegraphics[width=0.14\textwidth]{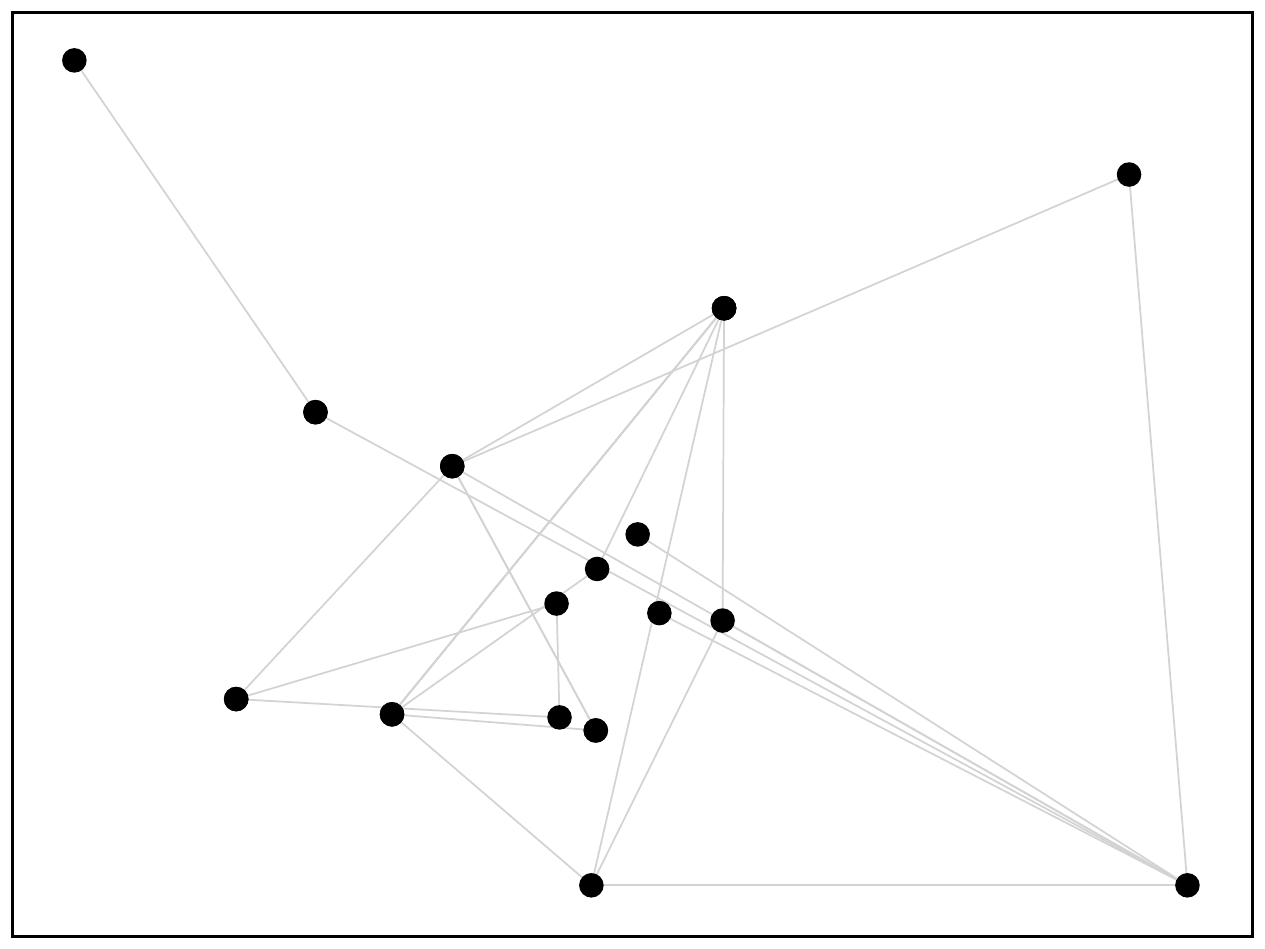}
    &
    \includegraphics[width=0.14\textwidth]{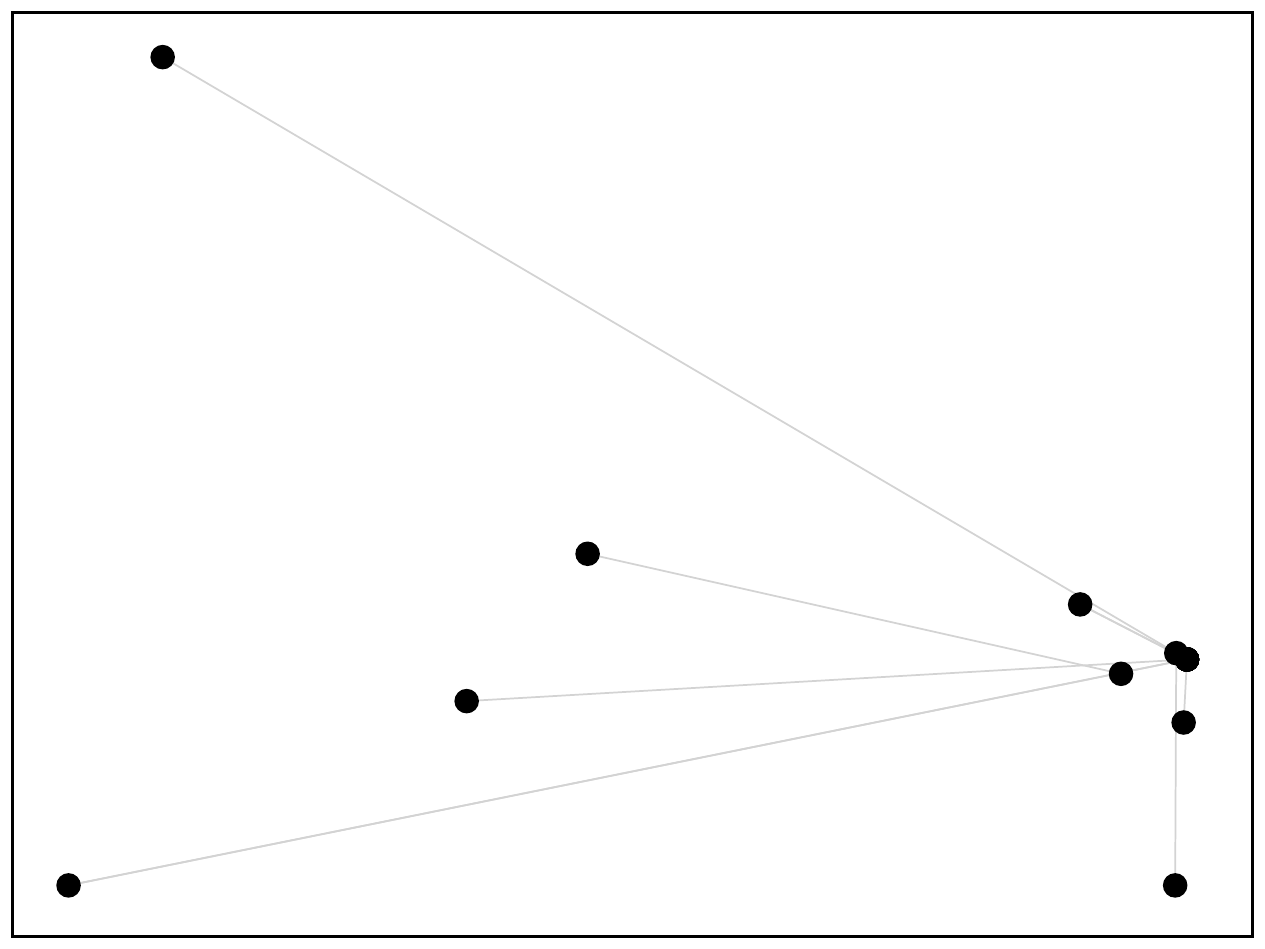}
    &
    \includegraphics[width=0.14\textwidth]{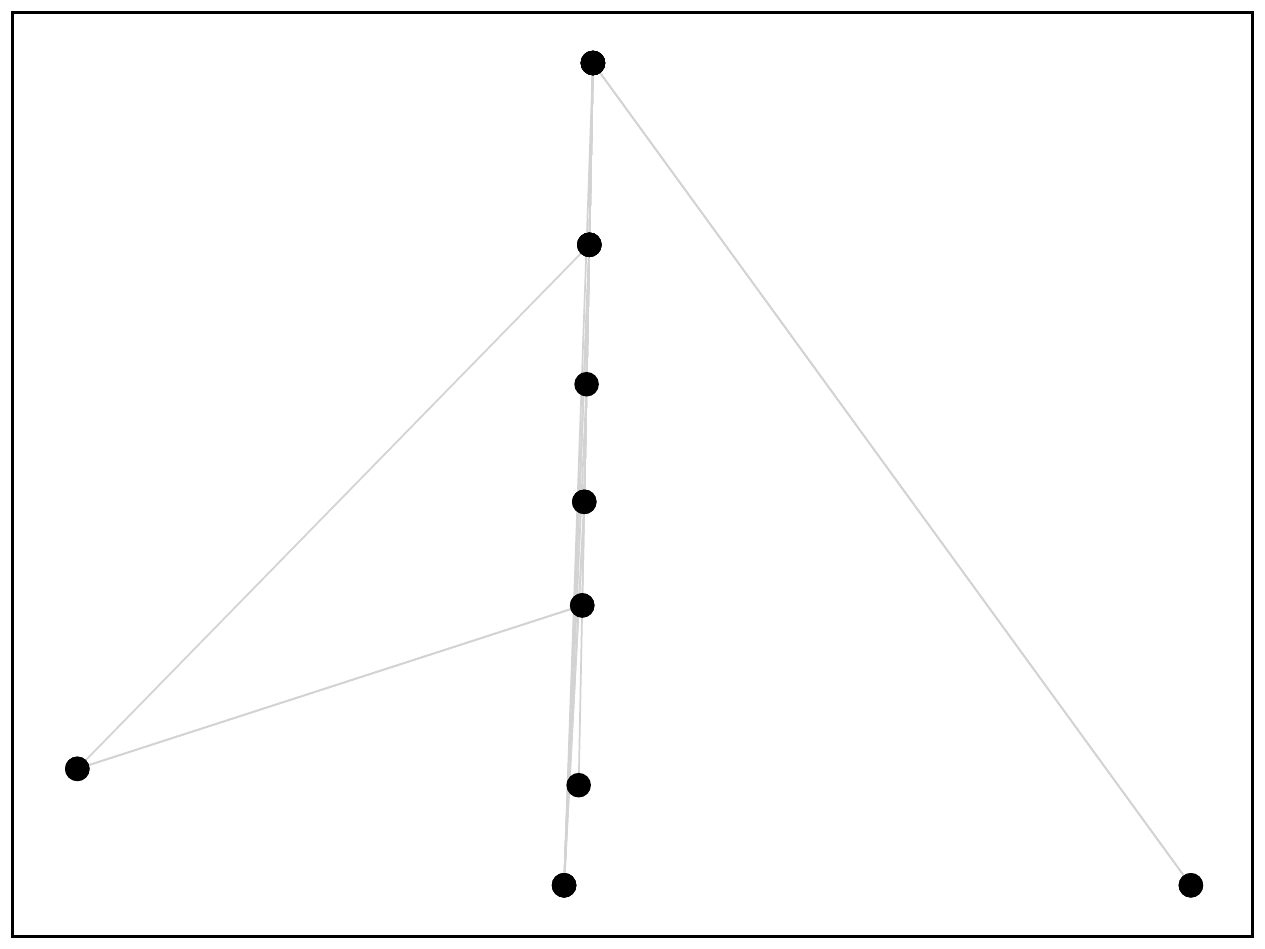}
    &
    \includegraphics[width=0.14\textwidth]{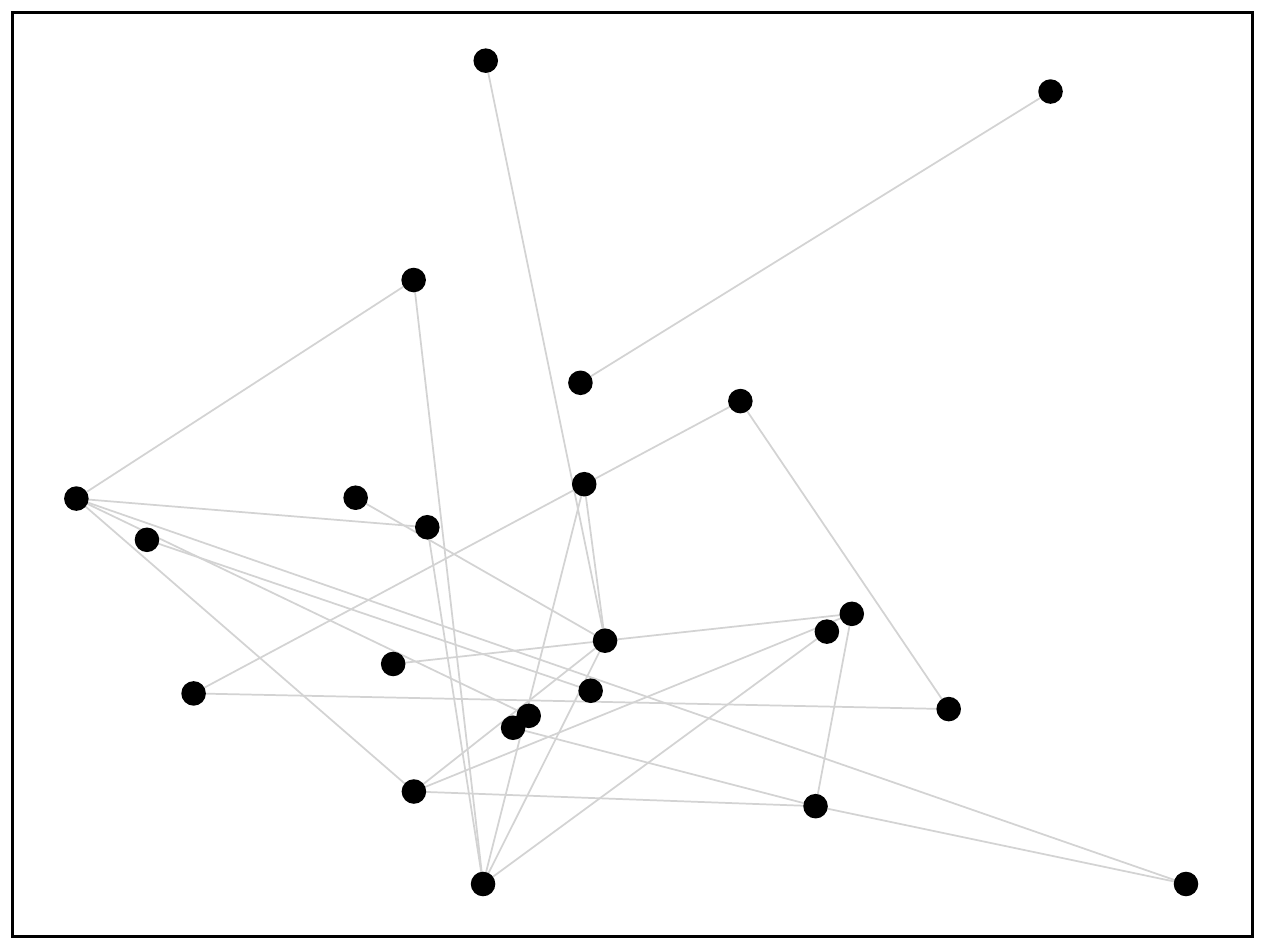}
    &
    \includegraphics[width=0.14\textwidth]{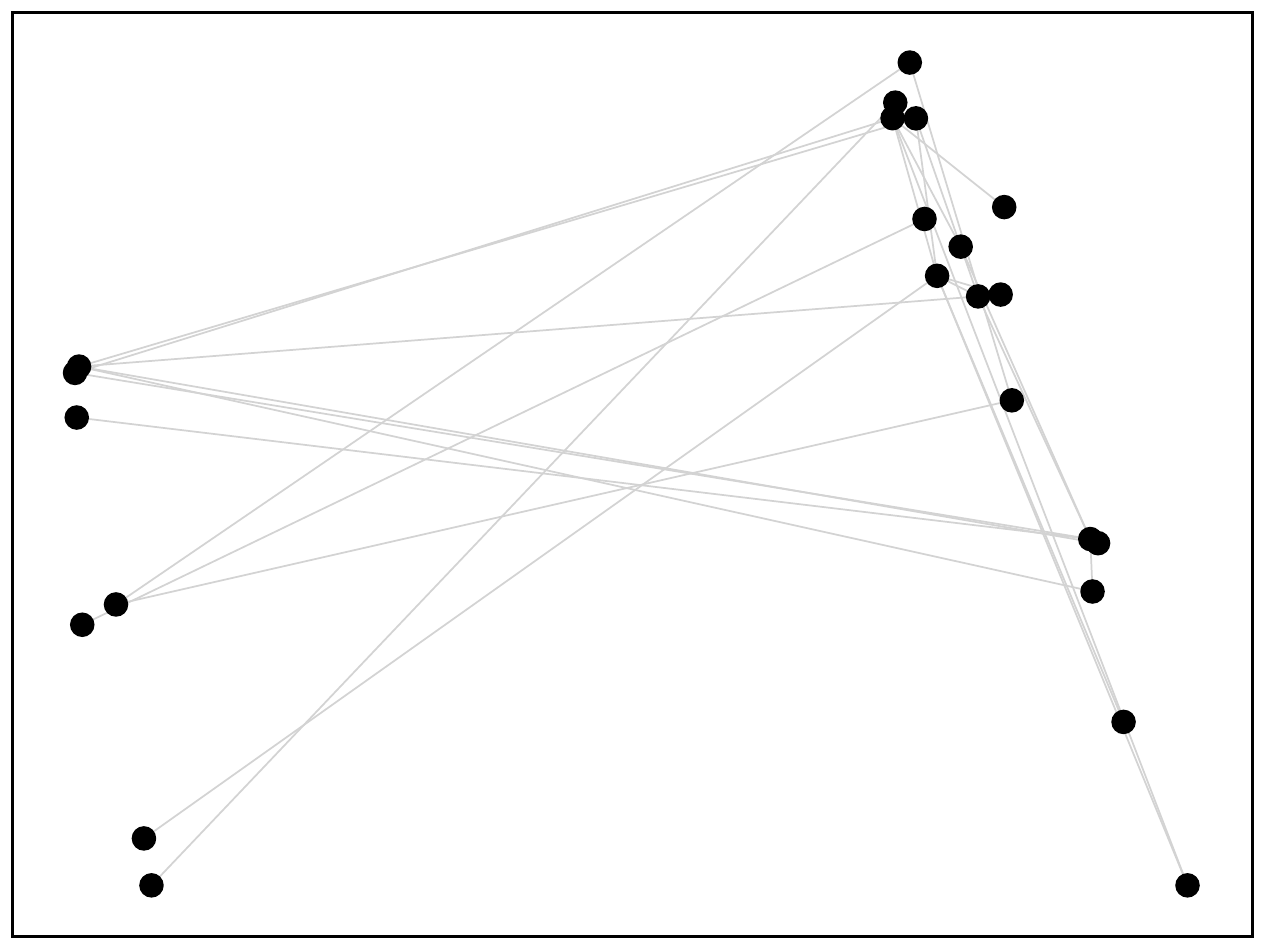}
    &   
    \includegraphics[width=0.14\textwidth]{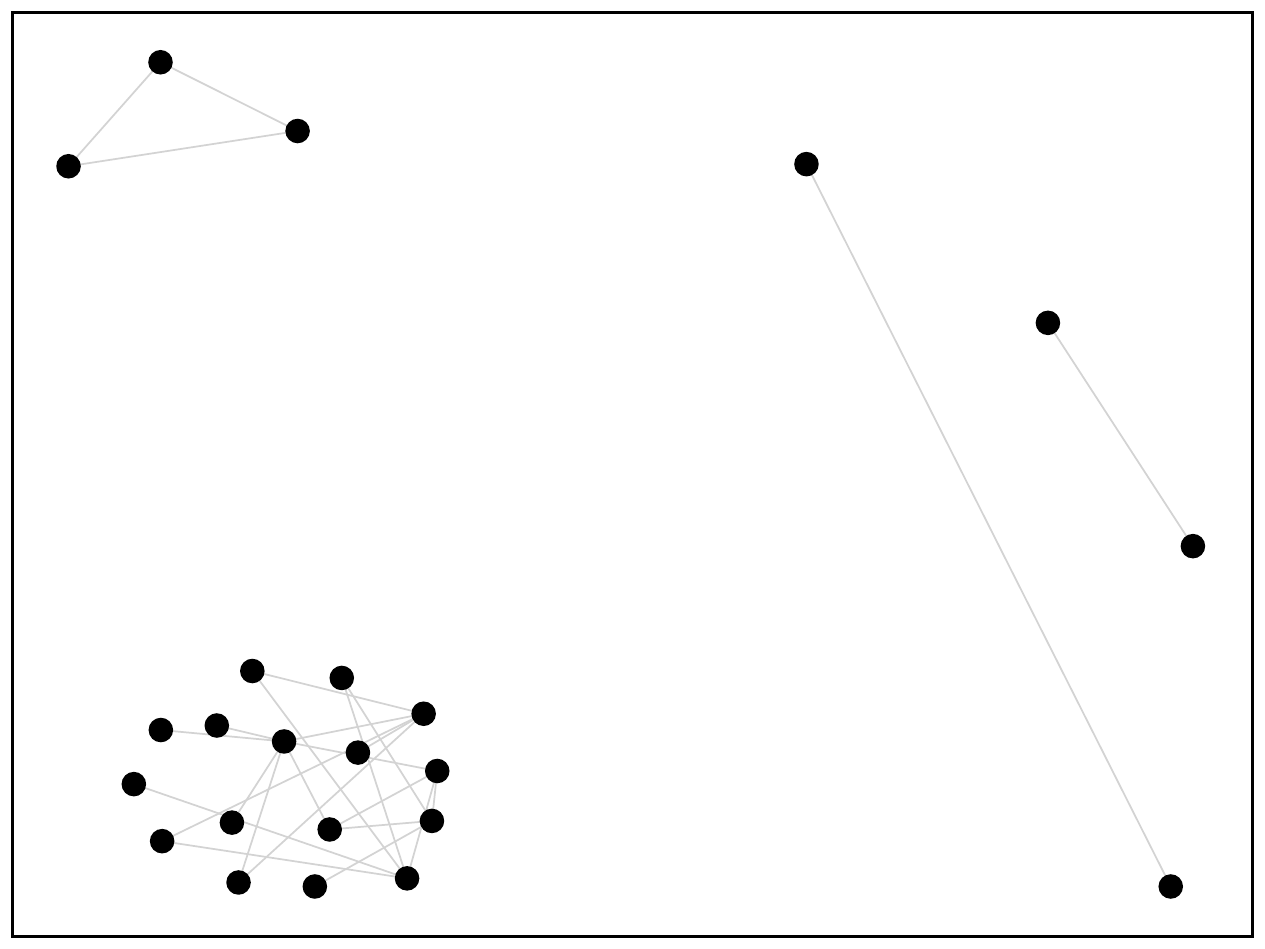}
    \\
     (g) $\star$~\gf~\cite{ahmed2013distributed} & 
     (h) $\star$~\sdne~\cite{wang2016structural} & 
     (i) $\star$~\leemb~\cite{belkin2003laplacian} & 
     (j) $\star$~\lle~\cite{roweis2000nonlinear} & 
     (k) $\star$~\nodevec~\cite{grover2016node2vec} & 
     (l) \simrank~\cite{jeh2002simrank}
    \\
  \end{tabular}
 \end{small}
  \vspace{-3mm}
  \caption{Visualization results and aesthetic metrics for the \twego graph: force-directed methods are marked with $\blacklozenge$; stress methods are marked with $\blacktriangle$; graph embedding methods are marked with $\star$.}\label{fig:tw-viz}
\end{figure*}

\begin{figure}[!t]
\centering
\begin{small}
\rc{
  \begin{tabular}{cc}
    \includegraphics[width=0.45\columnwidth]{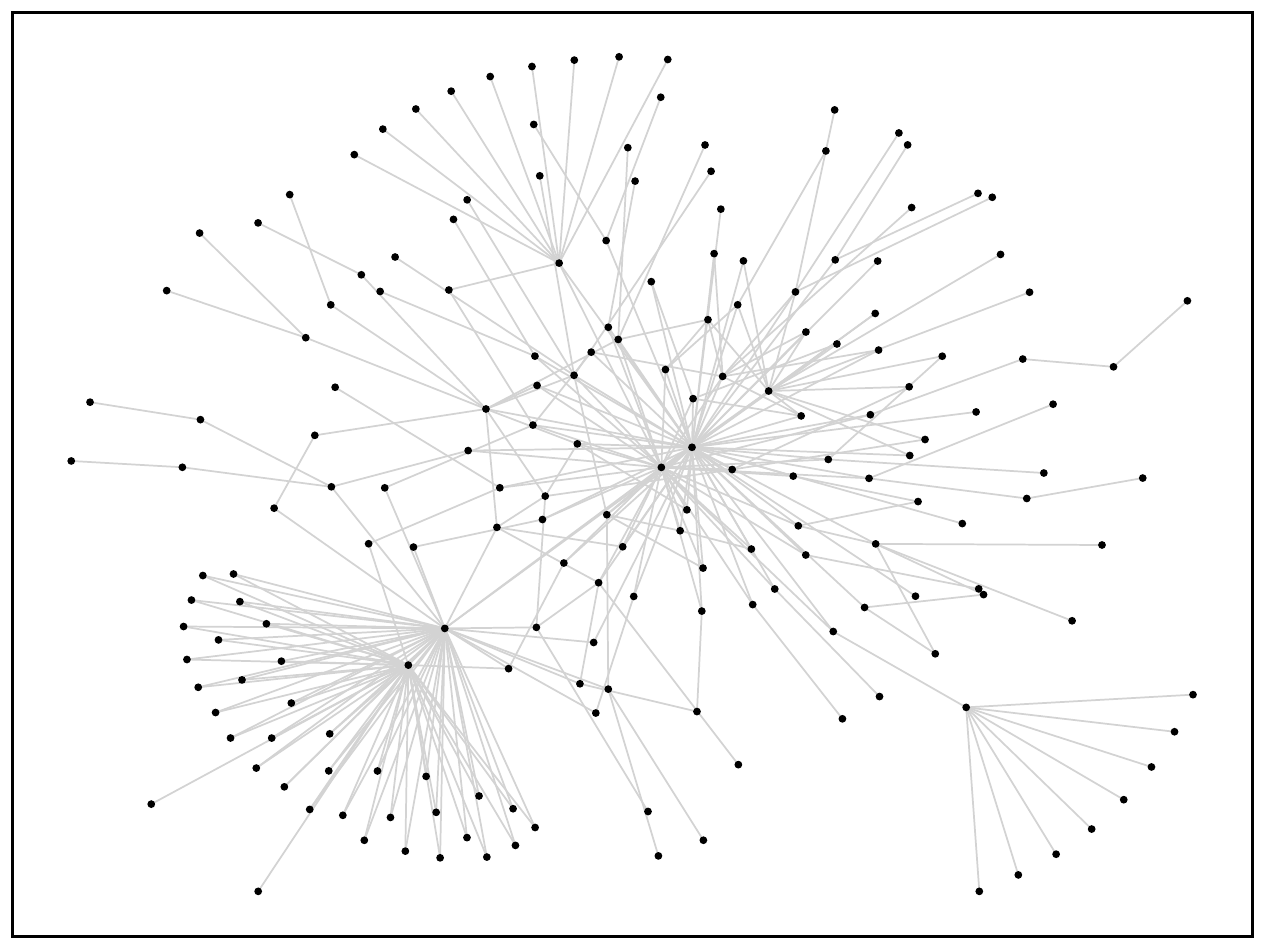}
     &
    \includegraphics[width=0.45\columnwidth]{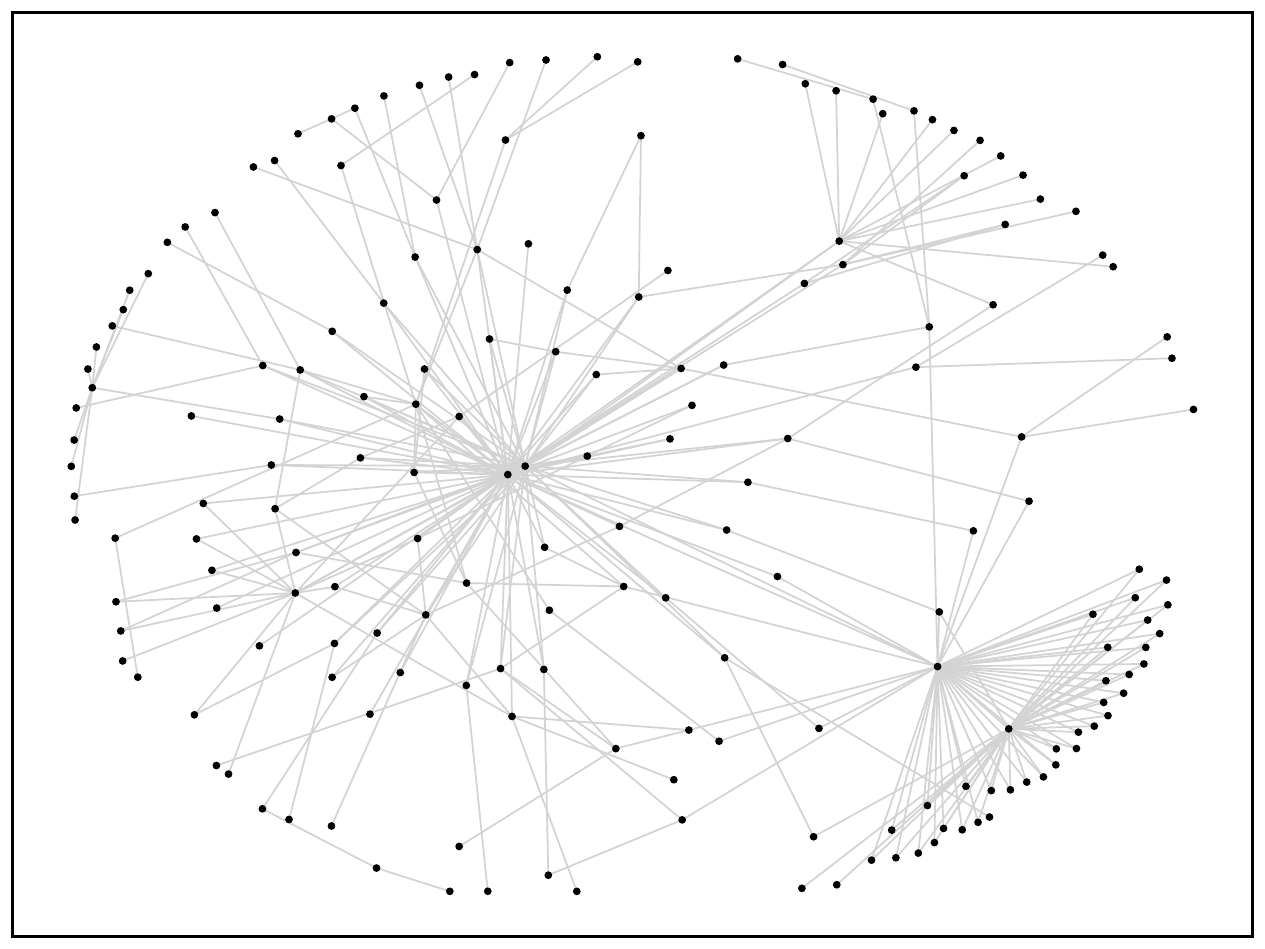}
    \\
    (a) \pprviz (Ours)  & (b) $\blacklozenge$~\fr~\cite{fruchterman1991graph} 
    \\
  \end{tabular}
}
\end{small}
\vspace{-3mm}
\caption{\rc{Visualization of \pprvizb and \frb on \wiki.}}\label{fig:wiki-viz-part}
\vspace{-2mm}
\end{figure}

\begin{figure}[!t]
\centering
\begin{small}
\rc{
  \begin{tabular}{cc}
    \includegraphics[width=0.45\columnwidth]{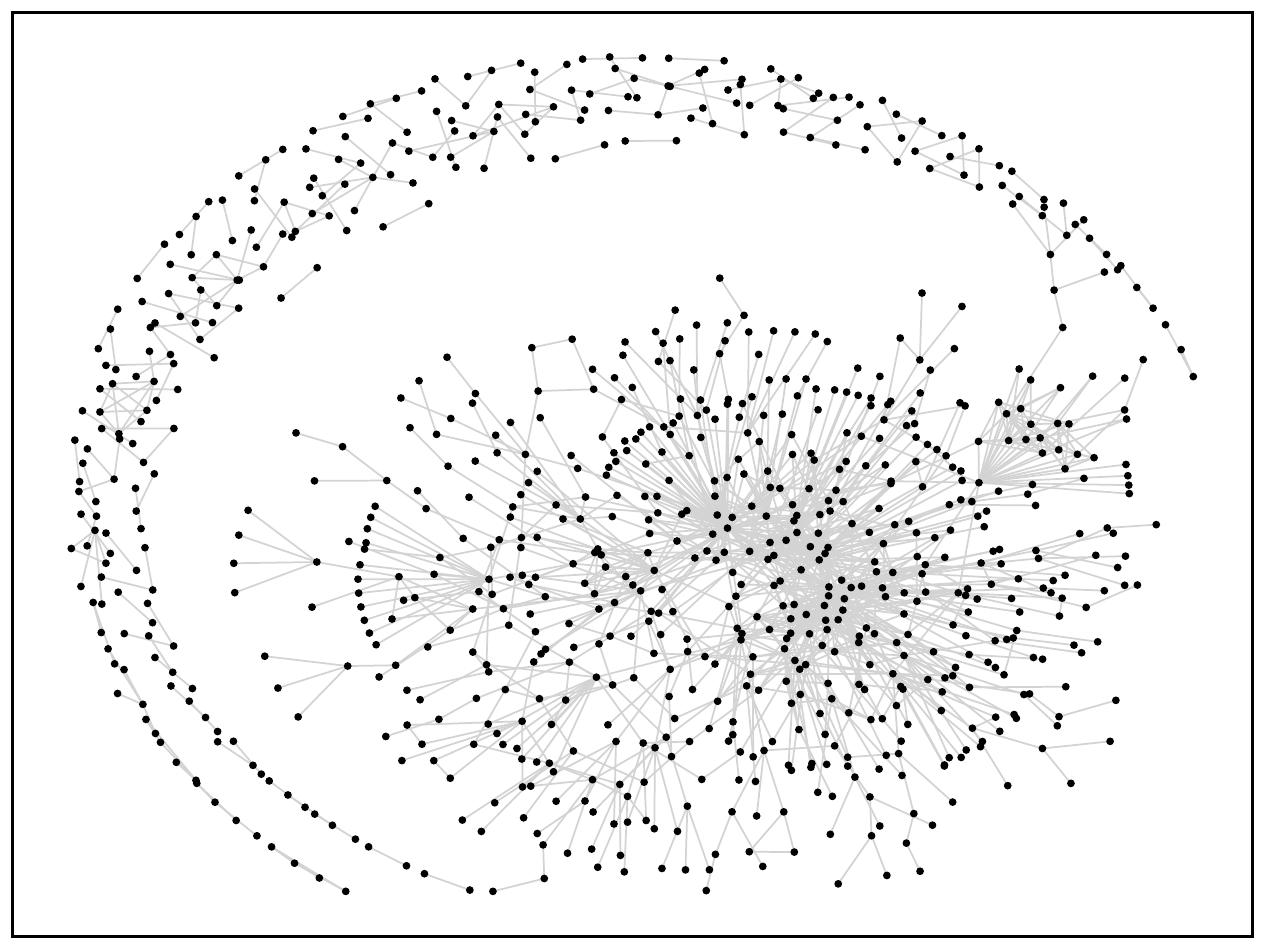}
     &
    \includegraphics[width=0.45\columnwidth]{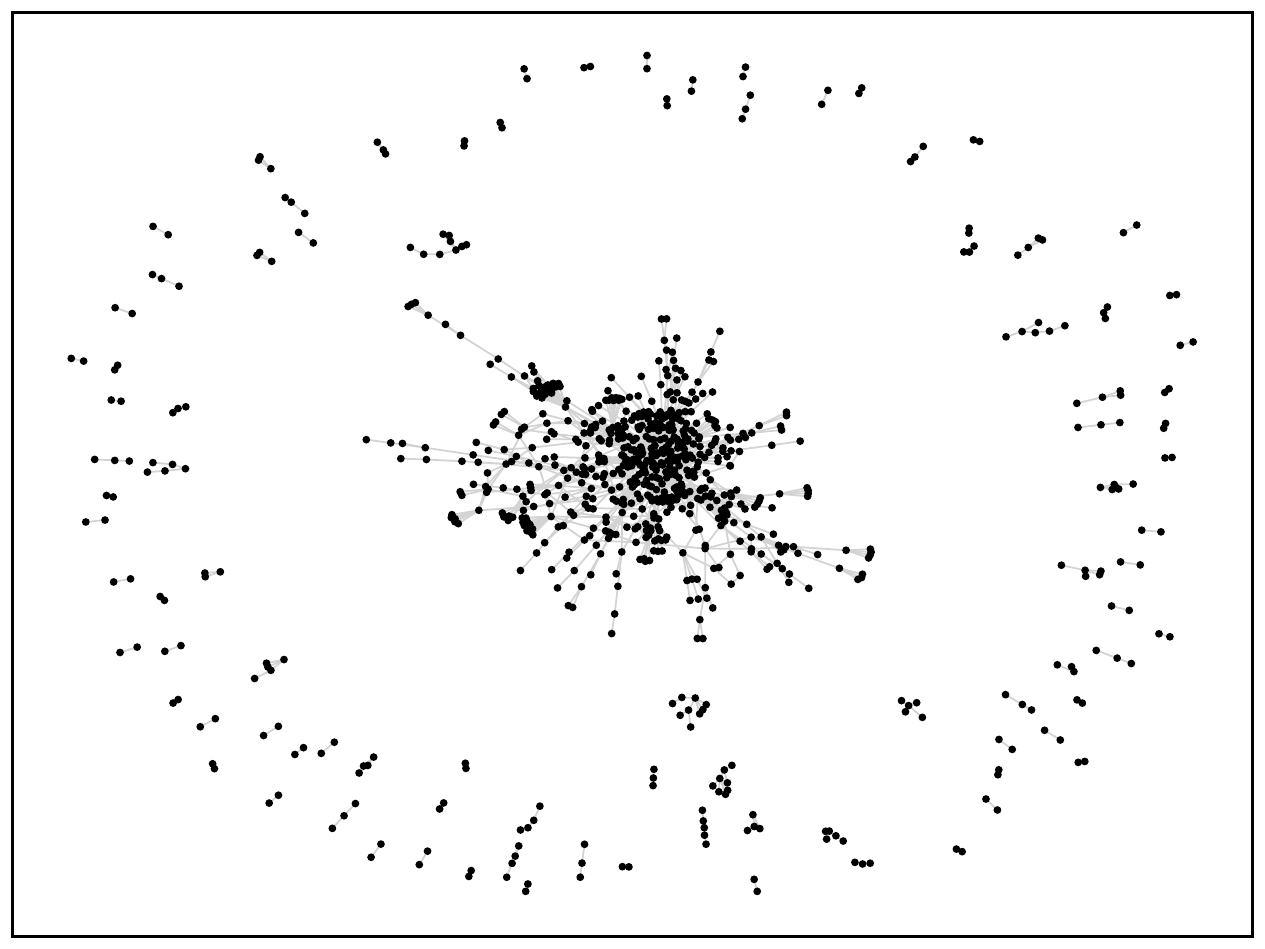}
    \\
    (a) \pprviz (Ours)  & (b) $\blacklozenge$~\fr~\cite{fruchterman1991graph} 
    \\
  \end{tabular}
 }
\end{small}  
\vspace{-3mm}
\caption{\rc{Visualization of \pprvizb and \frb on \filmtrust.}}\label{fig:filmtrust-viz-part}
\vspace{-2mm}
\end{figure}

\input{exps/exp-time}

\subsubsection{\bf User Study}\label{sec:exp-user-study}
In the second set of experiments, we conduct a user study to evaluate the visualization quality of \pprviz. This study aims to answer two questions: (i) does \pprviz output better visualization results than the competitors, and (ii) does the approximate level-$\ell$ \pprdistname computation in \pprviz affect visualization quality from the perspective of human observers? We recruited 30 participants with 6 females and 24 males, among which 28 individuals are aged 20 to 30 and 2 individuals are aged 30 to 40.

To answer the first question, we generate 6 groups of visualizations, each of which is obtained on one of the 6 small graphs using \pprviz and 4 representative competitors, including \fr, \mds, \nodevec, and \simrank, that achieve relatively good performance in Section~\ref{sec:exp-metrics}.
In each group, the 5 visualizations generated by different methods are randomly shuffled before presenting to the participants. Following the assessment paradigm in graph drawing~\cite{lee2006task,du2017isphere,wu2016evaluation}, the participants are asked to rank the visualizations in terms of readability and cluster structure, which are identified as two paramount tasks by~\cite{gibson2013survey,battista1998graph} and are described as follows.
\begin{itemize}[leftmargin=*]
\item Task 1 (T1): rank the 5 visualizations in each group from the highest (1) to the lowest readability (5), where high readability means that the graph elements are well displayed. 
\item Task 2 (T2): rank the 5 visualizations in each group from the best (1) to the worst (5) in terms of the structure layout, where a good layout means that clusters and strongly connected nodes can be clearly observed. 
\end{itemize}

To answer the second question, we compare the visualizations generated by employing \taupush in \pprviz with those by employing the near-exact solution \poweriter in \pprviz. Since \poweriter does not scale to large graphs, we consider two small graphs: \filmtrust and \scinet. For both methods, we vary $k$ in $\{15, 20, 25\}$.
For each supergraph, we generate 2 visualizations using \taupush and \poweriter and organize them as a group. The visualizations in each group are displayed to the participants without telling them the methods, and the participants are asked to conduct the following task.     
\begin{itemize}[leftmargin=*]
    \item Task 3 (T3): for the 2 visualizations in a group, select the one with superior visualization quality.
\end{itemize}

Note that for each task, we collected 180 instances (30 participants $\times$ 6 groups). Fig.~\ref{fig:user-study-t12}(a) (resp.\ Fig.~\ref{fig:user-study-t12}(b)) reports the ranking frequency of the visualization generated by each approach in T1 (resp.\ T2). In particular, the rankings of \pprviz are concentrated on top-1 and top-2 for both tasks. This result demonstrates \pprviz's sound quality in terms of readability and cluster inspection, which is consistent with our \nd{} and \ulcv{} results reported in Section \ref{sec:exp-metrics}. Regarding the results of T3, we find in Table~\ref{tab:user-study-t3} that the participants cannot tell the quality difference between the visualizations generated by \taupush and \poweriter in most cases, and the times the two methods are selected as the best are comparable, demonstrating that the approximate level-$\ell$ \pprdistname computation in \pprviz does not affect its visualization quality.

\subsubsection{\bf Case Study} \label{sec:exp-viz}
At last, we conduct a case study to compare the visualizations of \pprviz and other competitors.
In Fig.~\ref{fig:tw-viz}, we report the results of \pprviz and all competitors on \twego. Besides that, we also compare the results of \pprviz and the best competitor \fr on two large graphs \wiki and \filmtrust, which are shown in Fig.~\ref{fig:wiki-viz-part} and Fig.~\ref{fig:filmtrust-viz-part}, respectively.
We find that the comparison results of visualization are consistent with those of metrics in Section~\ref{sec:exp-metrics}.
More concrete, \pprviz yields a high-quality visualization, which clearly organizes the graph into a well-connected cluster and several cliques. In contrast, the competitors suffer from issues such as node overlapping and edge distortion. 
Furthermore, as shown in Fig.~\ref{fig:tw-viz}(d) and Fig.~\ref{fig:filmtrust-viz-part}(b), the structures of small-size clusters yielded by the best competitor \fr are hard to recognize. This is because nodes in the small-size clusters are more likely to be affected by the repulsive force from the large cluster, making all of them huddle together.

\subsection{Visualization Efficiency}\label{sec:exp-time}
\stitle{Response time}
Fig.~\ref{fig:query-time} shows the response time on the 6 larger graphs, including \amazon, \youtube, \orkut, \dblp, \itzerofour, and \twitter. We only report the results of \pprviz, \openord, \kadraw, \pivotmds, and \nodevec, and omit the rest of the competitors as they fail to terminate within 1000 seconds for all datasets.
We can observe that \pprviz consistently outperforms all competitors in terms of response time. Specifically, \pprviz outputs the visualization result within 1 second on all datasets. For example, \pprviz costs 0.63 seconds on the \twitter graph 3 billion edges, while all competing methods take more than 1000 seconds.
Regarding the competitors, \pivotmds only computes the positions of several pivot nodes and determines the positions of other nodes by linear combinations of the pivot nodes. However, it can only visualize the entire \amazon graph in 45 seconds.
In addition, \nodevec incurs costly random walk simulations, rendering it rather inefficient on large graphs and can only return the visualization for \amazon in 845 seconds.
The two multi-level methods, \ie \kadraw and \openord, have comparable performance to \pprviz on \amazon, \ie the smallest one among the 6 graphs, but turn to be markedly inferior to \pprviz when graph size increases. The reason is that \kadraw requires computing the forces between the leaf nodes inside a supernode which is rather costly for high-level supergraphs, and \openord takes many iterations to determine the layout in its five-stage design.

\stitle{Preprocessing time}
Fig.~\ref{fig:pre-time} plots the preprocessing time of the multi-level methods, \ie \pprviz, \openord, and \kadraw, as the single-level methods do not have the preprocessing phase. All three methods conduct hierarchical clustering for the input graph. However, \pprviz also computes the \dnpr vector and some single-target \pprdistname scores. The results show that the processing time of \pprviz is two to three orders of magnitude faster than \openord and one order of magnitude faster than \kadraw. Moreover, both \openord and \kadraw cannot finish preprocessing for the largest \twitter graph within 12 hours while \pprviz takes only 33 minutes. This is because \openord computes the layout of the entire graph first and then conducts hierarchical clustering on the two-dimensional layout.
For \kadraw, its clustering algorithm~\cite{meyerhenke2014partitioning} is more expensive than \louvainplus in \pprviz.

\stitle{Vary cluster size}
Table~\ref{tab:pprviz-varyk-time} reports the preprocessing time and response time of \pprviz by varying the maximum number of nodes (\ie the cluster size limit $k$) on the largest test graph \twitter. We exclude \kadraw and \openord from this experiment because configuring $k$ is difficult for them, as discussed in Section~\ref{sec:exp-setting}. \pprviz's preprocessing time drops when $k$ increases because \louvainplus organizes more nodes into a supernode with larger $k$, resulting in fewer supernodes in each level-$\ell$ supergraph and fewer levels of the supergraph hierarchy.  
For interactive visualization, \pprviz's response time increases with $k$, and the main reason is that more pairwise \pprdistname are computed with more nodes in each visualization. Furthermore, the response time of \pprviz is only 2.10 seconds even with $k=100$, above which the intra-structure of a supernode becomes dense and overburdens human perception~\cite{huang2009measuring}. 

\begin{table}[!t]
\centering
\caption{Time of \pprvizb on \twitter by varying $k$ (in seconds).} \label{tab:pprviz-varyk-time}
\vspace{-3mm}
\renewcommand{\arraystretch}{1.1}
\begin{small}
\begin{tabular}{|c|c|c|c|c|c|}
\hline
$k$                 & 5       & 10      & 25      & 50      & 100     \\ \hline
Preprocessing        & 2267.65   & 2114.39 & 1934.48 & 1842.45 & 1796.87 \\ \hline
Response            & 0.28    & 0.43    & 0.63    & 1.56    & 2.10    \\ \hline
\end{tabular}
\end{small}
\vspace{-3mm}
\end{table}

\subsection{\taupushb Performance}\label{sec:exp-dnppr}
\stitle{Comparing with other solutions}
To validate the efficiency of the \taupush algorithm for \pprdistname computation, we replace \taupush with 4 alternative solutions as mentioned in Section~\ref{sec:exp-setting}, and then compare \pprviz against these variants in terms of response time, preprocessing time, and index size on the 4 largest graphs including \youtube, \orkut, \itzerofour, and \twitter. 
In terms of response time, \rc{we use ``-'' to indicate that an approach fails to terminate within 1000 seconds.} Table~\ref{tab:variant-response} shows that all \pprviz variants incur more than 1000 seconds on all tested graphs. This is because they need to compute \pprdistname from $O(n)$ leaf nodes for the top-level supergraph.
As shown in Table~\ref{tab:variant-preprocess}, all methods achieve comparable performance regarding the preprocessing time, since supergraph hierarchy construction dominates preprocessing cost.
The two \pprviz variants using \poweriter and \resacc have slightly shorter preprocessing time as they only need to construct the supergraph hierarchies. 
For the same reason, they require much less space to store the indices as shown in Table~\ref{tab:variant-index}.
In contrast, \fora (resp.\ \taupush) precomputes random walk samples (resp.\ \dnpr values and results for \bwdpushsn) as indices.
Compared with \fora, our \taupush costs less space for indices because \taupush eliminates the need to store random walks by precomputing \dnpr to guide the termination of push operations. 
Specifically, besides $O(n)$ supernode partitions, \pprviz only requires storing $n$ \dnpr values and $O(k\cdot\sqrt{n\cdot k})$ extra \pprdistname values, which in total are insignificant compared with the size of the input graph.

\stitle{Ablation study}
In this set of experiments, we verify the effectiveness of the three techniques in \taupush, \ie the grouped push strategy, \dnpr-guided termination trick, and the filter-refinement optimization delineated in Section \ref{sec:mainidea}.
First, to demonstrate the effectiveness of the grouped push strategy, we replace \fwdpush in \fora with \fwdpushsn and call this variant \forasn.
In particular, for each level-$\ell$ supernode, \forasn first invokes \fwdpushsn to roughly estimate level-$\ell$ \pprdegname values w.r.t.\ the given source supernode and then refines the under-estimations by random walk samplings. After conducting a theoretical analysis, we generate a sufficient amount of random walks to ensure the correctness and optimize the time complexity by balancing the overhead of two stages for a fair comparison.
Akin to the analysis in Table~\ref{tab:complex-compare}, we can show that the time complexity of \forasn to compute all pairwise approximate level-$\ell$ \pprdistname in a supernode $\mathcal{S}$ is $O\left(\frac{k\cdot\log{n}\cdot\sqrt{m}}{\epsilon}\right)$, which improves the time complexity of \fora by ${n}/{\sqrt{k}}$.
As shown in Table~\ref{tab:variant-response}, the response time of \forasn is at least four orders of magnitude faster than \fora by adopting the grouped push strategy in our \taupush.

\input{exps/exp-dnppr}

Note that if we set $\tau = \max_{\mathcal{V}_j\in \mathcal{S}\backslash\mathcal{V}_i}{\tau_j}$ for Eq.\eqref{eq:rmax}, all values returned by \fwdpushsn are $(\epsilon,\delta)-$approximate level-$\ell$ \pprdegname.  Here, we call this variant \fwdpushsn ($\tau_{max}$) and compare \forasn against it. We find in Table~\ref{tab:variant-response} that \fwdpushsn ($\tau_{max}$) improves \forasn on the \orkut, \itzerofour, and \twitter graphs that have massive edges, with 3$\times$, 2.3$\times$, and 4$\times$ speedups, respectively. Furthermore, as shown in Table~\ref{tab:variant-index}, \fwdpushsn ($\tau_{max}$) incurs much less space overhead for index storage. The reason is that on such graphs, \forasn requires a multitude of excessive random walks, while \fwdpushsn ($\tau_{max}$) eliminates random walks and enables early termination by leveraging \dnpr as described in Section \ref{sec:mainidea}.

Finally, we evaluate the filter-refinement optimization by comparing \taupush with the initial solution \fwdpushsn ($\tau_{max}$).  Fig.~\ref{fig:abalation-time} shows that \taupush significantly reduces the response time of \fwdpushsn when the cluster contains nodes with large \dnpr values by employing \bwdpushsn. The speedup of \taupush varies since the skewness of \dnpr value is different on each graph. 
For the first cluster on \youtube, the largest \dnpr value is 3 orders of magnitude larger than the others, and thus \taupush speeds up \fwdpushsn by about $300\times$ by avoiding many rounds of push operations. For a similar reason, \taupush is over $1000\times$ faster than \fwdpushsn in the second cluster in \twitter. The speedup of \taupush is less significant on \orkut and \itzerofour graphs but can still reach 9$\times$ and 24$\times$, respectively.

\section{Conclusions}\label{sec:ccl}
This paper proposes a PPR-based node distance \pprdistname and its fast computation algorithm \taupush for massive graph visualization, whose performance is extensively evaluated by comparing with 13 competitors on 12 real-world graphs. The results show that our proposal achieves high effectiveness and efficiency.
Regarding future works, we plan to extend \pprdistname and \taupush to support dynamic or attributed graph visualization.

\balance
\bibliographystyle{ACM-Reference-Format}
\bibliography{main}


\begin{thebibliography}{88}


\ifx \showCODEN    \undefined \def \showCODEN     #1{\unskip}     \fi
\ifx \showDOI      \undefined \def \showDOI       #1{#1}\fi
\ifx \showISBNx    \undefined \def \showISBNx     #1{\unskip}     \fi
\ifx \showISBNxiii \undefined \def \showISBNxiii  #1{\unskip}     \fi
\ifx \showISSN     \undefined \def \showISSN      #1{\unskip}     \fi
\ifx \showLCCN     \undefined \def \showLCCN      #1{\unskip}     \fi
\ifx \shownote     \undefined \def \shownote      #1{#1}          \fi
\ifx \showarticletitle \undefined \def \showarticletitle #1{#1}   \fi
\ifx \showURL      \undefined \def \showURL       {\relax}        \fi
\providecommand\bibfield[2]{#2}
\providecommand\bibinfo[2]{#2}
\providecommand\natexlab[1]{#1}
\providecommand\showeprint[2][]{arXiv:#2}

\bibitem[\protect\citeauthoryear{Abello, Van~Ham, and Krishnan}{Abello
  et~al\mbox{.}}{2006}]%
        {abello2006ask}
\bibfield{author}{\bibinfo{person}{James Abello}, \bibinfo{person}{Frank
  Van~Ham}, {and} \bibinfo{person}{Neeraj Krishnan}.}
  \bibinfo{year}{2006}\natexlab{}.
\newblock \showarticletitle{Ask-graphview: A large scale graph visualization
  system}.
\newblock \bibinfo{journal}{\emph{TVCG}} \bibinfo{volume}{12},
  \bibinfo{number}{5} (\bibinfo{year}{2006}), \bibinfo{pages}{669--676}.
\newblock


\bibitem[\protect\citeauthoryear{Agapito, Guzzi, and Cannataro}{Agapito
  et~al\mbox{.}}{2013}]%
        {agapito2013visualization}
\bibfield{author}{\bibinfo{person}{Giuseppe Agapito},
  \bibinfo{person}{Pietro~Hiram Guzzi}, {and} \bibinfo{person}{Mario
  Cannataro}.} \bibinfo{year}{2013}\natexlab{}.
\newblock \showarticletitle{Visualization of protein interaction networks:
  problems and solutions}.
\newblock \bibinfo{journal}{\emph{BMC}} \bibinfo{volume}{14},
  \bibinfo{number}{1} (\bibinfo{year}{2013}), \bibinfo{pages}{1--30}.
\newblock


\bibitem[\protect\citeauthoryear{Ahmed, Shervashidze, Narayanamurthy,
  Josifovski, and Smola}{Ahmed et~al\mbox{.}}{2013}]%
        {ahmed2013distributed}
\bibfield{author}{\bibinfo{person}{Amr Ahmed}, \bibinfo{person}{Nino
  Shervashidze}, \bibinfo{person}{Shravan Narayanamurthy},
  \bibinfo{person}{Vanja Josifovski}, {and} \bibinfo{person}{Alexander~J
  Smola}.} \bibinfo{year}{2013}\natexlab{}.
\newblock \showarticletitle{Distributed large-scale natural graph
  factorization}. In \bibinfo{booktitle}{\emph{WWW}}. \bibinfo{pages}{37--48}.
\newblock


\bibitem[\protect\citeauthoryear{Andersen, Chung, and Lang}{Andersen
  et~al\mbox{.}}{2006}]%
        {andersen2006local}
\bibfield{author}{\bibinfo{person}{Reid Andersen}, \bibinfo{person}{Fan Chung},
  {and} \bibinfo{person}{Kevin Lang}.} \bibinfo{year}{2006}\natexlab{}.
\newblock \showarticletitle{Local graph partitioning using pagerank vectors}.
  In \bibinfo{booktitle}{\emph{FOCS}}. \bibinfo{pages}{475--486}.
\newblock


\bibitem[\protect\citeauthoryear{Archambault, Munzner, and Auber}{Archambault
  et~al\mbox{.}}{2008}]%
        {archambault2008grouseflocks}
\bibfield{author}{\bibinfo{person}{Daniel Archambault}, \bibinfo{person}{Tamara
  Munzner}, {and} \bibinfo{person}{David Auber}.}
  \bibinfo{year}{2008}\natexlab{}.
\newblock \showarticletitle{GrouseFlocks: Steerable exploration of graph
  hierarchy space}.
\newblock \bibinfo{journal}{\emph{TVCG}} \bibinfo{volume}{14},
  \bibinfo{number}{4} (\bibinfo{year}{2008}), \bibinfo{pages}{900--913}.
\newblock


\bibitem[\protect\citeauthoryear{Auber}{Auber}{2004}]%
        {auber2004tulip}
\bibfield{author}{\bibinfo{person}{David Auber}.}
  \bibinfo{year}{2004}\natexlab{}.
\newblock \showarticletitle{Tulip—A huge graph visualization framework}.
\newblock In \bibinfo{booktitle}{\emph{GDS}}. \bibinfo{pages}{105--126}.
\newblock


\bibitem[\protect\citeauthoryear{Bastian, Heymann, and Jacomy}{Bastian
  et~al\mbox{.}}{2009}]%
        {bastian2009gephi}
\bibfield{author}{\bibinfo{person}{Mathieu Bastian}, \bibinfo{person}{Sebastien
  Heymann}, {and} \bibinfo{person}{Mathieu Jacomy}.}
  \bibinfo{year}{2009}\natexlab{}.
\newblock \showarticletitle{Gephi: an open source software for exploring and
  manipulating networks}. In \bibinfo{booktitle}{\emph{ICWSM}}.
\newblock


\bibitem[\protect\citeauthoryear{Battista, Eades, Tamassia, and
  Tollis}{Battista et~al\mbox{.}}{1998}]%
        {battista1998graph}
\bibfield{author}{\bibinfo{person}{Giuseppe~Di Battista},
  \bibinfo{person}{Peter Eades}, \bibinfo{person}{Roberto Tamassia}, {and}
  \bibinfo{person}{Ioannis~G Tollis}.} \bibinfo{year}{1998}\natexlab{}.
\newblock \bibinfo{booktitle}{\emph{Graph drawing: algorithms for the
  visualization of graphs}}.
\newblock


\bibitem[\protect\citeauthoryear{Belkin and Niyogi}{Belkin and Niyogi}{2003}]%
        {belkin2003laplacian}
\bibfield{author}{\bibinfo{person}{Mikhail Belkin} {and}
  \bibinfo{person}{Partha Niyogi}.} \bibinfo{year}{2003}\natexlab{}.
\newblock \showarticletitle{Laplacian eigenmaps for dimensionality reduction
  and data representation}.
\newblock \bibinfo{journal}{\emph{Neural Comput.}} \bibinfo{volume}{15},
  \bibinfo{number}{6} (\bibinfo{year}{2003}), \bibinfo{pages}{1373--1396}.
\newblock


\bibitem[\protect\citeauthoryear{Bennett, Ryall, Spalteholz, and Gooch}{Bennett
  et~al\mbox{.}}{2007}]%
        {bennett2007aesthetics}
\bibfield{author}{\bibinfo{person}{Chris Bennett}, \bibinfo{person}{Jody
  Ryall}, \bibinfo{person}{Leo Spalteholz}, {and} \bibinfo{person}{Amy Gooch}.}
  \bibinfo{year}{2007}\natexlab{}.
\newblock \showarticletitle{The aesthetics of graph visualization.}
\newblock \bibinfo{journal}{\emph{CAe}} (\bibinfo{year}{2007}),
  \bibinfo{pages}{57--64}.
\newblock


\bibitem[\protect\citeauthoryear{Bhowmick, Huang, Chua, Yuan, Choi, and
  Zhou}{Bhowmick et~al\mbox{.}}{2020}]%
        {bhowmick2020aurora}
\bibfield{author}{\bibinfo{person}{Sourav~S Bhowmick}, \bibinfo{person}{Kai
  Huang}, \bibinfo{person}{Huey~Eng Chua}, \bibinfo{person}{Zifeng Yuan},
  \bibinfo{person}{Byron Choi}, {and} \bibinfo{person}{Shuigeng Zhou}.}
  \bibinfo{year}{2020}\natexlab{}.
\newblock \showarticletitle{AURORA: Data-driven construction of visual graph
  query interfaces for graph databases}. In \bibinfo{booktitle}{\emph{SIGMOD}}.
  \bibinfo{pages}{2689--2692}.
\newblock


\bibitem[\protect\citeauthoryear{Bikakis, Liagouris, Krommyda, Papastefanatos,
  and Sellis}{Bikakis et~al\mbox{.}}{2016}]%
        {bikakis2016graphvizdb}
\bibfield{author}{\bibinfo{person}{Nikos Bikakis}, \bibinfo{person}{John
  Liagouris}, \bibinfo{person}{Maria Krommyda}, \bibinfo{person}{George
  Papastefanatos}, {and} \bibinfo{person}{Timos Sellis}.}
  \bibinfo{year}{2016}\natexlab{}.
\newblock \showarticletitle{GraphVizdb: A scalable platform for interactive
  large graph visualization}. In \bibinfo{booktitle}{\emph{ICDE}}.
  \bibinfo{pages}{1342--1345}.
\newblock


\bibitem[\protect\citeauthoryear{Blondel, Guillaume, Lambiotte, and
  Lefebvre}{Blondel et~al\mbox{.}}{2008}]%
        {blondel2008fast}
\bibfield{author}{\bibinfo{person}{Vincent~D Blondel},
  \bibinfo{person}{Jean-Loup Guillaume}, \bibinfo{person}{Renaud Lambiotte},
  {and} \bibinfo{person}{Etienne Lefebvre}.} \bibinfo{year}{2008}\natexlab{}.
\newblock \showarticletitle{Fast unfolding of communities in large networks}.
\newblock \bibinfo{journal}{\emph{J. Stat. Mech. Theory Exp.}}
  \bibinfo{volume}{2008}, \bibinfo{number}{10} (\bibinfo{year}{2008}),
  \bibinfo{pages}{P10008}.
\newblock


\bibitem[\protect\citeauthoryear{Boldi and Vigna}{Boldi and Vigna}{2004}]%
        {BoVWFI}
\bibfield{author}{\bibinfo{person}{Paolo Boldi} {and}
  \bibinfo{person}{Sebastiano Vigna}.} \bibinfo{year}{2004}\natexlab{}.
\newblock \showarticletitle{The {W}eb{G}raph Framework {I}: {C}ompression
  Techniques}. In \bibinfo{booktitle}{\emph{WWW}}. \bibinfo{pages}{595--602}.
\newblock


\bibitem[\protect\citeauthoryear{Brandes and Pich}{Brandes and Pich}{2006}]%
        {brandes2006eigensolver}
\bibfield{author}{\bibinfo{person}{Ulrik Brandes} {and}
  \bibinfo{person}{Christian Pich}.} \bibinfo{year}{2006}\natexlab{}.
\newblock \showarticletitle{Eigensolver methods for progressive
  multidimensional scaling of large data}. In \bibinfo{booktitle}{\emph{GD}}.
  \bibinfo{pages}{42--53}.
\newblock


\bibitem[\protect\citeauthoryear{Chung and Lu}{Chung and Lu}{2006}]%
        {chung2006concentration}
\bibfield{author}{\bibinfo{person}{Fan Chung} {and} \bibinfo{person}{Linyuan
  Lu}.} \bibinfo{year}{2006}\natexlab{}.
\newblock \showarticletitle{Concentration inequalities and martingale
  inequalities: a survey}.
\newblock \bibinfo{journal}{\emph{Internet Math.}} \bibinfo{volume}{3},
  \bibinfo{number}{1} (\bibinfo{year}{2006}), \bibinfo{pages}{79--127}.
\newblock


\bibitem[\protect\citeauthoryear{Chung and Tsiatas}{Chung and Tsiatas}{2012}]%
        {chung2012finding}
\bibfield{author}{\bibinfo{person}{Fan Chung} {and} \bibinfo{person}{Alexander
  Tsiatas}.} \bibinfo{year}{2012}\natexlab{}.
\newblock \showarticletitle{Finding and visualizing graph clusters using
  PageRank optimization}.
\newblock \bibinfo{journal}{\emph{Internet Math.}} (\bibinfo{year}{2012}),
  \bibinfo{pages}{86--97}.
\newblock


\bibitem[\protect\citeauthoryear{Cohen, Erez, Ben-Avraham, and Havlin}{Cohen
  et~al\mbox{.}}{2001}]%
        {cohen2001breakdown}
\bibfield{author}{\bibinfo{person}{Reuven Cohen}, \bibinfo{person}{Keren Erez},
  \bibinfo{person}{Daniel Ben-Avraham}, {and} \bibinfo{person}{Shlomo Havlin}.}
  \bibinfo{year}{2001}\natexlab{}.
\newblock \showarticletitle{Breakdown of the internet under intentional
  attack}.
\newblock \bibinfo{journal}{\emph{Phys. Rev. Lett.}} \bibinfo{volume}{86},
  \bibinfo{number}{16} (\bibinfo{year}{2001}), \bibinfo{pages}{3682}.
\newblock


\bibitem[\protect\citeauthoryear{Davidson and Harel}{Davidson and
  Harel}{1996}]%
        {davidson1996drawing}
\bibfield{author}{\bibinfo{person}{Ron Davidson} {and} \bibinfo{person}{David
  Harel}.} \bibinfo{year}{1996}\natexlab{}.
\newblock \showarticletitle{Drawing graphs nicely using simulated annealing}.
\newblock \bibinfo{journal}{\emph{TOG}} \bibinfo{volume}{15},
  \bibinfo{number}{4} (\bibinfo{year}{1996}), \bibinfo{pages}{301--331}.
\newblock


\bibitem[\protect\citeauthoryear{De~Nooy, Mrvar, and Batagelj}{De~Nooy
  et~al\mbox{.}}{2018}]%
        {de2018exploratory}
\bibfield{author}{\bibinfo{person}{Wouter De~Nooy}, \bibinfo{person}{Andrej
  Mrvar}, {and} \bibinfo{person}{Vladimir Batagelj}.}
  \bibinfo{year}{2018}\natexlab{}.
\newblock \bibinfo{booktitle}{\emph{Exploratory social network analysis with
  Pajek: Revised and expanded edition for updated software}}.
  Vol.~\bibinfo{volume}{46}.
\newblock \bibinfo{publisher}{Cambridge university press}.
\newblock


\bibitem[\protect\citeauthoryear{Du, Cao, Lin, Xu, and Tong}{Du
  et~al\mbox{.}}{2017}]%
        {du2017isphere}
\bibfield{author}{\bibinfo{person}{Fan Du}, \bibinfo{person}{Nan Cao},
  \bibinfo{person}{Yu-Ru Lin}, \bibinfo{person}{Panpan Xu}, {and}
  \bibinfo{person}{Hanghang Tong}.} \bibinfo{year}{2017}\natexlab{}.
\newblock \showarticletitle{isphere: Focus+ context sphere visualization for
  interactive large graph exploration}. In \bibinfo{booktitle}{\emph{CHI}}.
\newblock


\bibitem[\protect\citeauthoryear{Duncan, Goodrich, and Kobourov}{Duncan
  et~al\mbox{.}}{1998}]%
        {duncan1998balanced}
\bibfield{author}{\bibinfo{person}{Christian~A Duncan},
  \bibinfo{person}{Michael~T Goodrich}, {and} \bibinfo{person}{Stephen~G
  Kobourov}.} \bibinfo{year}{1998}\natexlab{}.
\newblock \showarticletitle{Balanced aspect ratio trees and their use for
  drawing very large graphs}. In \bibinfo{booktitle}{\emph{GD}}.
  \bibinfo{pages}{111--124}.
\newblock


\bibitem[\protect\citeauthoryear{Eades}{Eades}{1984}]%
        {eades1984heuristic}
\bibfield{author}{\bibinfo{person}{Peter Eades}.}
  \bibinfo{year}{1984}\natexlab{}.
\newblock \showarticletitle{A heuristic for graph drawing}.
\newblock \bibinfo{journal}{\emph{Congr. Numer.}}  \bibinfo{volume}{42}
  (\bibinfo{year}{1984}), \bibinfo{pages}{149--160}.
\newblock


\bibitem[\protect\citeauthoryear{Fogaras, R{\'a}cz, Csalog{\'a}ny, and
  Sarl{\'o}s}{Fogaras et~al\mbox{.}}{2005}]%
        {fogaras2005towards}
\bibfield{author}{\bibinfo{person}{D{\'a}niel Fogaras},
  \bibinfo{person}{Bal{\'a}zs R{\'a}cz}, \bibinfo{person}{K{\'a}roly
  Csalog{\'a}ny}, {and} \bibinfo{person}{Tam{\'a}s Sarl{\'o}s}.}
  \bibinfo{year}{2005}\natexlab{}.
\newblock \showarticletitle{Towards scaling fully personalized pagerank:
  Algorithms, lower bounds, and experiments}.
\newblock \bibinfo{journal}{\emph{Internet Math.}} \bibinfo{volume}{2},
  \bibinfo{number}{3} (\bibinfo{year}{2005}), \bibinfo{pages}{333--358}.
\newblock


\bibitem[\protect\citeauthoryear{Fruchterman and Reingold}{Fruchterman and
  Reingold}{1991}]%
        {fruchterman1991graph}
\bibfield{author}{\bibinfo{person}{Thomas~MJ Fruchterman} {and}
  \bibinfo{person}{Edward~M Reingold}.} \bibinfo{year}{1991}\natexlab{}.
\newblock \showarticletitle{Graph drawing by force-directed placement}.
\newblock \bibinfo{journal}{\emph{SP\&E}} \bibinfo{volume}{21},
  \bibinfo{number}{11} (\bibinfo{year}{1991}), \bibinfo{pages}{1129--1164}.
\newblock


\bibitem[\protect\citeauthoryear{Gallo, Lissandrini, and Velegrakis}{Gallo
  et~al\mbox{.}}{2020}]%
        {gallo2020personalized}
\bibfield{author}{\bibinfo{person}{Denis Gallo}, \bibinfo{person}{Matteo
  Lissandrini}, {and} \bibinfo{person}{Yannis Velegrakis}.}
  \bibinfo{year}{2020}\natexlab{}.
\newblock \showarticletitle{Personalized page rank on knowledge graphs:
  Particle Filtering is all you need!}. In \bibinfo{booktitle}{\emph{EDBT}}.
  \bibinfo{pages}{447--450}.
\newblock


\bibitem[\protect\citeauthoryear{Gansner, Hu, and North}{Gansner
  et~al\mbox{.}}{2012}]%
        {gansner2012maxent}
\bibfield{author}{\bibinfo{person}{Emden~R Gansner}, \bibinfo{person}{Yifan
  Hu}, {and} \bibinfo{person}{Stephen North}.} \bibinfo{year}{2012}\natexlab{}.
\newblock \showarticletitle{A maxent-stress model for graph layout}.
\newblock \bibinfo{journal}{\emph{TVCG}} \bibinfo{volume}{19},
  \bibinfo{number}{6} (\bibinfo{year}{2012}), \bibinfo{pages}{927--940}.
\newblock


\bibitem[\protect\citeauthoryear{Gansner, Koren, and North}{Gansner
  et~al\mbox{.}}{2004}]%
        {gansner2004graph}
\bibfield{author}{\bibinfo{person}{Emden~R Gansner}, \bibinfo{person}{Yehuda
  Koren}, {and} \bibinfo{person}{Stephen North}.}
  \bibinfo{year}{2004}\natexlab{}.
\newblock \showarticletitle{Graph drawing by stress majorization}. In
  \bibinfo{booktitle}{\emph{GD}}. \bibinfo{pages}{239--250}.
\newblock


\bibitem[\protect\citeauthoryear{Gibson, Faith, and Vickers}{Gibson
  et~al\mbox{.}}{2013}]%
        {gibson2013survey}
\bibfield{author}{\bibinfo{person}{Helen Gibson}, \bibinfo{person}{Joe Faith},
  {and} \bibinfo{person}{Paul Vickers}.} \bibinfo{year}{2013}\natexlab{}.
\newblock \showarticletitle{A survey of two-dimensional graph layout techniques
  for information visualisation}.
\newblock \bibinfo{journal}{\emph{Inf. Vis.}} \bibinfo{volume}{12},
  \bibinfo{number}{3-4} (\bibinfo{year}{2013}), \bibinfo{pages}{324--357}.
\newblock


\bibitem[\protect\citeauthoryear{Goyal and Ferrara}{Goyal and Ferrara}{2018}]%
        {goyal2018graph}
\bibfield{author}{\bibinfo{person}{Palash Goyal} {and} \bibinfo{person}{Emilio
  Ferrara}.} \bibinfo{year}{2018}\natexlab{}.
\newblock \showarticletitle{Graph embedding techniques, applications, and
  performance: A survey}.
\newblock \bibinfo{journal}{\emph{Knowl.-Based Syst.}}  \bibinfo{volume}{151}
  (\bibinfo{year}{2018}), \bibinfo{pages}{78--94}.
\newblock


\bibitem[\protect\citeauthoryear{Grover and Leskovec}{Grover and
  Leskovec}{2016}]%
        {grover2016node2vec}
\bibfield{author}{\bibinfo{person}{Aditya Grover} {and} \bibinfo{person}{Jure
  Leskovec}.} \bibinfo{year}{2016}\natexlab{}.
\newblock \showarticletitle{node2vec: Scalable feature learning for networks}.
  In \bibinfo{booktitle}{\emph{SIGKDD}}. \bibinfo{pages}{855--864}.
\newblock


\bibitem[\protect\citeauthoryear{Hagberg, Swart, and S~Chult}{Hagberg
  et~al\mbox{.}}{2008}]%
        {hagberg2008exploring}
\bibfield{author}{\bibinfo{person}{Aric Hagberg}, \bibinfo{person}{Pieter
  Swart}, {and} \bibinfo{person}{Daniel S~Chult}.}
  \bibinfo{year}{2008}\natexlab{}.
\newblock \bibinfo{booktitle}{\emph{Exploring network structure, dynamics, and
  function using NetworkX}}.
\newblock \bibinfo{type}{{T}echnical {R}eport}. \bibinfo{institution}{Los
  Alamos National Lab.(LANL), Los Alamos, NM (United States)}.
\newblock


\bibitem[\protect\citeauthoryear{Herman, Melan{\c{c}}on, and Marshall}{Herman
  et~al\mbox{.}}{2000}]%
        {herman2000graph}
\bibfield{author}{\bibinfo{person}{Ivan Herman}, \bibinfo{person}{Guy
  Melan{\c{c}}on}, {and} \bibinfo{person}{M~Scott Marshall}.}
  \bibinfo{year}{2000}\natexlab{}.
\newblock \showarticletitle{Graph visualization and navigation in information
  visualization: A survey}.
\newblock \bibinfo{journal}{\emph{TVCG}} \bibinfo{volume}{6},
  \bibinfo{number}{1} (\bibinfo{year}{2000}), \bibinfo{pages}{24--43}.
\newblock


\bibitem[\protect\citeauthoryear{Hou, Chen, Wang, and Wei}{Hou
  et~al\mbox{.}}{2021}]%
        {hou2021massively}
\bibfield{author}{\bibinfo{person}{Guanhao Hou}, \bibinfo{person}{Xingguang
  Chen}, \bibinfo{person}{Sibo Wang}, {and} \bibinfo{person}{Zhewei Wei}.}
  \bibinfo{year}{2021}\natexlab{}.
\newblock \showarticletitle{Massively parallel algorithms for personalized
  PageRank}.
\newblock \bibinfo{journal}{\emph{PVLDB}} \bibinfo{volume}{14},
  \bibinfo{number}{9} (\bibinfo{year}{2021}), \bibinfo{pages}{1668--1680}.
\newblock


\bibitem[\protect\citeauthoryear{Hu}{Hu}{2005}]%
        {hu2005efficient}
\bibfield{author}{\bibinfo{person}{Yifan Hu}.} \bibinfo{year}{2005}\natexlab{}.
\newblock \showarticletitle{Efficient, high-quality force-directed graph
  drawing}.
\newblock \bibinfo{journal}{\emph{Mathematica}} \bibinfo{volume}{10},
  \bibinfo{number}{1} (\bibinfo{year}{2005}), \bibinfo{pages}{37--71}.
\newblock


\bibitem[\protect\citeauthoryear{Hu and Shi}{Hu and Shi}{2015}]%
        {hu2015visualizing}
\bibfield{author}{\bibinfo{person}{Yifan Hu} {and} \bibinfo{person}{Lei Shi}.}
  \bibinfo{year}{2015}\natexlab{}.
\newblock \showarticletitle{Visualizing large graphs}.
\newblock \bibinfo{journal}{\emph{Wiley Interdiscip. Rev. Comput. Stat.}}
  \bibinfo{volume}{7}, \bibinfo{number}{2} (\bibinfo{year}{2015}),
  \bibinfo{pages}{115--136}.
\newblock


\bibitem[\protect\citeauthoryear{Huang, Eades, and Hong}{Huang
  et~al\mbox{.}}{2009}]%
        {huang2009measuring}
\bibfield{author}{\bibinfo{person}{Weidong Huang}, \bibinfo{person}{Peter
  Eades}, {and} \bibinfo{person}{Seok-Hee Hong}.}
  \bibinfo{year}{2009}\natexlab{}.
\newblock \showarticletitle{Measuring effectiveness of graph visualizations: A
  cognitive load perspective}.
\newblock \bibinfo{journal}{\emph{Inf. Vis.}} \bibinfo{volume}{8},
  \bibinfo{number}{3} (\bibinfo{year}{2009}), \bibinfo{pages}{139--152}.
\newblock


\bibitem[\protect\citeauthoryear{Jacomy, Venturini, Heymann, and
  Bastian}{Jacomy et~al\mbox{.}}{2014}]%
        {jacomy2014forceatlas2}
\bibfield{author}{\bibinfo{person}{Mathieu Jacomy}, \bibinfo{person}{Tommaso
  Venturini}, \bibinfo{person}{Sebastien Heymann}, {and}
  \bibinfo{person}{Mathieu Bastian}.} \bibinfo{year}{2014}\natexlab{}.
\newblock \showarticletitle{ForceAtlas2, a continuous graph layout algorithm
  for handy network visualization designed for the Gephi software}.
\newblock \bibinfo{journal}{\emph{PloS one}} \bibinfo{volume}{9},
  \bibinfo{number}{6} (\bibinfo{year}{2014}), \bibinfo{pages}{e98679}.
\newblock


\bibitem[\protect\citeauthoryear{Jeh and Widom}{Jeh and Widom}{2002}]%
        {jeh2002simrank}
\bibfield{author}{\bibinfo{person}{Glen Jeh} {and} \bibinfo{person}{Jennifer
  Widom}.} \bibinfo{year}{2002}\natexlab{}.
\newblock \showarticletitle{Simrank: a measure of structural-context
  similarity}. In \bibinfo{booktitle}{\emph{SIGKDD}}.
  \bibinfo{pages}{538--543}.
\newblock


\bibitem[\protect\citeauthoryear{Jung, Park, Lee, and Kang}{Jung
  et~al\mbox{.}}{2017}]%
        {jung2017bepi}
\bibfield{author}{\bibinfo{person}{Jinhong Jung}, \bibinfo{person}{Namyong
  Park}, \bibinfo{person}{Sael Lee}, {and} \bibinfo{person}{U Kang}.}
  \bibinfo{year}{2017}\natexlab{}.
\newblock \showarticletitle{Bepi: Fast and memory-efficient method for
  billion-scale random walk with restart}. In
  \bibinfo{booktitle}{\emph{SIGMOD}}. \bibinfo{pages}{789--804}.
\newblock


\bibitem[\protect\citeauthoryear{Kamada, Kawai, et~al\mbox{.}}{Kamada
  et~al\mbox{.}}{1989}]%
        {kamada1989algorithm}
\bibfield{author}{\bibinfo{person}{Tomihisa Kamada}, \bibinfo{person}{Satoru
  Kawai}, {et~al\mbox{.}}} \bibinfo{year}{1989}\natexlab{}.
\newblock \showarticletitle{An algorithm for drawing general undirected
  graphs}.
\newblock \bibinfo{journal}{\emph{Inform. Process. Lett.}}
  \bibinfo{volume}{31}, \bibinfo{number}{1} (\bibinfo{year}{1989}),
  \bibinfo{pages}{7--15}.
\newblock


\bibitem[\protect\citeauthoryear{Klammler, Mchedlidze, and Pak}{Klammler
  et~al\mbox{.}}{2018}]%
        {klammler2018aesthetic}
\bibfield{author}{\bibinfo{person}{Moritz Klammler}, \bibinfo{person}{Tamara
  Mchedlidze}, {and} \bibinfo{person}{Alexey Pak}.}
  \bibinfo{year}{2018}\natexlab{}.
\newblock \showarticletitle{Aesthetic discrimination of graph layouts}. In
  \bibinfo{booktitle}{\emph{GD}}. \bibinfo{pages}{169--184}.
\newblock


\bibitem[\protect\citeauthoryear{Kunegis}{Kunegis}{2013}]%
        {konect}
\bibfield{author}{\bibinfo{person}{J\'{e}r\^{o}me Kunegis}.}
  \bibinfo{year}{2013}\natexlab{}.
\newblock \showarticletitle{{KONECT} -- {The} {Koblenz} {Network}
  {Collection}}. In \bibinfo{booktitle}{\emph{WWW}}.
\newblock


\bibitem[\protect\citeauthoryear{Kwak, Lee, Park, and Moon}{Kwak
  et~al\mbox{.}}{2010}]%
        {kwak2010twitter}
\bibfield{author}{\bibinfo{person}{Haewoon Kwak}, \bibinfo{person}{Changhyun
  Lee}, \bibinfo{person}{Hosung Park}, {and} \bibinfo{person}{Sue Moon}.}
  \bibinfo{year}{2010}\natexlab{}.
\newblock \showarticletitle{What is Twitter, a social network or a news
  media?}. In \bibinfo{booktitle}{\emph{WWW}}. \bibinfo{pages}{591--600}.
\newblock


\bibitem[\protect\citeauthoryear{Lee, Plaisant, Parr, Fekete, and Henry}{Lee
  et~al\mbox{.}}{2006}]%
        {lee2006task}
\bibfield{author}{\bibinfo{person}{Bongshin Lee}, \bibinfo{person}{Catherine
  Plaisant}, \bibinfo{person}{Cynthia~Sims Parr}, \bibinfo{person}{Jean-Daniel
  Fekete}, {and} \bibinfo{person}{Nathalie Henry}.}
  \bibinfo{year}{2006}\natexlab{}.
\newblock \showarticletitle{Task taxonomy for graph visualization}. In
  \bibinfo{booktitle}{\emph{BELIV}}.
\newblock


\bibitem[\protect\citeauthoryear{Leskovec and Krevl}{Leskovec and
  Krevl}{2014}]%
        {snapnets}
\bibfield{author}{\bibinfo{person}{Jure Leskovec} {and} \bibinfo{person}{Andrej
  Krevl}.} \bibinfo{year}{2014}\natexlab{}.
\newblock \bibinfo{title}{{SNAP Datasets}: {Stanford} Large Network Dataset
  Collection}.
\newblock \bibinfo{howpublished}{\url{http://snap.stanford.edu/data}}.
\newblock


\bibitem[\protect\citeauthoryear{Lin, Wong, Xie, and Wei}{Lin
  et~al\mbox{.}}{2020}]%
        {lin2020index}
\bibfield{author}{\bibinfo{person}{Dandan Lin}, \bibinfo{person}{Raymond
  Chi-Wing Wong}, \bibinfo{person}{Min Xie}, {and}
  \bibinfo{person}{Victor~Junqiu Wei}.} \bibinfo{year}{2020}\natexlab{}.
\newblock \showarticletitle{Index-free approach with theoretical guarantee for
  efficient random walk with restart query}. In
  \bibinfo{booktitle}{\emph{ICDE}}. \bibinfo{pages}{913--924}.
\newblock


\bibitem[\protect\citeauthoryear{Litvak, Scheinhardt, and Volkovich}{Litvak
  et~al\mbox{.}}{2007}]%
        {litvak2007degree}
\bibfield{author}{\bibinfo{person}{Nelly Litvak}, \bibinfo{person}{Werner~RW
  Scheinhardt}, {and} \bibinfo{person}{Yana Volkovich}.}
  \bibinfo{year}{2007}\natexlab{}.
\newblock \showarticletitle{In-degree and PageRank: why do they follow similar
  power laws?}
\newblock \bibinfo{journal}{\emph{Internet Math.}} \bibinfo{volume}{4},
  \bibinfo{number}{2-3} (\bibinfo{year}{2007}), \bibinfo{pages}{175--198}.
\newblock


\bibitem[\protect\citeauthoryear{Lofgren, Banerjee, and Goel}{Lofgren
  et~al\mbox{.}}{2016}]%
        {lofgren2016personalized}
\bibfield{author}{\bibinfo{person}{Peter Lofgren}, \bibinfo{person}{Siddhartha
  Banerjee}, {and} \bibinfo{person}{Ashish Goel}.}
  \bibinfo{year}{2016}\natexlab{}.
\newblock \showarticletitle{Personalized pagerank estimation and search: A
  bidirectional approach}. In \bibinfo{booktitle}{\emph{WSDM}}.
  \bibinfo{pages}{163--172}.
\newblock


\bibitem[\protect\citeauthoryear{Lofgren and Goel}{Lofgren and Goel}{2013}]%
        {lofgren2013personalized}
\bibfield{author}{\bibinfo{person}{Peter Lofgren} {and} \bibinfo{person}{Ashish
  Goel}.} \bibinfo{year}{2013}\natexlab{}.
\newblock \showarticletitle{Personalized pagerank to a target node}.
\newblock \bibinfo{journal}{\emph{arXiv}} (\bibinfo{year}{2013}).
\newblock


\bibitem[\protect\citeauthoryear{Luo, Xiao, Lin, and Kao}{Luo
  et~al\mbox{.}}{2019}]%
        {luo2019baton}
\bibfield{author}{\bibinfo{person}{Siqiang Luo}, \bibinfo{person}{Xiaokui
  Xiao}, \bibinfo{person}{Wenqing Lin}, {and} \bibinfo{person}{Ben Kao}.}
  \bibinfo{year}{2019}\natexlab{}.
\newblock \showarticletitle{Baton: Batch one-hop personalized pageranks with
  efficiency and accuracy}.
\newblock \bibinfo{journal}{\emph{TKDE}} \bibinfo{volume}{32},
  \bibinfo{number}{10} (\bibinfo{year}{2019}), \bibinfo{pages}{1897--1908}.
\newblock


\bibitem[\protect\citeauthoryear{Martin, Brown, Klavans, and Boyack}{Martin
  et~al\mbox{.}}{2011}]%
        {martin2011openord}
\bibfield{author}{\bibinfo{person}{Shawn Martin}, \bibinfo{person}{W~Michael
  Brown}, \bibinfo{person}{Richard Klavans}, {and} \bibinfo{person}{Kevin~W
  Boyack}.} \bibinfo{year}{2011}\natexlab{}.
\newblock \showarticletitle{OpenOrd: an open-source toolbox for large graph
  layout}. In \bibinfo{booktitle}{\emph{VDA}}, Vol.~\bibinfo{volume}{7868}.
  \bibinfo{pages}{786806}.
\newblock


\bibitem[\protect\citeauthoryear{Meyerhenke, N{\"o}llenburg, and
  Schulz}{Meyerhenke et~al\mbox{.}}{2017}]%
        {meyerhenke2017drawing}
\bibfield{author}{\bibinfo{person}{Henning Meyerhenke}, \bibinfo{person}{Martin
  N{\"o}llenburg}, {and} \bibinfo{person}{Christian Schulz}.}
  \bibinfo{year}{2017}\natexlab{}.
\newblock \showarticletitle{Drawing large graphs by multilevel maxent-stress
  optimization}.
\newblock \bibinfo{journal}{\emph{TVCG}} \bibinfo{volume}{24},
  \bibinfo{number}{5} (\bibinfo{year}{2017}), \bibinfo{pages}{1814--1827}.
\newblock


\bibitem[\protect\citeauthoryear{Meyerhenke, Sanders, and Schulz}{Meyerhenke
  et~al\mbox{.}}{2014}]%
        {meyerhenke2014partitioning}
\bibfield{author}{\bibinfo{person}{Henning Meyerhenke}, \bibinfo{person}{Peter
  Sanders}, {and} \bibinfo{person}{Christian Schulz}.}
  \bibinfo{year}{2014}\natexlab{}.
\newblock \showarticletitle{Partitioning complex networks via size-constrained
  clustering}. In \bibinfo{booktitle}{\emph{SEA}}. \bibinfo{pages}{351--363}.
\newblock


\bibitem[\protect\citeauthoryear{Newman}{Newman}{2006}]%
        {newman2006modularity}
\bibfield{author}{\bibinfo{person}{Mark~EJ Newman}.}
  \bibinfo{year}{2006}\natexlab{}.
\newblock \showarticletitle{Modularity and community structure in networks}.
\newblock \bibinfo{journal}{\emph{PNAS}} \bibinfo{volume}{103},
  \bibinfo{number}{23} (\bibinfo{year}{2006}), \bibinfo{pages}{8577--8582}.
\newblock


\bibitem[\protect\citeauthoryear{Niculescu and Persson}{Niculescu and
  Persson}{2006}]%
        {niculescu2006convex}
\bibfield{author}{\bibinfo{person}{Constantin Niculescu} {and}
  \bibinfo{person}{Lars-Erik Persson}.} \bibinfo{year}{2006}\natexlab{}.
\newblock \bibinfo{booktitle}{\emph{Convex functions and their applications}}.
\newblock \bibinfo{publisher}{Springer}.
\newblock


\bibitem[\protect\citeauthoryear{Noack}{Noack}{2005}]%
        {noack2005energy}
\bibfield{author}{\bibinfo{person}{Andreas Noack}.}
  \bibinfo{year}{2005}\natexlab{}.
\newblock \showarticletitle{Energy-based clustering of graphs with nonuniform
  degrees}. In \bibinfo{booktitle}{\emph{GD}}. \bibinfo{pages}{309--320}.
\newblock


\bibitem[\protect\citeauthoryear{Noack}{Noack}{2007}]%
        {noack2007unified}
\bibfield{author}{\bibinfo{person}{Andreas Noack}.}
  \bibinfo{year}{2007}\natexlab{}.
\newblock \showarticletitle{Unified quality measures for clusterings, layouts,
  and orderings of graphs, and their application as software design criteria}.
\newblock  (\bibinfo{year}{2007}).
\newblock


\bibitem[\protect\citeauthoryear{Page, Brin, Motwani, and Winograd}{Page
  et~al\mbox{.}}{1999}]%
        {page1999pagerank}
\bibfield{author}{\bibinfo{person}{Lawrence Page}, \bibinfo{person}{Sergey
  Brin}, \bibinfo{person}{Rajeev Motwani}, {and} \bibinfo{person}{Terry
  Winograd}.} \bibinfo{year}{1999}\natexlab{}.
\newblock \bibinfo{booktitle}{\emph{The PageRank citation ranking: Bringing
  order to the web.}}
\newblock \bibinfo{type}{{T}echnical {R}eport}. \bibinfo{institution}{Stanford
  InfoLab}.
\newblock


\bibitem[\protect\citeauthoryear{Perozzi, Al-Rfou, and Skiena}{Perozzi
  et~al\mbox{.}}{2014}]%
        {perozzi2014deepwalk}
\bibfield{author}{\bibinfo{person}{Bryan Perozzi}, \bibinfo{person}{Rami
  Al-Rfou}, {and} \bibinfo{person}{Steven Skiena}.}
  \bibinfo{year}{2014}\natexlab{}.
\newblock \showarticletitle{Deepwalk: Online learning of social
  representations}. In \bibinfo{booktitle}{\emph{SIGKDD}}.
  \bibinfo{pages}{701--710}.
\newblock


\bibitem[\protect\citeauthoryear{Purchase}{Purchase}{2002}]%
        {purchase2002metrics}
\bibfield{author}{\bibinfo{person}{Helen~C Purchase}.}
  \bibinfo{year}{2002}\natexlab{}.
\newblock \showarticletitle{Metrics for graph drawing aesthetics}.
\newblock \bibinfo{journal}{\emph{JVLC}} \bibinfo{volume}{13},
  \bibinfo{number}{5} (\bibinfo{year}{2002}), \bibinfo{pages}{501--516}.
\newblock


\bibitem[\protect\citeauthoryear{Purchase, Carrington, and Allder}{Purchase
  et~al\mbox{.}}{2002}]%
        {purchase2002empirical}
\bibfield{author}{\bibinfo{person}{Helen~C Purchase}, \bibinfo{person}{David
  Carrington}, {and} \bibinfo{person}{Jo-Anne Allder}.}
  \bibinfo{year}{2002}\natexlab{}.
\newblock \showarticletitle{Empirical evaluation of aesthetics-based graph
  layout}.
\newblock \bibinfo{journal}{\emph{Empir. Softw. Eng.}} \bibinfo{volume}{7},
  \bibinfo{number}{3} (\bibinfo{year}{2002}), \bibinfo{pages}{233--255}.
\newblock


\bibitem[\protect\citeauthoryear{Rodrigues, Tong, Traina, Faloutsos, and
  Leskovec}{Rodrigues et~al\mbox{.}}{2015}]%
        {rodrigues2015gmine}
\bibfield{author}{\bibinfo{person}{Jose Rodrigues}, \bibinfo{person}{Hanghang
  Tong}, \bibinfo{person}{Agma Traina}, \bibinfo{person}{Christos Faloutsos},
  {and} \bibinfo{person}{Jure Leskovec}.} \bibinfo{year}{2015}\natexlab{}.
\newblock \showarticletitle{Gmine: a system for scalable, interactive graph
  visualization and mining}.
\newblock \bibinfo{journal}{\emph{PVLDB}} \bibinfo{number}{4}
  (\bibinfo{year}{2015}), \bibinfo{pages}{1195–1198}.
\newblock


\bibitem[\protect\citeauthoryear{Roweis and Saul}{Roweis and Saul}{2000}]%
        {roweis2000nonlinear}
\bibfield{author}{\bibinfo{person}{Sam~T Roweis} {and}
  \bibinfo{person}{Lawrence~K Saul}.} \bibinfo{year}{2000}\natexlab{}.
\newblock \showarticletitle{Nonlinear dimensionality reduction by locally
  linear embedding}.
\newblock \bibinfo{journal}{\emph{Science}} \bibinfo{volume}{290},
  \bibinfo{number}{5500} (\bibinfo{year}{2000}), \bibinfo{pages}{2323--2326}.
\newblock


\bibitem[\protect\citeauthoryear{Sahu, Mhedhbi, Salihoglu, Lin, and
  {\"O}zsu}{Sahu et~al\mbox{.}}{2017}]%
        {sahu2017ubiquity}
\bibfield{author}{\bibinfo{person}{Siddhartha Sahu}, \bibinfo{person}{Amine
  Mhedhbi}, \bibinfo{person}{Semih Salihoglu}, \bibinfo{person}{Jimmy Lin},
  {and} \bibinfo{person}{M~Tamer {\"O}zsu}.} \bibinfo{year}{2017}\natexlab{}.
\newblock \showarticletitle{The ubiquity of large graphs and surprising
  challenges of graph processing}.
\newblock \bibinfo{journal}{\emph{PVLDB}} \bibinfo{volume}{11},
  \bibinfo{number}{4} (\bibinfo{year}{2017}), \bibinfo{pages}{420--431}.
\newblock


\bibitem[\protect\citeauthoryear{Shannon, Markiel, Ozier, Baliga, Wang, Ramage,
  Amin, Schwikowski, and Ideker}{Shannon et~al\mbox{.}}{2003}]%
        {shannon2003cytoscape}
\bibfield{author}{\bibinfo{person}{Paul Shannon}, \bibinfo{person}{Andrew
  Markiel}, \bibinfo{person}{Owen Ozier}, \bibinfo{person}{Nitin~S Baliga},
  \bibinfo{person}{Jonathan~T Wang}, \bibinfo{person}{Daniel Ramage},
  \bibinfo{person}{Nada Amin}, \bibinfo{person}{Benno Schwikowski}, {and}
  \bibinfo{person}{Trey Ideker}.} \bibinfo{year}{2003}\natexlab{}.
\newblock \showarticletitle{Cytoscape: a software environment for integrated
  models of biomolecular interaction networks}.
\newblock \bibinfo{journal}{\emph{Genome research}} \bibinfo{volume}{13},
  \bibinfo{number}{11} (\bibinfo{year}{2003}), \bibinfo{pages}{2498--2504}.
\newblock


\bibitem[\protect\citeauthoryear{Shi, Yang, Jin, Xiao, and Yang}{Shi
  et~al\mbox{.}}{2019}]%
        {shi2019realtime}
\bibfield{author}{\bibinfo{person}{Jieming Shi}, \bibinfo{person}{Renchi Yang},
  \bibinfo{person}{Tianyuan Jin}, \bibinfo{person}{Xiaokui Xiao}, {and}
  \bibinfo{person}{Yin Yang}.} \bibinfo{year}{2019}\natexlab{}.
\newblock \showarticletitle{Realtime top-k personalized pagerank over large
  graphs on gpus}.
\newblock \bibinfo{journal}{\emph{PVLDB}} \bibinfo{volume}{13},
  \bibinfo{number}{1} (\bibinfo{year}{2019}), \bibinfo{pages}{15--28}.
\newblock


\bibitem[\protect\citeauthoryear{Shi, Cao, Liu, Qian, Tan, Wang, Sun, and
  Lin}{Shi et~al\mbox{.}}{2009}]%
        {shi2009himap}
\bibfield{author}{\bibinfo{person}{Lei Shi}, \bibinfo{person}{Nan Cao},
  \bibinfo{person}{Shixia Liu}, \bibinfo{person}{Weihong Qian},
  \bibinfo{person}{Li Tan}, \bibinfo{person}{Guodong Wang},
  \bibinfo{person}{Jimeng Sun}, {and} \bibinfo{person}{Ching-Yung Lin}.}
  \bibinfo{year}{2009}\natexlab{}.
\newblock \showarticletitle{HiMap: Adaptive visualization of large-scale online
  social networks}. In \bibinfo{booktitle}{\emph{PacificVis}}.
  \bibinfo{pages}{41--48}.
\newblock


\bibitem[\protect\citeauthoryear{Shin, Jung, Lee, and Kang}{Shin
  et~al\mbox{.}}{2015}]%
        {shin2015bear}
\bibfield{author}{\bibinfo{person}{Kijung Shin}, \bibinfo{person}{Jinhong
  Jung}, \bibinfo{person}{Sael Lee}, {and} \bibinfo{person}{U Kang}.}
  \bibinfo{year}{2015}\natexlab{}.
\newblock \showarticletitle{Bear: Block elimination approach for random walk
  with restart on large graphs}. In \bibinfo{booktitle}{\emph{SIGMOD}}.
  \bibinfo{pages}{1571--1585}.
\newblock


\bibitem[\protect\citeauthoryear{Sokal}{Sokal}{1958}]%
        {sokal1958statistical}
\bibfield{author}{\bibinfo{person}{Robert~R Sokal}.}
  \bibinfo{year}{1958}\natexlab{}.
\newblock \showarticletitle{A statistical method for evaluating systematic
  relationships.}
\newblock \bibinfo{journal}{\emph{Univ. Kansas, Sci. Bull.}}
  \bibinfo{volume}{38} (\bibinfo{year}{1958}), \bibinfo{pages}{1409--1438}.
\newblock


\bibitem[\protect\citeauthoryear{Tang, Qu, Wang, Zhang, Yan, and Mei}{Tang
  et~al\mbox{.}}{2015}]%
        {tang2015line}
\bibfield{author}{\bibinfo{person}{Jian Tang}, \bibinfo{person}{Meng Qu},
  \bibinfo{person}{Mingzhe Wang}, \bibinfo{person}{Ming Zhang},
  \bibinfo{person}{Jun Yan}, {and} \bibinfo{person}{Qiaozhu Mei}.}
  \bibinfo{year}{2015}\natexlab{}.
\newblock \showarticletitle{Line: Large-scale information network embedding}.
  In \bibinfo{booktitle}{\emph{WWW}}. \bibinfo{pages}{1067--1077}.
\newblock


\bibitem[\protect\citeauthoryear{Taylor and Rodgers}{Taylor and
  Rodgers}{2005}]%
        {taylor2005applying}
\bibfield{author}{\bibinfo{person}{Martyn Taylor} {and} \bibinfo{person}{Peter
  Rodgers}.} \bibinfo{year}{2005}\natexlab{}.
\newblock \showarticletitle{Applying graphical design techniques to graph
  visualisation}. In \bibinfo{booktitle}{\emph{Inf. Vis.}}
  \bibinfo{pages}{651--656}.
\newblock


\bibitem[\protect\citeauthoryear{Tong, Faloutsos, and Pan}{Tong
  et~al\mbox{.}}{2006}]%
        {tong2006fast}
\bibfield{author}{\bibinfo{person}{Hanghang Tong}, \bibinfo{person}{Christos
  Faloutsos}, {and} \bibinfo{person}{Jia-Yu Pan}.}
  \bibinfo{year}{2006}\natexlab{}.
\newblock \showarticletitle{Fast random walk with restart and its
  applications}. In \bibinfo{booktitle}{\emph{ICDM}}.
  \bibinfo{pages}{613--622}.
\newblock


\bibitem[\protect\citeauthoryear{Torgerson}{Torgerson}{1952}]%
        {torgerson1952multidimensional}
\bibfield{author}{\bibinfo{person}{Warren~S Torgerson}.}
  \bibinfo{year}{1952}\natexlab{}.
\newblock \showarticletitle{Multidimensional scaling: I. Theory and method}.
\newblock \bibinfo{journal}{\emph{Psychometrika}} \bibinfo{volume}{17},
  \bibinfo{number}{4} (\bibinfo{year}{1952}), \bibinfo{pages}{401--419}.
\newblock


\bibitem[\protect\citeauthoryear{Van~der Maaten and Hinton}{Van~der Maaten and
  Hinton}{2008}]%
        {van2008visualizing}
\bibfield{author}{\bibinfo{person}{Laurens Van~der Maaten} {and}
  \bibinfo{person}{Geoffrey Hinton}.} \bibinfo{year}{2008}\natexlab{}.
\newblock \showarticletitle{Visualizing data using t-SNE.}
\newblock \bibinfo{journal}{\emph{JMLR}}  \bibinfo{volume}{9}
  (\bibinfo{year}{2008}), \bibinfo{pages}{2579--2605}.
\newblock


\bibitem[\protect\citeauthoryear{Von~Landesberger, Kuijper, Schreck,
  Kohlhammer, van Wijk, Fekete, and Fellner}{Von~Landesberger
  et~al\mbox{.}}{2011}]%
        {von2011visual}
\bibfield{author}{\bibinfo{person}{Tatiana Von~Landesberger},
  \bibinfo{person}{Arjan Kuijper}, \bibinfo{person}{Tobias Schreck},
  \bibinfo{person}{J{\"o}rn Kohlhammer}, \bibinfo{person}{Jarke~J van Wijk},
  \bibinfo{person}{J-D Fekete}, {and} \bibinfo{person}{Dieter~W Fellner}.}
  \bibinfo{year}{2011}\natexlab{}.
\newblock \showarticletitle{Visual analysis of large graphs: state-of-the-art
  and future research challenges}. In \bibinfo{booktitle}{\emph{CGF}},
  Vol.~\bibinfo{volume}{30}. \bibinfo{pages}{1719--1749}.
\newblock


\bibitem[\protect\citeauthoryear{Wang, Cui, and Zhu}{Wang
  et~al\mbox{.}}{2016a}]%
        {wang2016structural}
\bibfield{author}{\bibinfo{person}{Daixin Wang}, \bibinfo{person}{Peng Cui},
  {and} \bibinfo{person}{Wenwu Zhu}.} \bibinfo{year}{2016}\natexlab{a}.
\newblock \showarticletitle{Structural deep network embedding}. In
  \bibinfo{booktitle}{\emph{SIGKDD}}. \bibinfo{pages}{1225--1234}.
\newblock


\bibitem[\protect\citeauthoryear{Wang, Wei, Gan, Wang, and Huang}{Wang
  et~al\mbox{.}}{2020}]%
        {wang2020personalized}
\bibfield{author}{\bibinfo{person}{Hanzhi Wang}, \bibinfo{person}{Zhewei Wei},
  \bibinfo{person}{Junhao Gan}, \bibinfo{person}{Sibo Wang}, {and}
  \bibinfo{person}{Zengfeng Huang}.} \bibinfo{year}{2020}\natexlab{}.
\newblock \showarticletitle{Personalized PageRank to a Target Node, Revisited}.
  In \bibinfo{booktitle}{\emph{SIGKDD}}. \bibinfo{pages}{657--667}.
\newblock


\bibitem[\protect\citeauthoryear{Wang, Tang, Xiao, Yang, and Li}{Wang
  et~al\mbox{.}}{2016b}]%
        {wang2016hubppr}
\bibfield{author}{\bibinfo{person}{Sibo Wang}, \bibinfo{person}{Youze Tang},
  \bibinfo{person}{Xiaokui Xiao}, \bibinfo{person}{Yin Yang}, {and}
  \bibinfo{person}{Zengxiang Li}.} \bibinfo{year}{2016}\natexlab{b}.
\newblock \showarticletitle{HubPPR: effective indexing for approximate
  personalized pagerank}.
\newblock \bibinfo{journal}{\emph{PVLDB}} \bibinfo{volume}{10},
  \bibinfo{number}{3} (\bibinfo{year}{2016}), \bibinfo{pages}{205--216}.
\newblock


\bibitem[\protect\citeauthoryear{Wang, Yang, Wang, Xiao, Wei, Lin, Yang, and
  Tang}{Wang et~al\mbox{.}}{2019}]%
        {wang2019efficient}
\bibfield{author}{\bibinfo{person}{Sibo Wang}, \bibinfo{person}{Renchi Yang},
  \bibinfo{person}{Runhui Wang}, \bibinfo{person}{Xiaokui Xiao},
  \bibinfo{person}{Zhewei Wei}, \bibinfo{person}{Wenqing Lin},
  \bibinfo{person}{Yin Yang}, {and} \bibinfo{person}{Nan Tang}.}
  \bibinfo{year}{2019}\natexlab{}.
\newblock \showarticletitle{Efficient algorithms for approximate single-source
  personalized pagerank queries}.
\newblock \bibinfo{journal}{\emph{TODS}} \bibinfo{volume}{44},
  \bibinfo{number}{4} (\bibinfo{year}{2019}), \bibinfo{pages}{1--37}.
\newblock


\bibitem[\protect\citeauthoryear{Wang, Yang, Xiao, Wei, and Yang}{Wang
  et~al\mbox{.}}{2017b}]%
        {wang2017fora}
\bibfield{author}{\bibinfo{person}{Sibo Wang}, \bibinfo{person}{Renchi Yang},
  \bibinfo{person}{Xiaokui Xiao}, \bibinfo{person}{Zhewei Wei}, {and}
  \bibinfo{person}{Yin Yang}.} \bibinfo{year}{2017}\natexlab{b}.
\newblock \showarticletitle{FORA: simple and effective approximate
  single-source personalized pagerank}. In \bibinfo{booktitle}{\emph{SIGKDD}}.
  \bibinfo{pages}{505--514}.
\newblock


\bibitem[\protect\citeauthoryear{Wang, Wang, Sun, Zhu, Lu, Fu, Sedlmair,
  Deussen, and Chen}{Wang et~al\mbox{.}}{2017a}]%
        {wang2017revisiting}
\bibfield{author}{\bibinfo{person}{Yunhai Wang}, \bibinfo{person}{Yanyan Wang},
  \bibinfo{person}{Yinqi Sun}, \bibinfo{person}{Lifeng Zhu},
  \bibinfo{person}{Kecheng Lu}, \bibinfo{person}{Chi-Wing Fu},
  \bibinfo{person}{Michael Sedlmair}, \bibinfo{person}{Oliver Deussen}, {and}
  \bibinfo{person}{Baoquan Chen}.} \bibinfo{year}{2017}\natexlab{a}.
\newblock \showarticletitle{Revisiting stress majorization as a unified
  framework for interactive constrained graph visualization}.
\newblock \bibinfo{journal}{\emph{TVCG}} (\bibinfo{year}{2017}).
\newblock


\bibitem[\protect\citeauthoryear{Wu, Gan, Wei, and Zhang}{Wu
  et~al\mbox{.}}{2021}]%
        {wu2021unifying}
\bibfield{author}{\bibinfo{person}{Hao Wu}, \bibinfo{person}{Junhao Gan},
  \bibinfo{person}{Zhewei Wei}, {and} \bibinfo{person}{Rui Zhang}.}
  \bibinfo{year}{2021}\natexlab{}.
\newblock \showarticletitle{Unifying the Global and Local Approaches: An
  Efficient Power Iteration with Forward Push}. In
  \bibinfo{booktitle}{\emph{SIGMOD}}. \bibinfo{pages}{1996--2008}.
\newblock


\bibitem[\protect\citeauthoryear{Wu, Cao, Archambault, Shen, Qu, and Cui}{Wu
  et~al\mbox{.}}{2016}]%
        {wu2016evaluation}
\bibfield{author}{\bibinfo{person}{Yanhong Wu}, \bibinfo{person}{Nan Cao},
  \bibinfo{person}{Daniel Archambault}, \bibinfo{person}{Qiaomu Shen},
  \bibinfo{person}{Huamin Qu}, {and} \bibinfo{person}{Weiwei Cui}.}
  \bibinfo{year}{2016}\natexlab{}.
\newblock \showarticletitle{Evaluation of graph sampling: A visualization
  perspective}.
\newblock \bibinfo{journal}{\emph{TVCG}} (\bibinfo{year}{2016}).
\newblock


\bibitem[\protect\citeauthoryear{Xu, Williams, Hong, Liu, and Zhang}{Xu
  et~al\mbox{.}}{2009}]%
        {xu2009semi}
\bibfield{author}{\bibinfo{person}{Kai Xu}, \bibinfo{person}{Rohan Williams},
  \bibinfo{person}{Seok-Hee Hong}, \bibinfo{person}{Qing Liu}, {and}
  \bibinfo{person}{Ji Zhang}.} \bibinfo{year}{2009}\natexlab{}.
\newblock \showarticletitle{Semi-bipartite graph visualization for gene
  ontology networks}. In \bibinfo{booktitle}{\emph{GD}}.
\newblock


\bibitem[\protect\citeauthoryear{Yang, Shi, Xiao, Yang, and Bhowmick}{Yang
  et~al\mbox{.}}{2020}]%
        {yang2020homogeneous}
\bibfield{author}{\bibinfo{person}{Renchi Yang}, \bibinfo{person}{Jieming Shi},
  \bibinfo{person}{Xiaokui Xiao}, \bibinfo{person}{Yin Yang}, {and}
  \bibinfo{person}{Sourav~S Bhowmick}.} \bibinfo{year}{2020}\natexlab{}.
\newblock \showarticletitle{Homogeneous network embedding for massive graphs
  via reweighted personalized PageRank}.
\newblock \bibinfo{journal}{\emph{PVLDB}} \bibinfo{volume}{13},
  \bibinfo{number}{5} (\bibinfo{year}{2020}), \bibinfo{pages}{670--683}.
\newblock


\bibitem[\protect\citeauthoryear{Yoon, Jung, and Kang}{Yoon
  et~al\mbox{.}}{2018}]%
        {yoon2018tpa}
\bibfield{author}{\bibinfo{person}{Minji Yoon}, \bibinfo{person}{Jinhong Jung},
  {and} \bibinfo{person}{U Kang}.} \bibinfo{year}{2018}\natexlab{}.
\newblock \showarticletitle{Tpa: Fast, scalable, and accurate method for
  approximate random walk with restart on billion scale graphs}. In
  \bibinfo{booktitle}{\emph{ICDE}}. \bibinfo{pages}{1132--1143}.
\newblock


\bibitem[\protect\citeauthoryear{Zhuang, Chang, and Li}{Zhuang
  et~al\mbox{.}}{2019}]%
        {zhuang2019dynamo}
\bibfield{author}{\bibinfo{person}{Di Zhuang}, \bibinfo{person}{J~Morris
  Chang}, {and} \bibinfo{person}{Mingchen Li}.}
  \bibinfo{year}{2019}\natexlab{}.
\newblock \showarticletitle{DynaMo: Dynamic community detection by
  incrementally maximizing modularity}.
\newblock \bibinfo{journal}{\emph{TKDE}} (\bibinfo{year}{2019}).
\newblock


\end{thebibliography}

\balance
\appendix
\section{Appendix}\label{sec:appendix}

\subsection{Algorithmic Details}\label{sec:reproducibility}

\stitle{\louvainplusb}
To construct supergraph hierarchy, a solution is directly using multilevel community detection algorithms such as \louvain~\cite{blondel2008fast}, which merges well-connected nodes based on modularity optimization~\cite{newman2006modularity}. However, \louvain boils down to two defects for visualization: (i) the number of communities (supernodes) in the highest level is too large as there is no merge that increases the modularity after some point, which causes visual clutter; (ii) the number of nodes in the communities are imbalanced. Specifically, low-level supernodes tend to contain many children, which leads to visual cluster while high-level supernodes usually contain only a few children and thus provide very limited structural information about the graph. To fix these, we extend \louvain to \louvainplus.
Here we ignore the direction in the raw graph and take the undirected graph as the input for community detection.
To generate a level-$(\ell+1)$ supergraph $G_{\ell+1}$, the detailed clustering strategy is that
either (i) directly merge supernode $\mathcal{S}$ to its neighboring supernode $\mathcal{T}$ if $\mathcal{T}$ is the only neighbor; or (ii) merge $\mathcal{S}$ to its neighboring supernode $\mathcal{T}$ with the largest modularity gain $\widehat{Q} (\mathcal{S},\mathcal{T})$ if the size of $\mathcal{T}$ after this merge is less than $k$. The modularity gain $\widehat{Q} (\mathcal{S},\mathcal{T})$ after merging level-$(\ell+1)$ supernodes $\mathcal{S}$ and $\mathcal{T}$ is defined as
\begin{equation*} \label{eq:deltaQsuper}
    \widehat{Q} (\mathcal{S},\mathcal{T}) = \sum_{\mathcal{V}_{i}\in \mathcal{S}} q(\mathcal{V}_i,\mathcal{T}). 
\end{equation*}
Note that $q(\mathcal{V}_i,\mathcal{T})$ is the modularity change of $G_{\ell}$ after moving a node $\mathcal{V}_i$ into supernode $\mathcal{T}$, which is defined by
\begin{equation*}\label{eq:deltaQ}
\begin{split}
 q(\mathcal{V}_i,\mathcal{T}) = & \left[\frac{w(\mathcal{T})+w_{cr}(\mathcal{V}_i,\mathcal{T})}{2m}-\left(\frac{w_{in}(\mathcal{T})+w_{in}(\mathcal{V}_i)}{2m}\right)^2\right]\\
- & \left[\frac{w(\mathcal{T})}{2m}-\left(\frac{w_{in}(\mathcal{T})}{2m}\right)^2-\left(\frac{w_{in}(\mathcal{V}_i)}{2m}\right)^2\right],
\end{split}
\end{equation*}
where $w(\mathcal{T})$ is the number of leaf edges with both endpoints within $\mathcal{T}$, $w_{cr}(\mathcal{V}_i,\mathcal{T})$ is the number of leaf edges crossing supernodes $\mathcal{V}_i$ and $\mathcal{T}$, and $w_{in}(\mathcal{T})$ is the number of leaf edges incident to $\mathcal{T}$. 

\stitle{Stress Majorization}{
Stress majorization~\cite{gansner2004graph} is adapted to enable efficient optimization, Particularly, Eq.~\eqref{eq:loss} is transformed via the following steps.
First, the expansion of Eq.~\eqref{eq:loss} is rewritten as follows:
\begin{equation}\label{eq:loss-expan}
   Loss(\XM|\pprdist) = \sum\limits_{i<j}1+\sum\limits_{i<j}\frac{||\XM[i]-\XM[j]||^2}{\pprdist[i,j]^2}-2\sum\limits_{i<j}\frac{||\XM[i]-\XM[j]||}{\pprdist[i,j]}.
\end{equation}
In Eq.~\eqref{eq:loss-expan}, the first term is a constant and the second term can be represented by $\textrm{trace}(\XM^T\LM^w\XM)$, where $\LM^w$ is the weighted Laplacian matrix and defined as
\begin{equation*}\label{eq:lw}
    \LM^w[i,j] =
    \begin{cases}
       -\frac{1}{\pprdist[i,j]^2} & \text{if $i\neq j$}\\
       \sum_{k\neq i}\frac{1}{\pprdist[i,k]^2} & \text{if $i = j$}
    \end{cases}.
\end{equation*}
According to the Cauchy-Schwartz inequality, we know that for any position matrix $\YM$, $||\XM[i]-\XM[j]|| ||\YM[i]-\YM[j]|| \geq (\XM[i]-\XM[j])^T(\YM[i]-\YM[j])$ holds.
Hence, the third term can be bounded by
$$ -2\sum\limits_{i<j}\frac{||\XM[i]-\XM[j]||}{\pprdist[i,j]} \leq -2\sum\limits_{i<j}\frac{(\XM[i]-\XM[j])^T(\YM[i]-\YM[j])}{\pprdist[i,j]||\YM[i]-\YM[j]||}. $$
Then the third term can be written as $-\textrm{trace}(2\XM^T\LM^Y\YM)$, where the matrix $\LM^Y$ is defined as
\begin{equation*}\label{eq:ly}
    \LM^Y[i,j] =
    \begin{cases}
       -\frac{1}{\pprdist[i,j]||\YM[i]-\YM[j]||} & \text{if $i\neq j$ and $\YM[i]\neq \YM[j]$}\\
       -\sum_{k\neq i}\LM^Y[i,k] &  \text{if $i = j$}\\
        0 & \text{otherwise}
    \end{cases}.
\end{equation*}
Hence, the loss function in Eq.~\eqref{eq:loss} can be bounded by
\begin{equation}\label{eq:loss-g}
       \sum\limits_{i<j}1+\textrm{trace}(\XM^T\LM^w\XM)-\textrm{trace}(2\XM^T\LM^Y\YM).
\end{equation}

To optimize Eq.~\eqref{eq:loss-g}, let $\YM$ be the position matrix from the previous iteration and $\XM$ be the position matrix to be optimized in the current iteration.
By setting the derivative of Eq.~\eqref{eq:loss-g} to zero with respect to $\XM$, the minimizer of the loss function satisfies
\begin{equation}\label{eq:pprlayout-leq}
    \XM=(\LM^w)^{-1}\LM^Y\YM,
\end{equation}
by which position matrix $\XM$ is iteratively optimized and finally returned for visualization. 
}


\subsection{\forab and \forasnb}
\stitle{\forab}
\citet{wang2017fora} propose to invoke \fora from each leaf node $v_s \in \mathcal{V}_i$. In particular, \fora first utilizes \fwdpush~\cite{andersen2006local} to derive rough approximations of the \pprdegname values, and then estimates the error term in Eq.~\eqref{eq:fwd-invariant} by exploiting random walk samplings~\cite{fogaras2005towards}.
Adapting the conclusions in \cite{wang2017fora}, we can show that \fora yields approximate \pprdegname for each $v_s \in \mathcal{V}_i$.
{In particular, given a source $v_s\in V$, by setting the initial residue value as $r(v_s,v_s)=d(v_s)$ and performing $r_{sum}\cdot W$ random walks, \fora returns $(\epsilon,\delta)-$approximate \pprdegname $\widehat{\pi}_d(v_s,v_t)$ with probability at least $1-p_f$, 
where $r_{sum} =\sum_{v_j\in V}r(v_s,v_j)$ and $W = \frac{(2+2\epsilon/3)\cdot \log{(1/p_{f})}}{\epsilon^2\delta}$. Accordingly,} plugging the approximate \pprdegname into Eq.~\eqref{eq:spprdeg} yields $(\epsilon,\delta)-$approximate level-$\ell$ \pprdegname. 
However, \fora has high computation complexity. Specifically, the \fwdpush phase costs $O(d(v_s)/r_{max})$~\cite{andersen2006local}, and the random walk phase costs $O(\omega)=O(m\cdot r_{max}\cdot W)$ as $r_{sum}\leq m\cdot r_{max}$ when push phase finishes. 
Following~\cite{wang2017fora}, we set ${r_{max}=\sqrt{{d(v_s)/(m\cdot W)}}}$ to balance the time complexities of two phases.
Thus, \fora costs $O\left(\sqrt{d(v_s)\cdot m\cdot W}\right)$ for a source leaf node $v_s \in \mathcal{V}_i$. By summarizing the time complexities for each $v_s \in \mathcal{V}_i$ and $\mathcal{V}_i\in\mathcal{S}$, the time complexity for computing all pairwise approximate level-$\ell$ \pprdegname in a supernode $\mathcal{S}$ is ${O\left(\sum_{\mathcal{V}_i \in \mathcal{S}} \sum_{v_s \in \leaf(\mathcal{V}_i)}\sqrt{d(v_s)\cdot m\cdot W}\right)}$, which is prohibitively high, as a high-level supernode $\mathcal{V}_i$ can contain many leaf nodes.

\begin{algorithm}[!t]
\KwIn{Graph $G$, supernode $\mathcal{S}$, constants $r_{max}$, $\omega$.}
\KwOut{Estimated $\widehat{\pprdist}[i,j],$\ $ \forall{\mathcal{V}_i,\mathcal{V}_j}\in \mathcal{S}$.}

\For{each supernode $\mathcal{V}_i \in \mathcal{S}$}{
$\forall{\mathcal{V}_j \in \mathcal{S}},\ \widehat{\pi}_d(\mathcal{V}_i,\mathcal{V}_j) \gets$\fwdpushsn$(G,\mathcal{S}, \mathcal{V}_i, r_{max})$\;    
$r_{sum} \gets \sum_{v_j\in V}r(\mathcal{V}_i,v_j)$\;
\Repeat{$\omega$ times}{
    Perform a random walk from a node $v_k$ with probability $\frac{r(\mathcal{V}_i,v_k)}{r_{sum}}$. Let $v_t$ be the ending node of the random walk\;
    $\widehat{\pi}_d(\mathcal{V}_i,\mathcal{V}_j)$ increases by $\frac{r_{sum}}{\omega\cdot|\leaf(\mathcal{V}_j)|}$ for $v_t\in \leaf(\mathcal{V}_j)$\;
}
}
Lines 8-9 in Algorithm~\ref{alg:taupush}\;
\caption{\forasn}
\label{alg:forasn}
\end{algorithm}

\stitle{\forasnb}
To improve the efficiency of the above \fora adaptation, we propose \forasn, which conducts push and sampling from the perspective of supernodes. 
The pseudo-code of \forasn is illustrated in Algorithm~\ref{alg:forasn}. Particularly, \forasn first performs \fwdpushsn of \taupush, which returns the estimated value $\widehat{\pi}_d(\mathcal{V}_i,\mathcal{V}_j)$ for each $\mathcal{V}_j \in \mathcal{S}$ and the residue value $r(\mathcal{V}_i,v_k)$ for each $v_k\in V$ (Lines 2-3). After that, $\omega$ times of random walk samplings are invoked. For each sampling, it starts from node $v_k$, where $v_k$ is selected based on its residue $r(\mathcal{V}_i,v_k)$ (Line 5). Suppose that the sampling stops at $v_t\in \leaf(\mathcal{V}_j)$, then the estimated value $\widehat{\pi}_d(\mathcal{V}_i,\mathcal{V}_j)$ is increased by $\frac{r_{sum}}{\omega\cdot|\leaf(\mathcal{V}_j)|}$ (Line 6).
Given a single source supernode $\mathcal{V}_i$, \forasn reduces $O(\leaf(\mathcal{V}_i))$ times of calls of vanilla \fora to only once by utilizing \fwdpushsn. 
Based on Lemma~\ref{lem:fwdpushag}, the following theorem establishes the correctness of \forasn.
\begin{theorem}\label{thm:forasn-correct}
For any user-selected supernode $\mathcal{S}$, by performing $\omega = \frac{r_{sum}}{\minleaf}\cdot W$ random walks where $\minleaf = \min_{\mathcal{V}_i\in \mathcal{S}}|\leaf(\mathcal{V}_i)|$ and $W = \frac{(2+2\epsilon/3)\cdot \log{(1/p_{f})}}{\epsilon^2\delta}$, Algorithm~\ref{alg:forasn}
returns $(\epsilon,\delta)-$approximate level-$\ell$ \pprdegname $\widehat{\pi}_d(\mathcal{V}_i,\mathcal{V}_j)$ for $\mathcal{V}_i,\mathcal{V}_j\in\mathcal{S}$ with probability at least $1-p_f$.
\end{theorem}


Following the time complexity analysis of vanilla \fora, we can show that \forasn costs $ O\left(\sum_{\mathcal{V}_i \in \mathcal{S}} \frac{d(\mathcal{V}_i)}{|\leaf(\mathcal{V}_i)|\cdot r_{max}}+\frac{m\cdot r_{max}}{\minleaf}\cdot W\right)$ to approximate all-pair  level-$\ell$ \pprdegname in $\mathcal{S}$.
For the setting of $r_{max}$, we balance the complexity of \taupush from all source nodes $\mathcal{V}_i \in \mathcal{S}$ with that of the random walk phase.
Thus, we have ${r_{max}=\sqrt{\minleaf\cdot \sum_{\mathcal{V}_i\in \mathcal{S}}{d(\mathcal{V}_i)}/{|\leaf(\mathcal{V}_i)|}/(m\cdot W)}}$, and the time complexity becomes ${O\left(\sqrt{\sum_{\mathcal{V}_i\in \mathcal{S}}{(d(\mathcal{V}_i)}/{|\leaf(\mathcal{V}_i)|})\cdot({m\cdot W}/{\minleaf})}\right)}$.

\subsection{Proofs}\label{sec:proofs}

\header
{\bf Proof of Theorem~\ref{thm:pprd-nd}.}
We use $z_{ij}$ to denote $\pi_d(v_i,v_j)+\pi_d(v_j,v_i)$. Thus,
\begin{align}
\sum_{v_i,v_j\in V}{z_{ij}}&=\rc{\sum_{v_i,v_j\in V}{\pi_d(v_i,v_j)}+\sum_{v_i,v_j\in V}{\pi_d(v_j,v_i)}}=2\sum_{v_i,v_j\in V}{\pi_d(v_i,v_j)}\notag \\
&=2\sum_{v_i\in V}{d(v_i)\cdot\sum_{v_j\in V}{\pi(v_i,v_j)}}=2m \label{eq:z-bound}.
\end{align}
Define function $f(z)$ as $f(z)=\frac{1}{\min(\max(2,1-\log{(z)}),2\log n)^2}$. It is easy to show that $f(z)$ is monotonically increasing when $z\in [2,2\log n]$ and is always smaller than function $h(z)=0.215e\cdot z + 0.035$.
By Eq. \eqref{eq:pprdist}, we have $\pprdist[i,j]=\min(\max(2,1-\log{(z_{ij})}),2\log n)$. As we assume that $\pprdist[i,j]=||X[i]-X[j]||$, we can obtain
\begin{align*}
\textrm{ND}(\XM)&=\sum_{i< j}{\frac{1}{\pprdist[i,j]^2}}=\sum_{i< j}{f(z_{ij})}\le \frac{1}{2}\sum_{1\le i,j\le n}{f(z_{ij})}\\
&\le \frac{1}{2}\sum_{1\le i,j\le n}{h(z_{ij})}=\frac{1}{2}\left(0.215e\sum_{v_i,v_j\in V}{z_{ij}}+0.035n^2\right).
\end{align*}
Plugging Eq. \eqref{eq:z-bound} into the above inequality gives ND$(\XM)\le 0.215e\cdot m+0.0175n^2$, which completes the proof.

\header
{\bf Proof of Theorem~\ref{thm:pprd-ulcv}.}
For any random walk starting from $v_i$, the probability that the walk moves to a certain one-hop neighbor $v_j$ is $(1-\alpha)/d(v_i)$. Since the walk stops at $v_j$ with probability $\alpha$, then the PPR value of a neighbor is bounded by
$\pi(v_i,v_j) \geq \frac{\alpha(1-\alpha)}{d(v_i)}.$
As we assume that $\pprdist[i,j]=||X[i]-X[j]||$, by Definition~\ref{def:pprdist}, the length of any edge $l(v_i,v_j)$ is bounded by
$$l(v_i,v_j)=\pprdist[i,j]\leq 1-\log{\left(2\alpha(1-\alpha)\right)}.$$
Based on the range of \pprdistname, the edge length is in the range
of $[2,1-\log{2\alpha(1-\alpha)}]$,
where $\textstyle \alpha \leq \tfrac{1}{2}-\sqrt{\tfrac{1}{4}-\frac{1}{2e}}$. 
According to Popoviciu's inequality~\cite{niculescu2006convex},
\begin{equation*}
\textrm{ULCV}(\XM)=\frac{l_{\sigma}}{l_{\mu}}\leq\frac{{\left(\log{\tfrac{1}{{2\alpha(1-\alpha)}}}-1\right)}/{2}}{2}=\frac{\left(\log{\tfrac{1}{{2\alpha(1-\alpha)}}}-1\right)}{4}.
\end{equation*}

\header
{\bf Proof of Lemma \ref{lem:appro-relation}.}
We call the approximate \pprdistname mentioned in Lemma~\ref{lem:appro-relation} $(\theta,\sigma)$-approximate \pprdistname.
We first focus on the relative error part in Lemma~\ref{lem:appro-relation}. Denote $c = \left(\frac{\pi_d(\mathcal{V}_i, \mathcal{V}_j)+\pi_d(\mathcal{V}_j, \mathcal{V}_i)}{e}\right)^{\theta}$. There exist the following two cases.
\begin{itemize}[leftmargin=*]
    \item To ensure $\widehat{\pprdist}[i,j]-\pprdist[i,j]\leq \theta\cdot\pprdist[i,j]$, the following approximation should be satisfied
    $$\pi_d(\mathcal{V}_i, \mathcal{V}_j)-\widehat{\pi}_d(\mathcal{V}_i, \mathcal{V}_j)\leq (1-c)\cdot \pi_d(\mathcal{V}_i, \mathcal{V}_j).$$
    \item To ensure $\pprdist[i,j]-\widehat{\pprdist}[i,j] \leq \theta\cdot\pprdist[i,j]$, the following approximation should be satisfied
    $$\widehat{\pi}_d(\mathcal{V}_i, \mathcal{V}_j)-\pi_d(\mathcal{V}_i, \mathcal{V}_j)\leq (\frac{1}{c}-1)\cdot \pi_d(\mathcal{V}_i, \mathcal{V}_j).$$
\end{itemize}
Since $\frac{1}{c}-1\geq 1-c$ and each $\Delta[i,j]$ is in range $[2,2\log{n}]$ by definition,  by setting $$1-c=1-\left(\frac{\pi_d(\mathcal{V}_i, \mathcal{V}_j)+\pi_d(\mathcal{V}_j, \mathcal{V}_i)}{e}\right)^{\theta}\geq 1-\left(\frac{1}{e^{2}}\right)^{\theta} = \epsilon_1,$$ 
the relative error part of $(\theta,\sigma)$-approximate \pprdistname holds.
Akin to the above analysis, we can derive that by setting 
$$\epsilon_2=1-\left(\frac{1}{e^{\sigma}}\right)^{\theta},$$
the absolute error part $|\pprdist[i,j]-\widehat{\pprdist}[i,j]|\leq \theta\cdot\sigma$ holds.
Since $\sigma\geq 2$ and $\epsilon_1\leq \epsilon_2$, $(\theta,\sigma)$-approximate \pprdistname holds by setting $\epsilon=\epsilon_1=1-\left(\frac{1}{e^{2}}\right)^{\theta}$.

\header
{\bf Proof of Lemma \ref{lem:fpsn-correct}.}
We first need a crucial property of Algorithm \ref{alg:fwdpushsn} in Lemma \ref{lem:fwdpushag}.
\begin{lemma}\label{lem:fwdpushag}
{By initializing the residue values as Line 2,}
Algorithm~\ref{alg:fwdpushsn} satisfies the following invariant
\begin{equation}\label{eq:fwdpushsn-invariant}
    \pprdeg(\mathcal{V}_i,\mathcal{V}_j) =\widehat{\pi}_d(\mathcal{V}_i,\mathcal{V}_j)+\sum\limits_{v_t\in \leaf(\mathcal{V}_j)}\sum\limits_{v_k\in V}\frac{r(\mathcal{V}_i,v_k)}{|\leaf(\mathcal{V}_j)|\cdot d(v_k)}\cdot\pi_d(v_k,v_t).
\end{equation}
\end{lemma}
Based on Lemma~\ref{lem:fwdpushag}, for any supernode $\mathcal{V}_i,\mathcal{V}_j \in \mathcal{S}$, the approximation error is bounded by 
$${\sum\limits_{v_t\in \leaf(\mathcal{V}_j)}\sum\limits_{v_k\in V}\frac{r(\mathcal{V}_i,v_k)}{|\leaf(\mathcal{V}_j)|\cdot d(v_k)}\cdot\pi_d(v_k,v_t)}.$$
As \fwdpushsn stops when each residue value $r(\mathcal{V}_i,v_k)\leq r_{max}\cdot d(v_k)$, the error bound turns to
$$\tfrac{r_{max}}{|\leaf(\mathcal{V}_j)|}\cdot\sum\limits_{v_t\in \leaf(\mathcal{V}_j)}\sum\limits_{v_k\in V}\pi_d(v_k,v_t).$$
Thus, if we set $ r_{max}=\tfrac{|\leaf(\mathcal{V}_j)|\cdot\epsilon\cdot\delta}{\sum\limits_{v_t\in \leaf(\mathcal{V}_j)}\sum\limits_{v_k\in V}\pi_d(v_k,v_t)}$, $\widehat{\pi}_d(\mathcal{V}_i,\mathcal{V}_j)$ satisfies $(\epsilon,\delta)-$approximate. 
Based on Eq.~\eqref{eq:dnpr-dnppr}, for a target $\mathcal{V}_j\in \mathcal{S}$, the correctness of \fwdpushsn is guaranteed by setting $ r_{max} = \tfrac{\epsilon\cdot \delta}{m\cdot\tau_{j}}$. Hence, with $r_{max} = \tfrac{\epsilon\cdot \delta}{m\cdot\tau}$, \fwdpushsn is $(\epsilon,\delta)-$approximate for target supernode $\mathcal{V}_j \in \mathcal{S}$ with $\tau_j\leq \tau$.

\header
{\bf Proof of Lemma \ref{lem:fwdpushag}.}
For each $v_s\in\leaf(\mathcal{V}_{i})$, denote $\widehat{\pi}_d(\mathcal{V}_{i},v_s,v_k)$ and $r(\mathcal{V}_{i},v_s,v_k)$ as the lower bound and residue value, which is firstly initialized from $v_s$ and then distributed to $v_k$. 
Initially, $r(\mathcal{V}_{i},v_s,v_s)$ is set to $\frac{d(v_s)}{\leaf(\mathcal{V}_{i})}$ (Line 2). According to the proof in~\cite{andersen2006local}, the following equation holds for each $v_t\in V$ in graph traversal,
\begin{equation*}
        \frac{\pprdeg(v_s,v_t)}{|\leaf(\mathcal{V}_{i})|} = \widehat{\pi}_d(\mathcal{V}_{i},v_s,v_t)+\sum_{v_k\in V}r(\mathcal{V}_{i},v_s,v_k)\cdot \pi(v_k,v_t).
\end{equation*}
Note that for each $v_t\in V$, $\sum_{v_s\in\leaf(\mathcal{V}_{i})}\widehat{\pi}_d(\mathcal{V}_{i},v_s,v_t) = \widehat{\pi}_d(\mathcal{V}_{i},v_t)$ and residue value holds the same relationships. By summing up all $v_s\in\leaf(\mathcal{V}_{i})$, the above equation turns to
$$\pprdeg(\mathcal{V}_{i},v_t) = \widehat{\pi}_d(\mathcal{V}_{i},v_t)+\sum_{v_k\in V}r(\mathcal{V}_{i},v_k)\cdot \pi(v_k,v_t).$$
By Eq.~\eqref{eq:spprdeg}, the \pprdegname between supernodes $\mathcal{V}_{i}$ and $\mathcal{V}_{j}$ is $\pprdeg (\mathcal{V}_{i},\mathcal{V}_{j})=\frac{\sum_{v_t\in \leaf(\mathcal{V}_{j})}\pprdeg(\mathcal{V}_{i},v_t)}{|\leaf(\mathcal{V}_{j})|}$. Then the above equation can be transformed to
$$\pprdeg(\mathcal{V}_i,\mathcal{V}_j) =\widehat{\pi}_d(\mathcal{V}_i,\mathcal{V}_j)+\sum\limits_{v_t\in \leaf(\mathcal{V}_j)}\sum\limits_{v_k\in V}\frac{r(\mathcal{V}_i,v_k)}{|\leaf(\mathcal{V}_j)|\cdot d(v_k)}\cdot\pi_d(v_k,v_t).$$

\header
{\bf Proof of Lemma \ref{lem:bpsn-correct}.}
Similar to the invariant property of conventional \bwdpush~\cite{lofgren2013personalized}, \bwdpushsn has the following property.
\begin{proposition}\label{prop:bwdpushsn}
\bwdpushsn satisfies the following invariant:
\begin{equation}\label{eq:bwdpushsn-property}
     \pi_d(v_s,\mathcal{V}_j) = \widehat{\pi}_d(v_s,\mathcal{V}_j)+d(v_s)\cdot\sum\limits_{v_k\in V}\pi(v_s,v_k)\cdot r(v_k,\mathcal{V}_j).
\end{equation}
\end{proposition}
Based on Eq.\eqref{eq:bwdpushsn-property} and $\widehat{\pi}_d(\mathcal{V}_i,\mathcal{V}_j)=\sum_{v_s \in \leaf(\mathcal{V}_i)}{\frac{\widehat{\pi}_d(v_{s},\mathcal{V}_j)}{|\leaf(\mathcal{V}_i)|}}$, the approximation error is
$$ \pi_d(\mathcal{V}_i,\mathcal{V}_j)-\widehat{\pi}_d(\mathcal{V}_i,\mathcal{V}_j)=\sum\limits_{v_s \in \leaf(\mathcal{V}_i)}\frac{d(v_s)}{|\leaf(\mathcal{V}_i)|}\cdot{\sum\limits_{v_k\in V}\pi(v_s,v_k)\cdot r(v_k,\mathcal{V}_j)}.$$
Note that $\sum_{v_k\in V}\pi(v_s,v_k)=1$ and \bwdpushsn stops when $ r(v_k,\mathcal{V}_j)\leq r^b_{max}$ for all $v_k \in V$. Hence, the error term on the r.h.s. is upper-bounded by $$\sum_{v_s \in \leaf(\mathcal{V}_i)}\frac{d(v_s)\cdot{r^b_{max}}}{|\leaf(\mathcal{V}_i)|}.$$
For a source supernode $\mathcal{V}_i$, the approximation quality is met with $$ r^b_{max}=\frac{\epsilon\cdot\delta}{d_i},$$
where $d_i=\sum_{v_s \in \leaf(\mathcal{V}_i)}\frac{d(v_s)}{|\leaf(\mathcal{V}_i)|}$. Hence, correctness is guaranteed to all source supernodes $\mathcal{V}_i\in \mathcal{S}\backslash \mathcal{V}_j$ by setting $r^b_{max}=\frac{\epsilon\cdot\delta}{d_{max}}$, where $d_{max}=\max_{\mathcal{V}_i\in \mathcal{S}\backslash \mathcal{V}_j}d_i$.

\header
{\bf Proof of Theorem \ref{thm:forasn-correct}.}
Recall that the input values of sampling phase satisfy (see Lemma~\ref{lem:fwdpushag})
$$\pprdeg(\mathcal{V}_i,\mathcal{V}_j) =\widehat{\pi}_d(\mathcal{V}_i,\mathcal{V}_j)+\sum\limits_{v_t\in \leaf(\mathcal{V}_j)}\sum\limits_{v_k\in V}\frac{r(\mathcal{V}_i,v_k)}{|\leaf(\mathcal{V}_j)|}\cdot\pi(v_k,v_t).$$
For any sampled node $v_k$, let $X_{k,j}$ be a Bernoulli variable that takes value 1 if the random walk starting from $v_k$ stops at any node of $\leaf(\mathcal{V}_{j})$. 
By definition, $$\mathbb{E}[X_{k,j}]=\sum\limits_{v_t\in \leaf(\mathcal{V}_j)} \pi(v_k,v_{t}).$$
Recall that $v_k$ is selected with probability of $\frac{r(\mathcal{V}_{i},v_k)}{r_{sum}}$ and the estimation is refined by $\frac{r_{sum}}{\omega\cdot|\leaf(\mathcal{V}_{j})|}$, hence,
$$\mathbb{E}[\frac{1}{\omega}\sum^{\omega}\frac{r_{sum}}{|\leaf(\mathcal{V}_{j})|}\cdot \frac{r(\mathcal{V}_{i},v_k)}{r_{sum}}\cdot X_{k,j}]=\sum\limits_{v_t\in \leaf(\mathcal{V}_j)}\frac{r(\mathcal{V}_{i},v_k)}{|\leaf(\mathcal{V}_{j})|}\cdot \pi(v_k,v_t).$$
Denote $\psi_k = \frac{r(\mathcal{V}_{i},v_k)}{|\leaf(\mathcal{V}_{j})|}\cdot \pi(v_k,v_t)$, then
$$\mathbb{E}[\sum_{v_k\in V}\psi_k]=\sum\limits_{v_t\in \leaf(\mathcal{V}_j)}\sum_{v_k\in V}\frac{r(\mathcal{V}_{i},v_k)}{|\leaf(\mathcal{V}_{j})|}\cdot \pi(v_k,v_t).$$
which is exactly the second term of Lemma~\ref{lem:fwdpushag}. Thus, it is an unbiased estimator of $\pi_d(\mathcal{V}_{i},\mathcal{V}_{j})$. Next, by applying Chernoff Bound~\cite{chung2006concentration}, we can derive that, for any $\mathcal{V}_{i},\mathcal{V}_{j}\in \mathcal{S}$,
\begin{itemize}[leftmargin=*]
    \item when $\pi_d(\mathcal{V}_{i},\mathcal{V}_{j})< \delta$, by setting $\omega=\frac{r_{sum}}{\minleaf}\cdot W$, we have 
        \begin{align*}
        & \mathsf{Pr}[|\widehat{\pi}_d(\mathcal{V}_{i},\mathcal{V}_{j})-\pi_d(\mathcal{V}_{i},\mathcal{V}_{j})| \geq \epsilon \cdot \delta] \\
        &\leq \exp(-\frac{\omega\cdot (\epsilon \cdot \delta)^2}{\frac{r_{sum}}{|\leaf(\mathcal{V}_{j})|}(\frac{2}{3}(\epsilon \cdot \delta)+2\delta)}) \leq p_f.
        \end{align*}
    \item when $\delta \leq \pi_d(\mathcal{V}_{i},\mathcal{V}_{j})$,  by setting $\omega=\frac{r_{sum}}{\minleaf}\cdot W$, we have 
        \begin{align*}
        & \mathsf{Pr}[|\widehat{\pi}_d(\mathcal{V}_{i},\mathcal{V}_{j})-\pi_d(\mathcal{V}_{i},\mathcal{V}_{j})| \geq \epsilon \cdot \pi_d(\mathcal{V}_{i},\mathcal{V}_{j})]\\
        &\leq \exp(-\frac{\omega\cdot (\epsilon \cdot \pi_d(\mathcal{V}_{i},\mathcal{V}_{j}))^2}{\frac{r_{sum}}{|\leaf(\mathcal{V}_{j})|}(\frac{2}{3}(\epsilon \cdot \pi_d(\mathcal{V}_{i},\mathcal{V}_{j}))+2\pi_d(\mathcal{V}_{i},\mathcal{V}_{j}))}) \leq p_f.    
        \end{align*}
\end{itemize}

\subsection{Complexity Analysis and Constant Setting}
For the following analysis about time complexity and $\tau$ setting, we focus on the level-0 children, \ie $v_i$ of a level-1 supernode $\mathcal{S}$, where the worst complexities of \fwdpushsn and \bwdpushsn are achieved. Take \fwdpushsn as an example. As the level increases, each supernode $\mathcal{V}_i$ contains more leaf nodes and $\tau_{i}$ tends to $1/n$ (\ie the average \dnpr of the entire graph). Thus, the largest $\tau_{i}$ and worst time complexity occur in the level-0.

\begin{table*}[t]
\vspace{-2mm}
\centering
\caption{\ar{} of \pprvizb and the competitors. Smaller value indicates better visualization quality, the best in bold and the second best in italic, and ``-'' indicates undefined.}
\label{tab:metrics-ar}
\vspace{-2mm}
\renewcommand{\arraystretch}{1.1}
\begin{small}
\resizebox{1\linewidth}{!}{%
\begin{tabular}{|c|c|c|c|c|c|c|c|c|c|c|c|c|}
\hline
           & \pprviz  & \openord/\fr & \linlog  & \forceatlas & \mds     & \pivotmds & \gf      & \sdne    & \leemb   & \lle     & \nodevec & \simrank \\ \hline
\twego     & \textbf{0.00E+00} & 1.95E+00     & 1.21E+00 & \textit{9.26E-03}    & \textbf{0.00E+00} & 5.00E+00  & 1.40E+01 & 6.00E+01 & 2.51E+01 & 3.87E+00 & 1.27E+01 & 3.45E+00 \\
\fbego     & \textbf{4.09E+01} & 5.60E+01     & 6.83E+01 & \textit{4.54E+01}    & \textit{4.54E+01} & 9.81E+01  & 2.71E+02 & 9.58E+02 & 6.44E+02 & 7.41E+02 & 2.97E+02 & 6.43E+01 \\
\wiki      & \textit{4.08E+02} & \textbf{4.01E+02}     & 4.70E+02 & 5.20E+02    & 5.14E+02 & 3.15E+03  & 6.14E+02 & 2.81E+03 & 2.70E+03 & 2.25E+03 & 1.01E+03 & 1.46E+03 \\
\physic    & 4.60E+02 & 8.87E+02     & 4.34E+02 & \textbf{3.87E+02}    & \textit{4.19E+02} & 5.83E+02  & 1.38E+03 & 6.69E+03 & 4.74E+03 & 5.34E+03 & 2.40E+03 & 7.82E+02 \\
\filmtrust & \textbf{6.72E+02} & 8.03E+02     & \textit{7.89E+02} & 8.09E+02    & 1.15E+03 & 2.02E+03  & 1.51E+03 & 8.80E+03 & 6.63E+03 & 3.73E+03 & 3.44E+03 & 1.02E+03 \\
\scinet    & \textbf{7.51E+02} & 1.41E+03     & 1.57E+03 & 1.29E+03    & 1.20E+03 & 7.52E+03  & 3.60E+03 & 1.14E+04 & 1.00E+04 & -        & 5.17E+03 & \textit{1.10E+03} \\
\hline
\end{tabular}
}
\end{small}
\vspace{-3mm}
\end{table*}

\stitle{Time for the worst case}
The threshold $\tau$ determines how quickly \fwdpushsn can be terminated and how many \bwdpushsn are performed. A small $\tau$ engenders a low cost for \fwdpushsn and a high cost for \bwdpushsn, and vice versa. Therefore, the appropriate setting of $\tau$ should balance the workloads of \fwdpushsn and \bwdpushsn in \taupush.
First, note that the compleixty of \fwdpushsn for a given leaf source ${v}_i\in \mathcal{S}$ is $ O\left(d({v}_i)\cdot\frac{m\tau}{\epsilon\delta}\right)$~\cite{andersen2006local}, by which the worst time complexity of \fwdpushsn occurs when the source out-degree $d(v_i)$ is the largest.
Assume that there is only one largest out-degree in the scale-free network. Following the proof by ~\cite{cohen2001breakdown}, the largest out-degree is $O\left(n^{\frac{1}{b-1}}\right)$, where $b\in[2,3]$ is the exponent of degree distribution and $b=2$ on the \twitter graph. Hence, the worst time complexity of Lines 1 - 2 of Algorithm~\ref{alg:taupush} is $O\left(\frac{knm\tau}{\epsilon\delta}\right)$.
Next, for a leaf target $v_j$, the worst time complexity of \bwdpushsn is ${O\left(\frac{\sum_{v_i\in V}d(v_i)\cdot \pi(v_i,v_j)}{r^b_{max}}\right)}$~\cite{wang2020personalized} and can be simplified as ${O\left(\frac{m\cdot\tau_j}{r^b_{max}}\right)}$ using Eq.~\eqref{eq:dnpr-dnppr}.
As $\tau_j=O(1)$ and $r^b_{max}=\frac{\epsilon\delta}{d_{max}}$, the time complexity of \bwdpushsn can be re-written as $O\left(\frac{d_{max}\cdot m}{\epsilon\delta}\right)$.
Since there are at most $O(1/\tau)$ target nodes with average \dnpr larger than $\tau$, the time complexity of Lines 3 - 4 of Algorithm~\ref{alg:taupush} is $O\left(\frac{1}{\tau}\cdot\frac{d_{max}\cdot m}{\epsilon\delta}\right)$.

\stitle{$\boldsymbol \tau$ setting}
Based on the aforementioned analysis, the worst-case time complexity of \taupush is ${O\left(\frac{knm\tau}{\epsilon\delta} + \frac{d_{max}\cdot m}{\tau\epsilon\delta}\right)}$. Since the degree and \dnpr follows the same power law~\cite{litvak2007degree}, the probability that a random-select node has out-degree larger than $d_{max}$ is $\frac{1/\tau}{n}$.
Hence, following the proof by~\cite{cohen2001breakdown}, $d_{max}=O(\tau n)$ for the scale-free networks. Therefore, the time complexity turns into ${O\left(\frac{knm\tau}{\epsilon\delta} + \frac{nm}{\epsilon\delta}\right)}$.
For the single-source approximation in \fwdpushsn, we set $\delta$ to $\frac{1}{10\cdot k}=O(\frac{1}{k})$~\cite{shi2019realtime} as nodes in the same supernode have good connectivity (\ie large \pprdegname value) and we focus on the top-$O(k)$ \pprdegname values for a source.
However, the above setting is not suitable for single-target approximation in \bwdpushsn because the source nodes and their degrees are unknown with a given target node. Hence, we set $\delta=O({n\tau}/{k})$ for \bwdpushsn as empirically ${\sum_{v_i\in \leaf(\mathcal{S})}\pi_d(v_i,v_j)}/\leaf(\mathcal{S})=n\cdot\tau_j/{k}$ for a random level-1 supernode $\mathcal{S}$ and $\tau_j$ should be comparable to $\tau$.
With these configurations, by setting
$\tau={1}/{\sqrt{kn}}$, the worst-case time complexity of \taupush in Algorithm~\ref{alg:taupush} is minimized to $ O\left(\frac{km}{\epsilon}\cdot \sqrt{kn}\right)$. Note that, by employing \fwdpushsn only, the worst-case complexity is $O\left(\frac{k^2nm}{\epsilon}\right)$, which is $\sqrt{kn}$ times slower than \taupush. 

\stitle{Time for a random supernode $\boldsymbol{\mathcal{S}}$}
In scale-free networks, both \dnpr and \pprdegname values follow the power law~\cite{lofgren2016personalized,shi2019realtime,litvak2007degree}, hence, the i-th largest \dnpr value is $\frac{1}{i\cdot \log{n}}$~\cite{shi2019realtime}. As there are $n/k$ supernodes at level-1 and $\mathbb{E}[\tau]$ is upper-bounded by the average of the $(n/k)$ largest \dnpr values, we have $$\mathbb{E}[\tau]=\frac{1}{n/k}\cdot\sum_{i=1}^{n/k}\frac{1}{i\cdot \log{n}}=\frac{k\cdot\log{(n/k)}}{n\cdot \log{n}}\leq \frac{k}{n}.$$
Recall that $\tau$ is set to $1/\sqrt{kn}$. Hence, given a random query supernode $\mathcal{S}$, \taupush will only conduct \fwdpushsn if $\mathbb{E}[\tau]\leq 1/\sqrt{kn}$, \ie $n\geq k^3$.
Note that \fwdpushsn costs $O\left(\frac{d(\mathcal{V}_i)}{|\leaf(\mathcal{V}_i)|}\cdot\frac{m\tau}{\epsilon\delta}\right)$~\cite{andersen2006local} from a supernode $\mathcal{V}_i \in \mathcal{S}$, by setting $r_{init}$ as Line 1 in Algorithm~\ref{alg:fwdpushsn} and $r_{max}$ as Lemma~\ref{lem:fpsn-correct}. Therefore, by plugging $\mathbb{E}[\tau]$ into the above complexity, we can derive that the complexity of \taupush for a random selected supernode $\mathcal{S}$ is $O\left(\sum_{\mathcal{V}_i \in \mathcal{S}}\frac{d(\mathcal{V}_i)}{|\leaf(\mathcal{V}_i)|}\cdot\frac{k m}{\epsilon\delta n}\right)$. For easy presentation, we simplify the above complexity by setting $m/n=O(\log{n})$ and $\sum_{\mathcal{V}_i \in \mathcal{S}}\frac{d(\mathcal{V}_i)}{|\leaf(\mathcal{V}_i)|}=O(k \cdot \log{n})$, where $k$ is the number of supernodes in $\mathcal{S}$ and $O(\log{n})$ is the average node degree of scale-free networks.
Recall that $\delta=O({1}/{k})$~\cite{shi2019realtime}. Hence, the complexity is massaged into
$O\left(\frac{k^3\cdot(\log{n})^2}{\epsilon}\right)$.

\begin{figure*}[!t]
\centering
\begin{small}
  \begin{tabular}{cc}
    \includegraphics[width=0.5\textwidth]{pprvizs_output/fbego-pprvizs.pdf}
     &
    \includegraphics[width=0.5\textwidth]{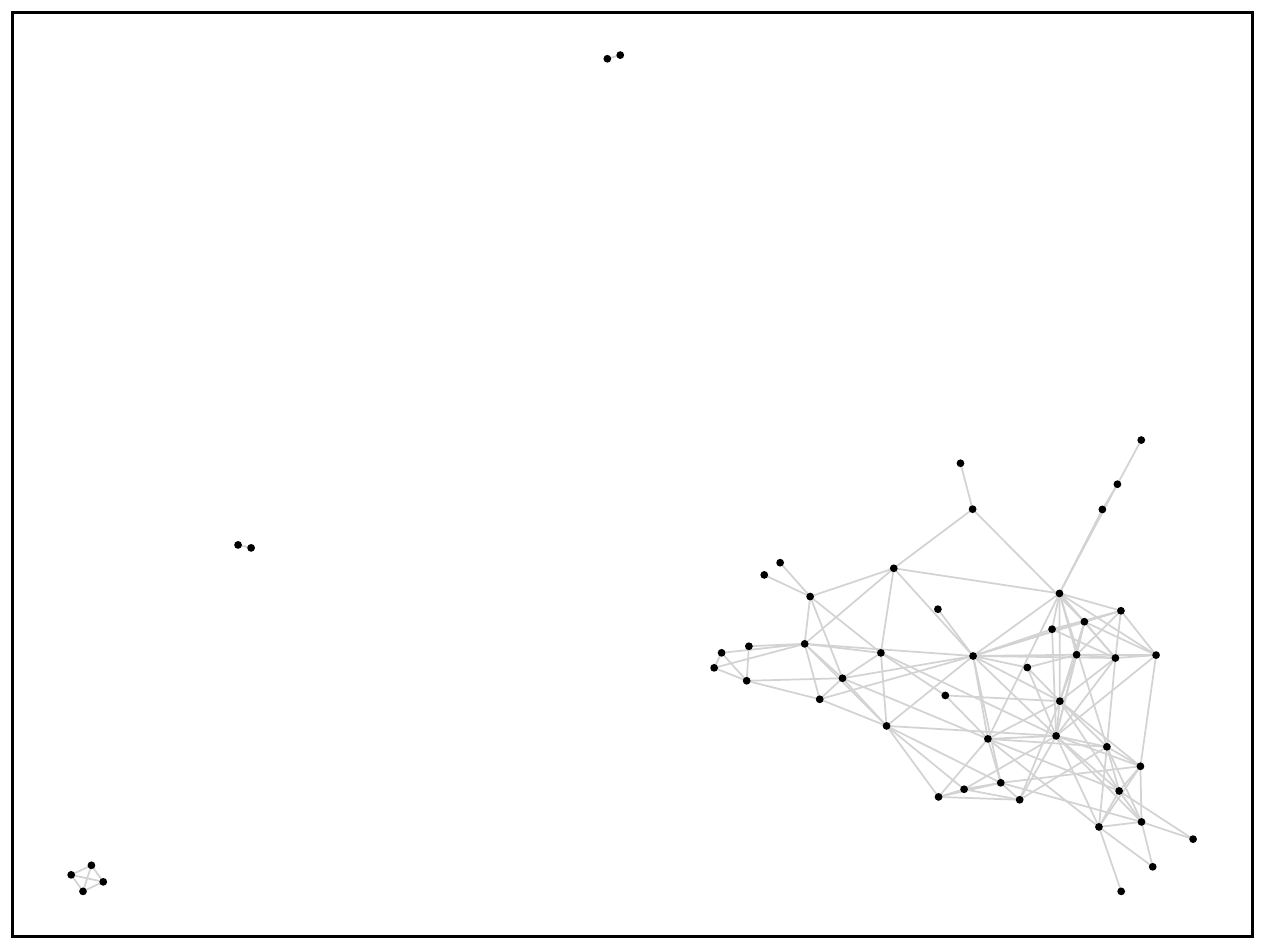}
    \\
    \footnotesize{\nd{}=2.4E+03, \ulcv{}=0.39} & \footnotesize{\nd{}=1.3E+04, \ulcv{}=0.49}
    \\
    (a) \pprviz (Ours)  & (b) $\blacklozenge$~\forceatlas~\cite{jacomy2014forceatlas2} 
    \\
  \end{tabular}
\end{small}
\vspace{-3mm}
\end{figure*}

\begin{figure*}[]
\centering
\begin{small}
  \begin{tabular}{cccc}
    \includegraphics[width=0.24\textwidth]{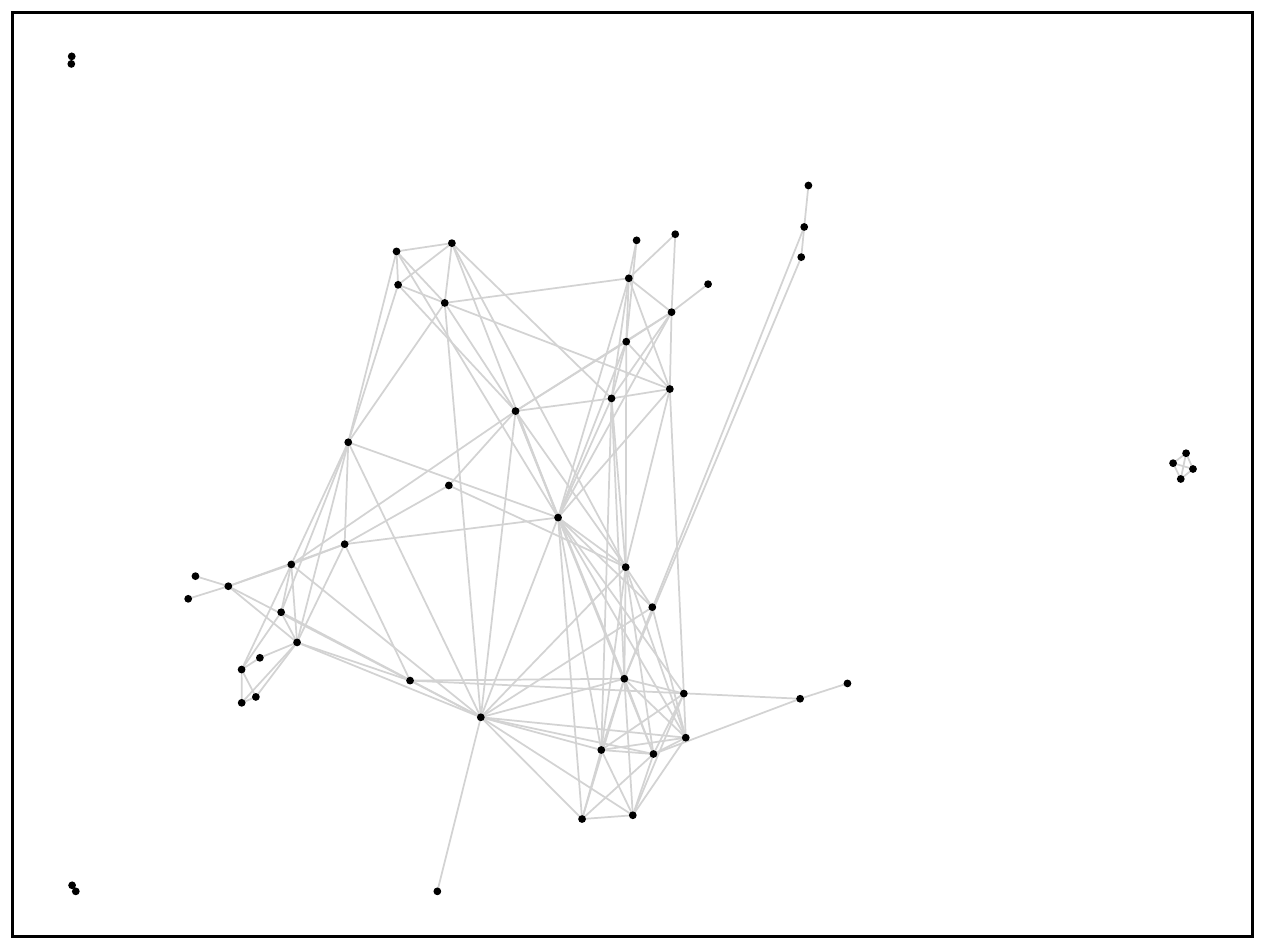}
    &
    \includegraphics[width=0.24\textwidth]{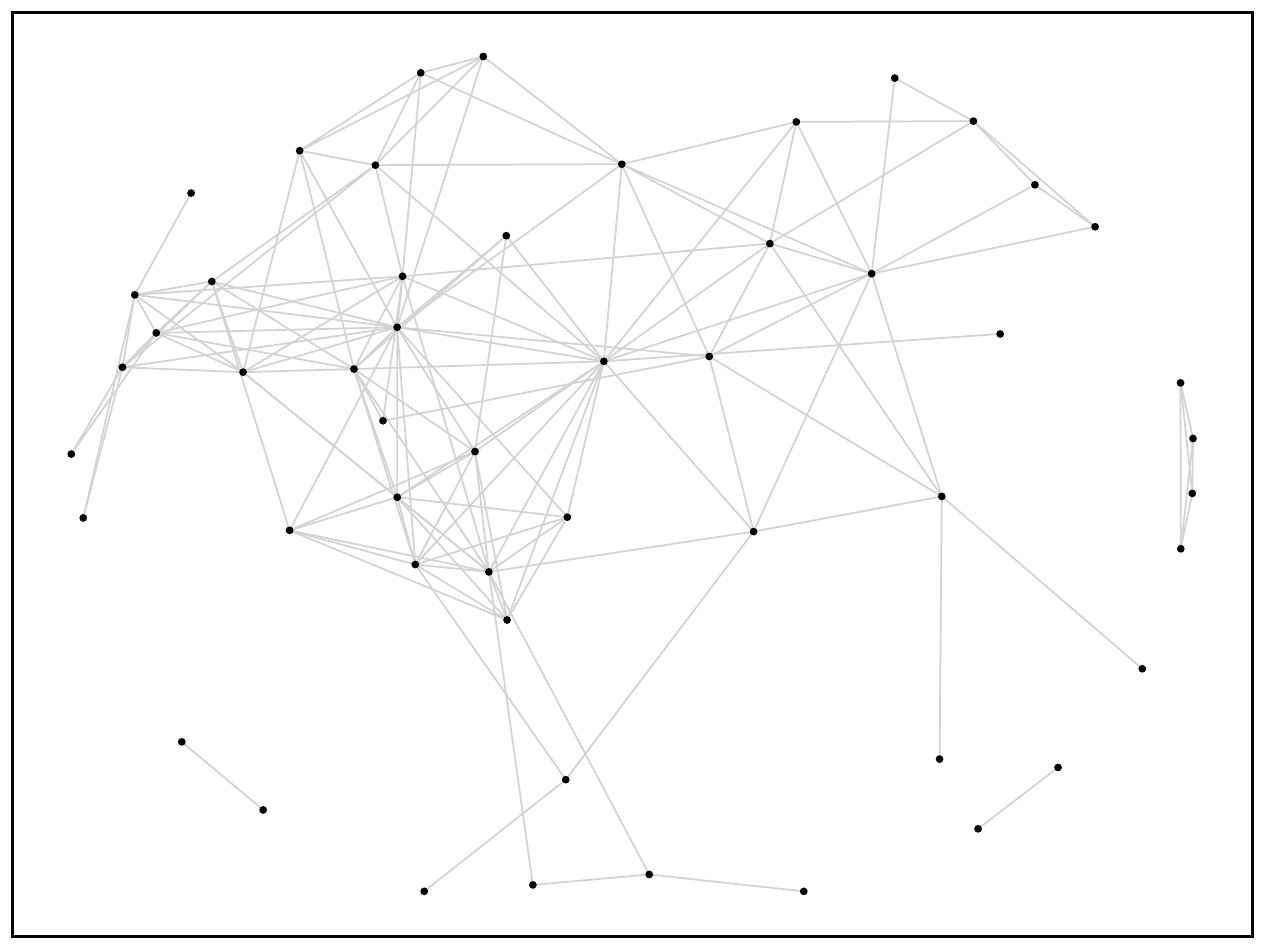}
    &
    \includegraphics[width=0.24\textwidth]{pprvizs_output/fbego-mds.pdf}
    &
    \includegraphics[width=0.24\textwidth]{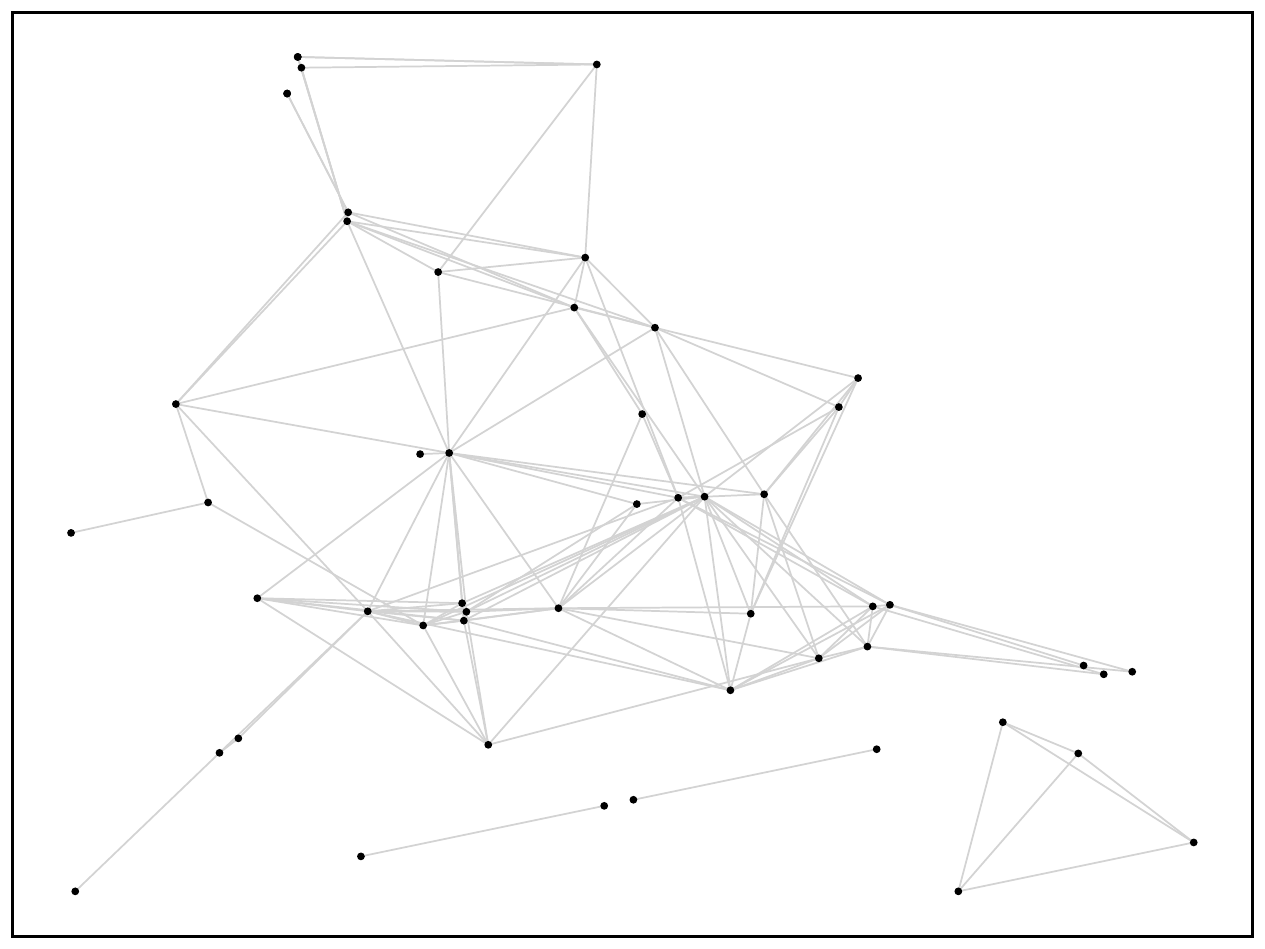}
    \\
    \footnotesize{\nd{}=9.5E+03, \ulcv{}=0.67} & \footnotesize{\nd{}=1.1E+03, \ulcv{}=0.42} & \footnotesize{\nd{}=2.0E+04, \ulcv{}=0.46} & \footnotesize{\nd{}=$\infty$, \ulcv{}=0.45}
    \\
    (c) $\blacklozenge$~\linlog~\cite{noack2005energy} 
    &
    (d) $\blacklozenge$~\fr~\cite{fruchterman1991graph} 
    & 
    (e) $\blacktriangle$~\mds~\cite{gansner2004graph}
    &
    (f) $\blacktriangle$~\pivotmds~\cite{brandes2006eigensolver} 
    \\
    \includegraphics[width=0.24\textwidth]{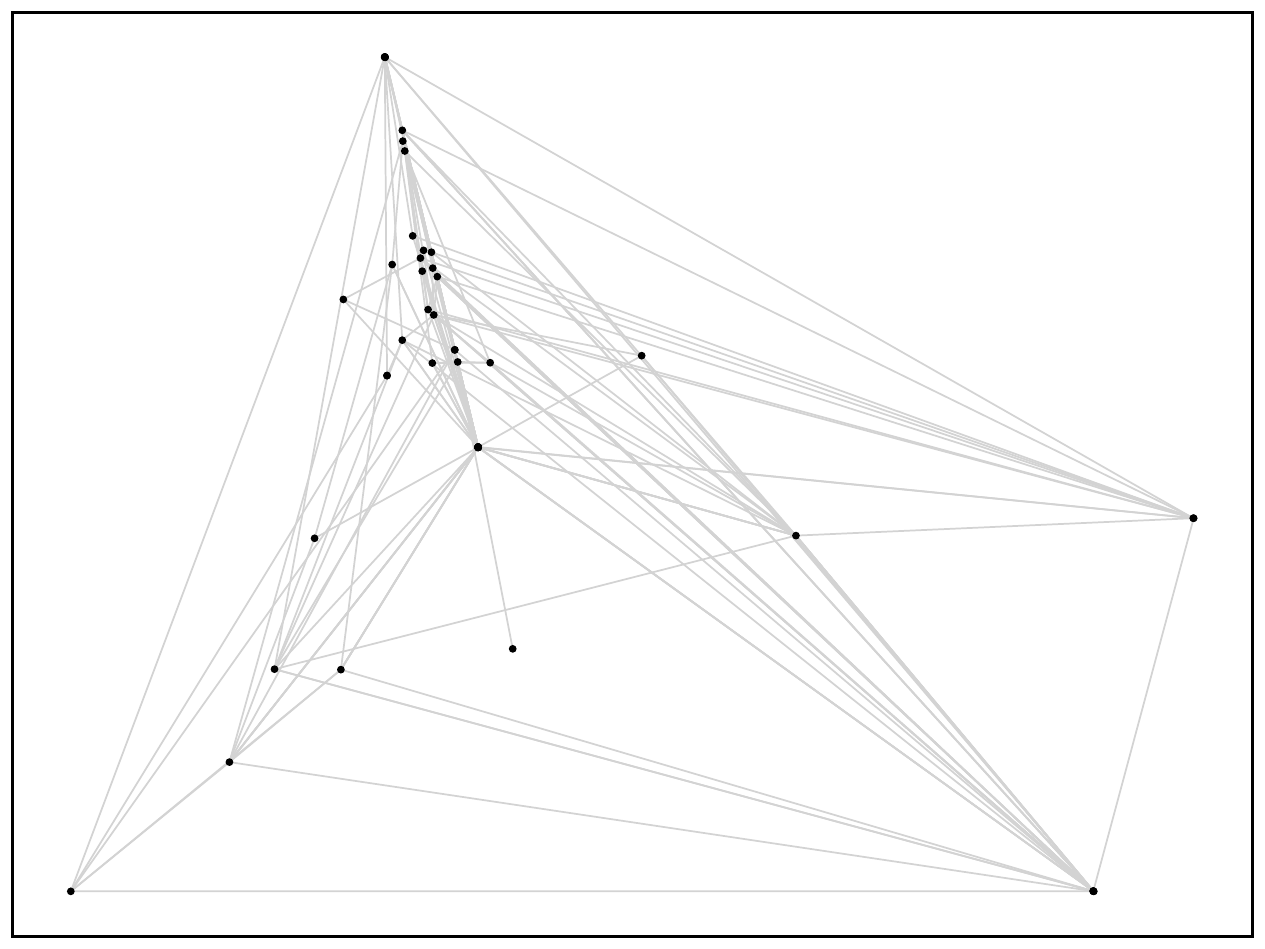}
    &
    \includegraphics[width=0.24\textwidth]{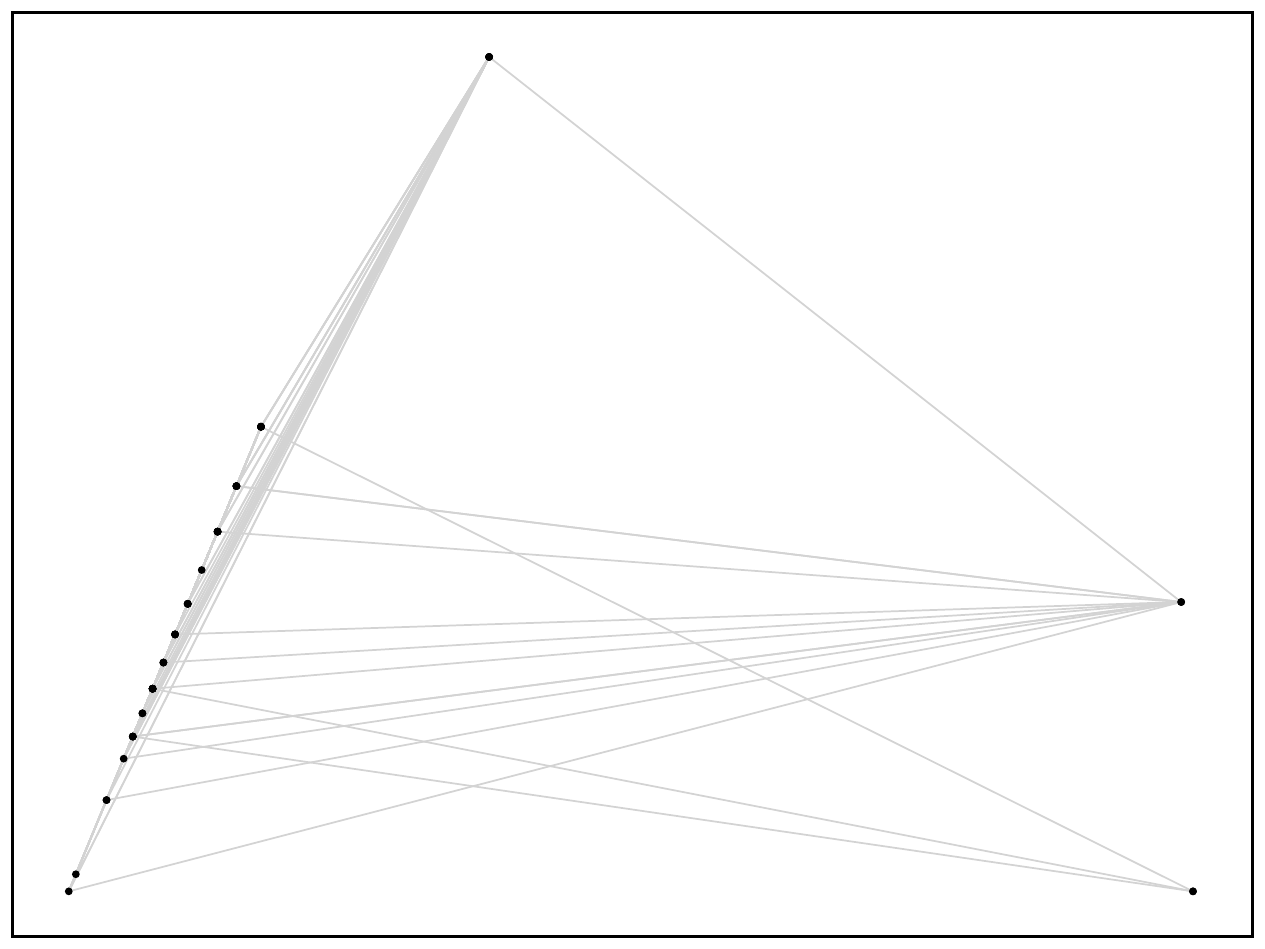}
    &
    \includegraphics[width=0.24\textwidth]{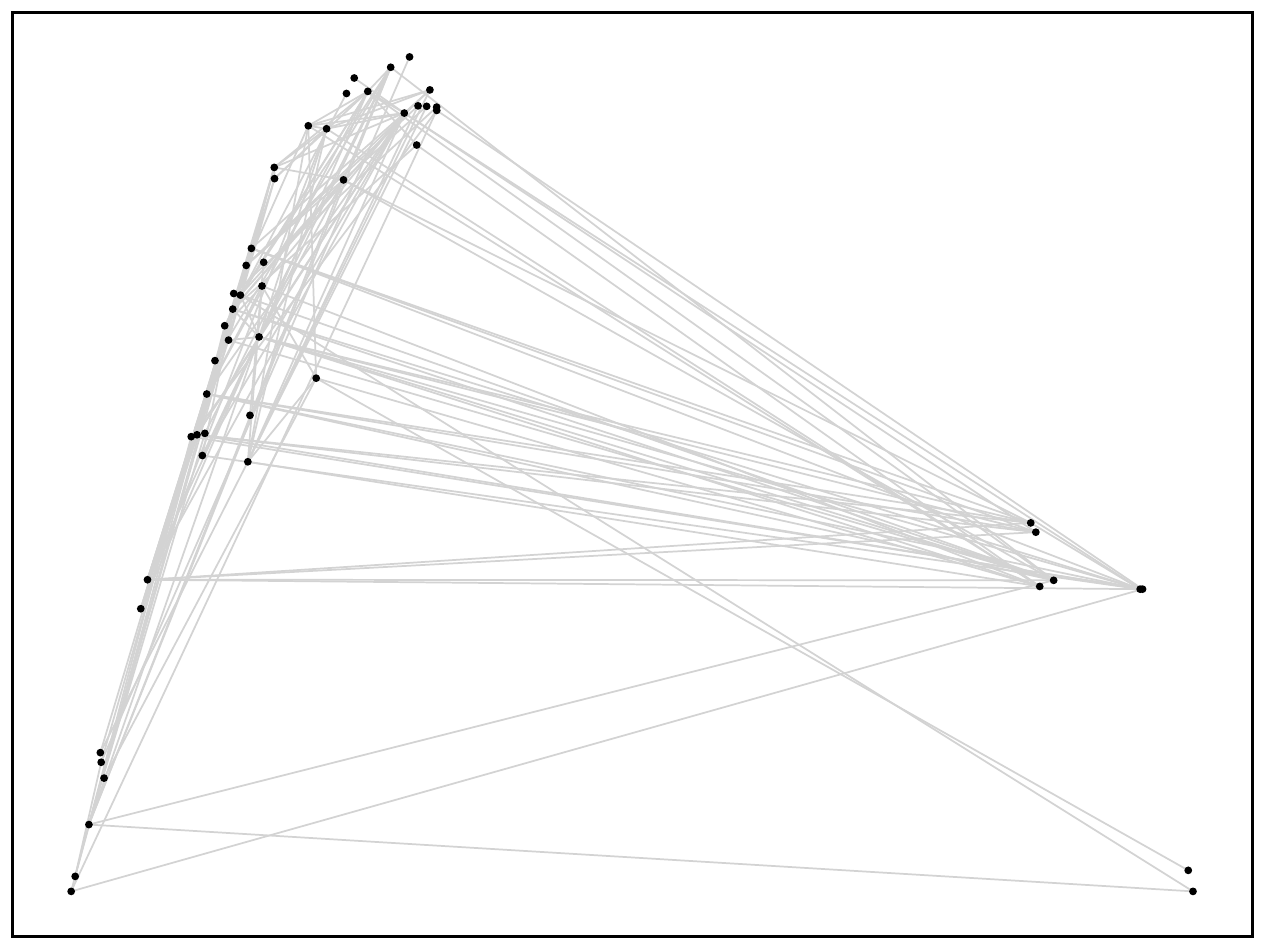}
    &
    \includegraphics[width=0.24\textwidth]{pprvizs_output/fbego-simrank.pdf}
    \\
    \footnotesize{\nd{}=3.6E+12, \ulcv{}=0.91} & \footnotesize{\nd{}=$\infty$, \ulcv{}=0.98} & \footnotesize{\nd{}=1.2E+05, \ulcv{}=0.96} & \footnotesize{\nd{}=6.2E+03, \ulcv{}=0.75}
    \\
    (g) $\star$~\gf~\cite{ahmed2013distributed} & 
    (h) $\star$~\leemb~\cite{belkin2003laplacian} &
    (i) $\star$~\nodevec~\cite{grover2016node2vec} & 
    (j) \simrank~\cite{jeh2002simrank}
    \\
  \end{tabular}
\end{small}
\caption{Visualization results for the \fbego graph: force-directed methods are marked with $\blacklozenge$; stress methods are marked with $\blacktriangle$; graph embedding methods are marked with $\star$.}\label{fig:fb-viz}
\end{figure*}

\begin{figure*}[]
\centering
\begin{small}
  \begin{tabular}{cc}
    \includegraphics[width=0.5\textwidth]{pprvizs_output/wikiedit-pprvizs.pdf}
     &
    \includegraphics[width=0.5\textwidth]{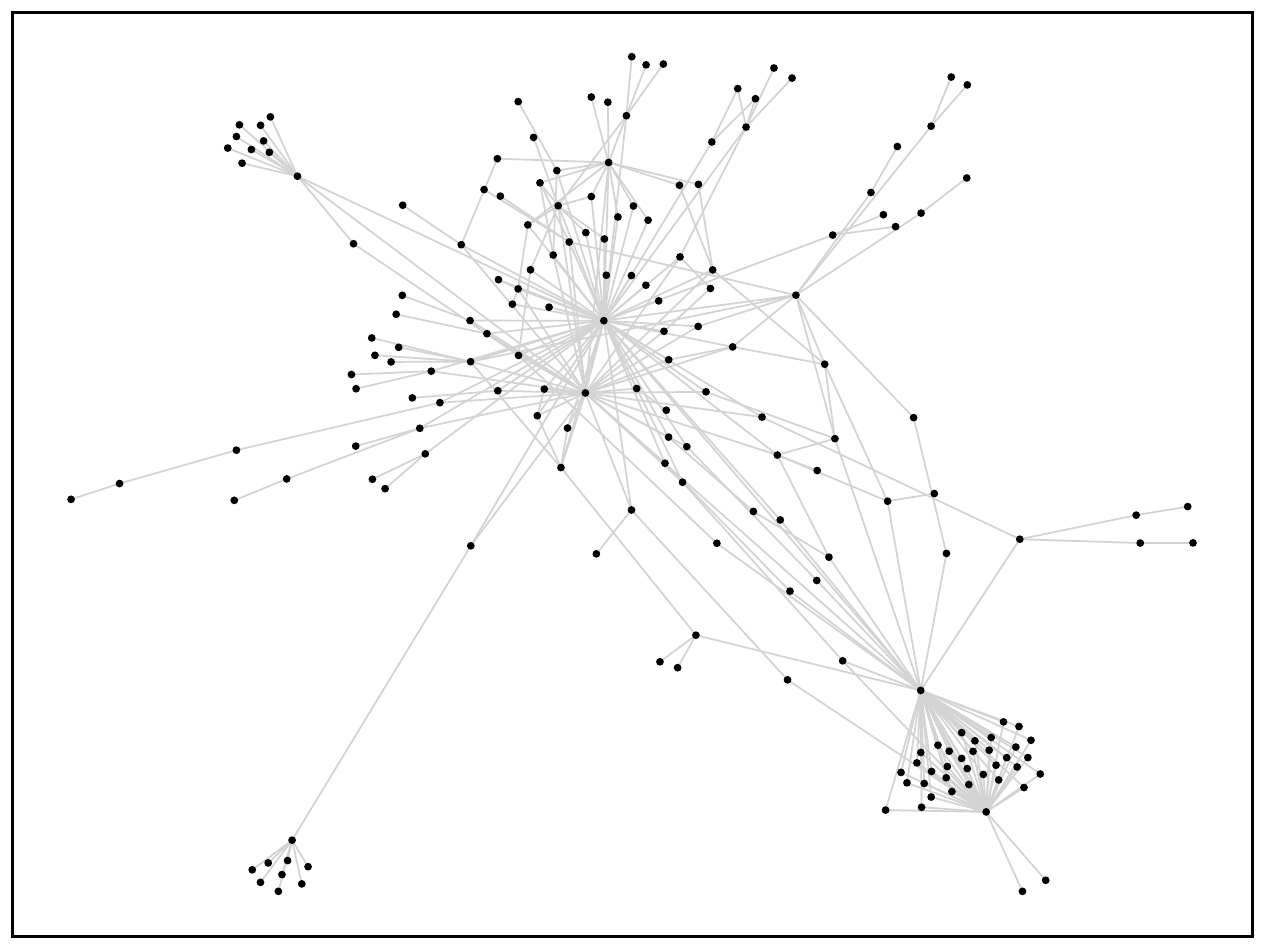}
    \\
    \footnotesize{\nd{}=2.7E+04, \ulcv{}=0.35} & \footnotesize{\nd{}=8.1E+04, \ulcv{}=0.64}
    \\
    (a) \pprviz (Ours)  & (b) $\blacklozenge$~\forceatlas~\cite{jacomy2014forceatlas2} 
    \\
  \end{tabular}
\end{small}
\vspace{-3mm}
\end{figure*}

\begin{figure*}[]
\centering
\begin{small}
  \begin{tabular}{cccc}
    \includegraphics[width=0.24\textwidth]{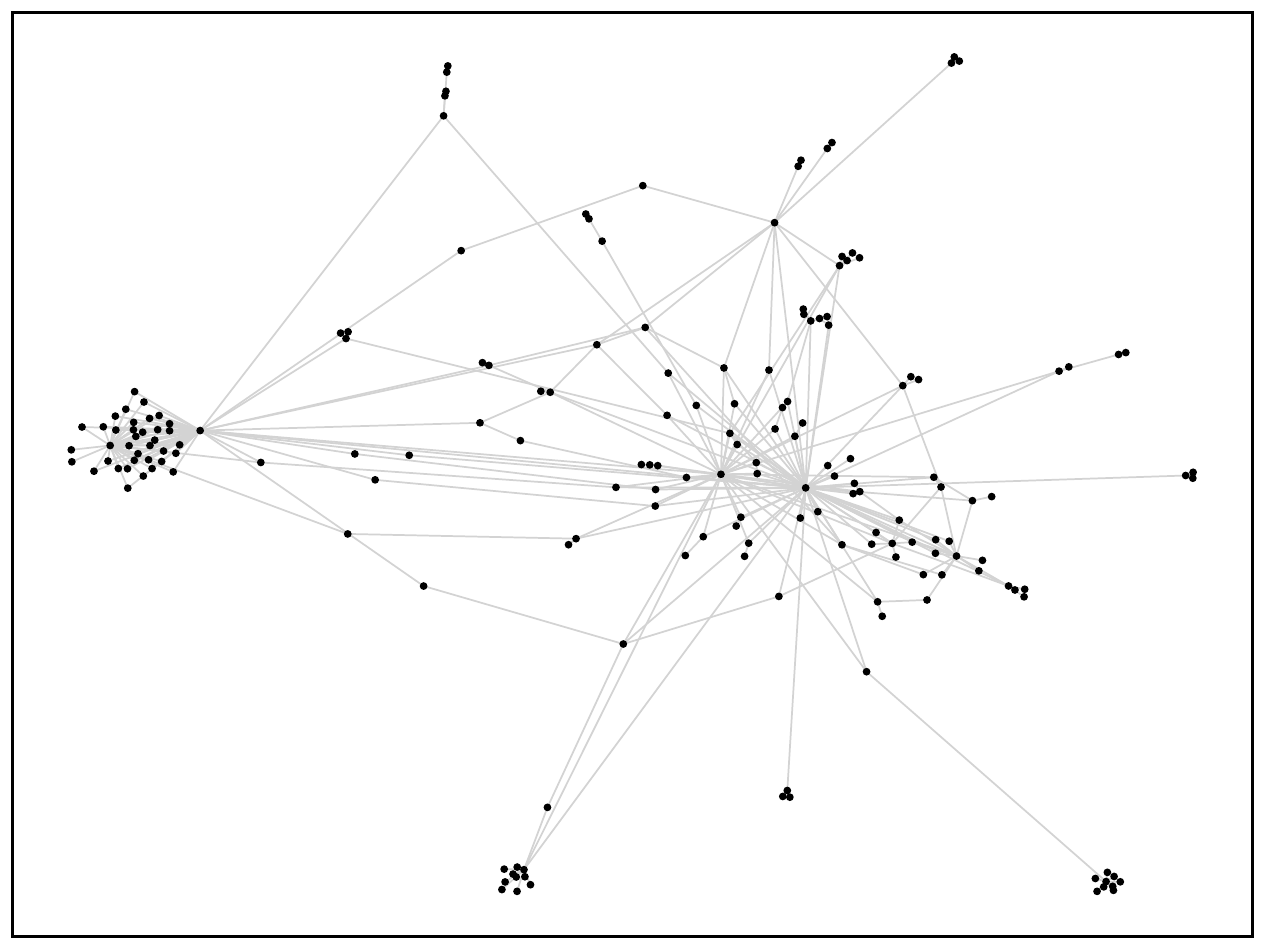}
    &
    \includegraphics[width=0.24\textwidth]{pprvizs_output/wikiedit-fr.pdf}
    &
    \includegraphics[width=0.24\textwidth]{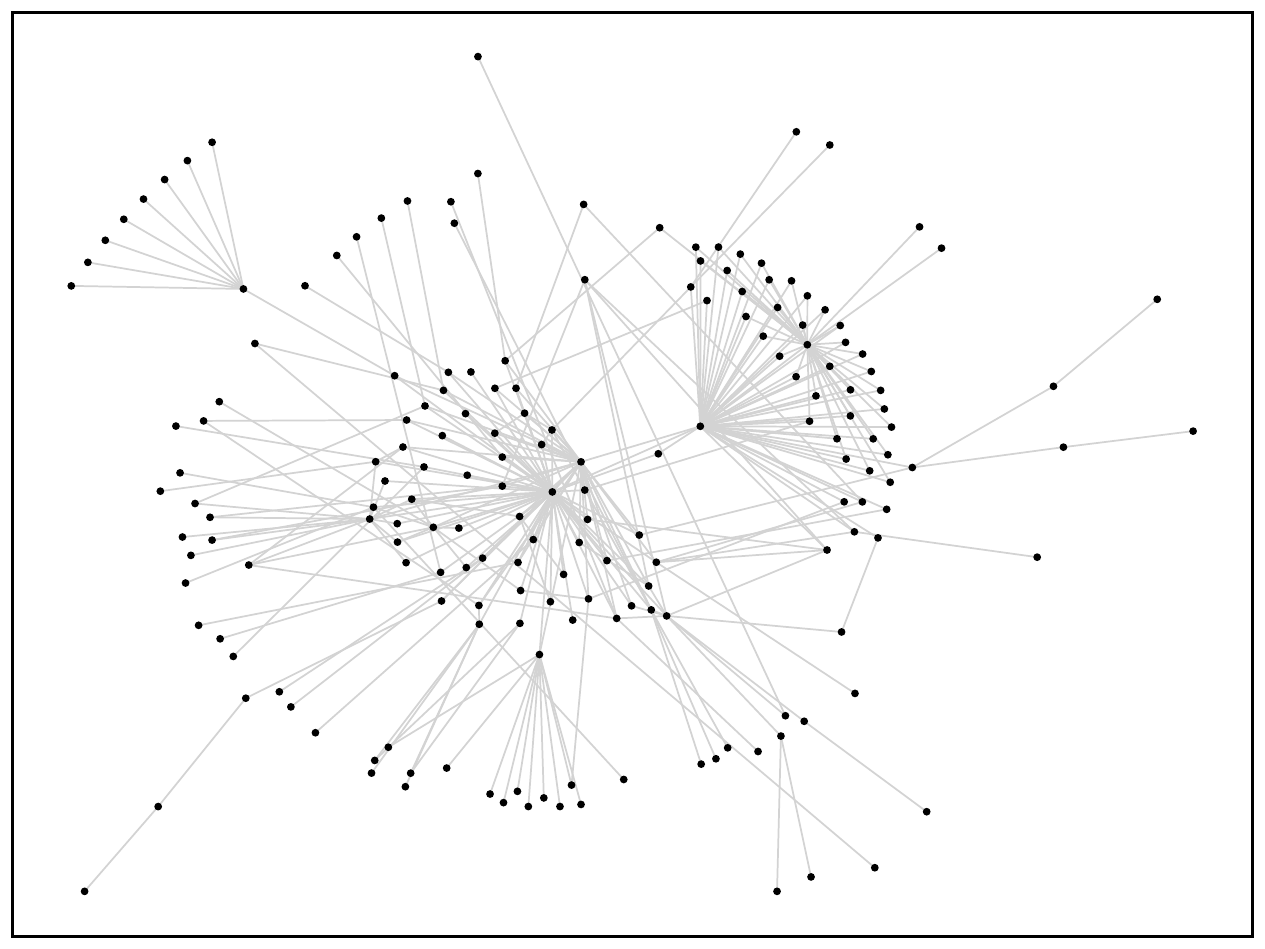}
    &
    \includegraphics[width=0.24\textwidth]{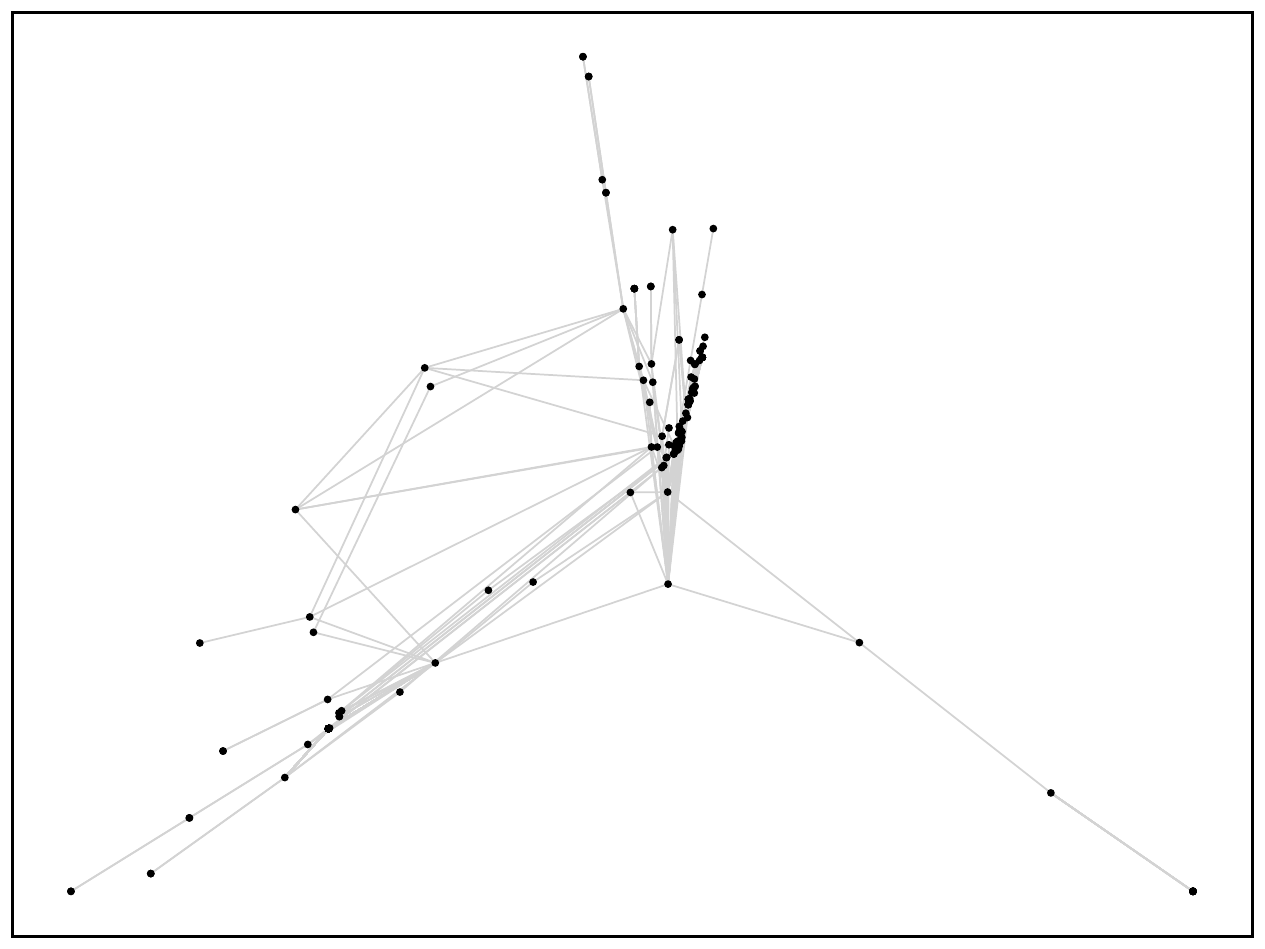}
    \\
    \footnotesize{\nd{}=1.4E+05, \ulcv{}=1.09} & \footnotesize{\nd{}=2.7E+04, \ulcv{}=0.41} & \footnotesize{\nd{}=4.9E+04, \ulcv{}=0.62} & \footnotesize{\nd{}=$\infty$, \ulcv{}=0.78}
    \\
    (c) $\blacklozenge$~\linlog~\cite{noack2005energy} 
    &
    (d) $\blacklozenge$~\fr~\cite{fruchterman1991graph} 
    & 
    (e) $\blacktriangle$~\mds~\cite{gansner2004graph}
    &
    (f) $\blacktriangle$~\pivotmds~\cite{brandes2006eigensolver} 
    \\
    \includegraphics[width=0.24\textwidth]{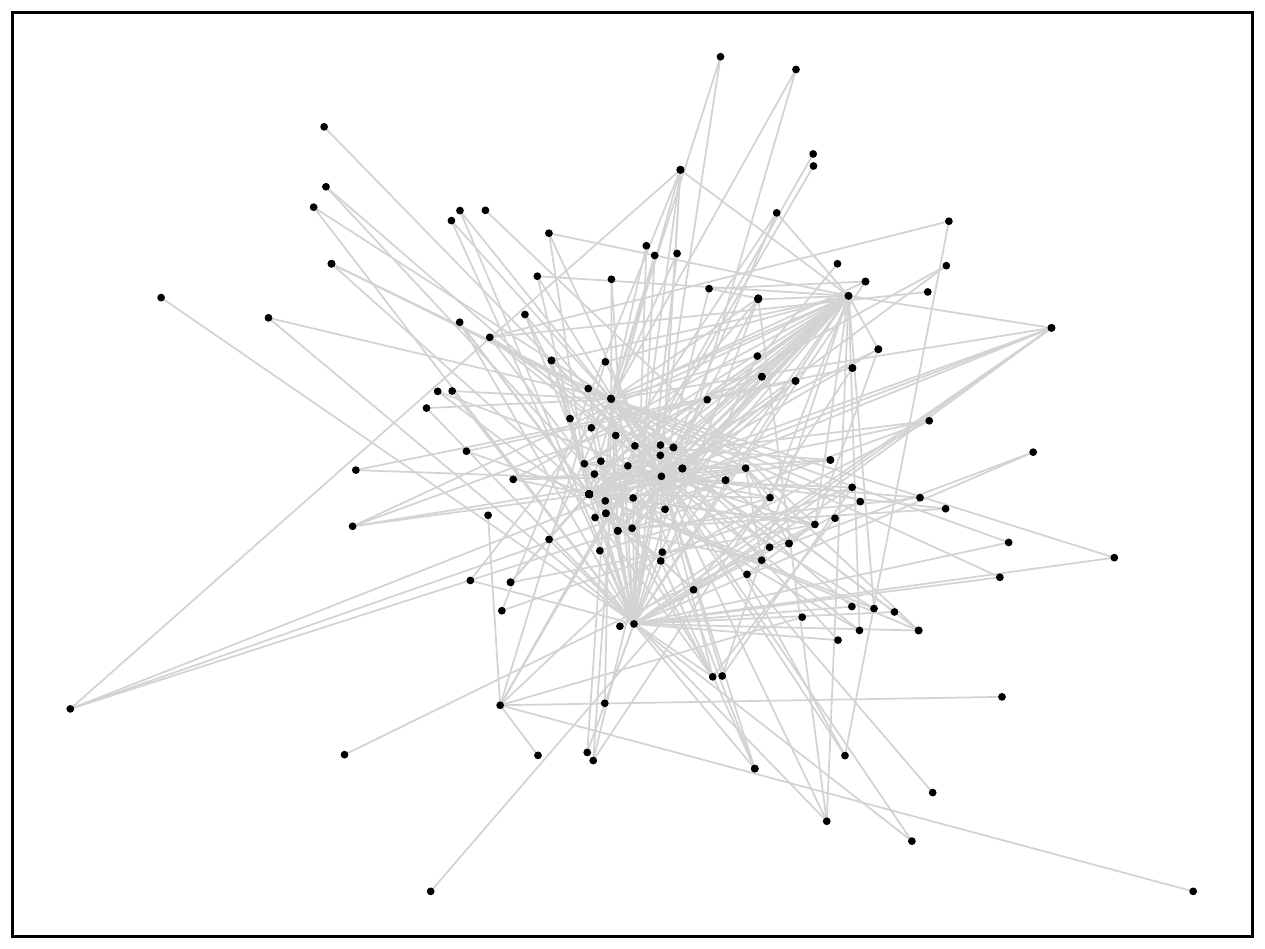}
    &
    \includegraphics[width=0.24\textwidth]{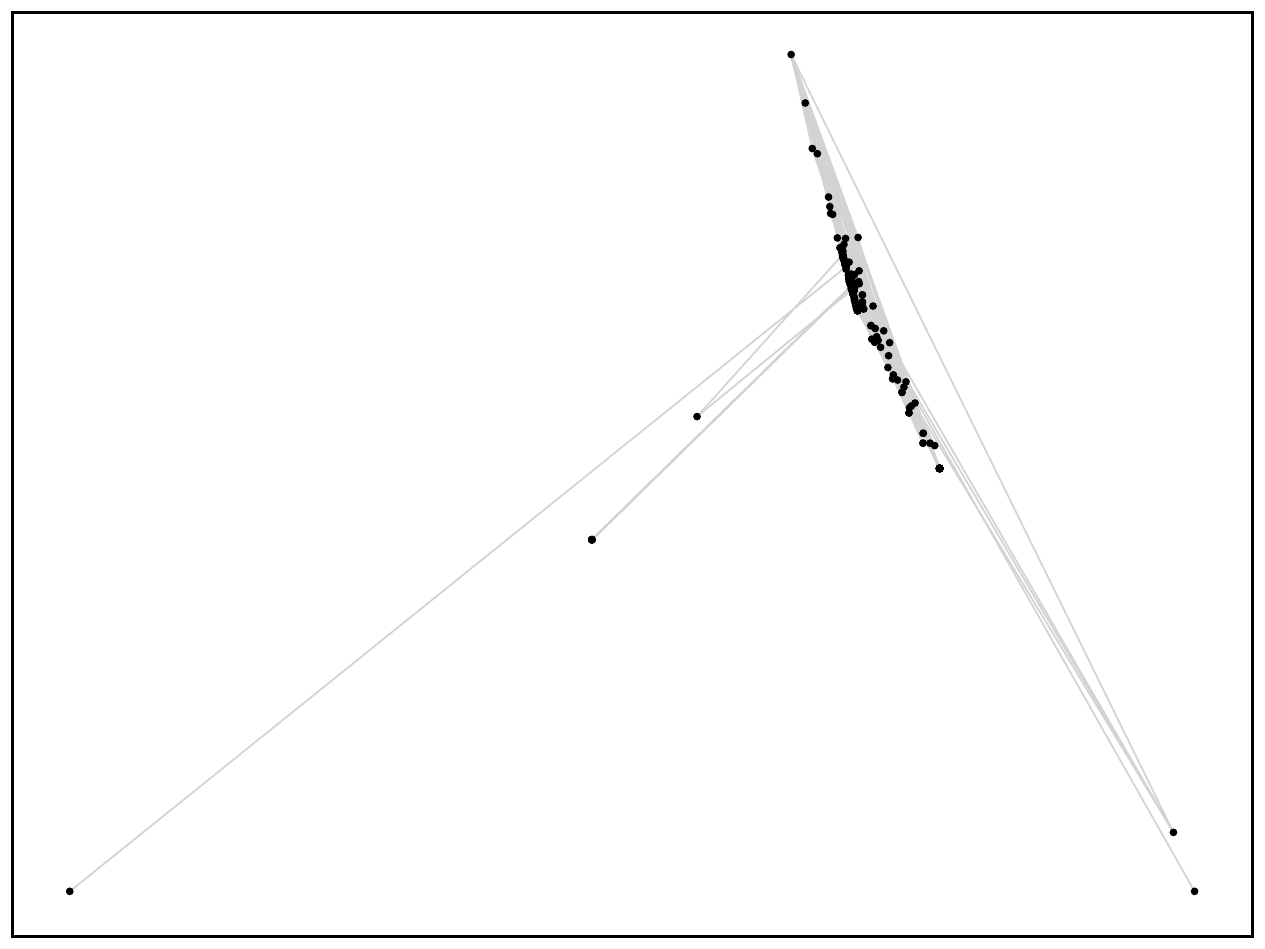}
    &
    \includegraphics[width=0.24\textwidth]{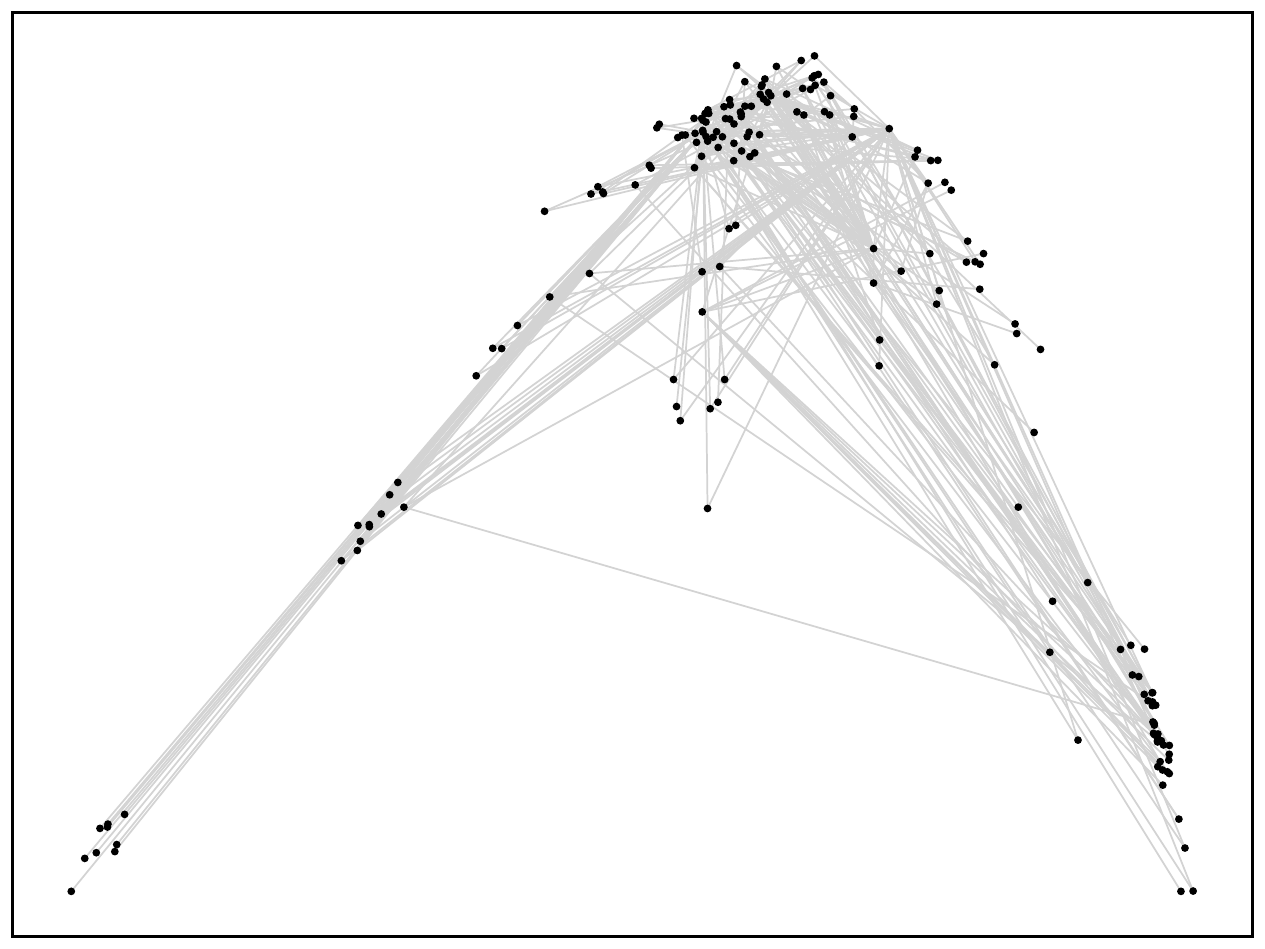}
    &
    \includegraphics[width=0.24\textwidth]{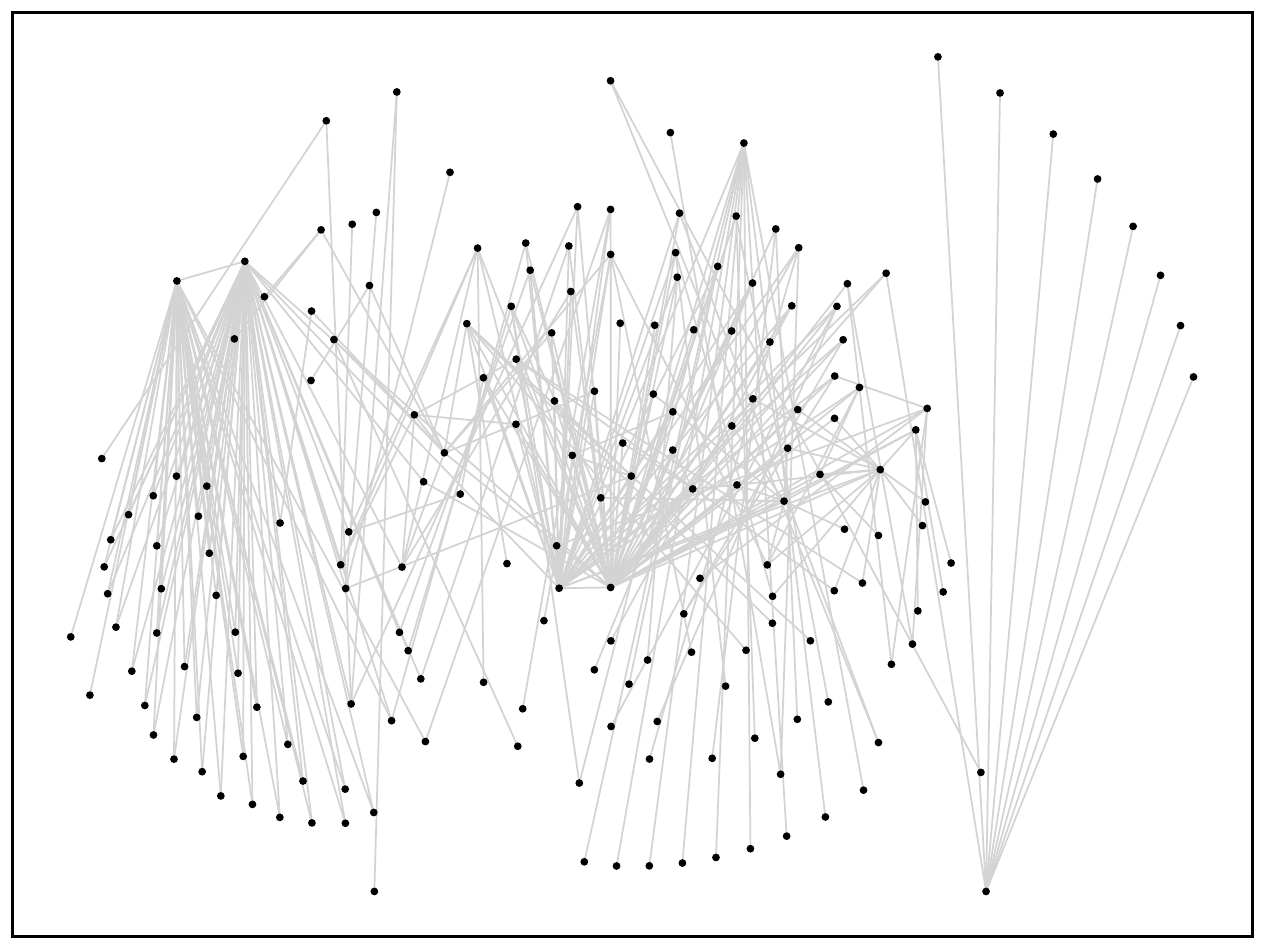}
    \\
    \footnotesize{\nd{}=9.2E+11, \ulcv{}=0.62} & \footnotesize{\nd{}=$\infty$, \ulcv{}=1.04} & \footnotesize{\nd{}=2.5E+06, \ulcv{}=0.86} & \footnotesize{\nd{}=2.7E+04, \ulcv{}=0.53}
    \\
    (g) $\star$~\gf~\cite{ahmed2013distributed} & 
    (h) $\star$~\leemb~\cite{belkin2003laplacian} &
    (i) $\star$~\nodevec~\cite{grover2016node2vec} & 
    (j) \simrank~\cite{jeh2002simrank}
    \\
  \end{tabular}
\end{small}
\caption{Visualization results for the \wiki graph: force-directed methods are marked with $\blacklozenge$; stress methods are marked with $\blacktriangle$; graph embedding methods are marked with $\star$.}\label{fig:trust-viz}
\end{figure*}

\begin{figure*}[]
\centering
\begin{small}
  \begin{tabular}{cc}
    \includegraphics[width=0.5\textwidth]{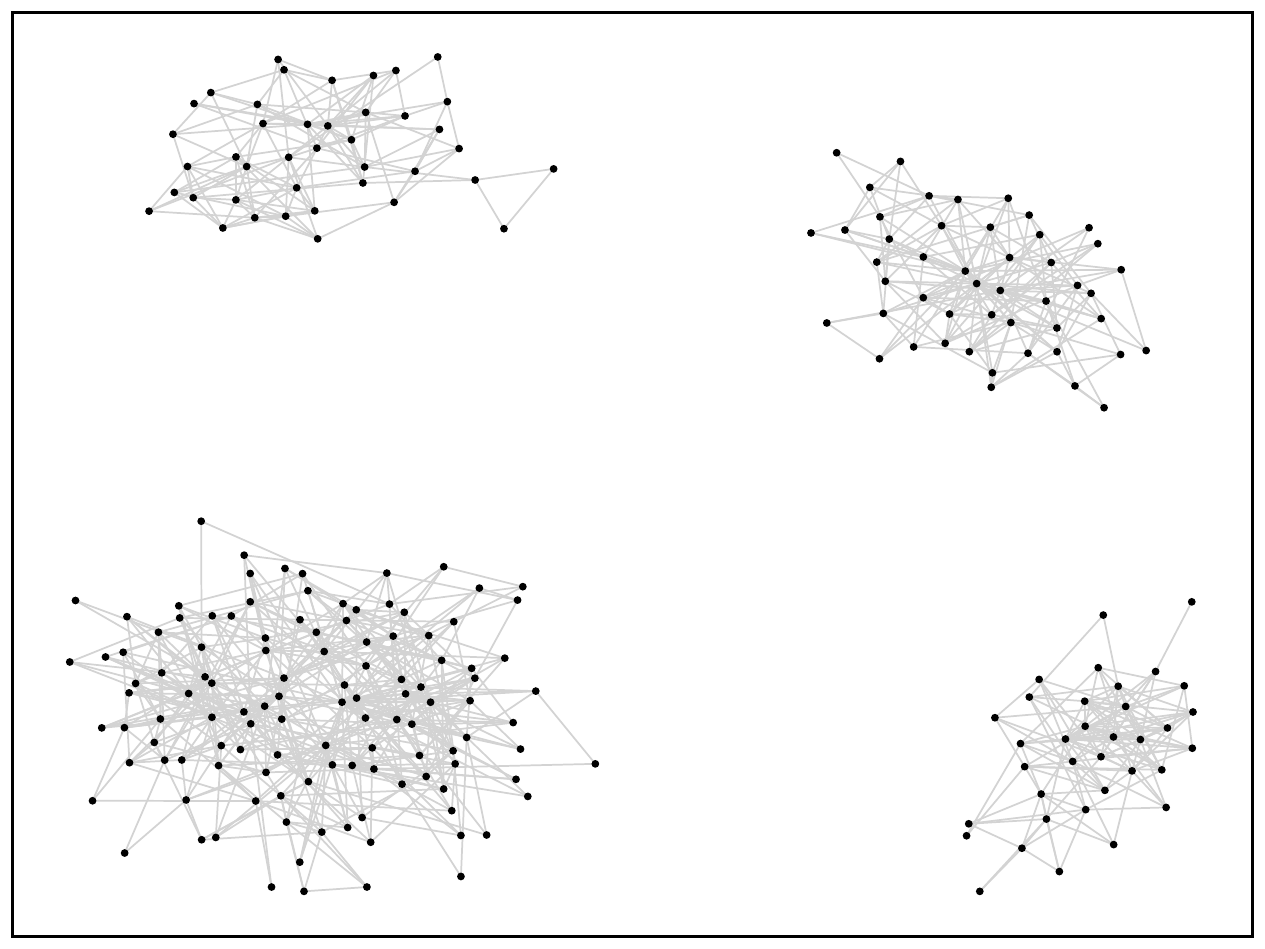}
     &
    \includegraphics[width=0.5\textwidth]{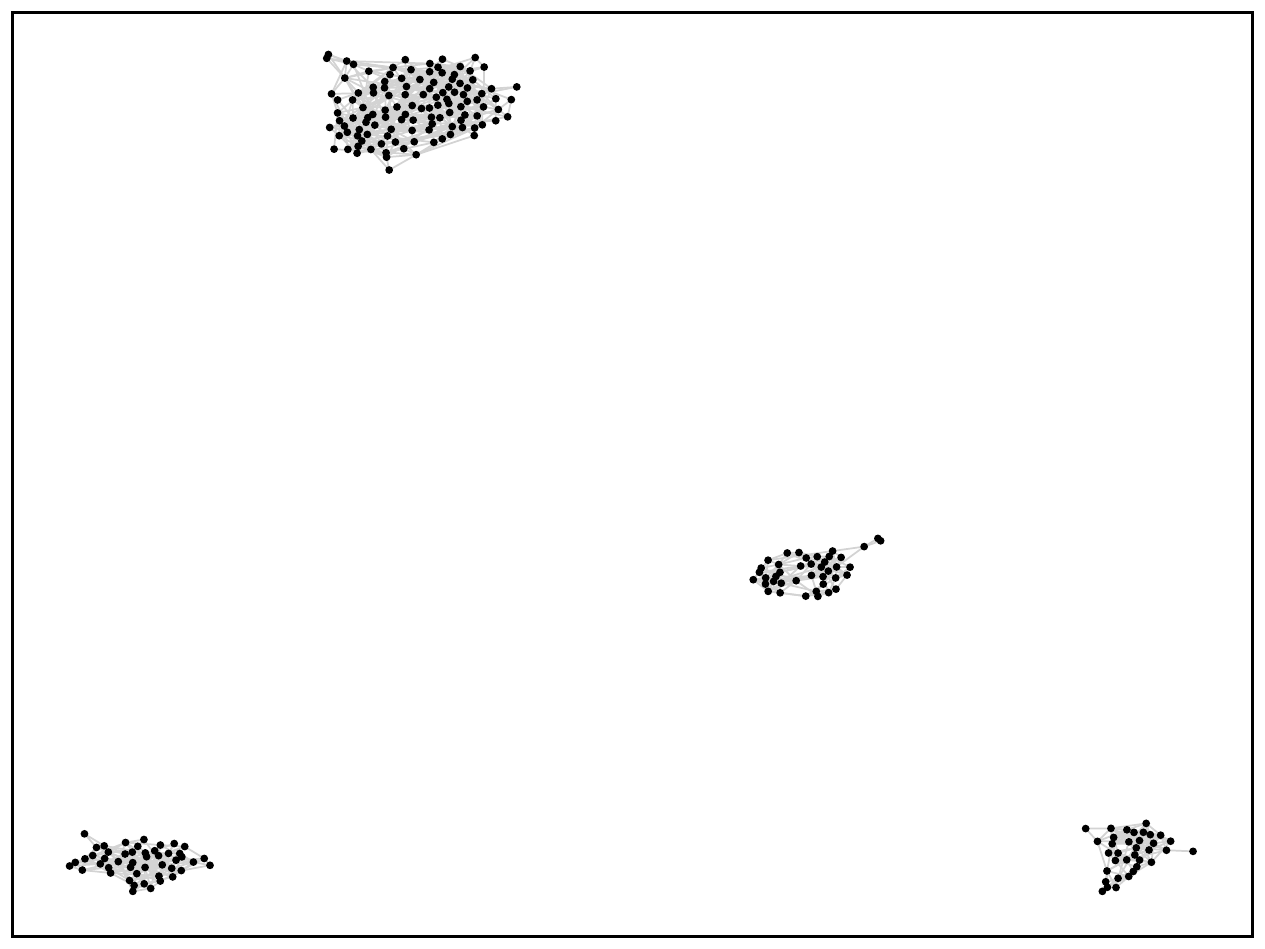}
    \\
    \footnotesize{\nd{}=6.7E+04, \ulcv{}=0.45} & \footnotesize{\nd{}=8.2E+05, \ulcv{}=0.55}
    \\
    (a) \pprviz (Ours)  & (b) $\blacklozenge$~\forceatlas~\cite{jacomy2014forceatlas2} 
    \\
  \end{tabular}
\end{small}
\vspace{-3mm}
\end{figure*}

\begin{figure*}[]
\centering
\begin{small}
  \begin{tabular}{cccc}
    \includegraphics[width=0.24\textwidth]{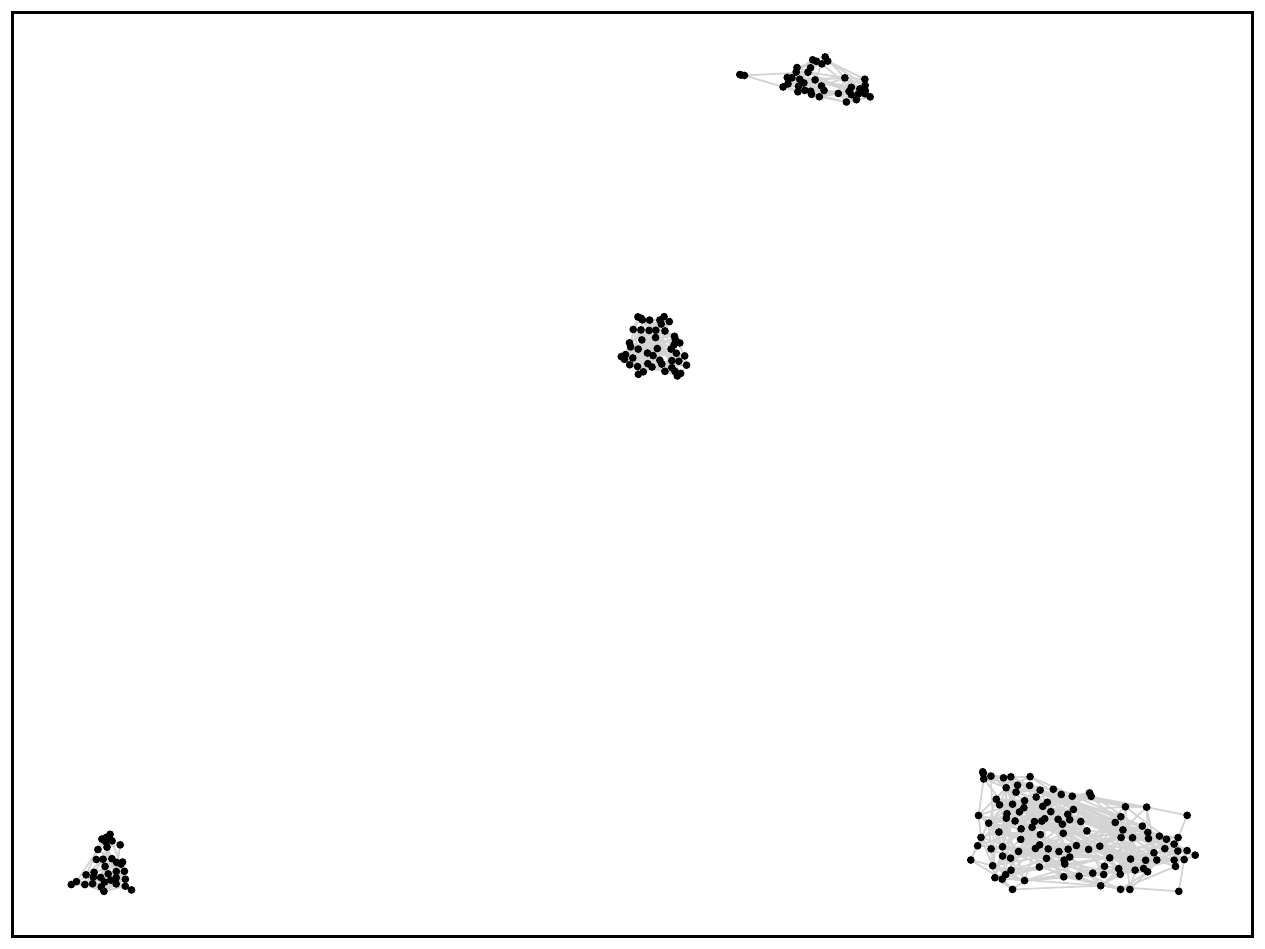}
    &
    \includegraphics[width=0.24\textwidth]{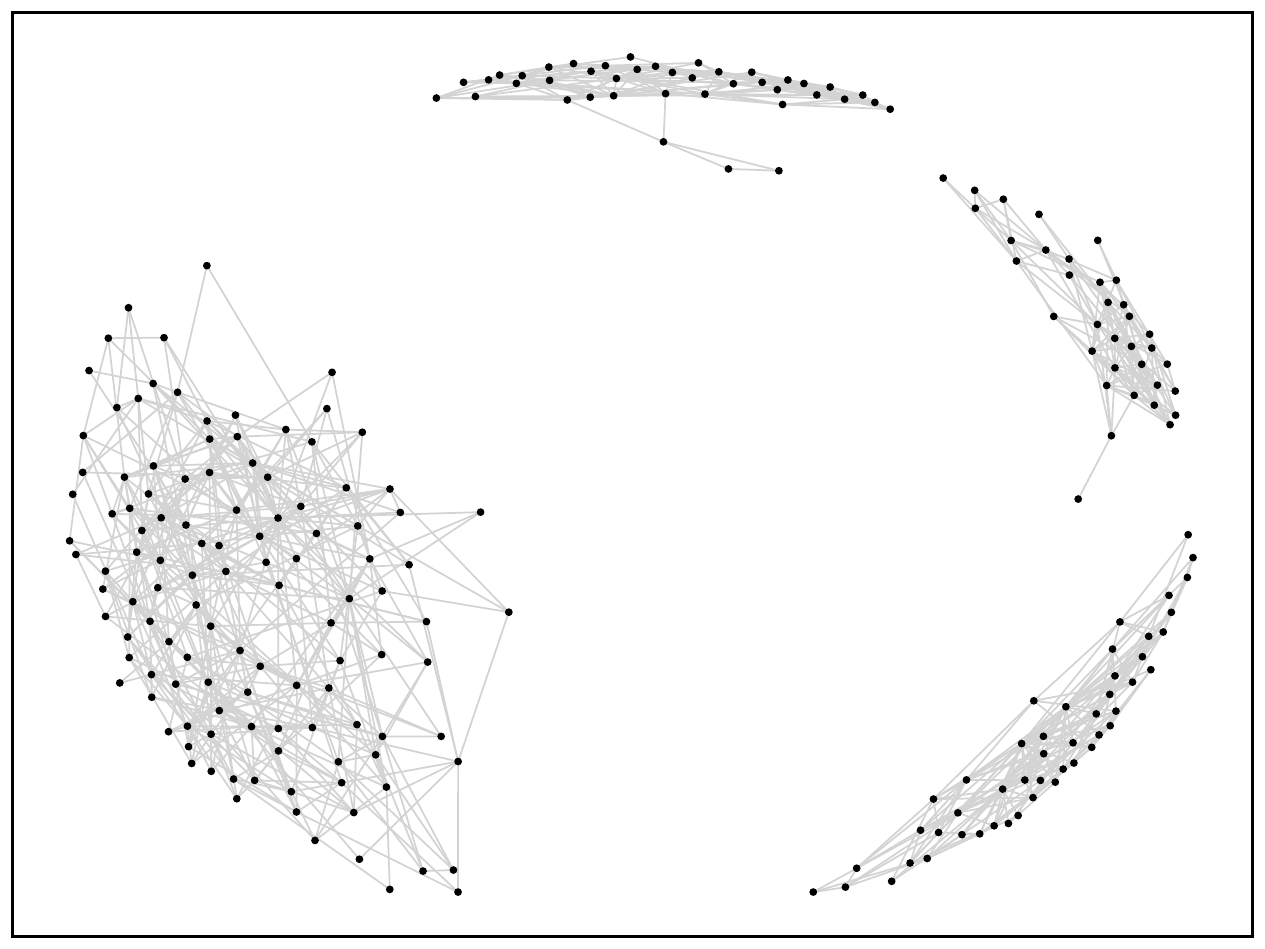}
    &
    \includegraphics[width=0.24\textwidth]{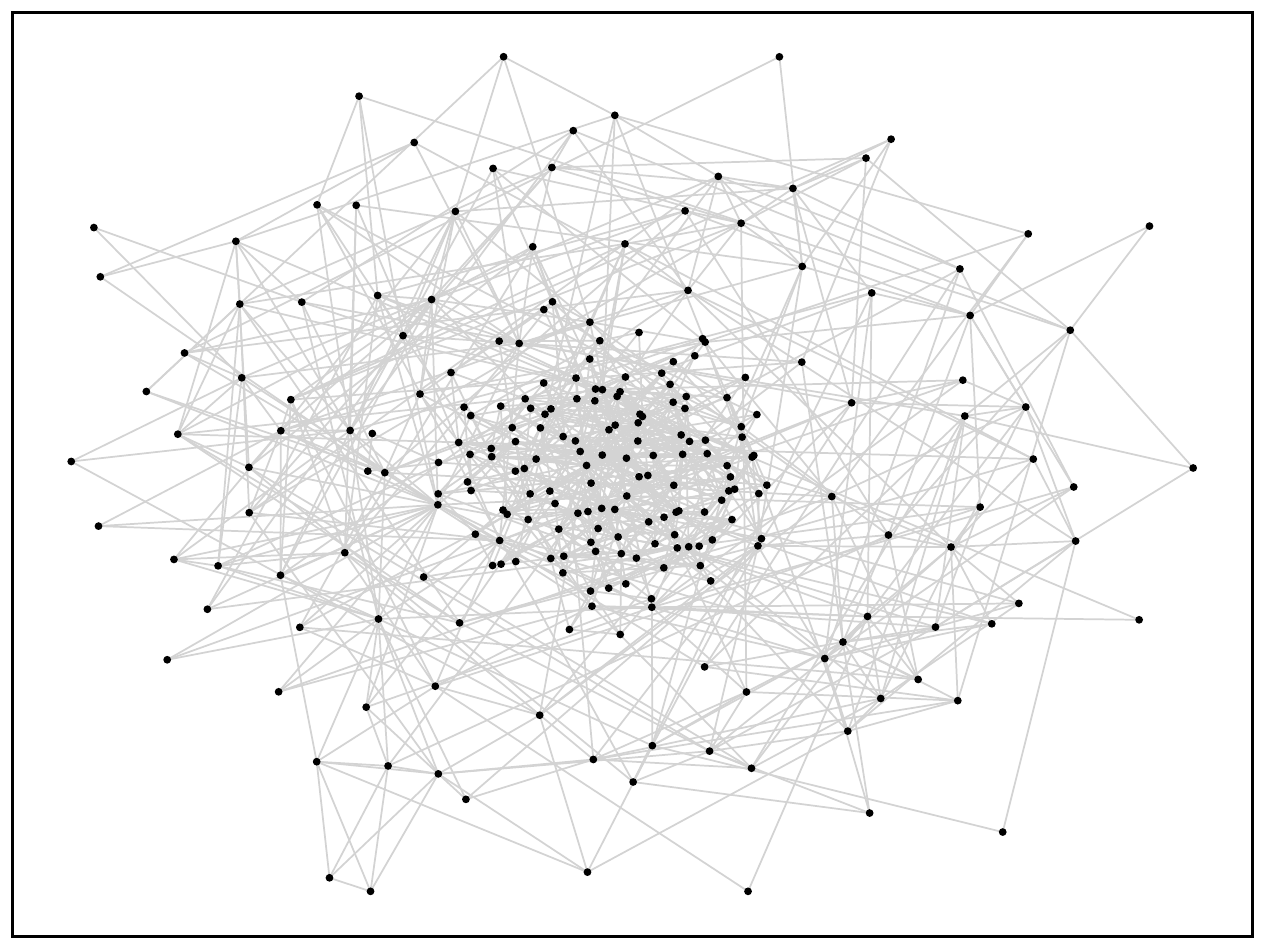}
    &
    \includegraphics[width=0.24\textwidth]{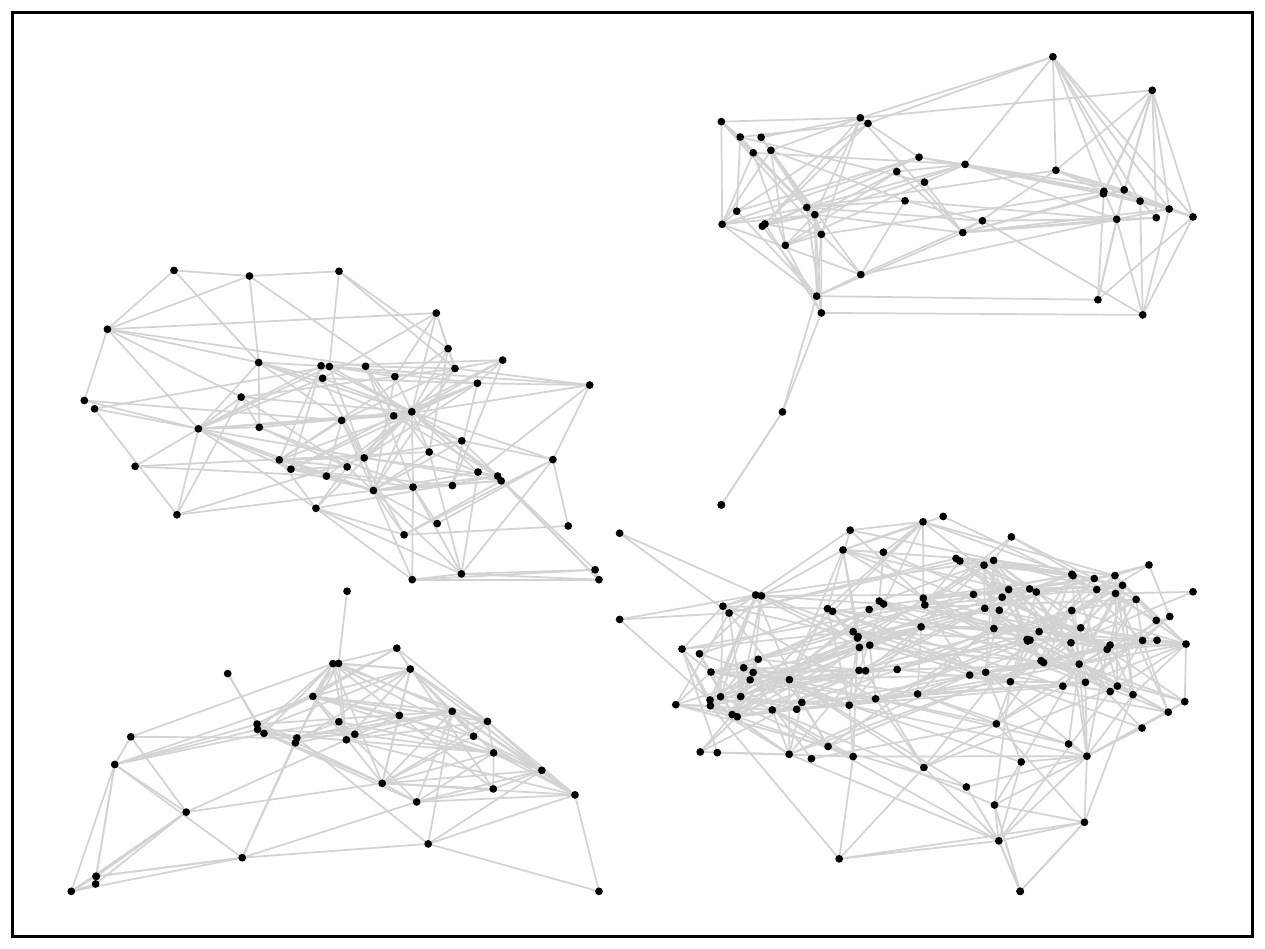}
    \\
    \footnotesize{\nd{}=7.6E+05, \ulcv{}=0.90} & \footnotesize{\nd{}=8.7E+04, \ulcv{}=0.53} & \footnotesize{\nd{}=1.5E+05, \ulcv{}=0.80} & \footnotesize{\nd{}=$\infty$, \ulcv{}=0.47}
    \\
    (c) $\blacklozenge$~\linlog~\cite{noack2005energy} 
    &
    (d) $\blacklozenge$~\fr~\cite{fruchterman1991graph} 
    & 
    (e) $\blacktriangle$~\mds~\cite{gansner2004graph}
    &
    (f) $\blacktriangle$~\pivotmds~\cite{brandes2006eigensolver} 
    \\
    \includegraphics[width=0.24\textwidth]{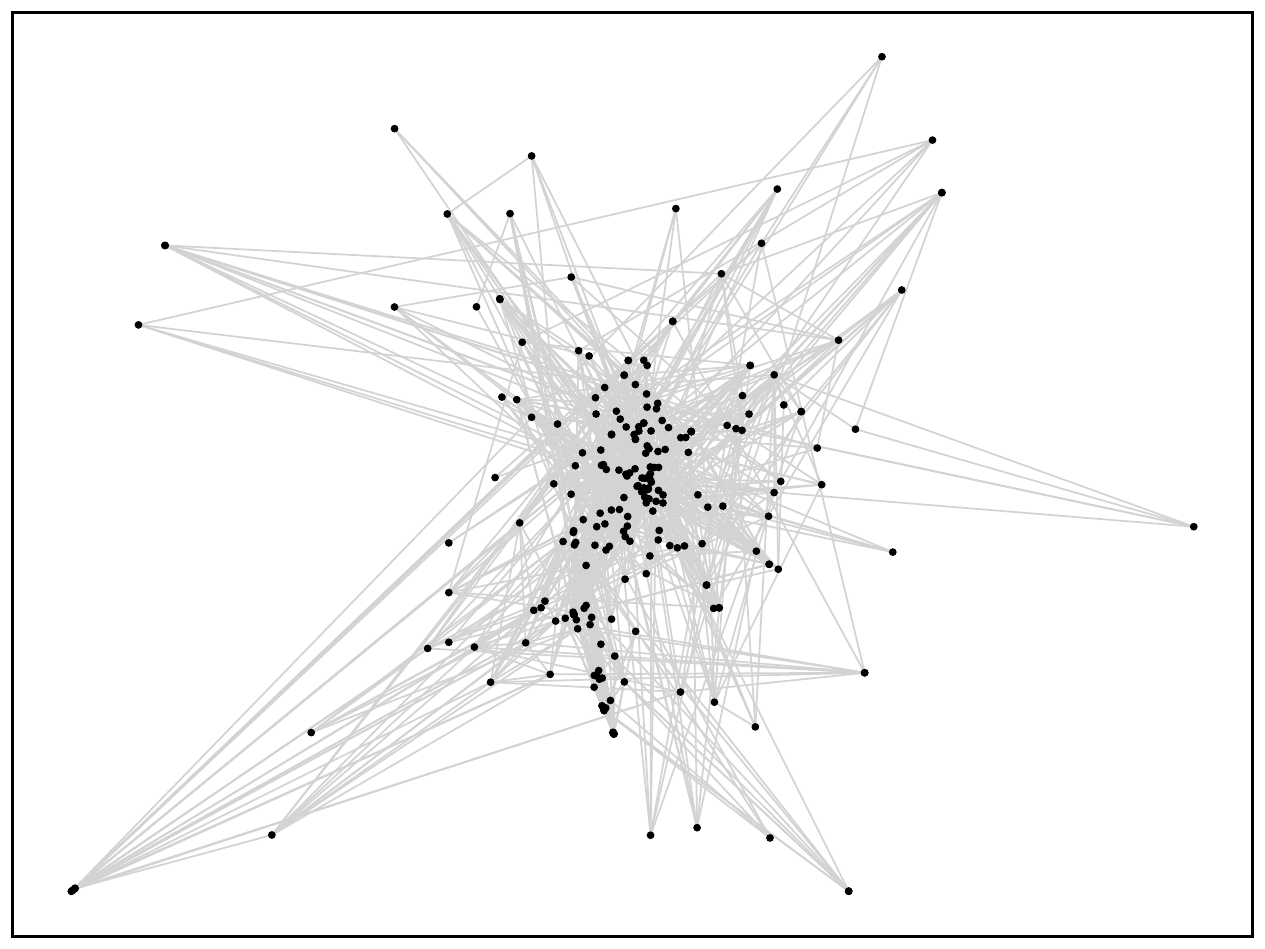}
    &
    \includegraphics[width=0.24\textwidth]{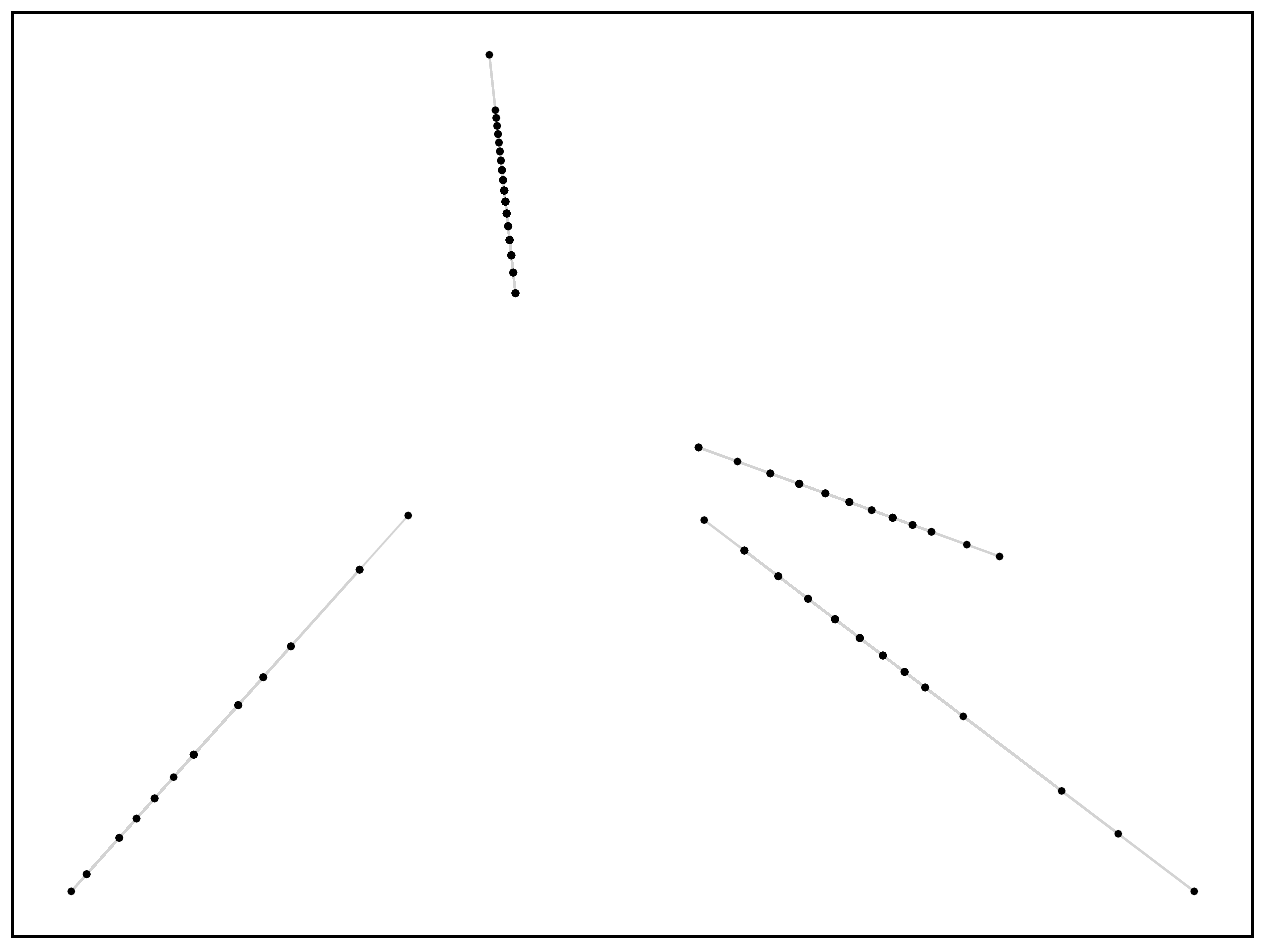}
    &
    \includegraphics[width=0.24\textwidth]{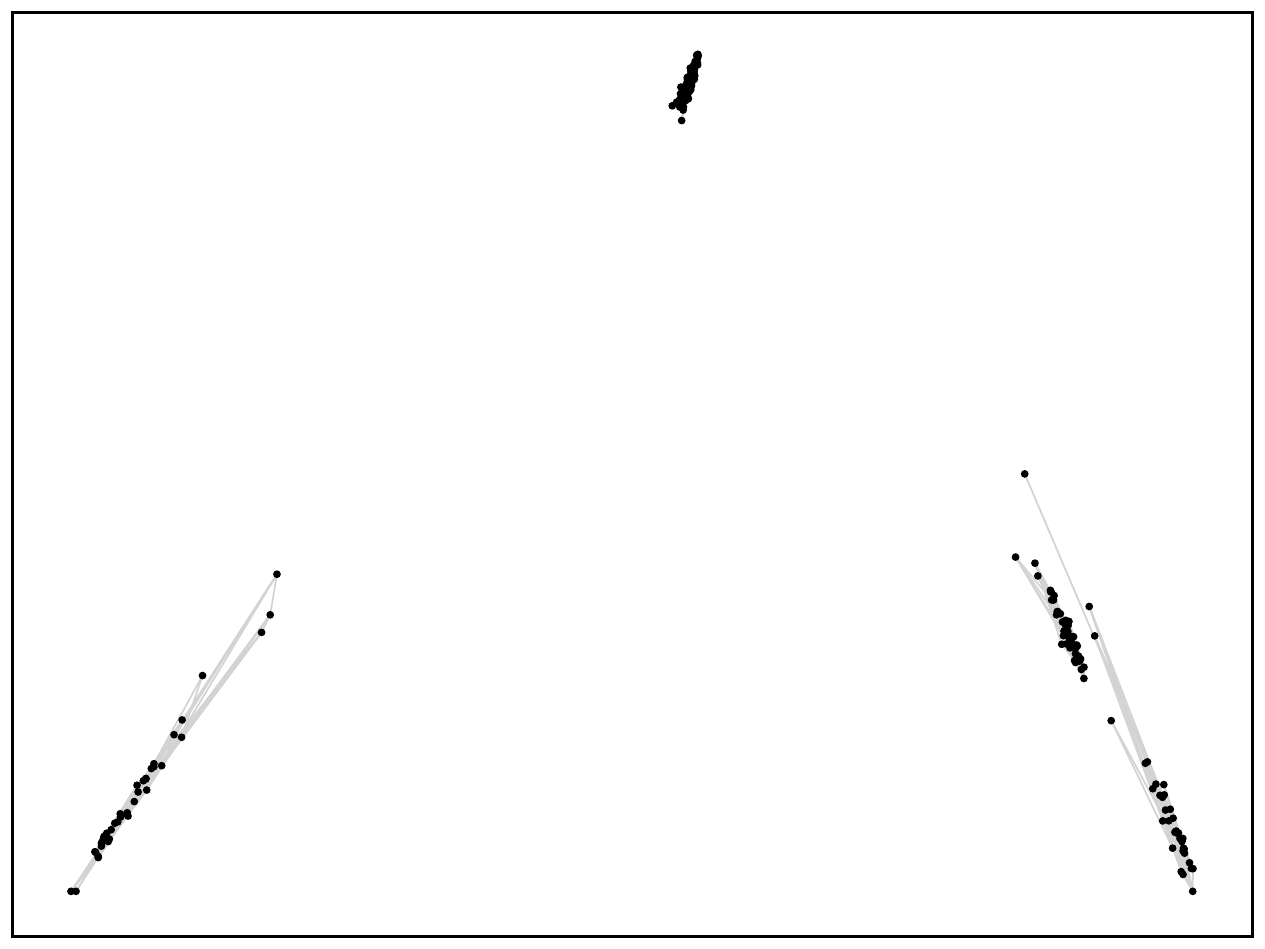}
    &
    \includegraphics[width=0.24\textwidth]{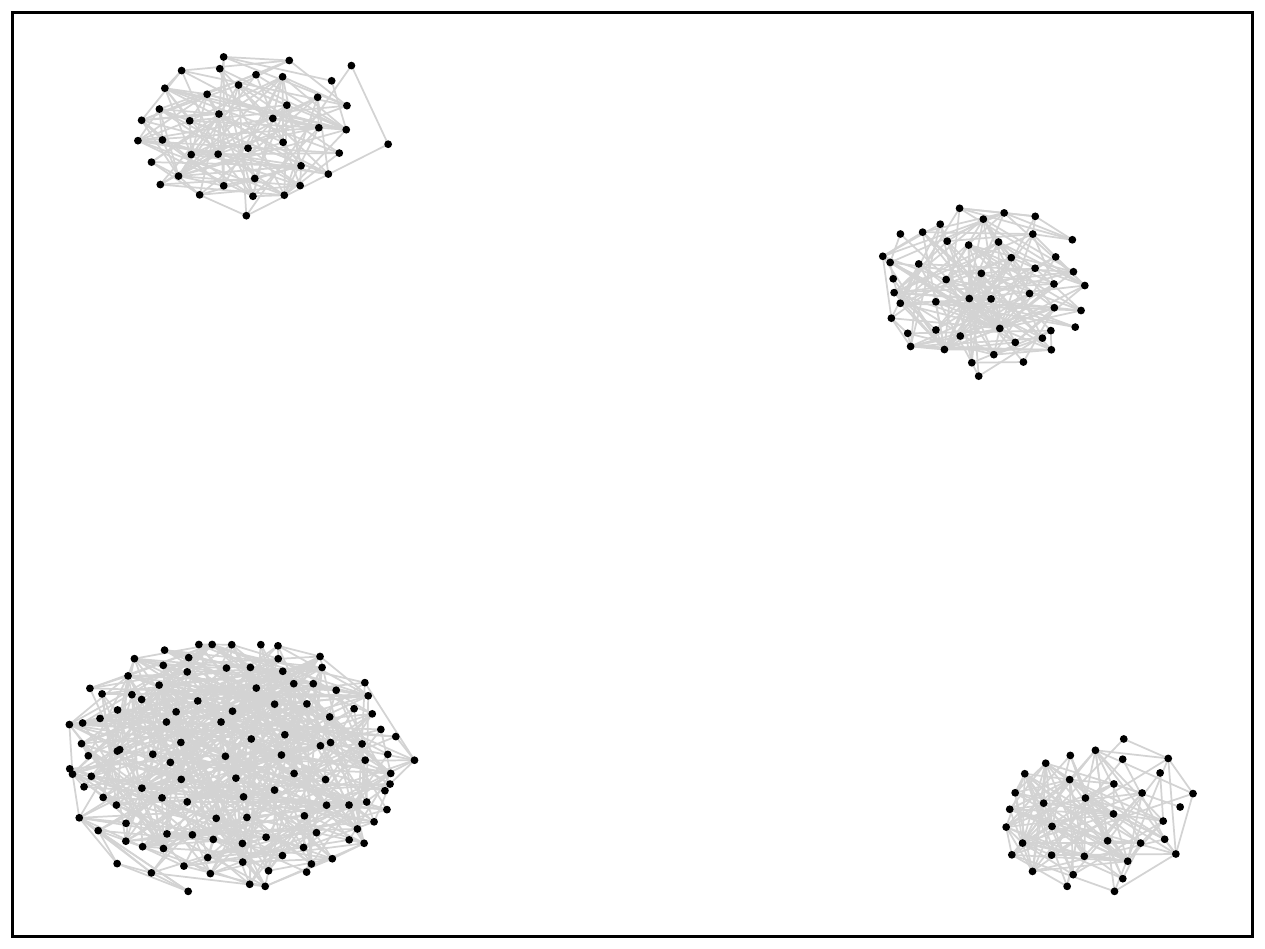}
    \\
    \footnotesize{\nd{}=2.5E+10, \ulcv{}=0.95} & \footnotesize{\nd{}=$\infty$, \ulcv{}=1.02} & \footnotesize{\nd{}=9.4E+07, \ulcv{}=1.41} & \footnotesize{\nd{}=1.1E+05, \ulcv{}=0.53}
    \\
    (g) $\star$~\gf~\cite{ahmed2013distributed} & 
    (h) $\star$~\leemb~\cite{belkin2003laplacian} &
    (i) $\star$~\nodevec~\cite{grover2016node2vec} & 
    (j) \simrank~\cite{jeh2002simrank}
    \\
  \end{tabular}
\end{small}
\caption{Visualization results for the \physic graph: force-directed methods are marked with $\blacklozenge$; stress methods are marked with $\blacktriangle$; graph embedding methods are marked with $\star$.}\label{fig:wiki-viz}
\end{figure*}

\begin{figure*}[]
\centering
\begin{small}
  \begin{tabular}{cc}
    \includegraphics[width=0.5\textwidth]{pprvizs_output/trust-pprvizs.pdf}
     &
    \includegraphics[width=0.5\textwidth]{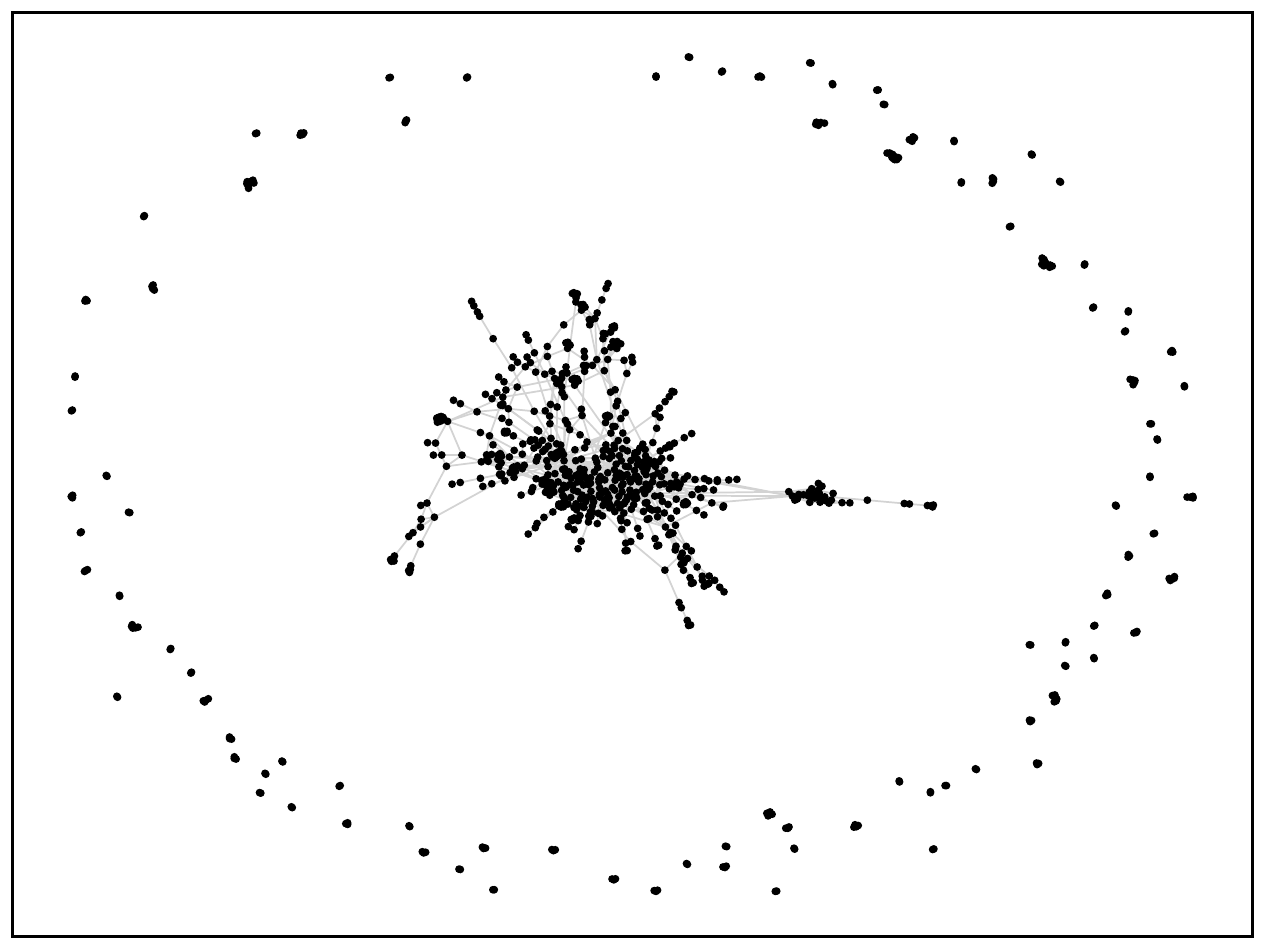}
    \\
    \footnotesize{\nd{}=9.1E+05, \ulcv{}=0.48} & \footnotesize{\nd{}=1.4E+07, \ulcv{}=0.96}
    \\
    (a) \pprviz (Ours)  & (b) $\blacklozenge$~\forceatlas~\cite{jacomy2014forceatlas2} 
    \\
  \end{tabular}
\end{small}
\vspace{-3mm}
\end{figure*}

\begin{figure*}[]
\centering
\begin{small}
  \begin{tabular}{cccc}
    \includegraphics[width=0.24\textwidth]{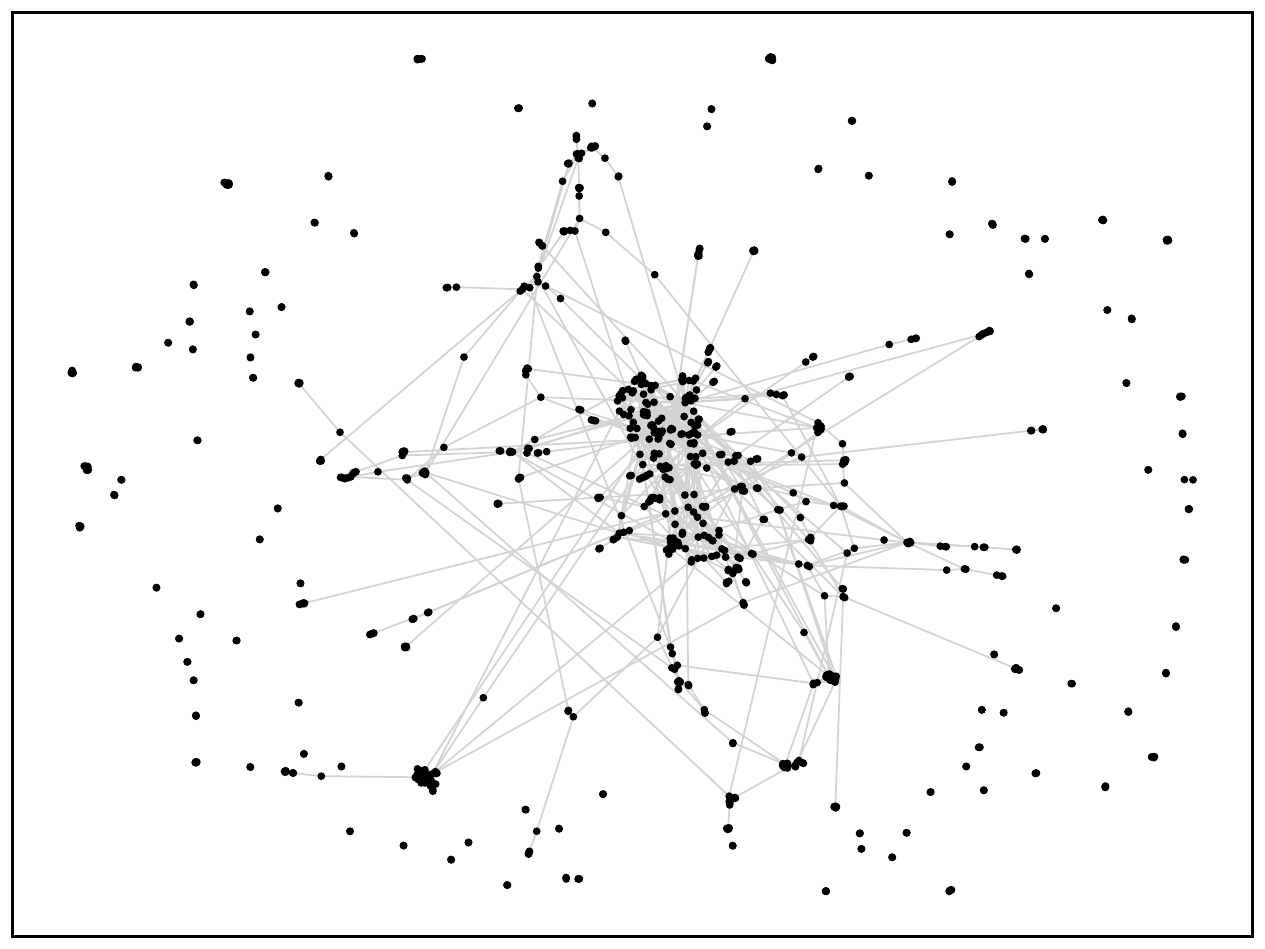}
    &
    \includegraphics[width=0.24\textwidth]{pprvizs_output/trust-fr.pdf}
    &
    \includegraphics[width=0.24\textwidth]{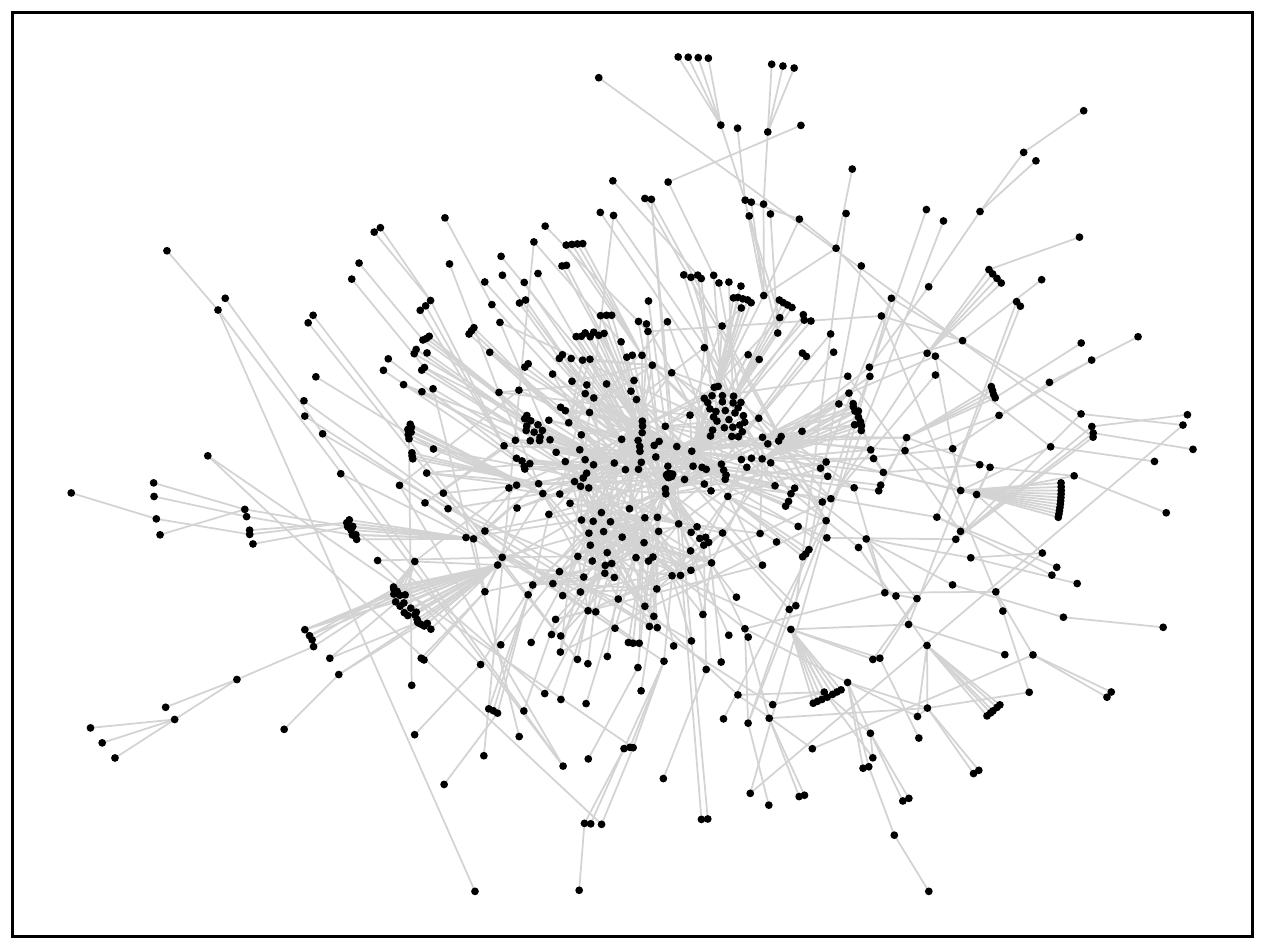}
    &
    \includegraphics[width=0.24\textwidth]{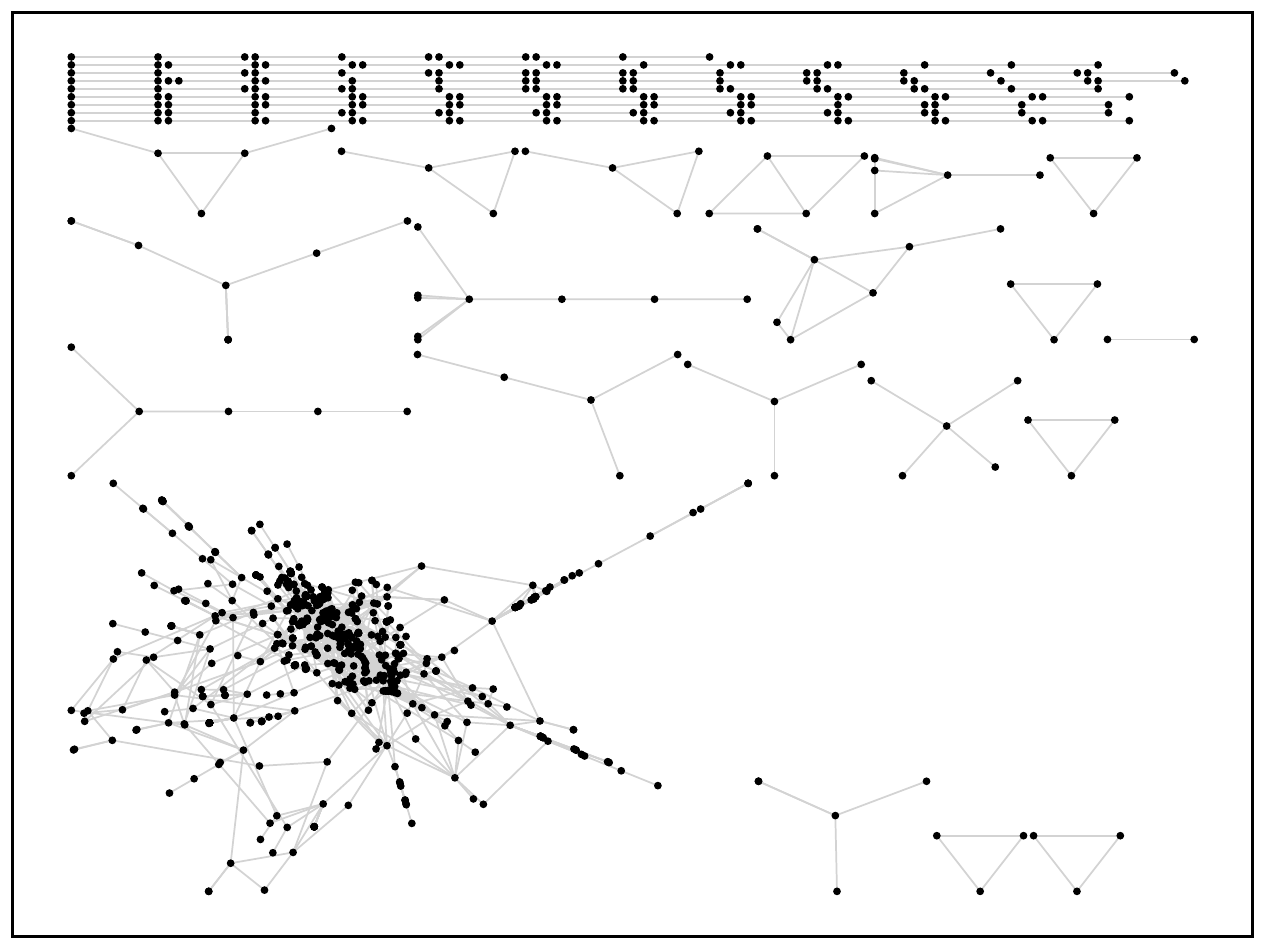}
    \\
    \footnotesize{\nd{}=3.2E+08, \ulcv{}=1.99} & \footnotesize{\nd{}=7.1E+06, \ulcv{}=0.54} & \footnotesize{\nd{}=$\infty$, \ulcv{}=1.05} & \footnotesize{\nd{}=$\infty$, \ulcv{}=0.69}
    \\
    (c) $\blacklozenge$~\linlog~\cite{noack2005energy} 
    &
    (d) $\blacklozenge$~\fr~\cite{fruchterman1991graph} 
    & 
    (e) $\blacktriangle$~\mds~\cite{gansner2004graph}
    &
    (f) $\blacktriangle$~\pivotmds~\cite{brandes2006eigensolver} 
    \\
    \includegraphics[width=0.24\textwidth]{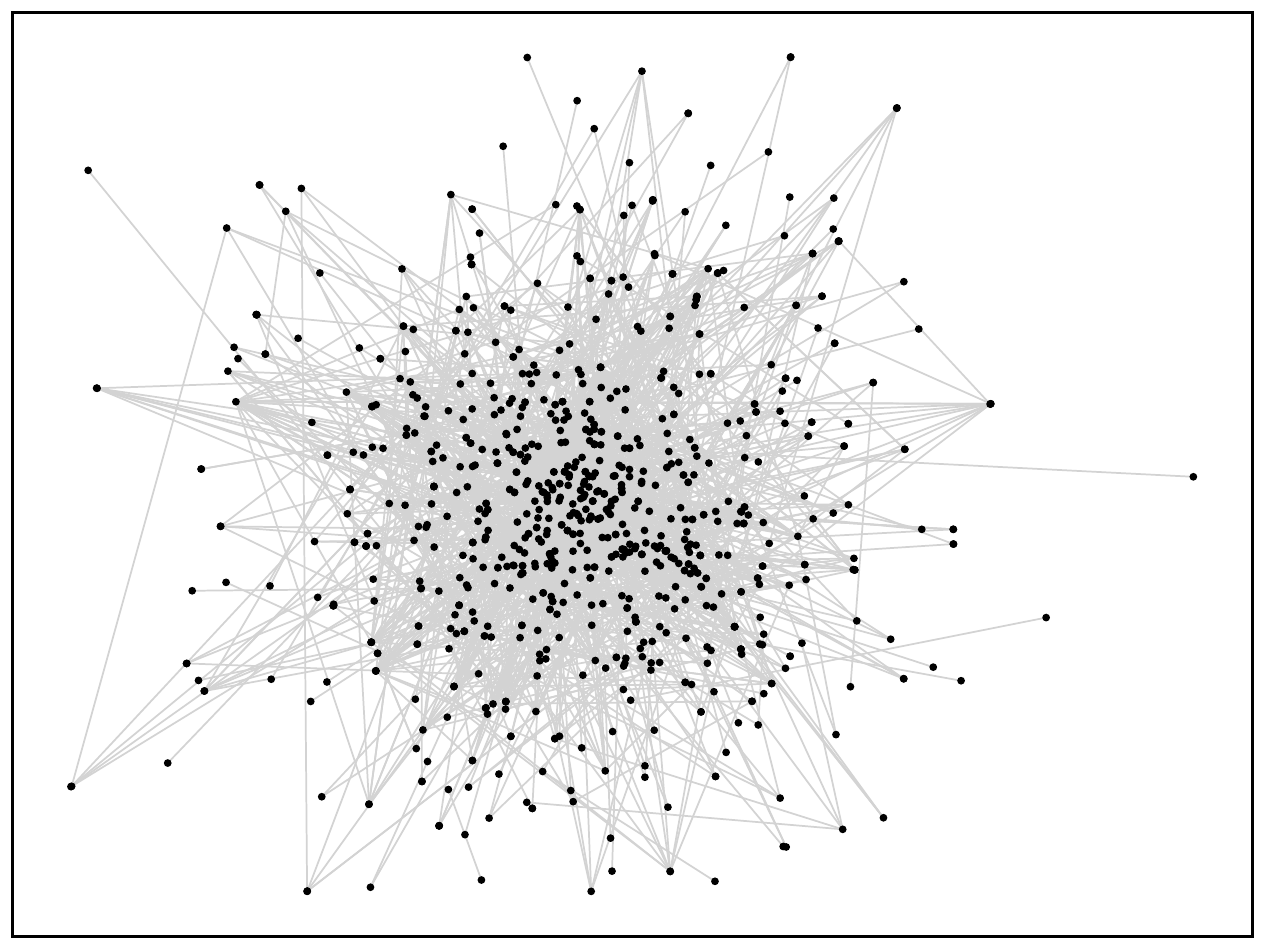}
    &
    \includegraphics[width=0.24\textwidth]{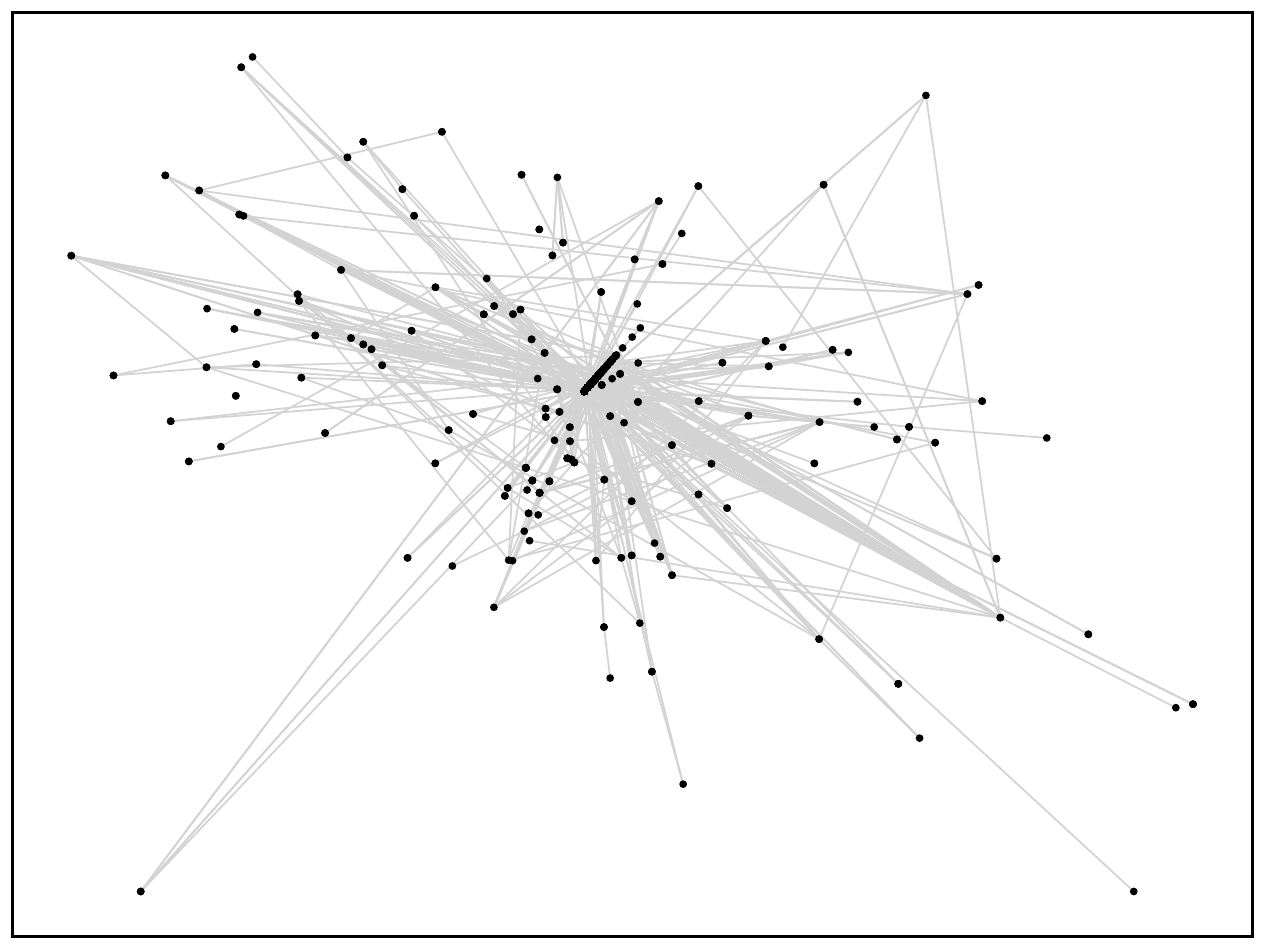}
    &
    \includegraphics[width=0.24\textwidth]{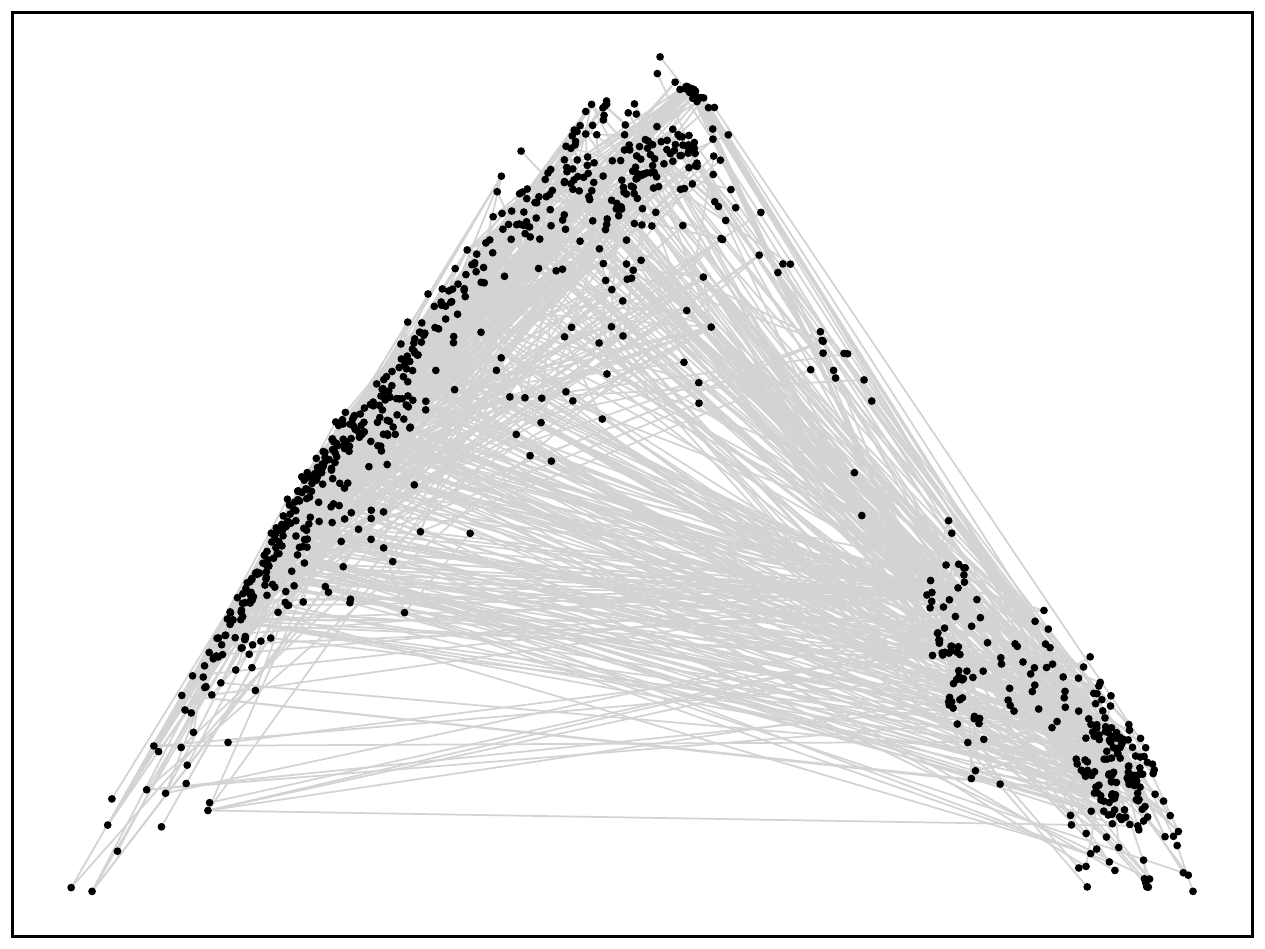}
    &
    \includegraphics[width=0.24\textwidth]{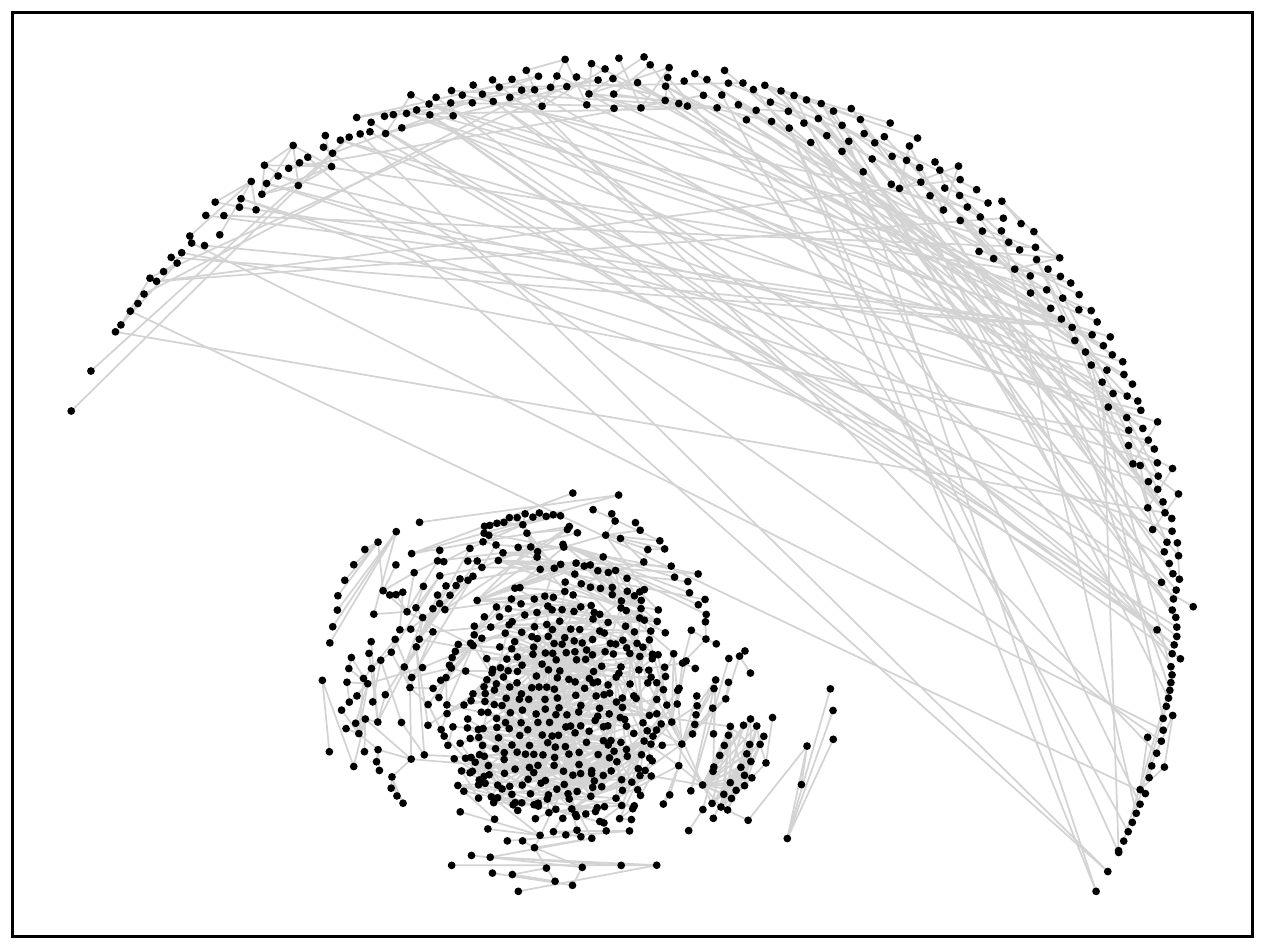}
    \\
    \footnotesize{\nd{}=1.2E+17, \ulcv{}=0.64} & \footnotesize{\nd{}=$\infty$, \ulcv{}=1.70} & \footnotesize{\nd{}=9.6E+07, \ulcv{}=0.89} & \footnotesize{\nd{}=2.9E+06, \ulcv{}=1.78}
    \\
    (g) $\star$~\gf~\cite{ahmed2013distributed} & 
    (h) $\star$~\leemb~\cite{belkin2003laplacian} &
    (i) $\star$~\nodevec~\cite{grover2016node2vec} & 
    (j) \simrank~\cite{jeh2002simrank}
    \\
  \end{tabular}
\end{small}
\caption{Visualization results for the \filmtrust graph: force-directed methods are marked with $\blacklozenge$; stress methods are marked with $\blacktriangle$; graph embedding methods are marked with $\star$.}\label{fig:trust-viz}
\end{figure*}

\begin{figure*}[]
\centering
\begin{small}
  \begin{tabular}{cc}
    \includegraphics[width=0.5\textwidth]{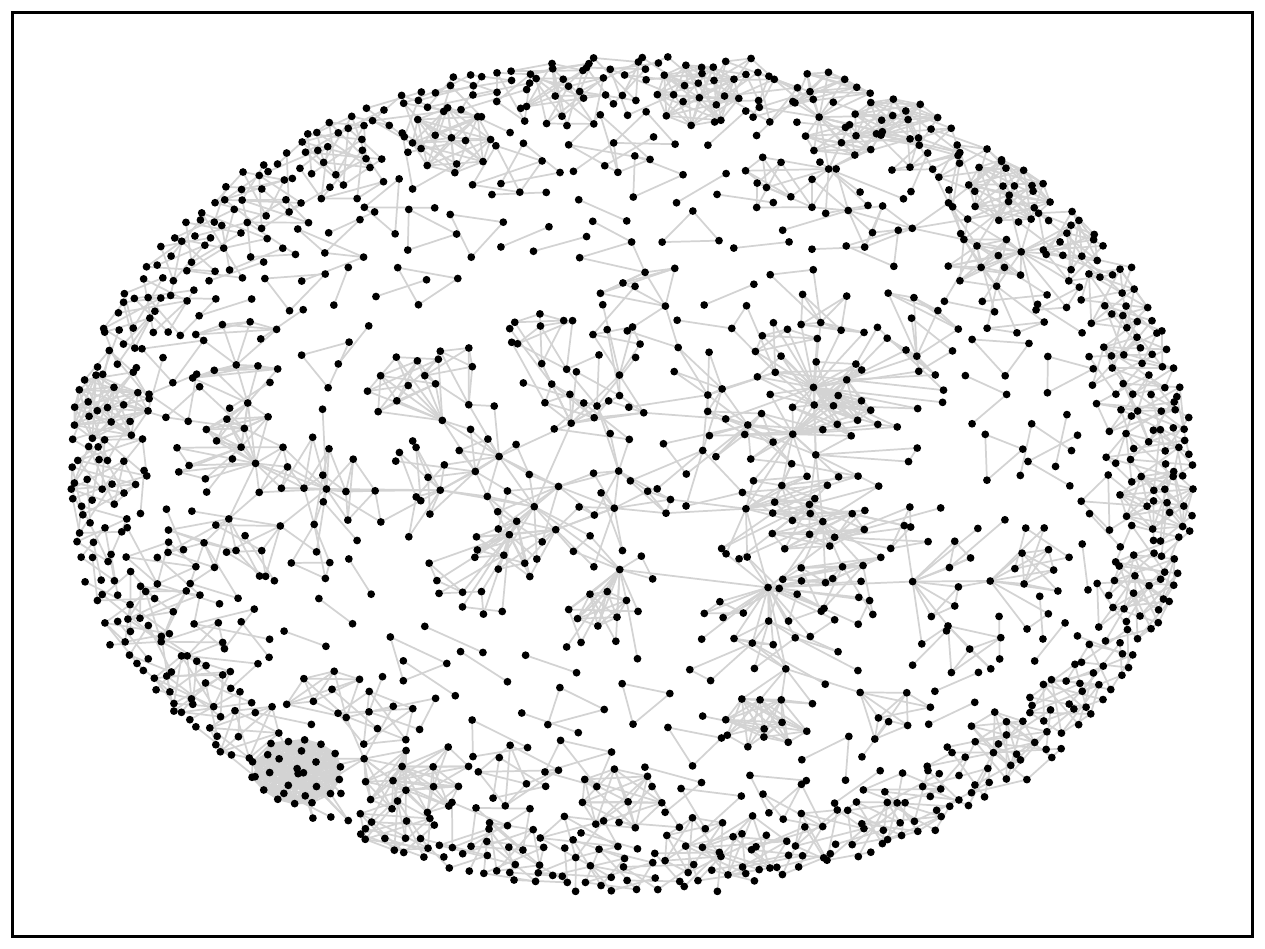}
     &
    \includegraphics[width=0.5\textwidth]{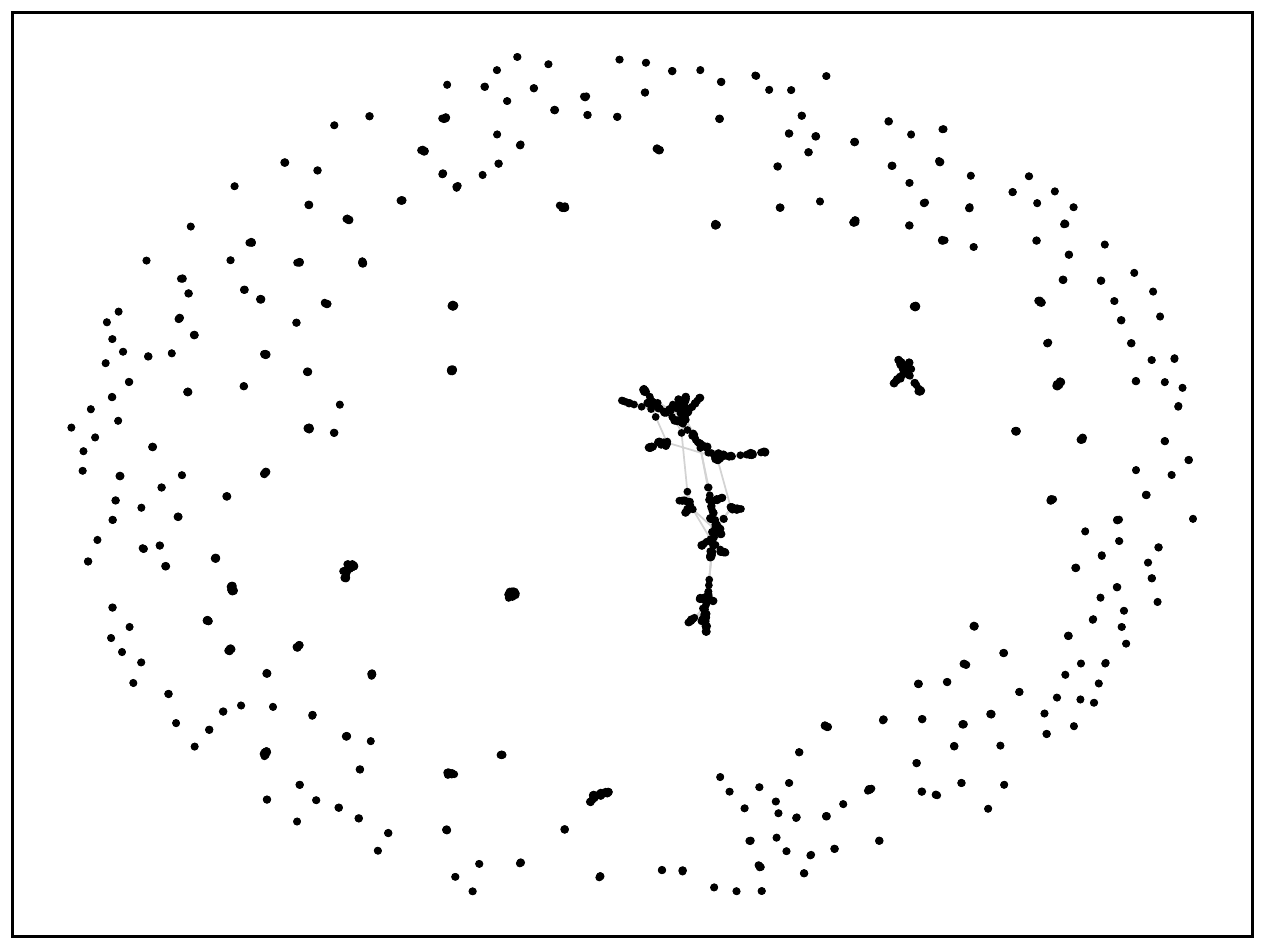}
    \\
    \footnotesize{\nd{}=2.0E+06, \ulcv{}=0.34} & \footnotesize{\nd{}=1.9E+08, \ulcv{}=1.52}
    \\
    (a) \pprviz (Ours)  & (b) $\blacklozenge$~\forceatlas~\cite{jacomy2014forceatlas2} 
    \\
  \end{tabular}
\end{small}
\vspace{-3mm}
\end{figure*}

\begin{figure*}[]
\centering
\begin{small}
  \begin{tabular}{cccc}
    \includegraphics[width=0.24\textwidth]{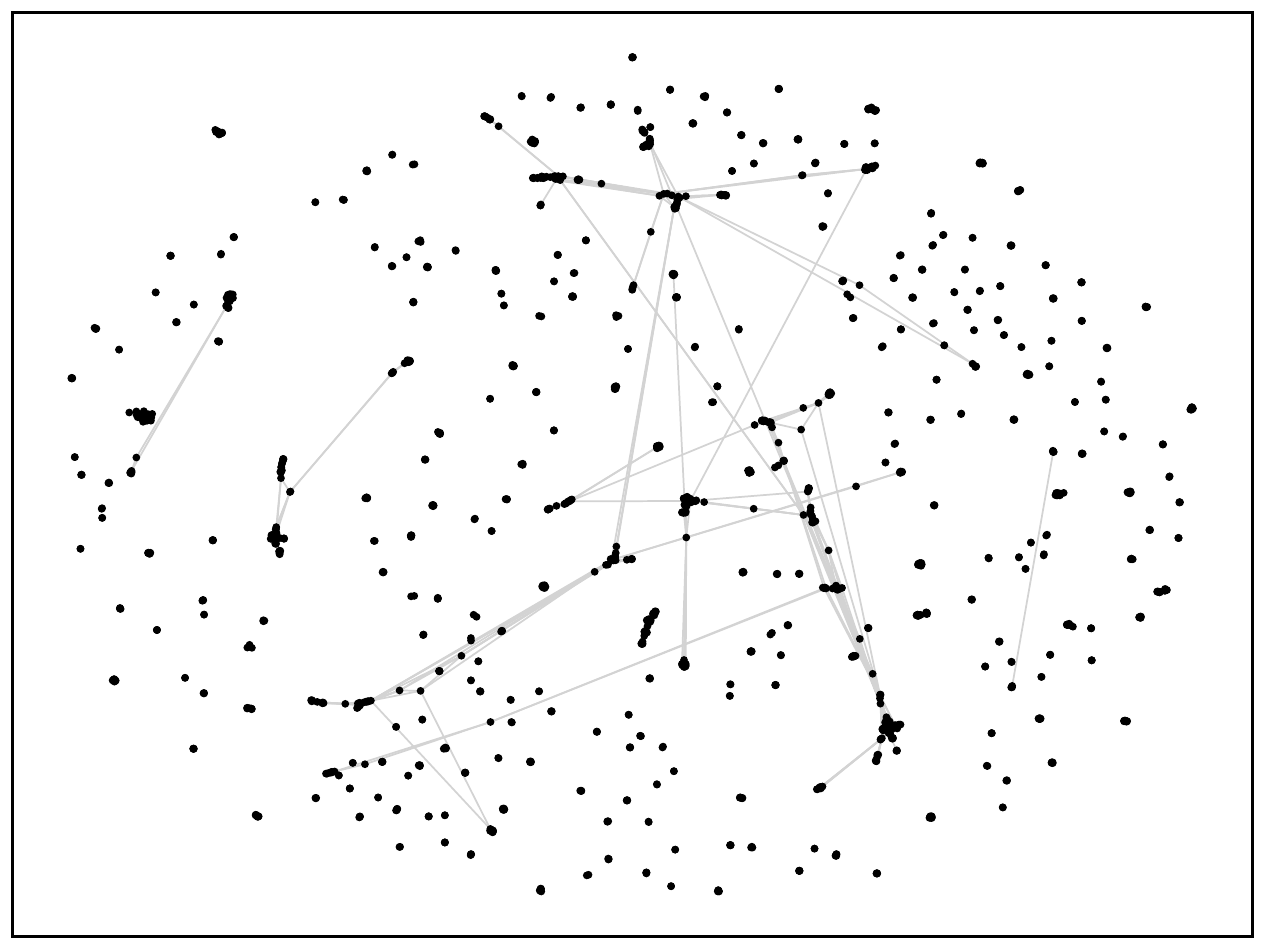}
    &
    \includegraphics[width=0.24\textwidth]{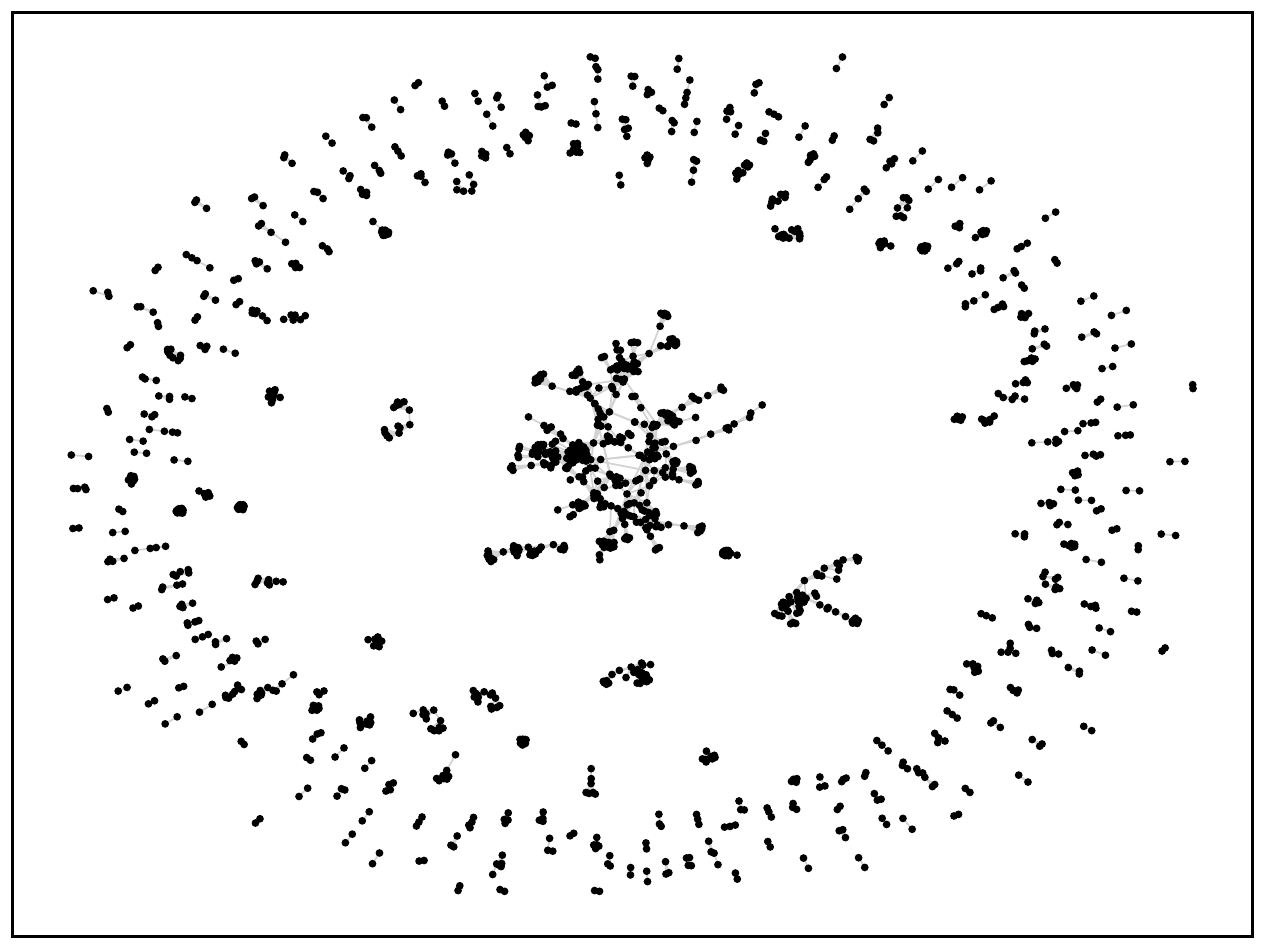}
    &
    \includegraphics[width=0.24\textwidth]{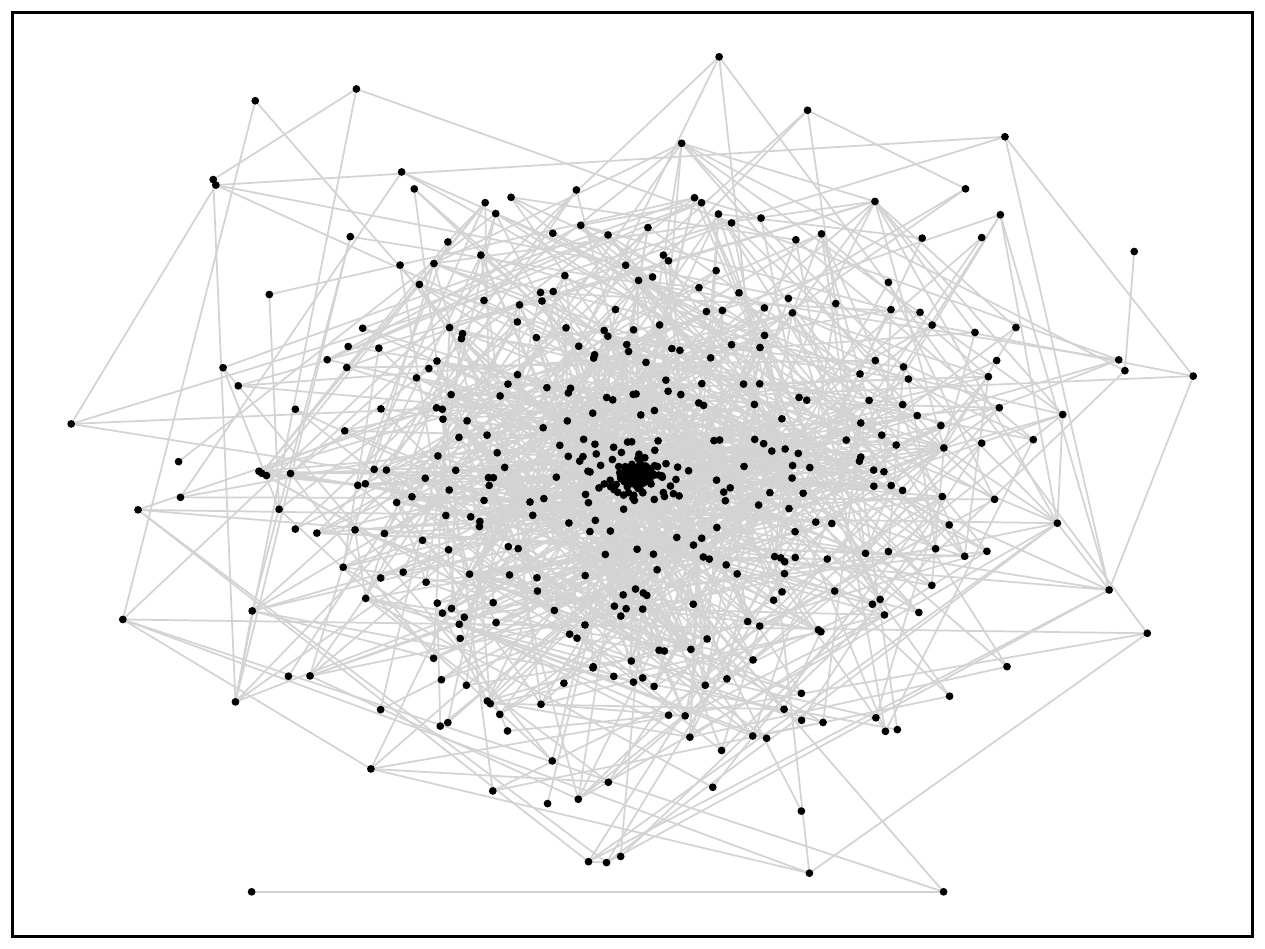}
    &
    \includegraphics[width=0.24\textwidth]{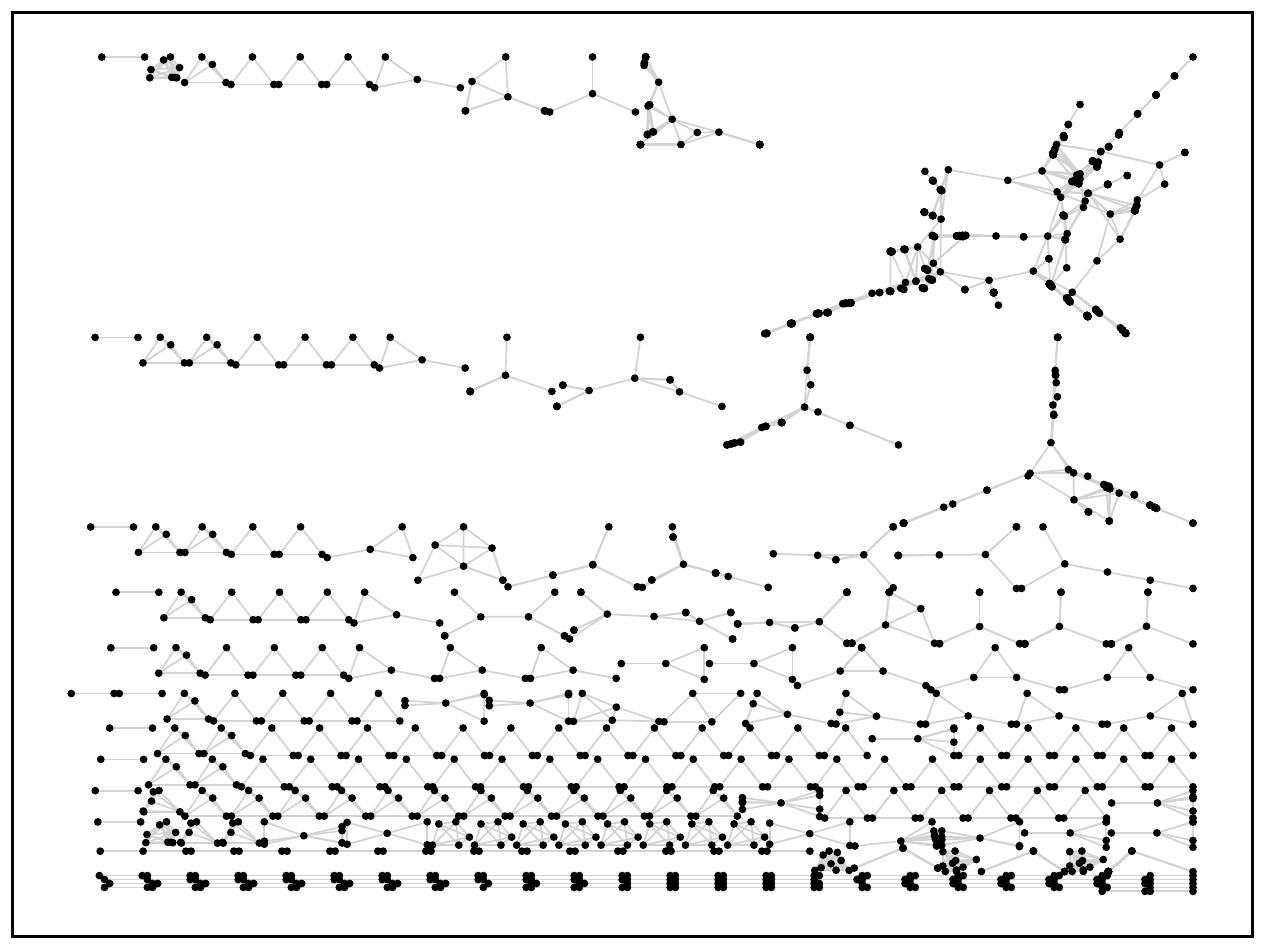}
    \\
    \footnotesize{\nd{}=2.3E+09, \ulcv{}=4.70} & \footnotesize{\nd{}=6.5E+12, \ulcv{}=0.77} & \footnotesize{\nd{}=9.9E+12, \ulcv{}=1.74} & \footnotesize{\nd{}=$\infty$, \ulcv{}=0.74}
    \\
    (c) $\blacklozenge$~\linlog~\cite{noack2005energy} 
    &
    (d) $\blacklozenge$~\fr~\cite{fruchterman1991graph} 
    & 
    (e) $\blacktriangle$~\mds~\cite{gansner2004graph}
    &
    (f) $\blacktriangle$~\pivotmds~\cite{brandes2006eigensolver} 
    \\
    \includegraphics[width=0.24\textwidth]{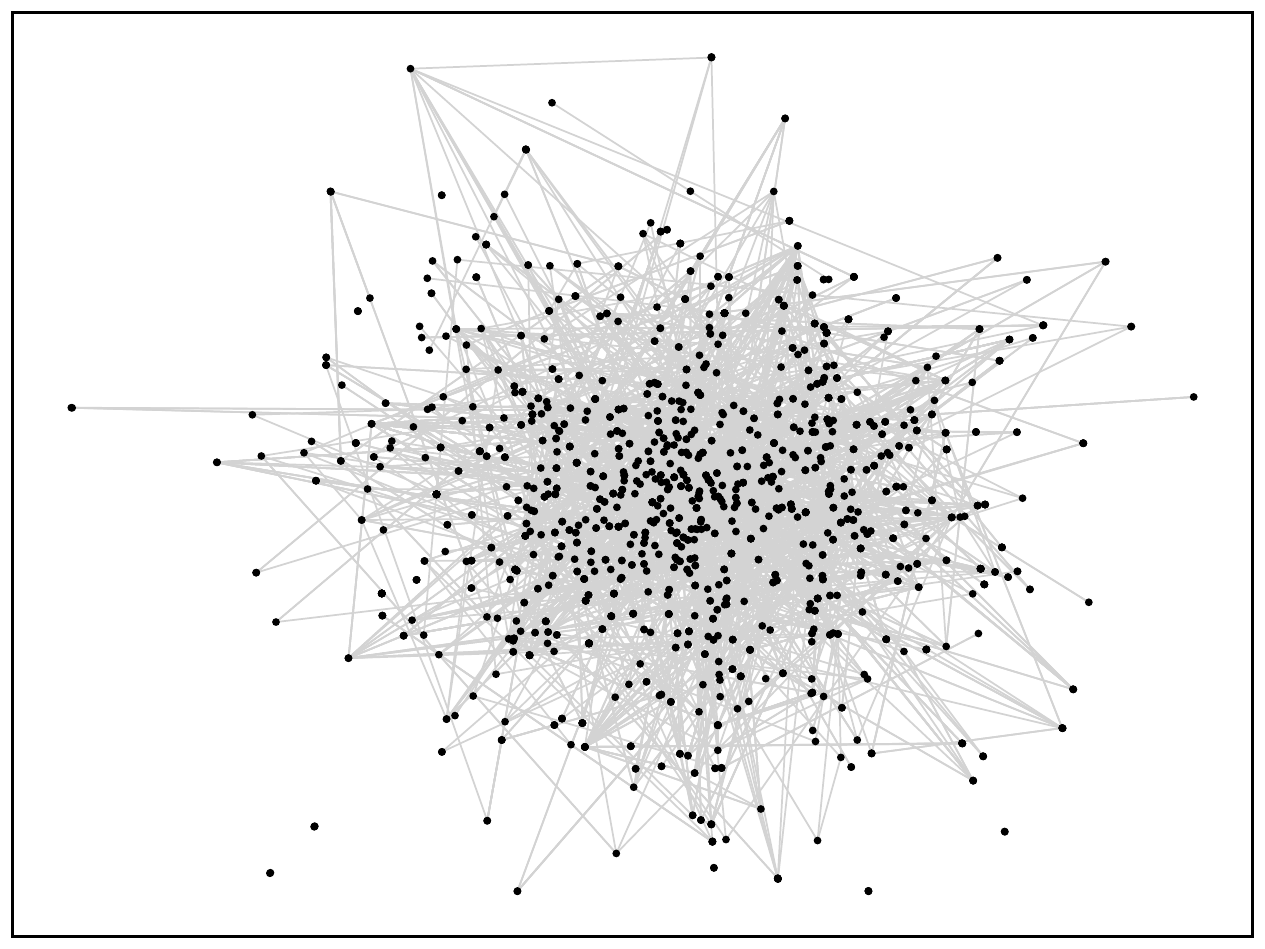}
    &
    \includegraphics[width=0.24\textwidth]{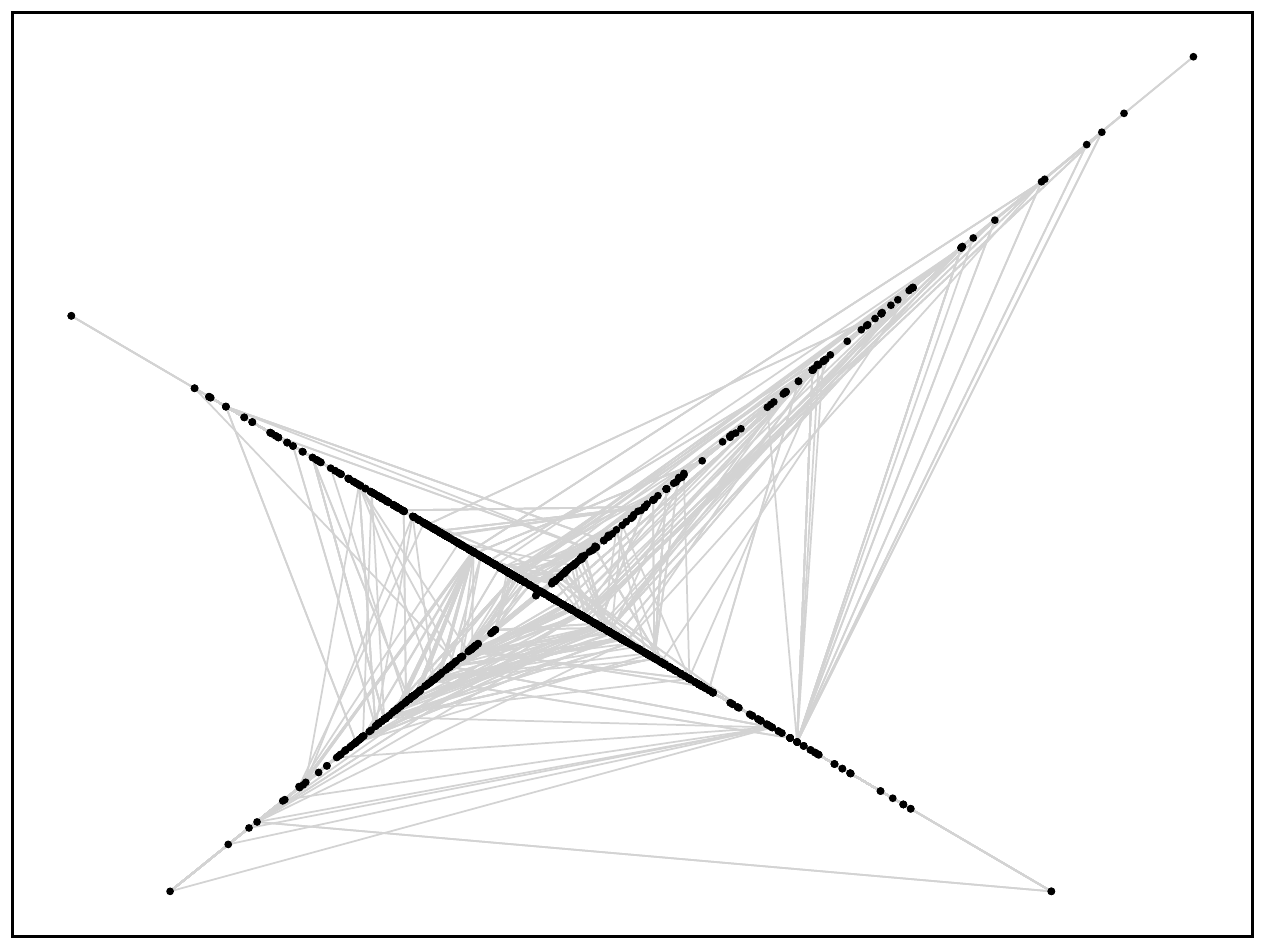}
    &
    \includegraphics[width=0.24\textwidth]{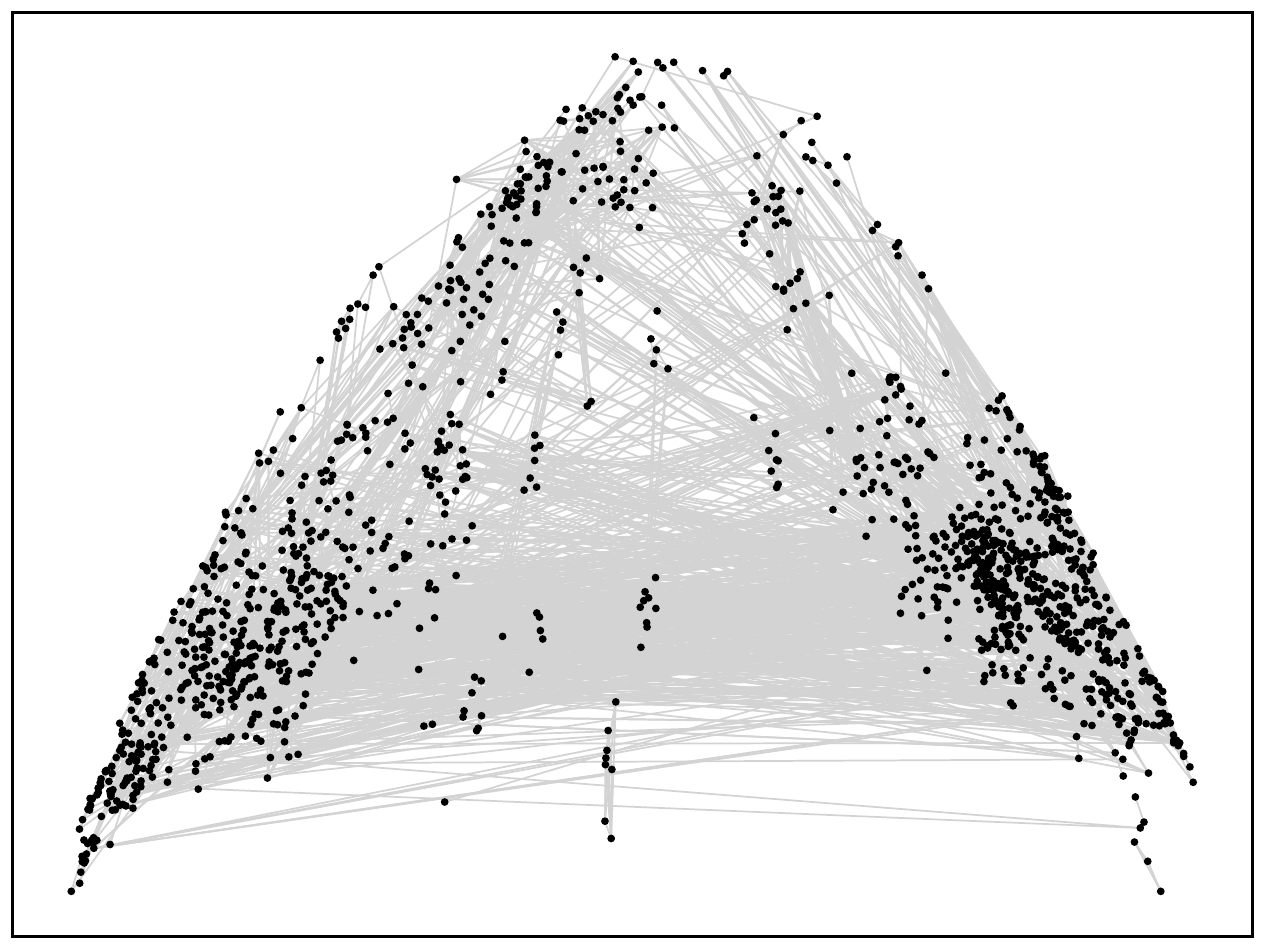}
    &
    \includegraphics[width=0.24\textwidth]{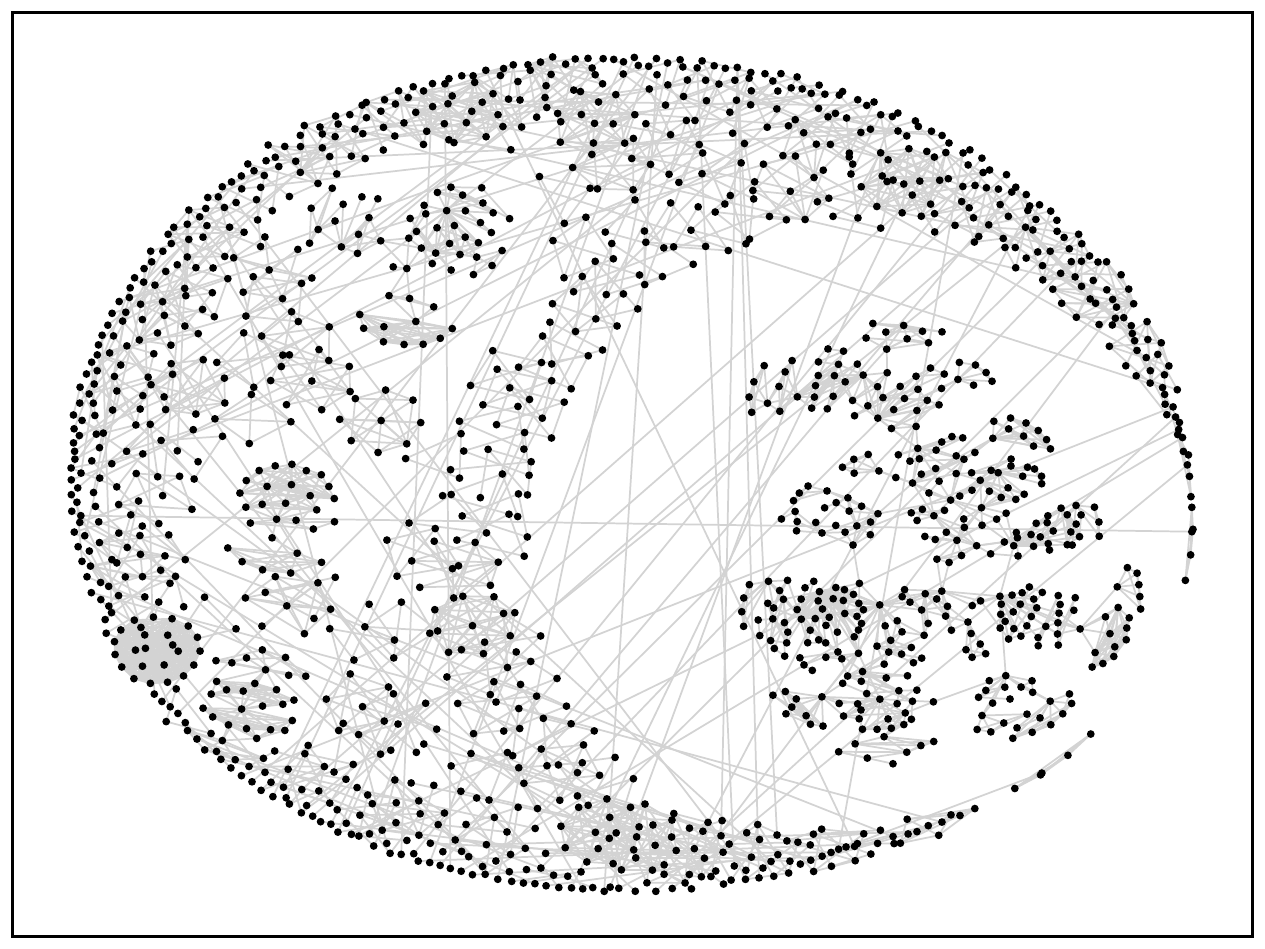}
    \\
    \footnotesize{\nd{}=1.1E+17, \ulcv{}=0.86} & \footnotesize{\nd{}=$\infty$, \ulcv{}=1.26} & \footnotesize{\nd{}=6.6E+07, \ulcv{}=1.32} & \footnotesize{\nd{}=2.2E+06, \ulcv{}=1.98}
    \\
    (g) $\star$~\gf~\cite{ahmed2013distributed} & 
    (h) $\star$~\leemb~\cite{belkin2003laplacian} &
    (i) $\star$~\nodevec~\cite{grover2016node2vec} & 
    (j) \simrank~\cite{jeh2002simrank}
    \\
  \end{tabular}
\end{small}
\caption{Visualization results for the \scinet graph: force-directed methods are marked with $\blacklozenge$; stress methods are marked with $\blacktriangle$; graph embedding methods are marked with $\star$.}\label{fig:scinet-viz}
\end{figure*}

\subsection{More Experiments}

\header
{\bf More metrics results.}
{In this part, we evaluate the performance of \pprviz and other competitors in terms of another aesthetic metric: angular resolution (\ar{}).  Here, we omit the formal definitions of \ar{}, and refer interested readers to \cite{taylor2005applying}. Roughly speaking, 
\ar{} measures the angles of adjacent edges and a smaller angle leads to a larger \ar{} value. Intuitively, larger adjacent edge angles are friendly to user perception. Therefore, a lower score of \ar{} indicates better quality.
We report the \ar{} scores of all methods in Table~\ref{tab:metrics-ar}.
As illustrated in Table~\ref{tab:metrics-ar}, \pprviz has the best or second best \ar{} scores on all test graphs except \physic.
}

\header
{\bf More visualization results.}
{In Figures~\ref{fig:fb-viz} to~\ref{fig:scinet-viz}, we report the visualization results of \pprviz and other competitors on \fbego, \wiki, \physic, \filmtrust and \scinet graphs. Specifically, we enlarge the results of \pprviz and a promising competitor \forceatlas for user comparison.}

\end{document}